\documentclass[preprint]{article}

\usepackage{blindtext}

\usepackage{amsthm}

\usepackage{jmlr2e}

\usepackage[T1]{fontenc}
\usepackage{lmodern}
\usepackage[english]{babel}
\usepackage{microtype}
\usepackage{amsfonts}

\usepackage{setspace}
\usepackage{subfiles}

\usepackage{mathtools}    
\usepackage{bm}
\usepackage{bbm}
\usepackage{dsfont}
\usepackage{mathrsfs}

\usepackage{algorithm}
\usepackage{algorithmic}
\usepackage{enumitem}
\usepackage{float}

\usepackage{array}
\usepackage{multirow}
\usepackage{booktabs}
\usepackage{longtable}
\usepackage{subcaption}

\usepackage[colorinlistoftodos]{todonotes}
\usepackage{comment}

\newcommand{\hilbert}{\mathcal{H}}

\DeclareMathOperator{\Bernoulli}{Bernoulli}
\DeclareMathOperator{\Multinomial}{Multinomial}

\DeclareMathOperator{\UPsystematic}{UP\text{-}systematic}

\newcommand{\E}{\mathbb{E}}

\newcommand{\ind}{\mathbbm{1}}
\newcommand{\independent}{\mathrel{\perp\!\!\!\perp}}

\usepackage{lastpage}
\jmlrheading{}{2026}{}{}{}{}{Zou, Matabuena, and Kosorok}
\ShortHeadings{Survey Distributional Random Forests}{Zou, Matabuena, and Kosorok}
\firstpageno{1}

\begin{document}

\title{Distributional Random Forests for Complex Survey
Designs}

\author{\name Yating Zou \email yating@live.unc.edu \\
       \addr Department of Biostatistics\\
       University of North Carolina at Chapel Hill\\
       Chapel Hill, NC, USA
       \AND
       \name Marcos Matabuena \email marcos.matabuena@mbzuai.ac.ae \\
       \addr Department of Epidemiology\\
       Mohamed bin Zayed University of Artificial Intelligence\\
       Abu Dhabi, UAE
       \AND
       \name Michael R.\ Kosorok\thanks{Corresponding author.} \email kosorok@bios.unc.edu \\
       \addr Department of Biostatistics\\
       University of North Carolina at Chapel Hill\\
       Chapel Hill, NC, USA}

\editor{}

\maketitle

\begin{abstract}
We study estimation of the conditional law ${P}(Y \mid X = \mathbf{x})$ and continuous measurable maps of it when $Y \in \mathcal{Y}$ takes values in a locally compact Polish space (e.g.\ $\mathbb{R}^d$), $X \in \mathbb{R}^p$, and the observations arise from a complex survey design---a single- or multi-stage sampling scheme that may involve unequal selection, stratification, and clustering.
We propose a survey-calibrated distributional random forest (SDRF) that incorporates complex-design features via the pseudo-population bootstrap, PSU-level honesty, and a Maximum Mean Discrepancy (MMD) split criterion computed from kernel mean embeddings of design-weighted node distributions.
We provide a framework for analyzing forest-based estimators under various survey designs; establish consistency for both finite- and super-population conditional laws under explicit conditions on the design, kernel, resampling multipliers, and tree partitions. 
As far as we are aware, these are the first results on model-free estimation of conditional distributions under survey designs.
Simulations under a stratified two-stage cluster design expose the systematic bias incurred by ignoring survey structure.
We illustrate the broad applicability of SDRF on NHANES, estimating the conditional joint tolerance regions for two diabetes biomarkers, revealing subgroup-level distributional heterogeneity relevant to diabetes risk profiling in the U.S.\ population.
\end{abstract}


\section{Introduction}
\label{sec:introduction}

We study nonparametric estimation of the conditional law of an outcome given covariates, when the data come from a probability sample drawn under a known sampling design. We work under a broad range of survey designs, possibly combining unequal inclusion probabilities, stratification, and clustering, in either single-stage or multiple-stage form. These are known as \emph{complex survey designs} \citep{Lohr2021-cm}. They govern the collection of many nationally representative datasets in health, economics, and social sciences, providing the official statistics that underpin public health policy, regulation, and resource allocation.

\paragraph{A public health motivation.}
A concrete decision problem makes the methodological need precise.
U.S.\ diabetes diagnosis relies on marginal thresholds of A1C or fasting plasma glucose \citep{american20252}, yet the two tests can be discordant \citep{hopham2017discordance}, and the joint distribution of these biomarkers varies with age and BMI \citep{lado2021, Matabuena2023}. Moving from the current one-size-fits-all thresholds toward subgroup-adapted reference values requires estimating the conditional joint distribution of these biomarkers as a function of a individual characteristics, on data representative of the target population.
The National Health and Nutrition Examination Survey (NHANES)
\citep{mozumdar2011persistent} supplies such data; what is missing is a flexible method that estimates maps (or functionals) of the conditional joint distribution while respecting the underlying design.

\paragraph{What survey design provides.}
Survey data come with features ML methods typically have to assume or estimate from data: a well-defined target population, a representative sample, and an exact correction for the gap between the sample and that target \citep{DeMatteis2025, Lohr2021-cm}.
Where ML methods recover analogous corrections through propensity scores, density ratios, or importance weights \citep{bradley2022addressing}, with attendant bias-variance trade-offs, surveys provide them by design. This guarantee, however, holds only if the learning algorithm respects the design that delivers it.

Respecting the design is crucial for statistical consistency, rather than any potential efficiency gain. In the setting of estimating the $P(Y \mid X = \mathbf{x})$, this is true whenever design variables are correlated with the covariates of interest but absent from the analysis dataset, as we illustrate in our simulations (Section~\ref{sec:sim}). Moreover, treating dependent observations as i.i.d.\ invalidates downstream uncertainty quantification approaches such as conformal prediction. Design-based conformal calibration is possible \citep{wieczorek2023designbased}, but requires a base estimator that is itself design-aware.

\paragraph{A Hilbert-space view.}
We build such a method around a representational observation. Kernel mean embeddings (KMEs) \citep{Smola_2007, sripernumbudur_hilbert_2010}
map probability measures into an RKHS, in which the super-population
conditional law, its finite-population analog, and any H\'ajek-type
design-weighted estimate of either live in a common Hilbert space and are compared by the common maximum mean discrepancy (MMD) metric. This is what allows a single algorithmic and theoretical scheme to cover a range of designs as opposed to a design-by-design approach, and the construction is not limited to forests.

The distributional random forest (DRF) of
\citet{cevid2022distributional} applies to the i.i.d.\ setting. It is an MMD-split forest that yields plug-in estimators for any continuous functional of the conditional law. 
Under a complex design, the split criterion, the
bootstrap, and the honesty mechanism each require redesign, as does the consistency analysis. 
Classical random-forest consistency theory rests on i.i.d.\ U-statistic and leaf-geometry tools \citep{Wager2015-nc, biau2008consistency, Wager2018-lu, liu2024randomization}, while design-based empirical-process theory for H{\'a}jek- and Horvitz--Thompson-type processes is well-developed for classical estimands \citep{boistard_functional_2017, Han_Weller_2021}. Neither toolkit applies directly to the setting we consider here.
To bridge this gap, two auxiliary results we proved may be of independent interest: a moment-based sub-gamma bound for pseudo-population bootstrap multipliers and a conditional multiplier maximal inequality under weak dependence. They combine to yield the empirical-process control our split consistency proofs require.

\paragraph{Contributions.} We propose the survey-calibrated distributional random forest (SDRF). We use `calibrated' in the broad sense of estimation adjusted for the sampling design, rather than the specific sense of calibration to known auxiliary totals \citep{deville1992calibration}. 
Our contributions are:
\begin{itemize}[itemsep=2pt, topsep=2pt, leftmargin=2em]
    \item \emph{A design-aware ML algorithm.} SDRF handles design by (i) a design-weighted MMD splitting criterion, (ii) pseudo-population bootstrap to respect dependency from design, and (iii) PSU-level honesty for split and populating the leaves. Only one hyperparameter, a node-level weight ratio cap, is added relative to standard random forests.
    \item \emph{A unified consistency theory.} We establish that the MMD between the SDRF-estimated conditional law and the target converges to zero in two senses: design consistency (with respect to the survey mechanism, conditional on the realized finite population) and joint consistency (with respect to the joint super-population/design law). A companion result establishes the consistency of the MMD split criterion itself, closing the gap between the empirical split and the population-oracle split. To our knowledge, these are the first consistency guarantees for distributional forests under non-i.i.d.\ sampling.
    \item \emph{A practical, deployable method.} SDRF is theoretically grounded yet computationally modest, training on a laptop in minutes on NHANES-scale data without GPU acceleration. This flexibility is our deliberate design goal aimed at practical use: one estimator, one set of guarantees, covering many designs and many downstream targets (conditional CDFs, means, quantiles, moments, and more).
\end{itemize}

\paragraph{Related work.} 
Within survey methodology, regression methods are predominantly parametric or semiparametric, based on (pseudo) probability weights and M- or Z-estimating equations \citep{van1996m, nan2013general, lumley2017fitting}, with regularized variants for high-dimensional auxiliary information \citep{iparragirre2023variable}. Classical nonparametric smoothers including local polynomials \citep{kikechi2017local}, Nadaraya--Watson estimators \citep{harms2010kernel}, and additive models have been adapted to surveys \citep{breidt2005model}, with recent extensions to functional responses \citep{koffman2025functionscalarregressioncomplex}, but focus on conditional means of continuous univariate outcomes. Design-based central limit theorems and empirical-process results for Horvitz--Thompson and H\'ajek-weighted measures \citep{horvitz1952generalization, bellhouse2001central, Lumley2005-em, boistard_functional_2017, schochet2022design} characterize the asymptotic behavior of design-weighted estimators in classical settings, but do not extend directly to processes indexed by the function class of data-adaptive tree partitions. 

Recent machine-learning methods for survey data include kernel methods such as kernel ridge regression and kernel smoothing under specific designs \citep{matabuena2023distributional}, including distributional and Fr\'echet-regression approaches for NHANES \citep{10.1093/biostatistics/kxaf013}. In the i.i.d.\ setting, distributional random forests \citep{cevid2022distributional} estimate the full conditional law but assume independent sampling (cf.\ the Hilbert-space discussion above). Tree- and forest-based methods for survey data have been surveyed recently \citep{bhaduri2025review}; concrete instances include \citet{toth2011building, mcconvilletoth2019automated, dagdoug2023model, toth2025bayesian}, with theoretical guarantees confined to specific designs and to low-dimensional, mean- or prediction-type targets. Deep generative approaches to conditional density estimation (normalizing flows \citep{winkler2019learning}, conditional GANs \citep{mirza2014conditional}, diffusion models \citep{ho2020denoising}) offer, to our knowledge, no native mechanism for design incorporation, despite the opportunity for design-aware extensions. Consequently, practitioners applying ML to survey data have often had to ignore part of the design information \citep{west2016analytic, ridgeway2015propensity}.

\paragraph{Outline.} Section~\ref{sec:setup} formalizes the complex-survey setup and briefly reviews kernel mean embeddings. Section~\ref{sec:algorithm} presents SDRF and the main consistency results. Section~\ref{sec:implementation} collects practical guidance for deployment. Section~\ref{sec:sim} reports simulations, including ablations isolating each design ingredient. Section~\ref{sec:casestudy} applies SDRF to NHANES. Section~\ref{sec:dis} discusses limitations and extensions. All proofs and a notation table are in the Supplementary Material.

\section{Set-up and Background}
\label{sec:setup}

In this section we formalize our problem setting and introduce the necessary background on survey design and kernel mean embeddings. We show how a sample relates to the finite population (e.g.\ a census) by design, and how the finite population relates to an infinite super-population, the usual underlying probability space in the design-free i.i.d.\ setting.

Suppose we observe a sample
$D=\{(X_i,Y_i,w_i)\}_{i \in [n_s]}$
drawn from a finite population $D^N = \{(X_i, Y_i, Z_i)\}_{i \in [N]}$ of size $N$ under a sampling design $\mathbf{p}$ that may depend on
auxiliary design variables $Z_i$, which we will formalize shortly.
Here, $Y_i\in\mathcal Y$ is the outcome, $X_i\in\mathbb R^p$ are covariates and $w_i\in\mathbb R^{+}$ are survey weights
(possibly incorporating nonresponse adjustment and calibration).
Let $P^N_{Y\mid X=\mathbf x}$ denote the finite-population conditional distribution induced by $D^N$, and let $P_{Y\mid X=\mathbf x}$ denote the super-population conditional law.
We consider a fixed covariate value $\mathbf x\in\mathcal X\subseteq\mathbb R^p$ and a Borel-measurable map $\Psi:(\mathcal M_b(\mathcal Y), d_{\mathbf k})\to(\mathcal V, d_v)$ into a metric space $(\mathcal V, d_v)$, continuous at the target laws introduced below.
Our objective is to construct an estimator $\widehat\Psi_N(\mathbf x)$ that is \textit{design-consistent} for the
finite-population target $\Psi(P^N_{Y\mid X=\mathbf x})$ and \textit{joint-consistent} for the
super-population target $\Psi(P_{Y\mid X=\mathbf x})$. We make these two notions precise in Section~\ref{subsec:survey-design}, once the underlying probability spaces are constructed.

Throughout, we assume the covariate space $\mathcal{X}\subseteq\mathbb{R}^p$ is bounded, so that without loss of generality $\mathcal{X}\subseteq[0,1]^p$ after affine rescaling, matching the standard convention in the random forest literature \citep{cevid2022distributional, Wager2018-lu}. 

We will denote $[a] := \{1, \dots, a\}$ for any $a \in \mathbb{Z}_{> 0}$, and `:=' will be used to emphasize definition.

\subsection{Survey Design}
\label{subsec:survey-design}
We adopt the finite-population framework of \cite{2005_RubinBleuer, Han_Weller_2021} and standard notations. Consider a sequence of finite populations $\{\mathcal{U}^N\}_{N \geq 1}$. Let $U^N := \{1,\dots,N\}$, the index set for each $\mathcal{U}^N$ and $\mathcal{S}^N := \{S: S \subset U^N\}$ the collection of subsets of $U^N$. Let $\sigma(\mathcal{S}^N)$ denote the $\sigma$-algebra generated by $\mathcal{S}^N$. We view the finite population $\mathcal U^N$ as a realization of i.i.d.\ random vectors $\{(X_i,Y_i,Z_i)\}_{i=1}^N$ in a common probability space $(\Omega,\mathcal A,P)$,
where $Z_i\in \mathcal{Z}$ denotes the auxiliary (design) information available in the frame.
Let $D^N:=\{(X_i,Y_i)\}_{i=1}^N$ denote the finite-population data and write $Z^N:=(Z_1,\dots,Z_N)$.

A \emph{sampling design} can be formalized as a transition kernel $\mathbf{p} : \mathcal{S}^N \times \mathcal{Z}^{\otimes N} \to [0,1]$ such that for each $z^N$, $\mathbf{p}(\cdot, z^N)$ is a probability measure on $(\mathcal{S}^N, \sigma(\mathcal{S}^N))$. Write $P_{\mathcal{S}^N \mid \omega}(\cdot) := \mathbf{p}(\cdot, Z^N(\omega))$ for the conditional design measure. The joint measure $P_{\Omega \times \mathcal{S}^N}$ on $(\Omega \times \mathcal{S}^N, \mathcal{A} \otimes \sigma(\mathcal{S}^N))$ is then uniquely determined by \cite[Def.~4.3]{2005_RubinBleuer}
\[
P_{\Omega \times \mathcal{S}^N}(E \times S) := \int_E \mathbf{p}(S, Z^N(\omega))\, P(d\omega), \qquad 
\forall E \in \mathcal{A},\ S \in \sigma(\mathcal{S}^N).
\]
This completes the construction of the two probability spaces that underlie our notions of model- and design-consistency.
Specifically, we require that for any $\varepsilon > 0$,
\[
P_{\mathcal{S}^N\mid \omega}\!\left(
d_{v}\!\left(\widehat{\Psi}_N(\mathbf{x}), \Psi(P^N_{Y \mid X = \mathbf x})\right) > \varepsilon
\right) \to 0
\quad \text{$P_{\Omega}$--a.e.\ (design consistency), and}
\]
\[
P_{\Omega \times \mathcal{S}^N}\!\left(
d_{v}\!\left(\widehat{\Psi}_N(\mathbf{x}), \Psi(P_{Y \mid X = \mathbf x})\right) > \varepsilon
\right) \to 0
\quad \text{(joint consistency).}
\]
Design consistency holds with respect to the design measure $P_{\mathcal{S}^N\mid\omega}$, for $P_\Omega$-almost every realized population $\omega$; joint consistency holds with respect to the joint model--design measure $P_{\Omega \times \mathcal{S}^N}$.
We write $\xrightarrow{p_d}$ and $\xrightarrow{p_\otimes}$ for convergence in 
probability under $P_{\mathcal{S}^N\mid\omega}$ and $P_{\Omega\times\mathcal{S}^N}$, respectively. See the Notation Table (Supplement, Table~S1) for a summary.

With a well-defined design $\mathbf{p}$, a random sample is obtained from realization of inclusion indicators $\xi_i := \ind(i \in S)$ for $S \sim P_{\mathcal{S}^N \mid \omega}$, with first- and second-order inclusion probabilities $\pi_i := \mathbb{E}[\xi_i \mid Z^N]$ and $\pi_{ij} := \mathbb{E}[\xi_i \xi_j \mid Z^N]$. We denote the realized sample $D = \{(X_i, Y_i, w_i) : \xi_i = 1\}$ and its size $n_s := \sum_{i \in [N]} \xi_i$, with sampling weights $w_i$ and possibly other design variables such as stratum and PSU number.

We impose the following regularity conditions on the design:
\medskip

\noindent\underline{Assumptions on the sampling design}
\begin{itemize}[leftmargin=3em]
    \item[(D1)] \emph{Conditionally non-informative design:} $\xi^N \independent Y^N \mid (X^N, Z^N)$.
    \item[(D2)] \emph{Stable sampling fraction:} $n_s / N \to_{p_d} f$ for some $f \in (0, 1]$, and there exist $\delta \in (0, f)$ and $C < \infty$ such that $P_{\mathcal{S}^N \mid \omega}(n_s / N \le \delta) \le C N^{-\kappa}$ for some $\kappa > 4$ and $N$ large enough.
    \item[(D3)] \emph{Bounded inclusion probabilities:} there exist $0 < \underline{\lambda} \le \bar{\lambda} < \infty$ such that for all $N$, $\underline{\lambda} n / N \le \inf_i \pi_i \le \sup_i \pi_i \le \bar{\lambda} n / N$, where $n := \mathbb{E}_{P_{\mathcal{S}^N \mid \omega}}[n_s]$ is the expected sample size.
    \item[(D4)] \emph{Controlled second-order dependence:} $\max_{i \le j} |\pi_{ij} - \pi_i \pi_j| \le C_1 / n_s$ a.s., for some $C_1 < \infty$ and all $N$.
\end{itemize}

\begin{remark}
(D1) separates super-population from design randomness, allowing design expectations to be taken conditionally on $(X^N, Z^N)$. (D2) rules out designs with vanishing sampling fraction. (D3) keeps the design weights bounded by ensuring inclusion probabilities are of order $n/N$. (D4), standard in the survey-process literature \cite{Boistard2012-gc, boistard_functional_2017, Cardot2010-cn}, controls the dependence among $\{\xi_i\}$ at order $1/n_s$. Together, (D2)--(D4) yield the design-based law of large numbers (Supplement, Lemma~S3), which is invoked in our latter consistency arguments.

These conditions are mild: simple random sampling without replacement (SRSWOR), Poisson sampling, probability-proportional-to-size (PPS) with replacement with bounded measure of size, and their stratified versions all satisfy (D2)--(D4) (Supplement, Lemma~S1). Stratified two-stage cluster designs satisfying analogous conditions at each stage induce SSU-level (D2)--(D3), while the pairwise dependence at the SSU level has an explicit primary-sampling-unit block structure that replaces (D4), see Section S3.1 in the Supplement.
\end{remark}

For any fixed region $A(\mathbf{x}) \subseteq \mathcal{X}$, a natural design-based estimator for $P^N_{Y \mid X \in A(\mathbf{x})}(\mathbf{x}) = \frac{1}{N(\mathbf{x})} \sum_{i \in [N]} \ind\;(X_i \in A(\mathbf{x}))\, \delta_{Y_i}$ with $N(\mathbf{x}) = \sum_i \ind\;(X_i \in A(\mathbf{x}))$, is the H\'ajek estimator. For a single-stage design, it is expressed as
\[
\widehat{P}^N_{\mathrm{HJ},\, Y \mid X \in A(\mathbf{x})}(\mathbf{x}) = \frac{1}{\widehat{N}(\mathbf{x})} \sum_{i \in [N]} \frac{\xi_i}{\pi_i}\, \ind\;(X_i \in A(\mathbf{x}))\, \delta_{Y_i}, \qquad \widehat{N}(\mathbf{x}) = \sum_{i \in [N]} \frac{\xi_i}{\pi_i}\, \ind\;(X_i \in A(\mathbf{x})).
\]
For multistage designs, $1/\pi_i$ is replaced by the overall unit weight aggregating inclusion probabilities across stages. The SDRF estimator (\S\ref{sec:algorithm}) adopts a H\'ajek-like form with design-compatible resampling and a data-adaptive partition $L^*(\mathbf{x})$ in place of $A(\mathbf{x})$.

\subsection{Kernel Mean Embeddings}
\label{subsec:kme}

Kernel mean embeddings (KMEs) represent probability measures as points in a reproducing kernel Hilbert space (RKHS) \citep{Aronszajn1950, Gretton_2007, Smola_2007}, reducing comparisons between distributions to distances between vectors. We use this representation both as the splitting criterion for SDRF and as the metric for the consistency theory in Section~\ref{sec:algorithm}.

Let $\mathbf{k} : \mathcal{Y} \times \mathcal{Y} \to \mathbb{R}$ be a bounded positive definite kernel with RKHS $(\mathcal{H}, \langle \cdot, \cdot \rangle_{\mathcal{H}})$ and canonical feature map $\Phi_{\mathbf{k}}(y) := \mathbf{k}(y, \cdot) \in \mathcal{H}$. For any $P \in \mathcal{M}_b(\mathcal{Y})$ in the space of finite signed Radon measures, the kernel mean embedding is
\[
\mu_{\mathbf{k}}(P) := \int_{\mathcal{Y}} \mathbf{k}(y, \cdot)\, dP(y) = \mathbb{E}_{Y \sim P}[\Phi_{\mathbf{k}}(Y)] \in \mathcal{H},
\]
and the maximum mean discrepancy (MMD) between $P, Q \in \mathcal{M}_b(\mathcal{Y})$ is
\[
d_{\mathbf{k}}(P, Q) := \| \mu_{\mathbf{k}}(P) - \mu_{\mathbf{k}}(Q) \|_{\mathcal{H}}.
\]
A kernel is \emph{characteristic} if $\mu_{\mathbf{k}}$ is injective, so that $d_{\mathbf{k}}(P, Q) = 0 \iff P = Q$, and \emph{universal} \cite{Steinwart2001-ja} if $d_{\mathbf{k}}$ metrizes weak convergence: $P_N \rightsquigarrow P \iff d_{\mathbf{k}}(P_N, P) \to 0$.

Three properties of this representation are central to SDRF.
First, it provides a distributional target. A split criterion based on $d_{\mathbf{k}}$ between the empirical KMEs of two candidate child nodes \cite{cevid2022distributional} discriminates conditional laws rather than conditional moments. Plug-in estimators for any continuous functional $\Psi$ of $P_{Y \mid X = \mathbf{x}}$ are then obtained from the same forest output; we exploit this in our application to construct multivariate tolerance regions.

Second, the MMD is computationally tractable.
Naive evaluation of an empirical MMD between $n$- and $m$-sample empirical measures requires $O((n+m)^2)$ kernel evaluations. When $\mathbf{k}$ is translation-invariant on $\mathbb{R}^d$, Bochner's theorem yields a random Fourier feature map $\varphi : \mathcal{Y} \to \mathbb{R}^K$ such that $\mathbf{k}(y, y') \approx \langle \varphi(y), \varphi(y') \rangle$ \citep{rahimi2007random}, and the empirical MMD reduces to
\[
\widehat{d}^{\,2}_{\mathbf{k}}(P_n, Q_m) = \Big\| \tfrac{1}{n} \sum_{i=1}^n \varphi(Y_i) - \tfrac{1}{m} \sum_{j=1}^m \varphi(Y_j') \Big\|_2^2,
\]
a squared Euclidean norm in $\mathbb{R}^K$ with per-split cost $O(K(n+m))$ and $K \ll n$ in practice. For multivariate $Y$ in moderate dimension, this typically outperforms estimators based on Wasserstein \cite{Fournier2015-mu} or bounded-Lipschitz distances both in computation and in concentration.

While the first two properties are inherited from the i.i.d.\ distributional random 
forest of \cite{cevid2022distributional}, the third is the structural 
foundation for our design-aware algorithm: 
because $\mu_{\mathbf{k}}$ is defined on $\mathcal{M}_b(\mathcal{Y})$ directly, 
the super-population conditional law $P_{Y \mid X = \mathbf{x}}$, its 
finite-population counterpart $P^N_{Y \mid X = \mathbf{x}}$, and any 
H\'ajek-type sample estimate live in the same Hilbert space and are compared 
by the same metric. This is what allows a single algorithmic and theoretical 
scheme to cover single-stage, stratified, clustered, and multistage designs 
without redoing empirical-process arguments design-by-design.

The conditions on $\mathcal{Y}$ and $\mathbf{k}$ that make all of the above well-defined are:
\medskip

\noindent\underline{Assumptions on the outcome space and kernel}
\begin{itemize}[leftmargin=3em]
    \item[(S1)] $\mathcal{Y}$ is a locally compact Polish space.
    \item[(K1)] $\mathbf{k}$ is uniformly bounded, positive definite, and $C_0$-continuous: $\mathbf{k}(\cdot, y) \in C_0(\mathcal{Y})$ for all $y \in \mathcal{Y}$, where $C_0(\mathcal{Y})$ is the Banach space of bounded continuous functions vanishing at infinity \cite[\S2]{Sriperumbudur2010-xr}.
    \item[(K2)] $\mathbf{k}$ is $c_0$-universal: the RKHS $\mathcal{H}$ is dense in $C_0(\mathcal{Y})$ under the uniform norm \cite[Definition~2, Theorem~1]{Carmeli2010-ps}.
    \item[(K3)] (Optional, for the random Fourier feature acceleration described above) When $\mathcal{Y}$ carries a compatible Abelian topological group structure, $\mathbf{k}$ is translation-invariant.
\end{itemize}

\begin{remark}
Assumptions (S1)--(K2) ensure that $\mu_{\mathbf{k}}(P)$ is well-defined on $\mathcal{M}_b(\mathcal{Y})$ and that $d_{\mathbf{k}}$ is both characteristic and metrizes weak convergence. (K3) is not needed for the statistical theory; when satisfied, it enables the random Fourier feature acceleration of (ii). When $\mathcal{Y} = \mathbb{R}^d$, the empirical-MMD concentration rate can be made independent of $d$ \cite{Gretton_2007}.
\end{remark}

Assumptions (S1) and (K1)--(K3) cover many common outcome types: $\mathcal{Y} = \mathbb{R}^d$, graph representations in $[0,1]^{m(m-1)/2}$ or $\mathbb{R}^{m \times m}_{\mathrm{sym}}$, torus-valued data $\mathbb{T}^d$ (via $\theta \mapsto (\cos\theta, \sin\theta)$), and the simplex $\Delta^{p-1}$, equipped with standard kernels such as the Gaussian RBF $\mathbf{k}(y, y') = \exp\{-\|y - y'\|^2 / (2\sigma^2)\}$ or its restriction after embedding.

\section{The SDRF Algorithm and Theoretical Results}
\label{sec:algorithm}

Random forests combine a CART-style \citep{1984_Breiman} tree ensemble with bagging \citep{breiman1996bagging} and can equivalently be viewed as kernel-type estimators \citep{lin2006random, qiu2024random}: each tree induces a data-adaptive neighborhood, and the forest averages outcomes over these neighborhoods with data-dependent weights. Classical formulations target the conditional mean for regression and class probabilities for classification \citep{breiman2001random}; \citet{cevid2022distributional} extend the construction to the full conditional law for i.i.d.\ data by replacing the mean-based splitting criterion with one based on kernel mean embeddings.

SDRF adapts this distributional forest to complex survey designs through three modifications: a design-weighted MMD splitting criterion (Section~\ref{sec:split}), a pseudo-population bootstrap that preserves the dependence and unequal-probability structure the design induces (Section~\ref{subsec:resampling}), and PSU-level honesty (Section~\ref{sec:honesty}). Together these yield a kernel estimator $\widehat{\omega}_{\mathbf{x}}(\cdot)$ whose splitting criterion targets distributional discrepancy in $\hilbert$ and whose leaf aggregation takes a H\'ajek form compatible with the design. The estimator is stated in Definition~\ref{def:sdrf}; its ingredients $L^*_b$ and $n^*_{b,i}$ are constructed in the subsections that follow, and the consistency theory is developed in Section~\ref{sec:prediction}.

We introduce the following notation. Let $B$ denote the number of trees and $T_b$ the $b$-th tree, fitted on resampled data $D^*_b = \{(X_i, Y_i, n^*_{b,i}/\pi_i)\}_{i=1}^{n_s}$, where $n^*_{b,i} \ge 0$ are resampling multipliers. Let $q \in (0,1)$ be the PSU-level selection probability for inclusion in the splitting subsample $D^*_{b,\mathrm{split}}$, with $D^*_{b,\mathrm{est}} = D^*_b \setminus D^*_{b,\mathrm{split}}$ used for honest leaf aggregation, and let $\mathcal{I}(D^*_{b,\mathrm{est}}) \subseteq [n_s]$ collect the corresponding subject indices. Write $L(\cdot)$ for the \emph{oracle partition} fitted on the full finite population $D^N$; under (D1) and the i.i.d.\ super-population assumption, this reduces to a standard random forest on i.i.d.\ data. Write $L^*(\cdot)$ for the partition fitted on the resampled training data. Both $L(\cdot)$ and $L^*(\cdot)$ map any $\mathbf{x} \in \mathcal{X}$ to a terminal leaf region in $\mathcal{X}$. Full implementation details are in Algorithm~\ref{alg:rf}.

\begin{definition}[SDRF estimator]
\label{def:sdrf}
Given the forest $\mathcal{T}=\{T_1,\dots,T_B\}$ of Algorithm~\ref{alg:rf}, the SDRF
weight for observation $i$ at covariate value $\mathbf{x}$ is
\begin{equation}
\label{eq:sdrf-weight}
\widehat{\omega}_i(\mathbf{x})
= \frac{1}{B}\sum_{b=1}^B
\frac{ n^*_{b,i}\,\big/\,\big(\pi_i(1-q)\big)\, \ind\!\big(X_i \in L^*_b(\mathbf{x})\big)}
{\sum_{j \in \mathcal{I}(D^*_{b,\mathrm{est}})} n^*_{b,j}\,\big/\,\big(\pi_j(1-q)\big)\, \ind\!\big(X_j \in L^*_b(\mathbf{x})\big)},
\end{equation}
and the induced conditional-distribution estimator and plug-in estimator are
\begin{equation}
\label{eq:sdrf-plugin}
\widehat{P}^N_{Y \mid X = \mathbf{x}} := \sum_{i=1}^{n_s} \widehat{\omega}_i(\mathbf{x})\, \delta_{Y_i},
\qquad
\widehat{\Psi}_N(\mathbf{x}) := \Psi\!\big(\widehat{P}^N_{Y \mid X = \mathbf{x}}\big).
\end{equation}
\end{definition}

\begin{algorithm}[t]
{\fontsize{9.5}{13}\selectfont
\caption{Survey Distributional Random Forest (SDRF)}\label{alg:rf}
\begin{algorithmic}[1]
  \REQUIRE
    $D = \{X_i, Y_i, 1/\pi_i\}_{i=1}^{n_s}$.
    Setup hyper\textendash parameters: maximum depth, $B$ number of trees, minimum node size, maximum node-level weight ratio $\max_{i} \hat{w}_i / \min_{i} \hat{w}_i$, and a probability $q \in (0, \; 1)$ for selection of PSUs into $D^*_{b,\mathrm{split}}$, and parameters needed for the selected reproducing kernel $\mathbf{k}$. Denote $k(i)$ the mapping from subject index to PSU index.
  \ENSURE Random forest ensemble $\mathcal{T} = \{T_1, \dots, T_B\}$
  \STATE $\mathcal{T} \gets \emptyset$
  \FOR{$b = 1, \dots, B$}
    \STATE Draw resample $D^*_b = \{(X_i,Y_i, 1/\pi_i \times n^*_{b, i})\}_{i=1}^{n_s}$ respecting the design. \hfill\COMMENT{(i) Resampling, \S\ref{subsec:resampling}}
    \STATE Draw $\forall k(i)$, $\ind(k(i) \in D^*_{b,\mathrm{split}}) \sim Bernoulli(q)$. \hfill\COMMENT{(ii) Honesty, \S\ref{sec:honesty}}
    \STATE Split at the PSU level $D^*_b = D^*_{b,\mathrm{split}} \bigcup D^*_{b,\mathrm{est}}$.
    \STATE Train decision tree $T_b$ on $D^*_{b,\mathrm{split}} = \{(X_i, Y_i, \;  1/\pi_i \times n^*_{b,i} \times 1/q)\}$ by: \hfill\COMMENT{(iii) Splitting rule, \S\ref{sec:split}}
    \STATE Randomly select $mtry$ features.
    \STATE Choose the split that maximizes $M^*(\mathrm{split}; D^*_{\mathrm{split}}) = \frac{N^*_L N^*_R}{N^{*2}_{Pa}}\|\mu_{\mathbf k}(\hat{P}^*_L) - \mu_{\mathbf k}(\hat{P}^*_R)\|^2_{\hilbert}$ until reaching maximum depth, minimum node size, or maximum weight ratio.
    \STATE $\mathcal{T} \leftarrow \mathcal{T} \; \bigcup \;\{T_b\}$.
   \ENDFOR
  \STATE \textbf{return} For all $b$ with $i \in \mathcal{I}(D^*_{b,\mathrm{est}})$, \\
  $
  \widehat{\omega}_i(\cdot) = \frac{1}{B}\sum_{b=1}^B \frac{ n^*_{b,i} / (\pi_i (1 - q)) \, \ind(X_i \in L^*_b(\mathbf{x}))}{\sum_{j \in \mathcal{I}(D^*_{b,\mathrm{est}})} n^*_{b,j} / (\pi_j (1 - q)) \, \ind(X_j \in L^*_b(\mathbf{x}))}
  $.
  \STATE For $\Psi(\cdot)$, obtain plug-in estimator
  $\Psi(\widehat{P}^N_{Y\mid X = \mathbf{x}}) = \Psi(\sum_{i=1}^{n_s} \widehat{\omega}_i(\mathbf{x}) \; \mathbf{k}(Y_i, \, \cdot))$.
\end{algorithmic}
}
\end{algorithm}

\subsection{Splitting Rule}
\label{sec:split}
The splitting criterion of an RF algorithm guides the formation of tree structure by determining how the feature space should be partitioned into increasingly homogeneous child nodes.
Consider any node where a split is possible (has not met the stopping criteria). Denote $P_a \subseteq \mathcal{X}$ the feature space of that node. A candidate split can be parametrized as $\theta = (j, \; t) \in \Pi$, $j = 1, \dots, p$, $t \in Range(X_j) \subseteq [a_j, \; b_j]$. This partitions $P_a$ into child regions $C_L(\theta) := \{x \in P_a: x_j \leq t\}$ and $C_R(\theta) := \{x \in P_a:\; x_j > t\}$.
We find the split by maximizing the MMD between the KMEs of the (design--weighted) distribution estimates between the two child regions. Formally, $\hat{\theta} = \arg \max_{\theta \in \Pi} M^*_{n_s}(\theta)$, where
{\footnotesize
\begin{align}
\label{eq:mmd-split}
    M^*_{n_s}(\theta)
    =
    \frac{\hat{N}_{L}\hat{N}_{R}}{\hat{N}^2_{P_a}}\,
    \left\|
    \mu_{\mathbf k} \left( \frac{1}{\hat{N}_{L}} \sum_{\{i \in [n_s]: X_i \in C_L\}} \frac{n^*_{b,i}}{q \times \pi_i} \; \delta_{Y_i} \right)
    -
    \mu_{\mathbf k} \left( \frac{1}{\hat{N}_{R}} \sum_{\{i \in [n_s]: X_i \in C_R\}} \frac{n^*_{b,i}}{q \times \pi_i} \; \delta_{Y_i} \right)
    \right\|^2_{\hilbert},
\end{align}
}
with $\hat{N}_\ell = \sum_{i \in [n_s]} \ind(X_i \in C_\ell) \frac{n^*_{b,i}}{q \times \pi_i}$, $\ell \in \{L, R\}$.
Here the factor $1/q$ rescales the split subsample $D^*_{b,\mathrm{split}}$ back to the design weights; the complementary estimation set used for leaf aggregation in Definition~\ref{def:sdrf} carries the factor $1/(1-q)$ instead.
When under stratified two--stage designs, notice that the same form for $M^*_{n_s}$ holds, except that we replace the sum over subjects by sum $\sum_{h,k}\sum_{i \in \mathcal{U}_{hk}}$ over SSUs $(h,k)$ and then over subjects within SSUs, if they exist. In practice, the framework extends to designs with more than two stages,  
see Section~\ref{sec:implementation}.

The following theorem shows that the empirical split score in \eqref{eq:mmd-split} consistently recovers its finite-population and super-population counterparts, justifying its use as a split criterion.

\begin{theorem}[Local design/joint consistency of the MMD split]
\label{thm:split-consistency}
For a parent region $P_a \subseteq \mathcal{X}$ at depth $d$ generated by the
tree-growing process, consider axis-aligned potential splits
$\theta=(j,t)\in\Pi:=\{(j,t): j\in[p],\ t\in[a_j,b_j]\}$ which define child regions.
Let $M^*_{n_s}(\theta)$ denote the (design-resampled) split score in \eqref{eq:mmd-split}, and 
$M^N(\theta)$ be its finite-population analog.
Assume (R1)--(R3), (K1), and (A2)--(A3) of Section~\ref{sec:prediction}, and that $\mathcal{X}\subseteq\mathbb{R}^p$ is bounded. 
Assume further the survey design satisfies either 
(i) (D1)--(D4) for a single-stage design, with sample index $i\in[n_s]$ and inclusion probability $\pi_i$, or
(ii) (D-TS1)--(D-TS4) (Supplement, Section~S3.1) for a stratified two-stage cluster design, with SSU index $u=(h,k,j)\in[n_s]$ and inclusion probability $\pi_u=\pi^{(1)}_ {hk}\pi^{(2)}_{hkj\mid hk}$.
Then, conditional on the realized finite population,
\[
\sup_{\theta\in\Pi}\bigl|M^*_{n_s}(\theta)-M^N(\theta)\bigr| \xrightarrow{p_d} 0.
\]
If, in addition, the finite population is an i.i.d.\ realization from a super-population law $P$ under which $X$ admits a Lebesgue density $f_X$ on $\mathcal{X}$ with $\|f_X\|_\infty\le M_P<\infty$, and $M(\theta)$ denotes the corresponding population score, then
\[
\sup_{\theta\in\Pi}\bigl|M^*_{n_s}(\theta)-M(\theta)\bigr| \xrightarrow{p_{\otimes}} 0.
\]
\end{theorem}

Theorem~\ref{thm:split-consistency} provides a {local} result. For a fixed parent node $P_a$, the empirical MMD criterion $M^*_{n_s}(\theta)$ uniformly approximates  $M^N(\theta)$ and $M(\theta)$.
The uniformity over $\theta\in\Pi$ is essential for split selection, as it permits an argmax continuous-mapping argument and yields consistency of the selected split $\hat\theta_{n_s}$ whenever the oracle maximizer $\theta_0$ is unique \citep{van-der-Vaart2023-ur}. Under additional uniqueness and near-maximization conditions on the maximizers of $M^N$ and $M$, the split argmax $\hat\theta_{n_s}$ is design-consistent and jointly consistent: $\hat\theta_{n_s} \xrightarrow{p_d} \theta_0(\omega)$ for $P$-a.s.\ $\omega$, and $\hat\theta_{n_s} \xrightarrow{p_\otimes} \theta^\dagger$. The precise statement is Corollary~S4 in Section~S6 of the Supplement.

Although extending from a fixed node to an entire data-adaptive tree requires additional control of the growing process, this theorem sheds light on why one might believe the partitions found via resampled data to be close to those obtained from using the full finite population, which is in particular assumption (B2) of Theorem~\ref{thm:consistency} that is required for consistency in the MMD metric.

The split score \eqref{eq:mmd-split} is computed from a single resampling draw $n^*$. When the design weights are highly variable, this can make the score noisy; one may then average $\bar{n}_i^* := M^{-1}\sum_{b=1}^M n_{b,i}^*$ over $M$ independent draws to stabilize it, at the cost of reduced tree-to-tree diversity. Proposition~\ref{prop:avg-resample-effect} quantifies when this is worthwhile.

\begin{proposition}[Effect of averaging $M$ resampling draws on split-score stability]
\label{prop:avg-resample-effect}
Under the same setting of Theorem~\ref{thm:split-consistency},
denote
${N}_{P_a}:=|\{i \in [N]: X_i\in P_a\}|$, $N_\ell:=|\{i \in [N]: X_i\in C_\ell\}|$.
Let $M^N(\theta)$ be the finite-population split score, and $M_{n_s}^{*(M)}(\theta)$ the score using
averaged bootstrap multipliers $\bar{n}_i^* := M^{-1}\sum_{b=1}^M n_{b,i}^*$.
Assume (R1)--(R3), (A3), (K1), and (D1)--(D4) for single-stage designs or (D-TS1)--(D-TS4) for stratified two-stage cluster designs, then
\[
\bigl|M_{n_s}^{*(M)}(\theta)-M^N(\theta)\bigr|
=
O_{p_d}\!\left(
N_{P_a}^{-1/2}
\;+\;
\frac{1}{\sqrt{M}}\sum_{\ell\in\{L,R\}} n_{\mathrm{eff},\ell}^{-1/2}\right),
\]
\noindent where
$
n_{\mathrm{eff},\ell}
=
\frac{(\sum_{i \in S \bigcap C_\ell} \pi_i^{-1})^2}{\sum_{i \in S \bigcap C_\ell} \pi_i^{-2}}
$ is the design-weighted effective sample size \citep{Kish1995-aq} in child $\ell$.
\end{proposition}

\begin{remark}[Design--resampling cross-over]
Proposition~\ref{prop:avg-resample-effect} separates two sources of split-score fluctuation: an intrinsic \emph{design term} $|M_{n_s}(\theta)-M^N(\theta)|=O_{p_d}(N_{P_a}^{-1/2})$, present even at fixed $n^*$, and a \emph{resampling term} $|M^{*(M)}_{n_s}(\theta)-M_{n_s}(\theta)|=O_{p_d}\bigl(M^{-1/2}\sum_{\ell} n_{\mathrm{eff},\ell}^{-1/2}\bigr)$ from averaging over draws. The resampling term is the standard Monte Carlo $M^{-1/2}$ rate, with the design entering only through the constant via the effective sample sizes. 

They balance at the cross-over $\tilde M \asymp N_{P_a}\bigl(\sum_{\ell} n_{\mathrm{eff},\ell}^{-1/2}\bigr)^2$: once $M\gtrsim\tilde M$ the irreducible design term dominates and further averaging cannot stabilize the score. The threshold is small at deep nodes ($\tilde M\approx 8$ when $n_{\mathrm{eff},\ell}\approx 100$ and $N_{P_a}\approx 200$) so single-digit $M$ typically suffices, after which effort is better spent raising $n_{\mathrm{eff},\ell}$ (through minimum node size, depth, or subsampling fraction $q$) than adding resamples. This balances score fluctuation at a single node; split \emph{selection} depends on score differences across candidates $\theta$, so the threshold is a per-node stability heuristic rather than a guarantee on the selected split.
\end{remark}

\subsection{Resampling}
\label{subsec:resampling}
Bagging is a key ingredient of random forests that smooths the partition and injects tree-to-tree diversity \citep{breiman1996bagging, Buhlmann2002-vp}. Under a complex survey design, Efron's classical bootstrap \citep{Efron1993-eg}  can fail to preserve the survey-weighted empirical measure \citep{wang2022bootstrap, Mashreghi2016-nh}.

We adopt the pseudo-population bootstrap class of approaches, taking that of \citet{wang2022bootstrap} as our default scheme. For each resample, we first use the design weights $\{1/\pi_i\}_{i \in [n_s]}$ to reconstruct, via a multinomial draw, a pseudo finite population of expected size $\widehat{N} = \sum_{i \in [n_s]} 1/\pi_i$; the original sampling design $\mathbf{p}$ is then re-applied to this reconstructed population. The procedure yields multipliers $n^* = \{n_i^*\}_{i \in [n_s]}$ that up- or down-weight the sampled distribution while respecting the design structure; see \citet{Mashreghi2016-nh, Conti_2022} for a broader review of resampling under complex designs. In the forest, the multiplier attached to subject $i$ in tree $b$ is written $n^*_{b,i}$; we suppress the tree index here, as a single generic resample is described.
 
We require the following assumptions on the resampled multipliers, and describe specific bootstrapping procedures that satisfy them in the Supplement.
\begin{lemma}
\label{lem:Wangbs_Rassumptions}
For single-stage designs including Poisson, SRSWOR, and PPS-with-replacement, and two-stage clustering designs with bounded cluster sample size, possibly stratified, the pseudo-population bootstrap of Section~S5.1, Supplement, based on \citet{wang2022bootstrap} with modifications detailed therein, generates unit- (or SSU-level) multipliers $\{n_i^*\}_{i \in [n_s]}$ satisfying:
\medskip
\noindent\underline{On resample multipliers $n^*$}
\begin{enumerate}[leftmargin=3em]
  \item[(R1)] The conditional law of $n_i^*$ given $(\xi^N, Z^N)$ depends on the design only through these variables. Multipliers are conditionally independent across strata; for two-stage designs, they are additionally conditionally independent across distinct PSUs given the first-stage multipliers.
  \item[(R2)] $\sup_{i:\,\xi_i=1}\bigl|\mathbb{E}(n_i^* \mid \xi, Z^N) - 1\bigr| = o_{p_d}(1)$, and $\sup_{i:\,\xi_i=1} \mathbb{E}\bigl[|n_i^*|^{2+\epsilon}\mid\xi,Z^N\bigr] < \infty$ a.s.\ for some $\epsilon > 0$.
  \item[(R3)] There exists $C < \infty$ such that
  \[
    \max_i \sum_{j \ne i} \bigl|\mathrm{Cov}(n_i^*, n_j^* \mid \xi, Z^N)\bigr| \le C \sup_j \mathrm{Var}(n_j^* \mid \xi, Z^N) \quad \text{a.s.\ for all $N$.}
  \]
\end{enumerate}
\end{lemma}
\begin{remark}
(R1) ensures the multipliers are generated in a design-adapted, symmetric manner, and prevents the resampling step from introducing extraneous dependence on unobserved design features. (R2) enforces centering at the obtained sample and stable fluctuation, standard for bootstrap multipliers. (R3) is a weak-dependence requirement: it permits dependence across units while remaining strong enough for the concentration bounds used in Sections~\ref{sec:algorithm}--\ref{sec:prediction}.
\end{remark}
Conditions (R1)--(R3) are the only properties of the resampling scheme invoked in the consistency analysis of Section~\ref{sec:prediction}. For designs with more than one stage, the pseudo-population reconstruction is applied recursively, one stage at a time. The unit-level multiplier $n^*_{b,i}$ is then the product of all the per-stage multipliers. A counterpart of Lemma~\ref{lem:Wangbs_Rassumptions} is given in Section~S5.2, Supplementary Material.

\subsection{Honesty}
\label{sec:honesty}

Honesty requires the leaf structure and the leaf contents to be estimated from disjoint, independent data. $D_{\mathrm{split}}$ fits the partition $L$ and $D_{\mathrm{est}}$ populates the leaves forming $\widehat{\omega}(\mathbf{x})$ \citep{Athey2016-hs, Wager2018-lu, Athey_Tibshirani_Wager_2019}. Under an i.i.d.\ sample, any disjoint split of the observations automatically yields two independent set of observations. A clustered design violates the very assumption the guarantee rests on: subjects within a PSU are dependent, so a subject-level partition of the sample places dependent units on opposite sides of the $D_{\mathrm{split}}/D_{\mathrm{est}}$ divide and the independence fails. SDRF restores it by splitting at the PSU level, assigning whole PSUs to $D_{\mathrm{split}}$ or $D_{\mathrm{est}}$ via the inclusion probability $q \in (0,1)$ of Algorithm~\ref{alg:rf}, so that $D_{\mathrm{est}} \independent D_{\mathrm{split}}$ holds conditional on $\xi$. Because each PSU enters $D_{\mathrm{split}}$ with probability $q$ and $D_{\mathrm{est}}$ with probability $1-q$, the two subsamples are reweighted by $1/q$ and $1/(1-q)$ respectively, so that each recovers the design weights $1/\pi_i$ in expectation.
 
\begin{remark}
When data are available from a continuous survey conducted under the same design and $P_{Y \mid X}$ may be assumed stable between consecutive waves, one wave can serve as $D_{\mathrm{split}}$ and the other as $D_{\mathrm{est}}$, yielding honesty without sacrificing within-sample data to the split.
\end{remark}

\subsection{Consistency Result}
\label{sec:prediction}
We now establish the main result, namely, consistency of the SDRF plug-in estimator at a fixed covariate value $\mathbf x \in \mathcal X$. Two targets are natural. In the design space the estimand is the finite-population conditional law $P_{\mathcal S^N\mid\omega}(\mathbf x)$ of $Y\mid X=\mathbf x$; in the super-population the estimand is $P_{Y\mid X=\mathbf x}$.

The argument proceeds in two steps. Theorem~\ref{thm:consistency}, under additional conditions (A1)--(A3) and (B1)--(B2) introduced below, shows that the MMD between the SDRF estimate $\widehat{P}_{\mathcal S^N\mid\omega}(\mathbf x)$ and each target vanishes in the design measure. Corollary~\ref{cor:consistency_functional} then lifts these MMD statements to consistency of $\Psi$ at fixed $\mathbf{x}$.

\begin{theorem}
\label{thm:consistency}
Suppose (S1) for the space $\mathcal{Y}$, (D1)--(D4) (or (D-TS1)--(D-TS4)) for the survey design $\mathbf{p}$, (R1)--(R3) for the resample multiplier $n^*$, and (K1)--(K2) for the reproducing kernel $\mathbf{k}$. Suppose the true super-population conditional kernel mean embedding is Lipschitz. That is, there exists $L_{\mathrm{lip}} > 0$ such that for all $\mathbf{x}, \mathbf{x}' \in \mathcal{X} \subseteq \mathbb{R}^p$,
\[
    \big\| \E_P[\mathbf{k}(Y, \cdot) \mid X = \mathbf{x}] - \E_P[\mathbf{k}(Y, \cdot) \mid X = \mathbf{x}'] \big\|_{\hilbert}
    \leq L_{\mathrm{lip}} \|\mathbf{x} - \mathbf{x}'\|.
\]
Fix an evaluation point $\mathbf{x} \in \mathcal{X}$ that is an atom of the covariate law, $p_{\mathbf{x}} := P(X = \mathbf{x}) > 0$, so that $N_{\mathbf{x}} := \sum_{i=1}^N \ind(X_i = \mathbf{x}) = \Theta_{p_d}(N) \to \infty$. Suppose further:\\

\underline{Algorithm regularization}
\begin{itemize}[leftmargin=3em]
    \item[(A1)] Shape regularity: For a.s. $L(\mathbf{x})$ for any $\mathbf{x}$ in the sampled covariate space satisfying $diam(L(\mathbf{x})) < k_N$, $\exists \; 0 < c_1 \leq 1$ s.t. $\bar{B}_{r_N}(\mathbf{x}) \subseteq L(\mathbf{x})$, $r_N = c_1 k_N$.
    \item[(A2)] There exist constants $0 < m < M < \infty$ such that for all radius r, of some $L(\mathbf{x})$ for the given $\mathbf{x}$ such that $0 < r \leq k_N$,
    $m r^p \leq \#\{i : X_i \in B_r(\mathbf{x})\}/N \leq M r^p$, where $B_r(\mathbf{x})$ is the ball centered at $\mathbf{x}$ with radius $r$.
    \item[(A3)] There exists $\alpha \in (0, 1/2]$ such that every candidate split $\theta$ at every internal node with parent region $P_a$ satisfies
    $\alpha \le N_L(\theta)/N_{P_a} \le 1-\alpha$ and $\alpha \le N_R(\theta)/N_{P_a} \le 1-\alpha$. At estimation time every leaf receives at least $n_{L^*,\min} > 0$ subjects: $n_{L^*(\mathbf{x})} = \sum_{i=1}^N \xi_i\, \ind(X_i \in L^*(\mathbf{x})) \geq n_{L^*,\min}$.
\end{itemize}

\underline{Algorithmic behavior}
\begin{itemize}[leftmargin=3em]
    \item[(B1)] For any $\mathbf{x}$ and some deterministic chosen sequence $k_N \to 0$ such that $k^p_N N \to \infty$  as $N \to \infty$, there exists $t_N \to 0$ such that
    $P_{L}(diam(L(\mathbf{x})) \geq k_N) \leq t_N$,
    where $diam(L(\mathbf{x})) = \sup_{x_1, x_2 \in L(\mathbf{x})}\rho(x_1, x_2)$, and
    $\rho$ is some metric on $\mathcal{X}$.
    \item[(B2)] $\sup_{x \in \mathcal{X}_1 \cup \mathcal{X}_2} \mathbb{P}\{X \in L^*(\mathbf{x})\,\Delta\,L(\mathbf{x})\} = o_p(d_N)$ for some $d_N = o(k_N^p)$, where $\Delta$ denotes symmetric difference: $A \Delta B := (A \setminus B)\;\cup\; (B \setminus A)$.
\end{itemize}
Let $\widehat{P}_{\mathcal{S}^N|\omega}(\mathbf{x})$ denote the SDRF estimate of the finite-population conditional law $P_{\mathcal{S}^N|\omega}(\mathbf{x})$ of $Y \mid X = \mathbf{x}$ in the design space.
Then $d_{\mathbf k}(\widehat{P}_{\mathcal{S}^N|\omega}(\mathbf x), {P}_{\mathcal{S}^N|\omega}(\mathbf x)) \xrightarrow{p_d} 0$ and $d_{\mathbf k}(\widehat{P}_{\mathcal{S}^N|\omega}(\mathbf x), P_{Y \mid X = \mathbf{x}})$ $ \xrightarrow{p_d} 0$ as $N, B \to \infty$.
\end{theorem}
We discuss these conditions below.
The Lipschitz condition on the conditional KME is a smoothness requirement on the target, ensuring that nearby covariate values induce nearby conditional laws in $\hilbert$.
Regularization--type conditions (A1)--(A3) can be controlled via the imposed stopping criteria. (A1) imposes a geometric control on the shape of $L$ to control the stretching. (A2) rules out clumping or sparsity of covariates near $\mathbf x$, keeping the local density bounded away from zero and infinity. (A3) forbids degenerate splits and empty leaves.
(B1)--(B2) formalize how the partitions produced by the algorithm behave as the finite population size grows.
(B1) is a shrinking diameter assumption that mirrors the adaptive concentration result proved for i.i.d. forests \citep{Wager2015-nc}.
(B2) requires that the thickness of the error band be negligible relative to the leaf's own diameter. In other words, the rate at which the leaf $L^*(\cdot)$ function estimated from the bootstrap sample approaches the leaf $L(\cdot)$ estimated when using the full finite population must be faster than the oracle leaf's own shrinking-volume rate. 
Under $d_N = o(k_N^p)$ and the shape regularity of (A1), the symmetric-difference region carries a vanishing fraction of the oracle leaf's mass, so the estimated-leaf tree and its conditional variance agree with their oracle-leaf counterparts asymptotically. 
Among all the assumptions, (B2) is the hardest to directly control. Thus we provide Theorem~\ref{thm:split-consistency}, which partially sheds light on conditions under which (B2) can be reasonably assumed to hold.

\begin{corollary}
\label{cor:consistency_functional}
Suppose $\{(X_i, Y_i, Z_i)\}_{i=1}^N$ are i.i.d. realizations from $P$, $\mathbf k$ satisfies (K1)--(K2), and $\mathcal{Y}$ is locally compact Polish (S1). Fix $\mathbf{x} \in \mathcal{X}$, and let $\Psi : (\mathcal{M}_b(\mathcal{Y}), d_{\mathbf{k}}) \to (\mathcal{V}, d_v)$ be a measurable map into a metric space $(\mathcal{V}, d_{v})$ that is continuous at both $P_{\mathcal{S}^N|\omega}(\mathbf{x})$ and $P_{Y \mid X = \mathbf{x}}$ for $P$-almost every $\omega \in \Omega$. If \,
$
d_{\mathbf k}(\widehat{P}_{\mathcal{S}^N|\omega}(\mathbf x), {P}_{\mathcal{S}^N|\omega}(\mathbf x)) \xrightarrow{p_d} 0, 
$
and 
$
d_{\mathbf k}(\widehat{P}_{\mathcal{S}^N|\omega}(\mathbf x), P_{Y \mid X = \mathbf{x}})$ $ \xrightarrow{p_d} 0
$ hold then
\begin{align*}
    \Psi(\widehat{P}_{\mathcal{S}^N|\omega}(\mathbf x)) &\xrightarrow{p_d} \Psi({P}_{\mathcal{S}^N|\omega}(\mathbf x))
    \;\;
    \text{ (design consistency) }, \text{ and }\\
    \Psi(\widehat{P}_{\mathcal{S}^N|\omega}(\mathbf x)) &\xrightarrow{p_\otimes} \Psi(P_{Y \mid X = \mathbf{x}})
    \quad\,\;
    \text{(joint consistency) }.
\end{align*}
\end{corollary}

Corollary~\ref{cor:consistency_functional} shows that controlling MMD convergence suffices to establish both types of consistency of $\Psi(\widehat{P}_{\mathcal{S}^N|\omega}(\mathbf{x}))$.
The continuity requirement on $\Psi$ is mild, and is met by most targets of practical interest, as the following illustrate.
When $\mathcal Y=\mathbb R^d$, linear functionals such as conditional means
$\Psi(P)=\int y\,dP(y)$ and, more generally, $\Psi_g(P)=\int g(y)\,dP(y)$ for $g\in C_b(\mathcal Y)$
are continuous under weak convergence.
For distributional functionals, $\Psi_t(P)=F_P(t)$ and $\Psi_t(P)=P(Y>t)$ are continuous at $P$
whenever $P(\{t\})=0$.
The quantile functional $\Psi_q(P)=F_P^{-1}(q)$ is continuous at $P$ whenever the $q$-quantile is
unique. Equivalently, that $F_P$ has no flat segment at level $q$.
The kernel mean element itself and its norm $\Psi(P) = \|\mu_{\mathbf{k}}(P)\|_{\hilbert}$, and the MMD between the true conditional $P$ and a reference distribution $Q$, $d_{\mathbf{k}}(P, Q)$, are also examples.
For higher-order continuous functionals such as covariances or variances of kernel embeddings, larger effective sample sizes may be needed to ensure adequate concentration of the estimator $\widehat{P}_{\mathcal{S}^N|\omega}$ in the $d_{\mathbf{k}}$-metric.

\section{Practical Considerations}
\label{sec:implementation}

This section collects guidance for deploying SDRF: the simplification available 
for deep multistage designs, kernel and bandwidth choice, hyperparameters, resampling and averaging choices, and optional post-processing.

\paragraph{Design reduction.}
When the survey has more than two stages, one collapses to a stratified two-stage skeleton via the ultimate-cluster approximation \citep{hansen1953sample}: first-stage PSUs are treated as the effective clusters, and later stages are absorbed as within-PSU sub-sampling. When the first-stage sampling fraction is small, first-stage PSU sampling is well-approximated as with-replacement and the between-PSU component dominates the design variance, together justifying the reduction. This treatment is standard in survey practice \citep{Sarndal1992-mo}. NHANES, though a four-stage design, is analyzed via its stratum-and-PSU structure alone per CDC analytic guidelines \citep{Johnson2013-nhanes}. 

\paragraph{Kernel and bandwidth.}
Any kernel satisfying (K1)--(K2) may be used. A translation-invariant RBF kernel $\mathbf{k}(y,y')=\exp\{-\|y-y'\|^2/(2\sigma^2)\}$ is a standard default and additionally satisfies (K3), enabling the random Fourier feature acceleration of Section~\ref{subsec:kme}. The bandwidth $\sigma$ may be fixed by the median heuristic: $\sigma$ equal to the median of pairwise distances among the sampled outcomes. This is simple and performs well in practice \citep{Garreau2017-ro}.

\paragraph{Tree-growth hyperparameters.}
The stopping criteria (maximum depth, minimum node size) play their usual role and 
additionally enforce (A1)--(A3); shallow trees keep local weights well-controlled 
in finite samples. The one new hyperparameter relative to the i.i.d. forest is 
the node-level weight-ratio cap $\max_{i\in\mathcal{I}(L)}\hat{w}_i/\min_{i\in\mathcal{I}(L)}\hat{w}_i$ (with $\mathcal{I}(L)$ indexing the subjects in node $L$), which bounds the influence of any single design weight.

\paragraph{Resampling choices.}
Generating one pseudo-population and drawing $B$ resamples from it is asymptotically equivalent to regenerating both the pseudo-population and the resample $B$ times \citep{Conti2020-mc}, so the cheaper single-reconstruction route may be used. When the number of PSUs is large but PSU sizes are small so that the second stage contributes little to the design variance, second-stage resampling may be omitted with mild loss of design fidelity.

\paragraph{Averaging resampling draws.}
Because resamples are determined by the nonnegative multipliers $n^*$, one may stabilize a noisy split by averaging $\bar n_i^*=M^{-1}\sum_{b=1}^M n_{b,i}^*$ over $M$ independent draws. This is useful when the design weights are highly variable, at the cost of reduced tree-to-tree diversity. A single-digit $M$ typically suffices, after which effort is better spent raising the effective sample size.

\paragraph{Post-processing.}
Because SDRF yields a kernel-type estimator, the conditional law $\widehat{P}(Y\mid X=\mathbf{x})$ or any functional $\widehat{\Psi}(\mathbf{x})$ may be further smoothed in $\mathbf{x}$ by an additional kernel-smoothing step \citep{verdinelli2021forest}, reducing finite-sample variability when the target is believed to vary smoothly in $\mathbf{x}$.

\section{Simulations}
\label{sec:sim}

We assess SDRF on synthetic data drawn from a stratified two-stage design that mimics the structure of the NHANES study analyzed in Section~\ref{sec:casestudy}: probability-proportional-to-size (PPS) selection of primary sampling units (PSUs) within strata, followed by simple random sampling without replacement (SRSWOR) of units within selected PSUs. The study has two goals:
\begin{enumerate}[itemsep=0pt, topsep=0pt]
\item Quantify the gain in accuracy from incorporating the design, by comparing SDRF with the distributional random forest (DRF) of \citet{cevid2022distributional}, which does not use survey weights in estimation, as the sample size grows.
\item Study how finite-sample performance depends on the number of trees $B$, the sample size, and the number of PSUs per stratum $N_{\mathrm{psu},h}$.
\end{enumerate}
Section~\ref{sec:sim-datagen} specifies the generative model and sampling design, Section~\ref{sec:sim-compute} the estimators and evaluation metrics, and Section~\ref{sec:sim-results} reports the results.

\subsection{Data Generation}
\label{sec:sim-datagen}

We begin by defining the generative model in the super--population framework. For each unit \(i=1,\dots,N\), let
\[
Y_i=(Y_{i1},Y_{i2})^\top\in\mathbb{R}^2,\qquad
X_i=(X_{i1},\dots,X_{ip})^\top\in \mathbb{R}^p,\qquad
Z_i \in \mathbb{R},
\]
where \(Y_i\) is the outcome vector, \(X_i\) the predictor vector and \(Z_i\) a design variable that influences sampling but is not observed at the unit level in the analysis.

\underline{The finite population \(\{(Y_i,X_i,Z_i):i=1,\dots,N\}\) is generated as follows}: For each \(i\),
\[
Z_i \overset{\mathrm{iid}}{\sim} \mathcal{N}(0,1),\quad
X_{i1} \overset{\mathrm{iid}}{\sim} \text{Uniform}(0,1),
\]
\[
B_i \overset{\mathrm{iid}}{\sim} \Bernoulli\!\left(\tfrac12\right),\quad
X_{i2}=2B_i-1,
\]
\[
X_{ij} \overset{\mathrm{iid}}{\sim} \mathcal{N}(0,1),\quad j=3,\dots,p,
\]
\noindent where all variables in the construction above are mutually independent across both units \(i\) and indices \(j\), unless otherwise stated. We then define the conditional mean of \(Y_i\) given \((X_i,Z_i)\) by
\[
Y_i \;=\; \mu(X_i,Z_i) + \Sigma^{1/2}\,\varepsilon_i,
\qquad
\varepsilon_i \overset{\mathrm{iid}}{\sim} \mathcal{N}(\mathbf{0}, I_2), \text{ where }
\]
\[
\mu(X_i,Z_i)
=
\begin{pmatrix}
\mu_1(X_i,Z_i)\\[2pt]
\mu_2(X_i,Z_i)
\end{pmatrix},
\quad
\Sigma=\begin{pmatrix}1&0.3\\[2pt]0.3&1\end{pmatrix}, \text{ and }
\]
\[
\mu_1(X_i,Z_i)=1.5\,X_{i2} + (2+50Z_i)\,X_{i1},\qquad
\mu_2(X_i,Z_i)=1.5\,X_{i2} - Z_i\,(3X_{i2}+1).
\]
\noindent \(\{\varepsilon_i\}^{N}_{i=1}\) are generated independently of \(\{(X_i,Z_i)\}\). Equivalently,
$
Y_i \mid (X_i,Z_i) \sim \mathcal{N}_2\bigl(\mu(X_i,Z_i),\,\Sigma\bigr)
$ \text{ i.i.d. across } i.

\underline{We now describe the sampling design.} Let \(H\ge 2\) be the number of strata, and introduce a stratum indicator
\[
\mathrm{Strat}_i \in \{1,\dots,H\},\qquad i=1,\dots,N.
\]
Each unit is assigned independently to a stratum with equal probability:
\[
\bigl(\ind\{\mathrm{Strat}_i=1\},\ldots,\ind\{\mathrm{Strat}_i=H\}\bigr)
\sim \Multinomial\!\left(1;\tfrac1H,\ldots,\tfrac1H\right).
\]

For each stratum \(h \in \{1,\dots,H\}\), let
\[
U_h = \{i : \mathrm{Strat}_i = h\},
\qquad
N_h = |U_h|
\]
\noindent denote the set and size of units in stratum \(h\). Within stratum \(h\), 
define the membership map
$g_h: U_h \to \{1,\ldots,N_h^{\mathrm{PSU}}\}$ by ordering units in $U_h$ by $X_{i1}$
(and breaking ties arbitrarily) and grouping the ordered list into $N_h^{\mathrm{PSU}}$
consecutive bins of approximately equal size. Write $g_h(i)=j$ if unit $i$ is assigned to PSU $j$,
and define
\[
\mathrm{PSU}_{hj} := \{ i \in U_h : g_h(i)=j\}, \qquad j=1,\ldots,N_h^{\mathrm{PSU}}.
\]

We define measure-of-size as a function of $Z$: For each PSU, \(\mathrm{PSU}_{hj}\),
\[
M_{hj} \;=\; 
8 \ind(\bar Z_{hj} > 0) + 2 \ind(\bar Z_{hj} \le 0),
\quad
\text{where  }
\bar Z_{hj} \;=\; \frac{1}{N_{hj}} \sum_{i \in \mathrm{PSU}_{hj}} Z_i
\]
is the PSU-level mean of \(Z_i\), \(N_{hj} = |\mathrm{PSU}_{hj}|\) the number of subjects within that PSU. 

\underline{First-stage sampling} selects PSUs within each stratum \(h\) with probabilities proportional to size. Fix an expected number \(n_h^{(1)}\) of PSUs to be selected in stratum \(h\), and define the inclusion probabilities of the first-stage
\[
\pi^{(1)}_{hj} \;=\; \min\!\Bigl(1,\; n^{(1)}_{h}\,\frac{M_{hj}}{\sum_{j'=1}^{N_h^{\mathrm{PSU}}} M_{hj'}}\Bigr),
\qquad j=1,\dots,N_h^{\mathrm{PSU}}.
\]
We then select PSUs by unequal-probability systematic sampling,
\[
\{\mathrm{PSU}_{hj}: \xi^{(1)}_{hj}=1\} \sim \UPsystematic\!\bigl(\{\pi^{(1)}_{hj}\}_{j=1}^{N_h^{\mathrm{PSU}}}\bigr),
\]
\noindent where \(\xi^{(1)}_{hj} \in \{0,1\}\) is the indicator that PSU \(j\) in stratum \(h\) is selected at the first stage.

\underline{In the second stage}, conditional on \(\xi^{(1)}_{hj}=1\), we sample units within PSU \(\mathrm{PSU}_{hj}\) by SRSWOR. Specifically, we select
\[
n^{(2)}_{hj} \;=\; \lceil 0.3\,N_{hj} \rceil
\]
\noindent units without replacement in each selected PSU. Thus, for any unit \(i \in \mathrm{PSU}_{hj}\),
\[
\pi^{(2)}_{hji} \;=\; \Pr(\xi^{(2)}_{hji}=1 \mid \xi^{(1)}_{hj}=1) \;=\; \frac{n^{(2)}_{hj}}{N_{hj}} \;=\; 0.3,
\]
\noindent where \(\xi^{(2)}_{hji} \in \{0,1\}\) is the second-stage selection indicator.

Under this two-stage design, the overall unit inclusion indicator is
$
\xi_{hji} = \xi^{(1)}_{hj}\,\xi^{(2)}_{hji},
$
with unit inclusion probability
$
\pi_{hji}
=
\Pr(\xi_{hji}=1)
=
\pi^{(1)}_{hj}\,\pi^{(2)}_{hji},
$
and corresponding survey weights
$
w_{hji} = 1/\pi_{hji}.
$
This completes the generation of the analysis dataset $D = \{(X_i, Y_i, w_{hji})\}_{\{i : \xi_{hji} = 1\}}$.
The sampling scheme satisfies the conditional non-informativeness requirement (D1).

\subsection{Computation Details}
\label{sec:sim-compute}

We run SDRF and DRF \citep{cevid2022distributional} on the above data-generating process for each finite-population size $N\in\{4000,5000,15000,22500\}$ and number of trees $B\in\{10,30,70,200\}$, with $200$ seeds per scenario. We set $p = 3$ to align with the case study (Section~\ref{sec:casestudy}). As $N$ increases, we also increase the total number of strata $H$ and the number of PSUs per stratum $N_{\mathrm{psu},h}$, so the expected sample size per stratum, and hence the overall sample size, grows with $N$. The sampling \emph{mechanism} (PPS at stage~1 and SRSWOR at stage~2, with fixed rates) is held fixed across the values of $N$, consistent with increasing-population survey asymptotics.

We report two evaluation criteria: (a) the maximum mean discrepancy (MMD) between the estimated and population kernel mean embeddings of the conditional law on a hold-out set, and (b) the mean squared error (MSE) of $\widehat{\Psi}(\mathbf x_1)$ for the regression functional
\[
\Psi(P_{Y \mid X = \mathbf{x}}) \;=\; \mathbb{E}\!\left[\,Y_1 \mid X_1 = \mathbf{x}_1\,\right].
\]
Unlike the training data, which are the complex-sample draws produced by the design, the testing data are generated directly from the super-population model; thus no design information (weights or $Z$) is required to evaluate $\Psi(\mathbf{x}_1)$ on the test set. We focus on $\Psi(\mathbf{x}_1)$ to enable visual inspection of pointwise performance across the range of $X_1$.

For both estimators we use a Gaussian kernel on the outcome space, with bandwidth set to the median of pairwise Euclidean distances among the observed $Y$'s. The maximum tree depth is fixed at $8$, and the minimum node size to $\max\{20, \lceil \sqrt{n_s}\,\rceil\}$, where $n_s$ is the training-sample size. We cap node-level effective sampling weights by $\lambda_{\max} = 5.5 \times (\max_i w_i / \min_i w_i)$, where $w_i$ are the survey weights. Although further tuning could improve performance, we keep the settings minimal, as the untuned configurations already reveal the main trends across $N$ and $B$.

We do not compare against alternative \emph{survey-calibrated} distributional methods because, to our knowledge, none are currently available in the survey-sampling literature; the unweighted DRF serves as the natural i.i.d.\ baseline. All code is implemented in R, version 4.4.2 \citep{R_cite}.

\subsection{Results}
\label{sec:sim-results}
Table~\ref{tab:mmd_mean_sd} reports the mean (SD) of the MMD between the true finite-population conditional law $P_{S^N|\omega}(Y\mid X)$ and its SDRF estimate $\widehat{P}_{S^N|\omega}(Y \mid X)$.
The mean MMD decreases as the finite-population size $N$ increases, while increasing the number of trees $B$ primarily reduces the SD (stabilizing forest performance), with comparatively smaller improvements in the mean and diminishing returns beyond moderate $B$.
These finite-sample patterns are consistent with the convergence asserted in Theorem~\ref{thm:consistency}.

\begin{table}[H]
\centering
\caption{Mean (SD) of $100 \times d_{\mathbf{k}}(\widehat{P}_{S^N|\omega}(Y\mid X),\; P_{S^N|\omega}(Y\mid X))$ across finite population size $N$ (hence different average sample sizes $\bar{n}_s$) and the number of trees $B$.}
\label{tab:mmd_mean_sd}
\begin{tabular}{c c c c c c}
\toprule
$N$ & $\bar{n}_s$ & $B = 10$ & $B = 30$ & $B = 70$ & $B = 200$ \\
\midrule
4{,}000  & 320 & 19.60 (2.52) & 18.60 (2.27) & 18.30 (2.19) & 18.30 (2.10) \\
5{,}000  & 400 & 19.30 (2.49) & 18.40 (2.14) & 18.00 (2.14) & 18.10 (2.07) \\
15{,}000 & 1{,}200 & 17.30 (1.86) & 16.80 (1.44) & 16.70 (1.30) & 16.80 (1.23) \\
22{,}500 & 1{,}800 & 17.40 (1.52) & 16.80 (1.23) & 16.60 (1.16) & 16.70 (1.08) \\
\bottomrule
\end{tabular}
\end{table}

\begin{table}[H]
\centering
\caption{Root mean squared error (RMSE) of the conditional mean estimator $\widehat{\mathbb{E}}[Y_1\mid X_1]$ averaged over $X_1$ for SDRF and benchmark DRF as finite population size $N$ and the number of trees $B$ grow.}
\label{tab:rmse_comparison}
\begin{tabular}{ccc ccccc}
\toprule
\multirow{2}{*}{{$N$}} &
\multirow{2}{*}{{$N_{\text{PSU},h}$}} &
\multirow{2}{*}{{$\bar{n}_s$}} &
\multirow{2}{*}{{Algorithm}} &
\multicolumn{4}{c}{{Number of Trees $B$}} \\
\cmidrule(lr){5-8}
& & & & 10 & 30 & 70 & 200 \\
\midrule
\multirow{2}{*}{4{,}000} &
\multirow{2}{*}{80} &
\multirow{2}{*}{320} &
DRF  & 7.80 & 7.49 & 7.32 & 7.33 \\
& & & SDRF & \textbf{4.11} & \textbf{3.44} & \textbf{3.29} & \textbf{3.27} \\
\midrule
\multirow{2}{*}{5{,}000} &
\multirow{2}{*}{100} &
\multirow{2}{*}{400} &
DRF  & 8.04 & 7.55 & 7.40 & 7.33 \\
& & & SDRF & \textbf{3.69} & \textbf{3.16} & \textbf{2.95} & \textbf{2.96} \\
\midrule
\multirow{2}{*}{15{,}000} &
\multirow{2}{*}{200} &
\multirow{2}{*}{1{,}200} &
DRF  & 7.64 & 7.45 & 7.45 & 7.44 \\
& & & SDRF & \textbf{2.48} & \textbf{2.11} & \textbf{2.00} & \textbf{1.95} \\
\midrule
\multirow{2}{*}{22{,}500} &
\multirow{2}{*}{300} &
\multirow{2}{*}{1{,}800} &
DRF  & 7.52 & 7.37 & 7.33 & 7.32 \\
& & & SDRF & \textbf{2.16} & \textbf{1.83} & \textbf{1.74} & \textbf{1.74} \\
\bottomrule
\end{tabular}
\end{table}

\begin{figure}[h]
  \centering
  \includegraphics[width=0.8\linewidth]{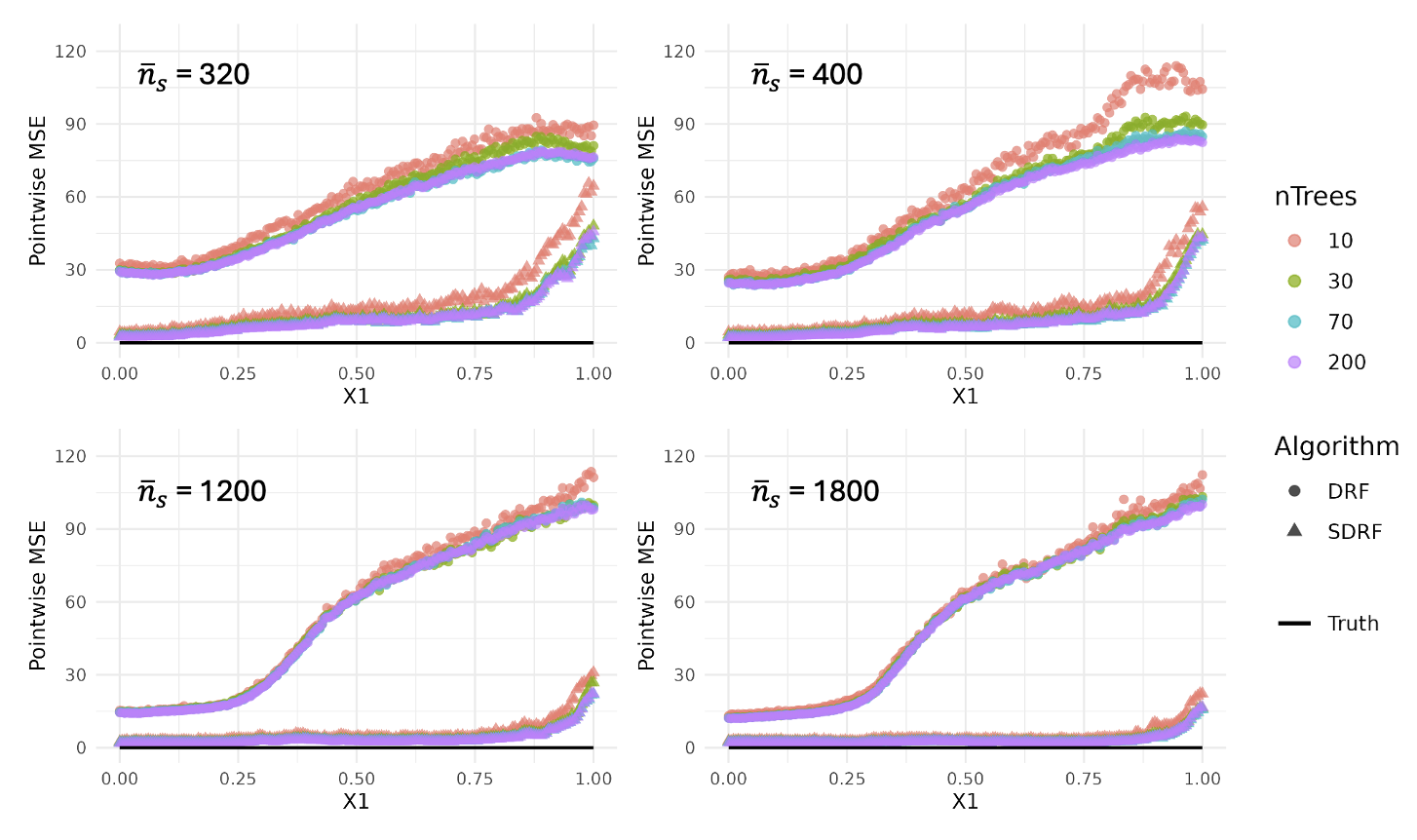}
  \caption{Pointwise mean square error (MSE) of the estimator $\widehat{\mathbb{E}}[Y_1\mid X_1 = \mathbf{x}_1]$ from SDRF and DRF. Each point represents an average over $200$ seeds.}
  \label{fig:mse}
\end{figure}

\begin{figure}[h]
  \centering
  \includegraphics[width=0.8\linewidth]{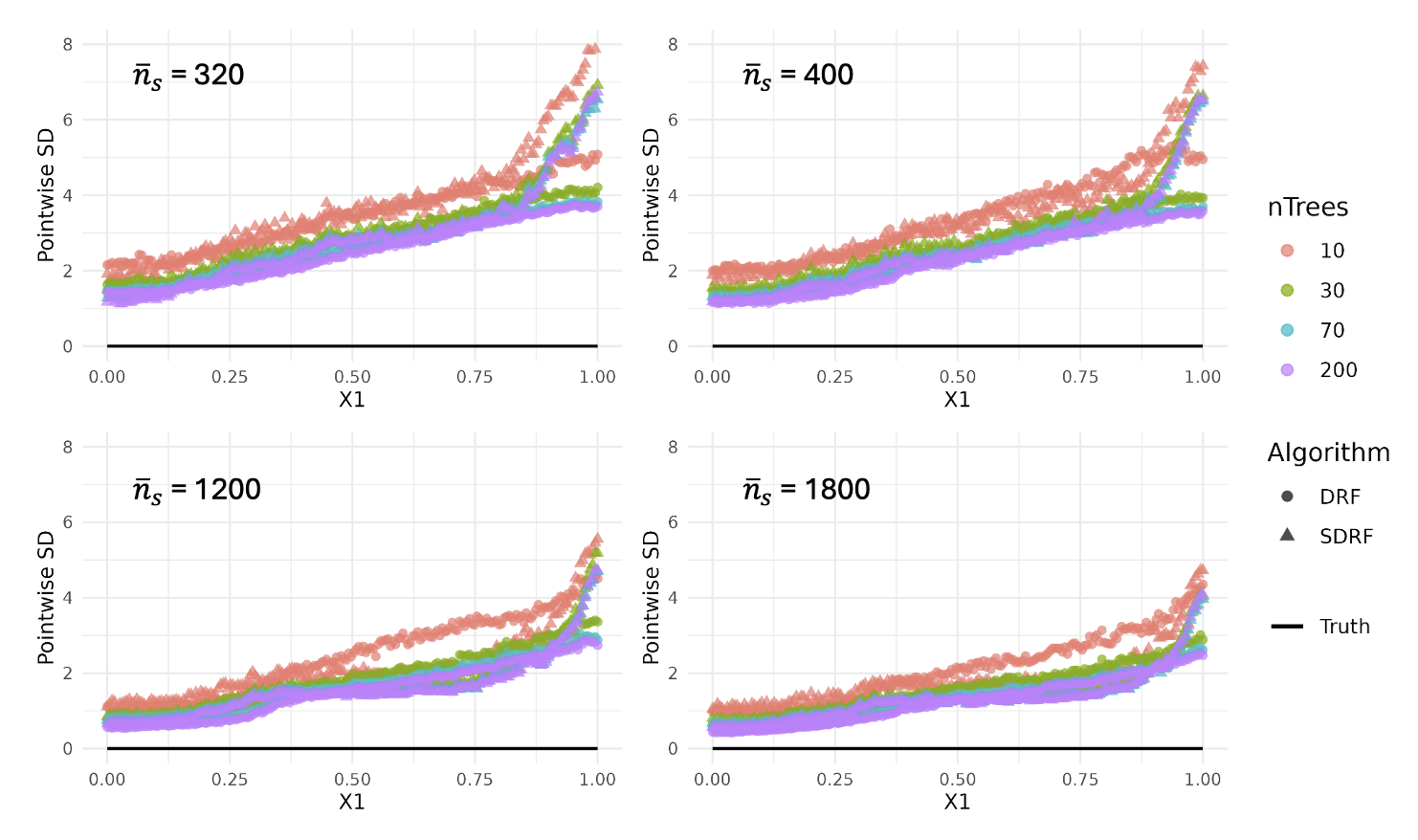}
   \caption{Pointwise standard error (SD) of the estimator $\widehat{\mathbb{E}}[Y_1\mid X_1 = \mathbf{x}_1]$ from SDRF and DRF. Each point represents an average over $200$ seeds.}
  \label{fig:sd}
\end{figure}

Table~\ref{tab:rmse_comparison} and Figures~\ref{fig:mse}--\ref{fig:sd} compare SDRF with the i.i.d.-based DRF for the regression functional $\Psi(\mathbf{x}_1) = \E[Y_1 \mid X_1 = \mathbf{x}_1]$. Averaged over a uniform grid of $200$ points in $[0,1]$, SDRF achieves smaller RMSE than DRF at every finite-population size $N$ and ensemble size $B$ considered (Table~\ref{tab:rmse_comparison}).

Figure~\ref{fig:mse} shows the corresponding pointwise MSE. For both methods MSE decreases with $B$, but DRF exhibits a clear error floor that persists as $N$ and $B$ increase, whereas SDRF's MSE decreases uniformly toward the truth, including in the upper range of $X_1$ where the sampling weights are more variable. Read together with Figure~\ref{fig:sd}, this indicates that the MSE gap is driven primarily by persistent bias in DRF, as anticipated in Section~\ref{sec:introduction}. Figure~\ref{fig:sd} further shows that pointwise variability decreases with $B$ for both methods and is comparable between SDRF and DRF; notably, SDRF shows no appreciable variance inflation despite the additional randomness introduced by its design-aware resampling step.

\section{Application: Diabetes Biomarkers in NHANES}
\label{sec:casestudy}

This section demonstrates the practical relevance of survey-calibrated distributional random forests (SDRF) for data analysis in nationally representative studies. We use data from the National Health and Nutrition Examination Survey (NHANES), which employs a multistage, stratified, cluster sampling design to represent the U.S.\ civilian, noninstitutionalized population: primary sampling units (PSUs) are selected at the county level, followed by smaller geographic areas and then individuals. All analyses incorporate the survey weights and design features provided by the Centers for Disease Control and Prevention (CDC).\footnote{\url{https://wwwn.cdc.gov/nchs/nhanes/Default.aspx}} We focus on the 2011--2012 NHANES cycle and apply an SDRF that incorporates the NHANES sampling mechanism during bagging and throughout tree construction (split selection and node estimation).

The objective is to characterize the conditional joint distribution of two glucose biomarkers used to diagnose and monitor diabetes---fasting plasma glucose (FPG) and glycated hemoglobin (HbA1c)---given demographic and anthropometric covariates: sex, age, and body mass index (BMI). Our primary clinical goal is to quantify how the population distribution of this bivariate outcome varies with these characteristics, supporting the detection of diabetes, the evaluation of glycemic control, and healthcare planning through the trends observed across population subgroups. A secondary objective is to assess whether the observed heterogeneity motivates individualized diabetes diagnostic thresholds for FPG and HbA1c, in contrast to the fixed cutoffs proposed by medical associations such as the American Diabetes Association (ADA) \citep{american20252}, which ignore individual characteristics. A total of $N=2732$ individuals, with and without diabetes, were included in the analysis.

We now articulate the analysis goal mathematically.
Let the clinical outcome be $Y=(Y_1,Y_2)^\top$, where $Y_1$ denotes FPG and $Y_2$ denotes HbA1c.
Let the covariate vector be $X=(X_1,X_2,X_3)^\top$, where $X_1=\mathrm{Age}$ and $X_2=\mathrm{BMI}$ are continuous and
$X_3=\mathrm{Sex}\in\{0,1\}$ is binary ($0=\text{male}$, $1=\text{female}$).

We run SDRF and estimate the conditional distribution of the bivariate outcome \(Y=(\mathrm{FPG},\mathrm{HbA1c})^{\top}\) given covariates \(X=\mathbf{x}\). Specifically, we compute the estimated conditional CDF
\[
\widehat F(y \mid \mathbf{x}) = \widehat{P}^{N}_{Y \le y \mid X=\mathbf{x}},
\qquad \text{for all } y\in\mathcal{Y},\ \mathbf{x}\in\mathcal{X},
\]
\noindent where the inequality \(Y \le y\) is understood component-wise. We evaluate \(\widehat F(\cdot \mid \mathbf{x})\) on a grid of covariate combinations,
\[
\mathrm{Age}\in\{30,50,70\},\quad \mathrm{BMI}\in\{20,30,40\},\quad \mathrm{Sex}\in\{0,1\}.
\]

Let $\widehat{\omega}_i(\mathbf{x})$ denote the SDRF distribution-regression weights for observation $i\in [N]$ at covariate value $\mathbf{x} \in \mathcal{X}$. Using these weights, we compute the estimated conditional mean and covariance,
\[
\widehat{\mu}(\mathbf{x})=\sum_{i=1}^N \widehat{\omega}_i(\mathbf{x})\,Y_i
\quad\text{and}\quad
\widehat{\Sigma}(\mathbf{x})=\sum_{i=1}^N \widehat{\omega}_i(\mathbf{x})\,\bigl(Y_i-\widehat{\mu}(\mathbf{x})\bigr)\bigl(Y_i-\widehat{\mu}(\mathbf{x})\bigr)^\top.
\]
We then define a weighted conditional Mahalanobis score for any pair $(\mathbf{x},y)\in \mathcal{X}\times \mathcal{Y}$ by
\[
S(\mathbf{x},y)=\bigl(y-\widehat{\mu}(\mathbf{x})\bigr)^\top \widehat{\Sigma}(\mathbf{x})^{-1}\bigl(y-\widehat{\mu}(\mathbf{x})\bigr).
\]
Let $\{(\mathbf{x}_a, y_a)\}_{a=1}^N$ denote the observed sample and let $\widehat{q}_{1-\alpha}$ be the (survey-weighted) empirical $(1-\alpha)$ quantile of $\{S(\mathbf{x}_a,y_a)\}_{a=1}^N$.
In the spirit of \citet{lado2021} for finite-dimensional Euclidean responses and \citet{lugosi2025conformal} for metric-space outcomes, we define the $(1-\alpha)$ tolerance region at covariate value $\mathbf{x} \in \mathcal{X}$ as
\[
\widehat{C}^{\,\alpha}(\mathbf{x})=\bigl\{\,y\in\mathcal{Y}:\; S(\mathbf{x},y)\le \widehat{q}_{1-\alpha}\,\bigr\}.
\]

A natural competitor in this setting is the method of \citet{lado2021}, which fits separate univariate local polynomial regressions for the conditional mean and uses residual-based estimates to approximate the conditional covariance and construct a tolerance region. Because it proceeds marginally through separate univariate fits, it is not designed for joint estimation; its smoothing parameters are moreover selected without an explicit global criterion, and it does not incorporate survey weights. SDRF instead requires a single model, avoids user-chosen smoothing parameters, and enables estimation under the complex NHANES design.

Figures~\ref{fig:example-age},~\ref{fig:example-bmi}, and~\ref{fig:example-sex} show the estimated tolerance regions for $1-\alpha \in \{0.10, 0.25, 0.50, 0.75,$ $ 0.90\}$ at the covariate combinations above; a larger $1-\alpha$ yields a wider region. Figure~\ref{fig:example-age} varies age, Figure~\ref{fig:example-bmi} varies BMI, and Figure~\ref{fig:example-sex} contrasts the sexes. FPG is plotted on a logarithmic scale, and the dashed red lines mark the ADA diagnostic thresholds, $\mathrm{FPG} \ge 126~\mathrm{mg/dL}$ and $\mathrm{HbA1c} \ge 6.5\%$.

We summarize the findings by covariate. The estimated joint distributions for women and men are similar in shape, but women exhibit more extreme observations and a higher prevalence of elevated FPG, especially in the wider regions. With increasing age, both FPG and HbA1c shift upward, indicating poorer glycemic control at older ages; the effect is most pronounced in the $1-\alpha = 0.9$ region. Higher BMI is moderately associated with higher glucose levels: individuals with $\mathrm{BMI}>30~\mathrm{kg/m}^2$ show higher values than those with $\mathrm{BMI}\in[20,30]~\mathrm{kg/m}^2$, though the regions change little between $30$ and $40~\mathrm{kg/m}^2$. Overall, glucose control differs between overweight or obese individuals and those in the normal BMI range.

Two implications follow. First, glycemic control in the U.S.\ population varies substantially with demographic and anthropometric characteristics, with consequences for public-health planning and potential policy interventions. Second, diagnostic cutoffs for diabetes based on FPG and HbA1c may warrant adjustment for at least age, and possibly BMI, rather than applying uniform rules to the entire population, in line with \citet{Berkovic2025}. From a modeling perspective, SDRF provides a principled framework to define reference values for specific subpopulations, reclassify individuals by biomarker profile, and forecast public-health risk using the full conditional distribution. Approaches that model only the conditional mean miss clinically important features such as quantiles, dispersion, and tail behavior, yielding less information for decision-making.

\begin{figure}[ht]
\centering
\begin{subfigure}{\linewidth}
\centering
\includegraphics[width=0.7\linewidth]{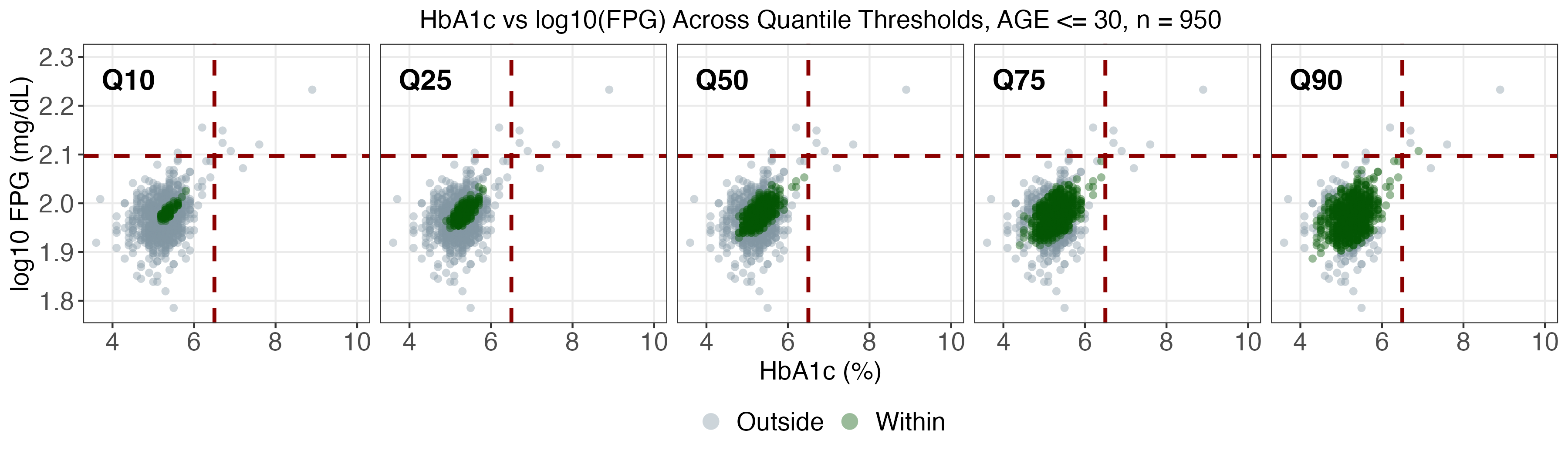}
\includegraphics[width=0.7\linewidth]{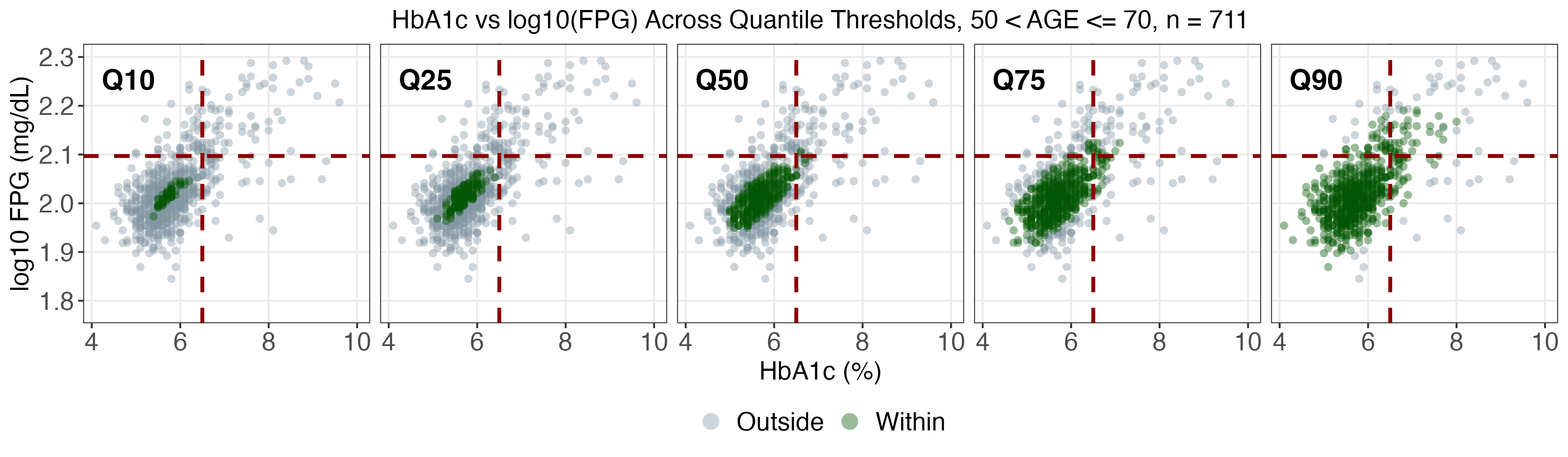}
\caption{Stratified by age.}
\label{fig:example-age}
\end{subfigure}
\caption{SDRF-estimated covariate-conditional tolerance regions for $\log_{10}(\mathrm{FPG})$ versus $\mathrm{HbA1c}$ at coverage levels $1-\alpha \in \{0.10, 0.25, 0.50, 0.75, 0.90\}$; a larger $1-\alpha$ yields a wider region. Green and gray points indicate observations inside and outside the estimated region, respectively; dashed red lines mark the ADA diagnostic thresholds ($\mathrm{FPG}\ge 126$~mg/dL, $\mathrm{HbA1c}\ge 6.5\%$). Panels~(a),~(b),~(c) stratify by age, BMI, and sex, respectively.}
\label{fig:example}
\end{figure}

\begin{figure}[ht]\ContinuedFloat
\centering
\begin{subfigure}{\linewidth}
\centering
\includegraphics[width=0.7\linewidth]{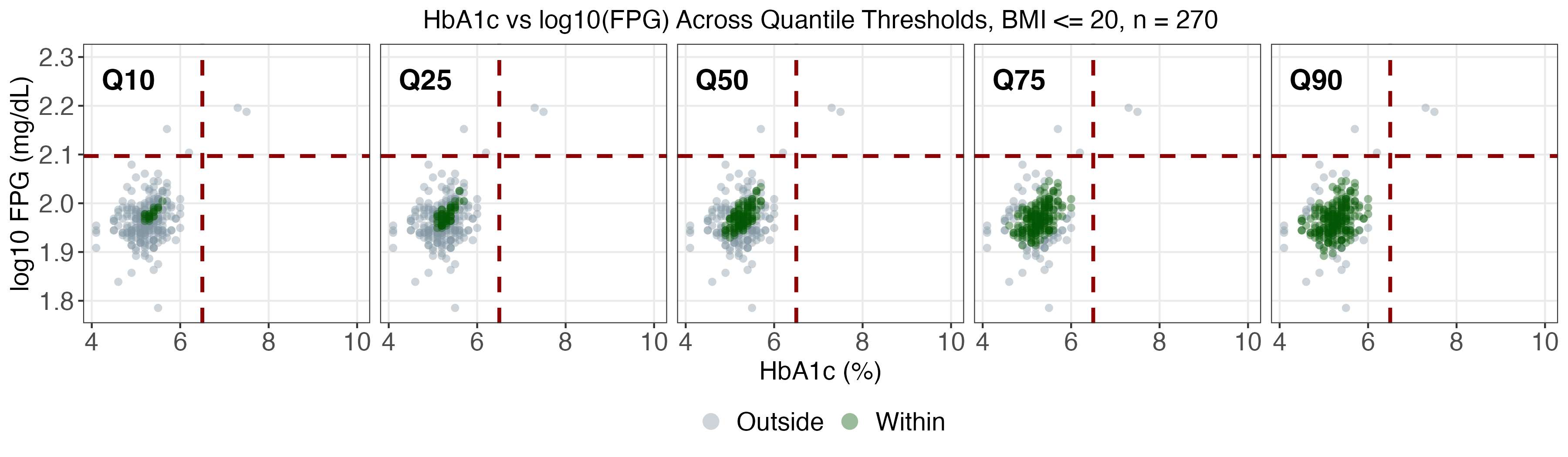}
\includegraphics[width=0.7\linewidth]{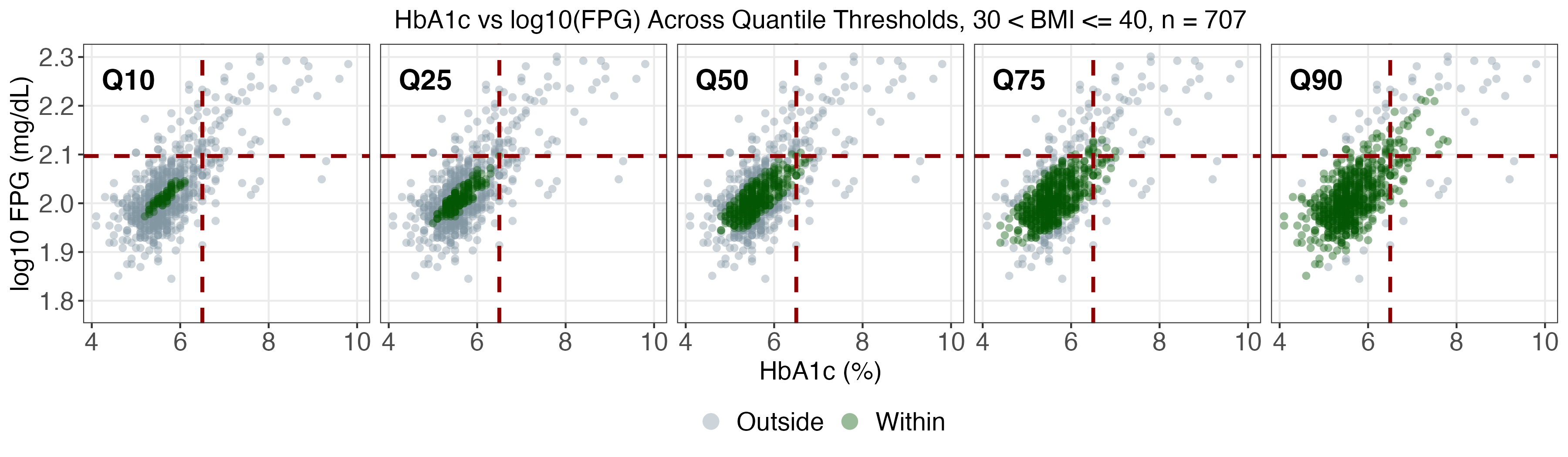}
\includegraphics[width=0.7\linewidth]{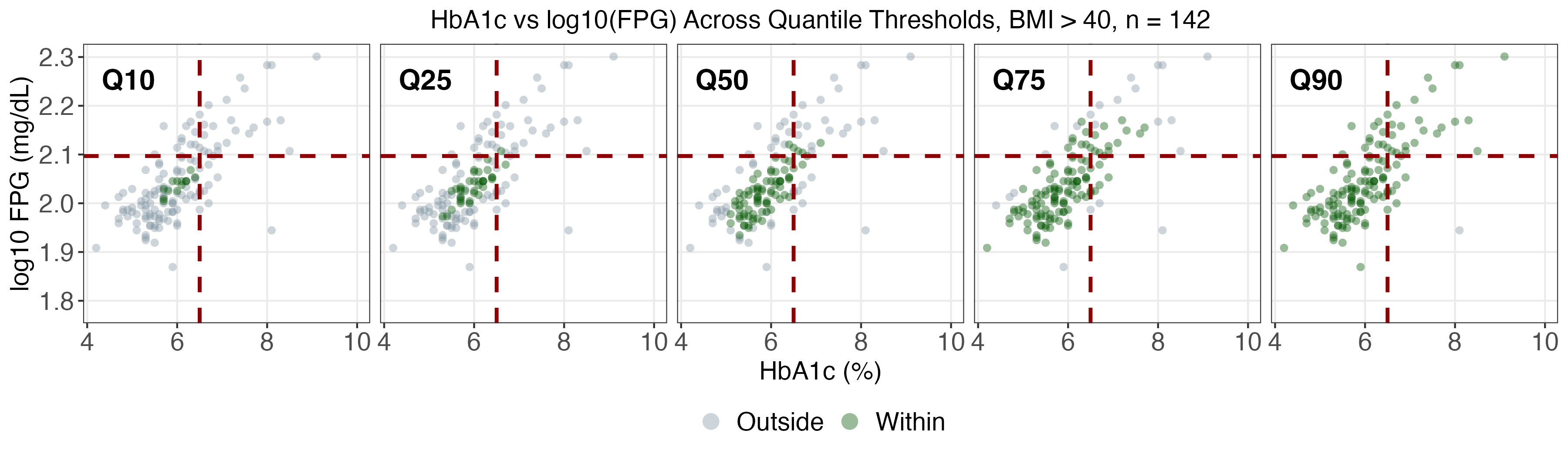}
\caption{Stratified by BMI.}
\label{fig:example-bmi}
\end{subfigure}
\end{figure}

\begin{figure}[ht]\ContinuedFloat
\centering
\begin{subfigure}{\linewidth}
\centering
\includegraphics[width=0.7\linewidth]{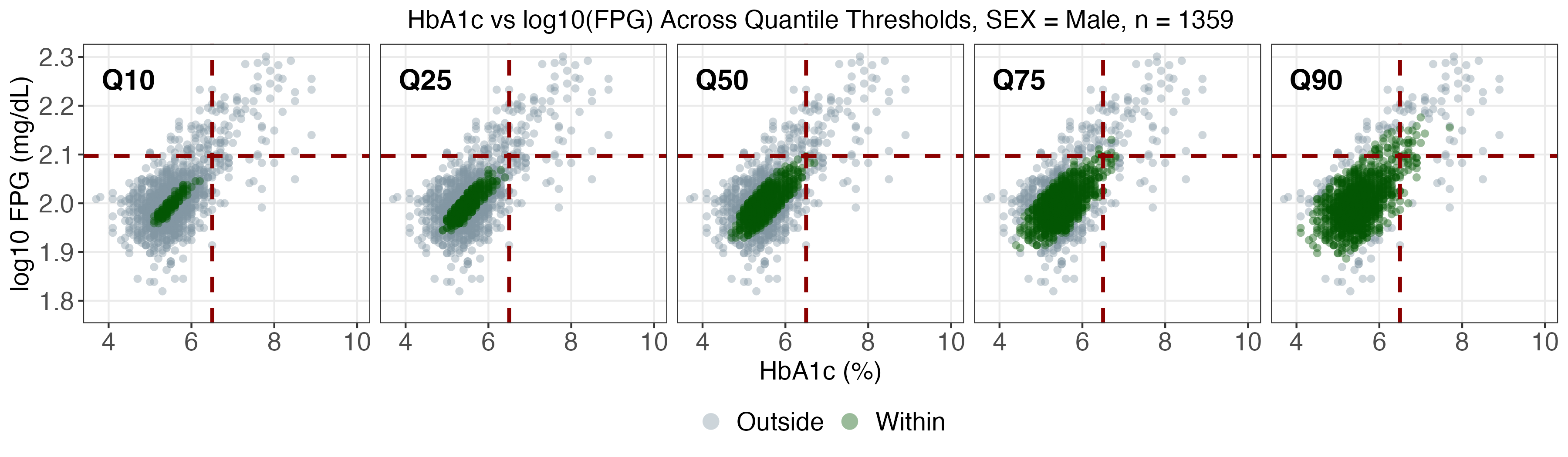}
\includegraphics[width=0.7\linewidth]{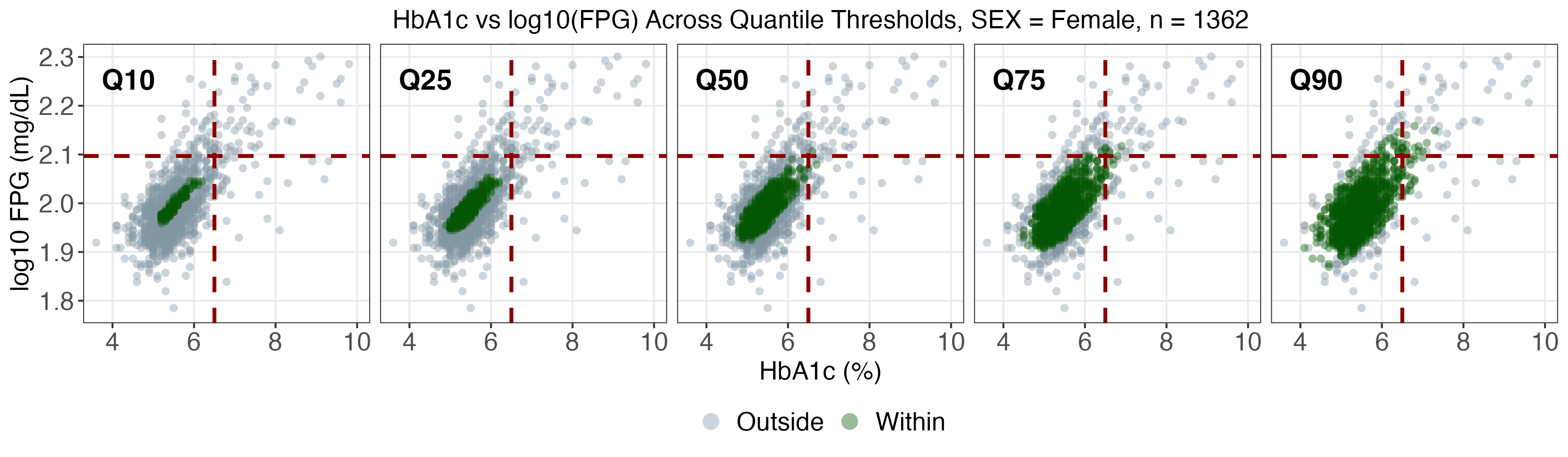}
\caption{Stratified by sex.}
\label{fig:example-sex}
\end{subfigure}
\end{figure}

\section{Discussion}
\label{sec:dis}

We have shown how the survey-statistical principle can be embedded inside a modern machine learning algorithm without sacrificing the tractable theory of the former or the function-class flexibility of the latter. The key structural observation is that kernel mean embeddings act on probability measures directly, so the super-population conditional law, its finite-population analog, and any H\'ajek-type design-weighted estimate all live in the same Hilbert space, allowing a single algorithmic and theoretical scheme to cover a wide variety of designs and achieve both design and joint consistency under a unified set of assumptions.

Like other MMD-based methods, SDRF's discriminative power degrades as the dimension of $Y$ grows. We also defer analysis of convergence rates. The standard route to forest rates \citep{Wager2015-nc} requires each leaf to shrink in diameter which is controlled by i.i.d. concentration arguments whose analogs under complex survey designs require non-trivial work.

Three future directions are natural. First, SDRF's H\'ajek-type leaf embeddings are ratio estimators, a design-weighted numerator divided by a design-weighted estimate of the subpopulation size. Thus, they inherit the standard finite-sample bias of such estimators \citep{cochran1977sampling, schochet2022design}. Jackknife and Taylor-linearization corrections are well-developed for individual ratio estimators \citep{tin1965comparison}, but whether per-leaf bias compounds, averages, or partially cancels across forest aggregation, and at what computational cost a forest-level correction can be implemented, remain open.

Second, SDRF's plug-in estimators support a wider range of population targets than the conditional means targeted by classical design-aware regression, and can serve as the base estimator for design-aware uncertainty wrappers such as conformal prediction \citep{wieczorek2023designbased}. The method does not, however, deliver inferential guarantees in the sense of confidence intervals for $\widehat{{P}}_{Y \mid X = \mathbf{x}}$ or its continuous maps. For i.i.d. random forests, the infinitesimal jackknife \citep{wager2014confidence} yields valid confidence intervals via the Hoeffding decomposition of the underlying U-statistic. Whether an analogous construction is available under complex sampling, where the i.i.d.\ Hoeffding decomposition no longer applies, is an open question.

Third, the principle of embedding design weights into the learning objective using kernel mean embeddings extends beyond forests. Generative alternatives based on design-weighted score matching, conditional diffusion, or normalizing flows are natural at larger scale, though their consistency under informative sampling remains open and presents structural obstacles distinct from those addressed here.

\textit{Data and Code Availability.} The NHANES data analyzed in Section~\ref{sec:casestudy} are publicly available from the CDC at \url{https://wwwn.cdc.gov/nchs/nhanes/Default.aspx}. R code implementing SDRF and reproducing all results is available at \href{https://github.com/yatingz205/SDRF}{\texttt{Github}}.


\acks{The last author was partly supported by the UNC Gillings Center for AI and Public Health. No other third-party funding or support was received for this work. None of the authors had financial relationships with entities that could be perceived to influence the submitted work.}


\appendix

\vskip 0.2in
\bibliography{reference}

\end{document}


\newpage
\setcounter{section}{0}
\setcounter{theorem}{0}
\setcounter{equation}{0}
\setcounter{table}{0}
\setcounter{figure}{0}
\renewcommand{\thesection}{S\arabic{section}}
\renewcommand{\thetheorem}{S\arabic{theorem}}
\renewcommand{\theequation}{S\arabic{equation}}
\renewcommand{\thetable}{S\arabic{table}}
\renewcommand{\thefigure}{S\arabic{figure}}

\section{Notation Table}
\label{sec:notation_table}
\renewcommand{\arraystretch}{0.85}
{\small
\begin{longtable}{%
  >{\raggedright\arraybackslash}p{0.35\textwidth}%
  >{\raggedright\arraybackslash}p{0.65\textwidth}}
\caption{Summary of notation}\label{tab:notation}\\
\toprule
\textbf{Symbol} & \textbf{Description} \\
\midrule
\endfirsthead

\toprule
\textbf{Symbol} & \textbf{Description} \\
\midrule
\endhead

\midrule
\endfoot

\bottomrule
\endlastfoot

\multicolumn{2}{l}{\textbf{Basic variables}}\\
$Z \in \mathcal{Z} \subseteq \mathbb{R}^q$       & Design variables (used by the sampling design) \\ 
$X \in \mathcal{X} \subseteq \mathbb{R}^p$       & Covariates / features in Euclidean space \\ 
$Y \in \mathcal{Y}$        & Outcome of interest on $\mathcal{Y}$ \\ 
$D^N = \{(X_i, Y_i, Z_i)\}_{i\in[N]}$    & Finite population \\[4pt]

\multicolumn{2}{l}{\textbf{Probability spaces, measures, and targets}}\\
$(\Omega,\mathcal{A},P)$ 
  & Super-population probability space for $(Z, X, Y)$ \\
$(\mathcal{S}^N,\sigma(\mathcal{S}^N),P_{\mathcal{S}^N\mid\omega})$ 
  & Design probability space\footnotemark \\
$(\Omega\times \mathcal{S}^N,\mathcal{A}\otimes\sigma(\mathcal{S}^N),\,P_{\Omega\times \mathcal{S}^N})$ 
  & Joint probability space \\
$P(Y\mid X=\mathbf{x})$ or $P_{Y\mid X=\mathbf{x}}$
  & Super-population conditional law of $Y\mid X=\mathbf{x}$ \\
$P_{\mathcal{S}^N\mid\omega}(Y\mid X=\mathbf{x})$ or $P^N_{Y\mid X=\mathbf{x}}$
  & Finite-population conditional law of $Y\mid X=\mathbf{x}$ \\
$\Psi(\mathbf{x})$           
  & Borel-measurable map into a metric space $(\mathcal{V},\, d_v)$, continuous at the relevant limits, for example  $\Psi(P_{Y\mid X=\mathbf{x}})$ \\
$\widehat{\Psi}_N(\mathbf{x})$  
  & SDRF plug-in estimator at covariate $\mathbf{x}$; $\widehat{\Psi}_N(\mathbf{x}) := \Psi(\widehat{P}^N_{Y\mid X=\mathbf{x}})$ \\[4pt]

\multicolumn{2}{l}{\textbf{Design quantities}}\\
$N \in \mathbb{Z}_{> 0}$      & Number of subjects in the finite population \\
$\mathbf{p}$                 & Sampling design transition function generating $\{\xi_i\}$ \\ 
$\xi_i \in \{0,1\}$          & Selection (inclusion) indicator for unit $i$ \\ 
$n_s = \sum_{i \in [N]} \xi_i$ & Realized (random) sample size \\ 
$\pi_i = \mathbb{E}[\xi_i \mid Z^N]$    & First–order inclusion probability of unit $i$ \\ 
$\pi_{ij} = \mathbb{E}[\xi_i \xi_j \mid Z^N]$ & Second–order joint inclusion probability \\  
$h \in [H]$                & Stratum index \\
$k \in [K_h]$              & Primary sampling unit (PSU) index within stratum $h$ \\
$j \in [M_{hk}]$           & Secondary sampling unit (SSU) index within PSU $(h,k)$ \\
$u = (h,k,j)$              & SSU index; sampled-unit index under two-stage designs \\
$K_h$                      & Number of PSUs in stratum $h$ \\
$M_{hk}$                   & Number of SSUs in PSU $(h,k)$ \\
$M_{\max}$                 & Uniform upper bound on cluster (PSU) size, $\max_{h,k} M_{hk} \le M_{\max}$ \\
$\pi^{(1)}_{hk}$           & First-stage (PSU-level) inclusion probability \\
$\pi^{(2)}_{hkj\mid hk}$   & Second-stage (SSU-level) conditional inclusion probability given PSU sampled \\
$\pi_u = \pi^{(1)}_{hk}\pi^{(2)}_{hkj\mid hk}$ & Overall SSU inclusion probability \\
$n_{\mathrm{eff}}$         & Effective sample size, $(\sum_i \pi_i^{-1})^2/\sum_i \pi_i^{-2}$ \\[4pt]

\multicolumn{2}{l}{\textbf{Design-aware bagging and honesty}}\\
$n_i^{*}$                  & Resampled multiplier (count) for unit $i$ used to start a tree \\ 
$G^*_{hk}, W^*_{hkj}$      & Stage-1 and stage-2 bootstrap multipliers in the two-stage pseudo-population bootstrap \\
$D_b^{*} = \{(X, Y, n_b^{*}/\pi)\}_{i=1}^{n_s}$ 
                           & Design-aware resampled dataset for tree $b$ \\
$\hat{w}_i = n_{b,i}^{*} / \pi_i$ 
                           & Tree-specific weight after resampling \\ 
$\omega_i(\mathbf x)$, $\widehat{\omega}_i(\mathbf x)$ 
                           & Final forest prediction weights at query $\mathbf x$ (averaged over trees) \\ 
$B$                        & Number of trees in the forest \\
$M$                        & Number of re-sample weights averaged to stabilize the split score (distinct from the split score $M^N(\theta)$ and the density bound $M$ of (A2), disambiguated by context) \\
$M^N(\theta),\ M_{n_s}(\theta),\ M^{*(M)}_{n_s}(\theta)$
                           & Finite-population, design-weighted sample, and $M$-averaged resampled split scores at candidate $\theta$ \\
$D_{\text{est}}, D_{\text{split}}$ 
                           & Estimation vs. splitting part of the sample (honesty) \\[4pt]

\multicolumn{2}{l}{\textbf{Partitions, leaves, and sets}}\\
$L(\cdot)$                 & (Oracle) partition map or leaf function on $\mathcal{X}$ if $D^N$ is used\\
$L_\xi(\cdot)$             & Partition map obtained from the \emph{whole sample} induced by $\xi$ \\ 
$L^{*}_{\xi}(\cdot)$ or $L^*(\cdot)$       & Partition map obtained from design-aware \emph{resampled} data \\ 
$\operatorname{diam}(L(\mathbf x))$& Diameter of a leaf $L(\mathbf x)$ under the Euclidean distance \\ 
$C_L(\theta), C_R(\theta)$ & Left/right child region induced by a split $\theta=(j,t)$ \\ 
$P_a \subseteq \mathcal{X}$   & Region on $\mathcal{X}$ (parent node) being split \\ 
$N_L, N_R, N_{P_a}$         & (Finite-population) sizes of left, right, and parent regions \\ 
$\widehat{N}_L, \widehat{N}_R, \widehat{N}_{P_a}$ 
                           & Sample / resample analogues used in split score \\ 
$k_N \to 0$                & Target (upper) diameter of oracle leaves (shrinking with $N$) \\ 
$r_N = c_1 k_N$            & Radius defining an inner ball inside $L(x)$ \\ 
$\overline{B}_{r_N}(\mathbf x)$
                           & Closed balls in $\mathcal{X}$ centered at $\mathbf x$ \\
$n_{\mathrm{eff},\ell}$
                           & Design-weighted effective sample size in child $\ell$, $\displaystyle n_{\mathrm{eff},\ell}=\frac{(\sum_{i\in S\cap C_\ell}\pi_i^{-1})^2}{\sum_{i\in S\cap C_\ell}\pi_i^{-2}}$ \parencite{Kish1995-aq} \\[4pt]

\multicolumn{2}{l}{\textbf{Kernel, RKHS, and MMD}}\\
$\mathbf{k}(\cdot,\cdot)$  
                           & Reproducing kernel, assumed characteristic/universal and bounded \\ 
$\mathcal{H}$              & Reproducing kernel Hilbert space (induced by $\mathbf k$) \\ 
$\mu_{\mathbf k}(P)$                 & Kernel mean embedding (KME) of some finite signed Radon measures $P$, an element in $\mathcal{H}$ \\ 
$K_{\max}$
                           & Uniform kernel bound $\sup_{y,y'\in\mathcal{Y}}|\mathbf{k}(y,y')|<\infty$ (finite by (K1)) \\
$d_{\mathbf{k}}(\cdot,\cdot)$ 
                           & MMD distance between embedded distributions, induced by $\mathbf k$ \\[4pt]

\multicolumn{2}{l}{\textbf{Convergence and order notation}}\\
$O_{p_d}(\cdot),\, o_{p_d}(\cdot)$ & Usual stochastic order w.r.t $P_{\mathcal{S}^N | \omega}$ \\ 
$X_n = \Theta_{p_d}(a_n)$      & $X_n=O_{p_d}(a_n)$ and $a_n=O_{p_d}(|X_n|)$ w.r.t $P_{\mathcal{S}^N | \omega}$ \\ 
$a_n \lesssim b_n$         & There exists $C>0$ with $a_n \le C b_n$ (equivalently, $a_n = O(b_n)$) \\ 
$X_n \xrightarrow{p_\omega} X$ & Convergence in super-population probability space $(\Omega, \mathcal A, P)$\\
$X_n \xrightarrow{p_d} X$  & Convergence in design probability space $(\mathcal{S}^N,\sigma(\mathcal{S}^N),P_{\mathcal{S}^N | \omega})$ \\ &  (Design consistency) \\
$X_n \xrightarrow{p_{\otimes}} X$    & Convergence in the joint probability space $(\Omega\times \mathcal{S}^N ,\mathcal{A}\otimes\sigma(\mathcal{S}^N ),\,P_{\Omega\times \mathcal{S}^N})$ \\ 
& (Joint consistency) \\[4pt]

\multicolumn{2}{l}{\textbf{Hyperparameters for SDRF}}\\
$q \in (0,1)$              & Base probability to send a PSU to the “split” part \\ 
$mtry \in \{1, \dots, p\}$                     & Number of candidate features considered at each split \\ 
$\lambda_{\max} := \max_{i \in L}\hat{w}_i / \min_{i \in L}\hat{w}_i$               & The maximum within-leaf max-to-min weight ratio \\
$n_{L,\min} \geq 1$               & Minimum number of (un-weighted) observations per leaf \\[4pt]
\end{longtable}
\footnotetext{Our $\mathcal S^N$ corresponds to the sample space $S$ of \textcite{2005_RubinBleuer}. The superscript $N$ emphasizes that the collection of possible samples is indexed by, and grows with, the population size $N$ along the asymptotic sequence.}
}
\renewcommand{\arraystretch}{1.0}

We use $O_p$ notation in the usual sense and $X_n = \Theta_p(a_n)$ if and only if 
$X_n = O_p(a_n)$ and $a_n = O_p(|X_n|)$.
We use \( a_n \lesssim b_n \) to denote that there exists \( C>0 \) such that 
\( a_n \leq C b_n \). In other words, $a_n \lesssim b_n \iff a_n = O(b_n)$; 
and similarly $a_n \lesssim_p b_n \iff a_n = O_p(b_n)$. 
We denote the index set $[a] := \{1, \dots, a\}$ for any positive integer $a \in \mathbb{Z}_{>0}$. Throughout, `$:=$' will be used to emphasize definition.\\

We use $X_n \xrightarrow{p_d} X$ to denote convergence in probability with respect to all sources of randomness conditional on the fixed finite population of size $N$. Specifically, $p_d$ encompasses randomness from the sampling design (indicators $\{\xi_i\}$), bootstrap resampling (multipliers $\{n^*_{b,i}\}$), and algorithmic randomness (PSU-level splitting indicators, random feature selection, and tree construction). We use $X_n \xrightarrow{p_\otimes} X$ to denote convergence in probability with respect to the joint measure $P_{\Omega \times \mathcal{S}^N}$ over the product space $(\Omega \times \mathcal{S}^N, \mathcal{A} \otimes \sigma(\mathcal{S}^N))$. This joint measure $p_\otimes$ includes all sources of randomness: super-population randomness (generating the finite population $\{(X_i, Y_i, Z_i)\}_{i=1}^N$ from $P_\Omega$), sampling design, bootstrap resampling, and algorithmic randomness. The ``$*$'' superscript emphasizes dependency on resampled weights ${n}^*_{b,i}$. We write $n_s$ or $n_{s,N}$ interchangeably for the (possibly random) sample size, with the subscript $N$ emphasizing dependency on the population size.\\

Under (stratified) two--stage cluster designs, let $h\in[H]$ index strata. In stratum $h$, there are $K_h$ primary sampling units (PSUs) indexed by $k\in[K_h]$, and PSU $(h,k)$ contains $M_{hk}$ secondary sampling units (SSUs) indexed by $j\in[M_{hk}]$. We therefore index an SSU by the triple $u=(h,k,j)$. Let $\xi^{(1)}_{hk}\in\{0,1\}$ denote the stage--1 inclusion indicator for PSU $(h,k)$, and, conditional on $\xi^{(1)}_{hk}=1$, let $\xi^{(2)}_{hkj}\in\{0,1\}$ denote the stage--2 inclusion indicator for SSU $(h,k,j)$ within PSU $(h,k)$. The overall SSU inclusion indicator is $\xi_u:=\xi^{(1)}_{hk}\xi^{(2)}_{hkj}$. Define the corresponding inclusion probabilities $\pi^{(1)}_{hk}:=\Pr(\xi^{(1)}_{hk}=1)$, $\pi^{(2)}_{hkj\mid hk}:=\Pr(\xi^{(2)}_{hkj}=1\mid \xi^{(1)}_{hk}=1)$, and $\pi_u:=\Pr(\xi_u=1)=\pi^{(1)}_{hk}\pi^{(2)}_{hkj\mid hk}$. For distinct SSUs $u\neq v$, write the joint inclusion probability as $\pi_{uv}:=\Pr(\xi_u=\xi_v=1)$. Finally, let $k(u):=(h,k)$ denote the PSU containing SSU $u$, so that $k(u)=k(v)$ iff $u$ and $v$ belong to the same PSU.

Under stratified two-stage cluster designs, the effective index driving asymptotic rates is the (deterministic) expected number of sampled primary sampling units,
\[
n_{\mathrm{eff}} \;:=\; \sum_{h\in[H]}\sum_{k\in[K_h]} \pi^{(1)}_{hk}
\;=\; \mathbb{E}_{P_{\mathcal{S}^N\mid\omega}}\!\left[\,\sum_{h\in[H]}\sum_{k\in[K_h]} \xi^{(1)}_{hk}\right],
\qquad
n_{\mathrm{eff}} \to \infty,
\]
where $H$ is fixed or satisfies $H = o(n_{\mathrm{eff}})$, and secondary sampling unit (SSU) counts per PSU are uniformly bounded from above:
\[
\sup_{h\in[H],\,k\in[K_h]} M_{hk} \;\le\; M_{\max} \;<\; \infty.
\]
The lower bound $M_{hk}\ge 1$ is automatic from the integer definition of PSU size (every PSU contains at least one SSU). The upper bound $M_{\max}<\infty$ is a regularity condition preventing a small number of oversized PSUs from dominating the design variance; it is the natural regime whenever PSUs are physical or institutional units of bounded capacity, such as households in a block, students in a school, patients per clinic, residents per census tract.

\section{Motivating Example}
\label{apsec:motivating_example}
When learning a conditional distribution, using design-aware algorithms is not only about potential efficiency gains; it is essential in certain settings.
Consider the following scenario. Denote by $Z \in \mathcal{Z}$ a design variable, $\xi \in \{0,1\}$ the inclusion indicator, and suppose that we are interested in $P(Y\mid X)$.
Design variables may be available only on the sampling frame and used to construct strata, define measures of size (MOS) for PPS sampling, or model nonresponse.
Examples include fine geographic indicators such as census blocks, facility size measures such as the number of providers, or other administrative flags.
Although $X$ and $Z$ may overlap, $Z$ often contains frame-only or confidential information and therefore must be removed from the released sampled data
$D=\{(X_i,Y_i,w_i):  i=1,\dots,N; \xi_i=1\}$ used for analysis.
Problem arise when $Z$ indexes subgroups with heterogeneous conditional outcome distributions and the design differentially samples these subgroups. Formally, when
\begin{align}
\begin{split}
   Y \not\!\perp\!\!\!\perp Z \mid X
   \quad
   \text{and}
    \quad
    \xi \not\!\perp\!\!\!\perp Z \mid X.
\end{split}
\end{align}
Even if the design is conditionally noninformative given $(X,Z)$, i.e., $Y \perp\!\!\!\perp \xi \mid X,Z$, we have
\[
P(Y \mid X , \xi = 1)
=
\int P(Y \mid X, Z = z)\, P(dz \mid X, \xi = 1)
\neq
P(Y \mid X), \text{ since }
\]
\[
P(dz \mid X, \xi = 1)
=
\frac{P(\xi = 1 \mid X, Z = z)}{\int_{\mathcal{Z}} P(\xi = 1 \mid X, Z = z)\,P(dz \mid X)}\, P(dz \mid X)
\neq P(dz \mid X).
\]
Thus, algorithms designed for i.i.d.\ data that implicitly learn $P(Y\mid X,\, \xi=1)$ lead to systematic bias relative to $P(Y\mid X)$, which does not vanish with increasing sample size or model complexity. This is the case even for non-parametric algorithms that free us from modeling error. Consistency for $P(Y\mid X)$ is impossible without incorporating the design. This phenomenon is showcased in Section~5 of the manuscript.

\section{Survey Design}
This section contains proofs for claims made in the Survey Design subsection of the manuscript.

\begin{lemma}
\label{lem:designs-D1-D4}
Each of the following first--stage designs are examples that satisfy assumptions \textup{(D2)}--\textup{(D4)}:

\begin{enumerate}[label=\textup{(\alph*)}, leftmargin=2.8em, itemsep=2pt, topsep=2pt]
\item \textbf{Poisson sampling.}
Conditionally independent indicators $\{\xi_i\}_{i=1}^N$ given $(X^N,Z^N)$ with
$\xi_i\sim\mathrm{Bern}(\pi_i)$, where the inclusion probabilities satisfy
\[
0<\pi_{\min}\le \pi_i \le \pi_{\max}<1\qquad\text{for all }i,N,
\]
and the expected sample size $n=\sum_{i=1}^N \pi_i$ satisfies $n/N\to f\in(0,1)$.

\item \textbf{SRSWOR (fixed size).}
Simple random sampling without replacement of a fixed size $n_N$ from $[N]$, where
$n_N/N\to f\in(0,1)$ as $N\to\infty$.

\item \textbf{PPS with replacement (distinct--unit sample).}
Let MOS values $\{m_i\}_{i=1}^N$ satisfy $0<m_{\min}\le m_i\le m_{\max}<\infty$ for all $i,N$.
Define per--draw probabilities $p_i:=m_i/\sum_{r=1}^N m_r$. Draw $m_N$ times with replacement,
independently across draws, with $\Pr(I_t=i)=p_i$, and define distinct--unit membership indicators
\[
\xi_i:=\ind\{i\in\{I_1,\dots,I_{m_N}\}\},\qquad n_{s,N}:=\sum_{i=1}^N \xi_i.
\]
Assume $m_N/N\to f_0\in(0,1)$ and that the expected number of distinct sampled units
$n:=\E(n_{s,N})=\sum_{i=1}^N \pi_i$ satisfies $n/N\to f\in(0,1)$.
\end{enumerate}
\end{lemma}

\begin{proof}[Proof of Lemma~\ref{lem:designs-D1-D4}]
We verify \textup{(D2)}--\textup{(D4)} case by case.

\medskip
\paragraph{Poisson sampling}
Let $n_{s,N}=\sum_{i=1}^N \xi_i$ with independent $\xi_i\sim\mathrm{Bern}(\pi_i)$ and
$n=\E(n_{s,N})=\sum_{i=1}^N \pi_i$. Since $n/N\to f\in(0,1)$, fix any $\delta\in(0,f)$ and set
$\eta:=1-\delta/f\in(0,1)$. For all large $N$, $n\ge (f/2)N$ and $\delta N \le (1-\eta/2)n$.
A standard Chernoff bound for Poisson--binomial sums yields
\[
\Pr\!\bigl(n_{s,N}\le (1-\eta/2)n\bigr)\le \exp\!\left(-\frac{\eta^2}{8}\,n\right).
\]
Consequently, for all large $N$,
\[
\Pr\!\left(\frac{n_{s,N}}{N}\le \delta\right)
\le \Pr\!\bigl(n_{s,N}\le (1-\eta/2)n\bigr)
\le \exp(-cN)
\]
for some $c>0$. Since $e^{-cN} = o(N^{-\kappa})$ for any fixed $\kappa>0$, there exists
$C<\infty$ such that $\exp(-cN)\le CN^{-\kappa}$ for all large $N$; in particular
$\Pr(n_{s,N}/N\le \delta)\le CN^{-\kappa}$ for any $\kappa>4$, proving \textup{(D2)}. Moreover,
$n_{s,N}/N\to_{p_d} f$ by the weak law of large numbers for independent bounded summands.

To verify \textup{(D3)}, note that $\pi_{\min}\le \pi_i\le \pi_{\max}$ for all $i,N$ implies
\[
\pi_{\min}\le \frac{n}{N}=\frac{1}{N}\sum_{i=1}^N \pi_i \le \pi_{\max}.
\]
Therefore, for all $i,N$,
\[
\pi_i \le \pi_{\max} \le \frac{\pi_{\max}}{\pi_{\min}}\cdot \frac{n}{N},\qquad
\pi_i \ge \pi_{\min} \ge \frac{\pi_{\min}}{\pi_{\max}}\cdot \frac{n}{N},
\]
so \textup{(D3)} holds with $\overline\lambda=\pi_{\max}/\pi_{\min}$ and
$\underline\lambda=\pi_{\min}/\pi_{\max}$.

Finally, independence gives $\pi_{ij}=\pi_i\pi_j$ for all $i\neq j$, hence \textup{(D4)} holds
with $C_1=0$.

\medskip
\paragraph{SRSWOR (SRS without Replacement)}
Here $n_{s,N}\equiv n_N$ deterministically, so $n_{s,N}/N\to f$ and, for any $\delta\in(0,f)$,
$\Pr(n_{s,N}/N\le \delta)=0$ for all large $N$, proving \textup{(D2)} with any $\kappa>0$.

Moreover $\pi_i=n_N/N$ for all $i$, so \textup{(D3)} holds with $\underline\lambda=\overline\lambda=1$
and $n=\E(n_{s,N})=n_N$.

For $i\neq j$,
\[
\pi_{ij}=\frac{n_N(n_N-1)}{N(N-1)},\qquad \pi_i\pi_j=\Big(\frac{n_N}{N}\Big)^2,
\]
hence
\[
\pi_{ij}-\pi_i\pi_j
= -\frac{n_N(N-n_N)}{N^2(N-1)}.
\]
Therefore
\[
\max_{1\le i<j\le N}|\pi_{ij}-\pi_i\pi_j|
=\frac{n_N(N-n_N)}{N^2(N-1)}
\le \frac{n_N}{N(N-1)}.
\]
Since $n_N/N\to f\in(0,1)$, there exists $N_0$ and $c>0$ such that $n_N\ge cN$ for all $N\ge N_0$.
Thus, for $N\ge N_0$,
\[
\frac{n_N}{N(N-1)} \le \frac{1}{c(N-1)}.
\]
Since $n_N\ge cN$ and $N-1\ge N/2$ for all large $N$, we have $n_N\ge cN\ge 2c(N-1)/2$,
hence $1/(N-1)\le 2/n_N \cdot (n_N/N) \le 2/n_N$. Therefore
\[
\frac{n_N}{N(N-1)} \le \frac{1}{c(N-1)} \le \frac{2}{c}\cdot\frac{1}{n_N}
= \frac{2}{c}\cdot\frac{1}{n_{s,N}},
\]
which is \textup{(D4)} with $C_1=2/c$.

\medskip
\paragraph{PPS with Replacement}
Define $p_i=m_i/\sum_{r=1}^N m_r$ and note that
\[
\frac{m_{\min}}{Nm_{\max}}\le p_i \le \frac{m_{\max}}{Nm_{\min}}
\qquad\text{for all }i,N.
\]
Let $\pi_i=1-(1-p_i)^{m_N}$ and $n_{s,N}=\sum_{i=1}^N \xi_i$ with $\xi_i=\ind\{i\in\{I_1,\dots,I_{m_N}\}\}$.
By assumption, $m_N/N\to f_0\in(0,1)$ and $n/N\to f\in(0,1)$, where $n=\E(n_{s,N})=\sum_{i=1}^N \pi_i$.

\emph{(D2).} Consider the function $F(I_1,\dots,I_{m_N})=n_{s,N}$. Changing one draw $I_t$ can change
$n_{s,N}$ by at most $1$, hence $F$ satisfies bounded differences with constants $c_t\equiv 1$.
McDiarmid's inequality yields, for all $t>0$,
\[
\Pr\!\left(n_{s,N}\le \E(n_{s,N})-t\right)\le \exp\!\left(-\frac{2t^2}{m_N}\right).
\]
Fix $\delta\in(0,f)$ and choose $\varepsilon:=(f-\delta)/2>0$. For all large $N$,
$\E(n_{s,N})/N=n/N\ge f-\varepsilon=\delta+\varepsilon$, hence
$\E(n_{s,N})\ge (\delta+\varepsilon)N$. 
Rearranging, we have $\delta N \le \E(n_{s,N}) - \varepsilon N$ for all large $N$.
Therefore, for all large $N$,
\[
\Pr\!\left(\frac{n_{s,N}}{N}\le \delta\right)
\le \Pr\!\left(n_{s,N}\le \delta N\right)
\le \Pr\!\left(n_{s,N}\le \E(n_{s,N})-\varepsilon N\right)
\le \exp\!\left(-\frac{2\varepsilon^2 N^2}{m_N}\right)
\le \exp(-cN)
\]
for some $c>0$ (since $m_N\asymp N$). As before, exponential decay implies
$\Pr(n_{s,N}/N\le \delta)\le CN^{-\kappa}$ for any $\kappa>4$ with suitable $C<\infty$,
proving \textup{(D2)}. Moreover $n_{s,N}/N\to_{p_d} f$ by the same concentration.

\emph{(D3).} Define
\[
a_N:=\min_{i\in[N]}\pi_i,\qquad b_N:=\max_{i\in[N]}\pi_i.
\]
Since $\pi_i$ is increasing in $p_i$ and $p_i\in[c_1/N,c_2/N]$ with
$c_1:=m_{\min}/m_{\max}$ and $c_2:=m_{\max}/m_{\min}$, we have
\[
1-\Big(1-\frac{c_1}{N}\Big)^{m_N}\le a_N \le \pi_i \le b_N \le 1-\Big(1-\frac{c_2}{N}\Big)^{m_N}.
\]
Also $n/N$ is the average of the $\pi_i$, hence $a_N\le n/N\le b_N$. Therefore, for all $i$,
\[
\pi_i \le b_N = \frac{b_N}{a_N}\,a_N \le \frac{b_N}{a_N}\,\frac{n}{N},\qquad
\pi_i \ge a_N = \frac{a_N}{b_N}\,b_N \ge \frac{a_N}{b_N}\,\frac{n}{N}.
\]
Since $m_N/N\to f_0\in(0,1)$ and $0<c_1\le c_2<\infty$ under the MOS bounds, we have
\[
a_N\to 1-e^{-f_0 c_1}\in(0,1),\qquad b_N\to 1-e^{-f_0 c_2}\in(0,1),
\]
with both limits strictly positive. Hence $b_N/a_N$ and $a_N/b_N$ are bounded away from
$0$ and $\infty$ for all large $N$, and there exist constants
$0<\underline\lambda\le \overline\lambda<\infty$ such that \textup{(D3)} holds for all
large $N$.

\emph{(D4).} For $i\neq j$,
\[
\pi_{ij}=1-(1-p_i)^{m_N}-(1-p_j)^{m_N}+(1-p_i-p_j)^{m_N}, \text{ and }
\]
\[
\pi_{ij}-\pi_i\pi_j
=(1-p_i-p_j)^{m_N}-\bigl(1-p_i-p_j+p_ip_j\bigr)^{m_N}.
\]
Let $a:=1-p_i-p_j$ and $b:=a+p_ip_j$. Since $p_i,p_j\le c_2/N$, we have
$p_i+p_j\le 2c_2/N\le 1$ and $p_ip_j\le (c_2/N)^2\le 1-a$ for all large $N$,
hence $a,b\in[0,1]$. By the mean value theorem and the bound
$\sup_{x\in[0,1]}|(x^{m_N})'|=m_N$,
\[
|\pi_{ij}-\pi_i\pi_j| = |a^{m_N}-b^{m_N}|
\le m_N|a-b| = m_N p_i p_j
\le m_N\left(\frac{c_2}{N}\right)^2
\le \frac{C}{N}
\]
for a constant $C<\infty$ and all $N$ (since $m_N/N$ is bounded).
Because $n_{s,N}\le m_N$ almost surely (the number of distinct sampled units cannot exceed the number of draws),
it follows that $1/m_N\le 1/n_{s,N}$ a.s., hence for all large $N$,
\[
\max_{1\le i<j\le N}|\pi_{ij}-\pi_i\pi_j|
\le \frac{C'}{m_N}
\le \frac{C'}{n_{s,N}}
\qquad\text{a.s.}
\]
for some $C'<\infty$, which proves \textup{(D4)}.
\end{proof}

\subsection{Stratified Two-stage Cluster Design}

The stagewise counterparts of (D1)--(D4) are as follows. Analogous to (D1)--(D4), these conditions are stated at the level of the design and are verified case by case for standard two-stage schemes (Poisson $\times$ SRSWOR, SRSWOR $\times$ SRSWOR, stratified PPSWR $\times$ SRSWOR).

\begin{enumerate}[label=\textup{(D-TS\arabic*)}, leftmargin=2.8em, itemsep=3pt, topsep=3pt]

\item \label{D-TS1} \emph{Cross-stratum and cross-PSU independence.} The collections $\{\xi^{(1)}_{hk}, \xi^{(2)}_{hkj} : k\in[K_h], j\in[M_{hk}]\}$ are mutually independent across $h\in[H]$. Conditional on $\{\xi^{(1)}_{hk}\}_{h,k}$, the stage-2 selections $\{\xi^{(2)}_{hkj}\}_j$ are mutually independent across distinct PSUs.

\item \label{D-TS2} \emph{Stable overall SSU sampling fraction.} There exist $f\in(0,1]$, $\delta\in(0,f)$, $\kappa>4$, and $C<\infty$ such that $n_s/N \xrightarrow{p_d} f$ and $P_{\mathcal{S}^N\mid\omega}(n_s/N \le \delta) \le C N^{-\kappa}$ for all $N\ge 1$, where $n_s := \sum_u \xi_u$ and $N := \sum_{h,k}M_{hk}$.

\item \label{D-TS3} \emph{Stagewise first-order inclusion bounds.} There exist $0 < \underline{\lambda}_1 \le \bar\lambda_1 < \infty$ and $0 < \underline{\lambda}_2 \le \bar\lambda_2 < 1$ such that, for all $h,k,j$,
\[
\underline{\lambda}_1\, \frac{n_{\mathrm{eff}}}{K}
\;\le\; \pi^{(1)}_{hk} \;\le\;
\bar\lambda_1\, \frac{n_{\mathrm{eff}}}{K},
\qquad
\underline{\lambda}_2 \;\le\; \pi^{(2)}_{hkj\mid hk} \;\le\; \bar\lambda_2,
\]
where $K := \sum_h K_h$.

\item \label{D-TS4} \emph{Stagewise controlled second-order dependence.} There exists $C<\infty$ such that:
\begin{enumerate}[label=(\alph*), leftmargin=2em, itemsep=1pt, topsep=1pt]
\item \emph{Between-PSU within stratum:} for each $h$ and $k\ne k'\in[K_h]$,
\(
\bigl|\pi^{(1)}_{hkk'} - \pi^{(1)}_{hk}\pi^{(1)}_{hk'}\bigr| \le C/n_{\mathrm{eff}}.
\)
\item \emph{Within-PSU:} for each $h,k$ and $j\ne j'\in[M_{hk}]$,
\(
\bigl|\pi^{(2)}_{hkj,hkj'\mid hk} - \pi^{(2)}_{hkj\mid hk}\pi^{(2)}_{hkj'\mid hk}\bigr| \le C.
\)
(Automatically satisfied with $C = 1/4$ since $\pi^{(2)} \in [0,1]$.)
\end{enumerate}
\end{enumerate}

These parallel (D1)--(D4): \ref{D-TS2} restates (D2) at the SSU level directly; \ref{D-TS3} splits (D3) into stagewise first-order bounds whose product recovers SSU-level $\pi_u \asymp n/N$; \ref{D-TS4} refines (D4) into between-PSU and within-PSU pairwise controls, since a single uniform bound on $|\pi_{uv}-\pi_u\pi_v|$ at rate $O(1/n_s)$ does \emph{not} hold at the SSU level under two-stage sampling (same-PSU pairs violate it; see Lemma~\ref{lem:designs-multistage} below).

\begin{lemma}[SSU-level order and PSU-block pairwise dependence]
\label{lem:designs-multistage}
Under the regime clause and \textup{\ref{D-TS1}--\ref{D-TS4}}:

\begin{enumerate}[label=\textup{(\roman*)}, leftmargin=2.5em, itemsep=3pt, topsep=3pt]

\item \emph{Order equivalence and SSU-level (D3).} $n \asymp n_{\mathrm{eff}} \asymp N$, and there exist $0<\underline{\lambda}\le\bar\lambda<\infty$ with
\[
\underline{\lambda}\,\frac{n}{N} \;\le\; \inf_{u\in[N]} \pi_u \;\le\; \sup_{u\in[N]} \pi_u \;\le\; \bar\lambda\,\frac{n}{N}.
\]

\item \emph{PSU-block pairwise dependence.} For distinct SSUs $u=(h,k,j)$ and $v=(h',k',j')$,
\[
\bigl|\pi_{uv} - \pi_u\pi_v\bigr|
\;\le\;
\begin{cases}
0, & h\ne h',\\[3pt]
C_1 / n_{\mathrm{eff}}, & h=h',\ k\ne k',\\[3pt]
C_2 \cdot n_{\mathrm{eff}}/N, & h=h',\ k=k',\ j\ne j',
\end{cases}
\]
for constants $C_1, C_2 < \infty$ depending only on $\underline{\lambda}_1,\bar\lambda_1,\underline{\lambda}_2,\bar\lambda_2,M_{\max}$, and the constant in \ref{D-TS4}\textup{(a)}.
\end{enumerate}
\end{lemma}

\begin{proof}[Proof of Lemma~\ref{lem:designs-multistage}]

\textbf{(i).} By $1\le M_{hk}\le M_{\max}$, $K\le N\le K M_{\max}$, so $K\asymp N$. From \ref{D-TS3},
\[
n \;=\; \sum_{h,k} \pi^{(1)}_{hk}\sum_j \pi^{(2)}_{hkj\mid hk}
\;\in\; \bigl[\underline{\lambda}_1\underline{\lambda}_2\, n_{\mathrm{eff}},\; \bar\lambda_1\bar\lambda_2 M_{\max}\, n_{\mathrm{eff}}\bigr],
\]
where the lower bound uses $\sum_j\pi^{(2)}_{hkj\mid hk}\ge \underline{\lambda}_2 M_{hk}\ge\underline{\lambda}_2$ and the upper bound uses $\sum_j\pi^{(2)}_{hkj\mid hk}\le \bar\lambda_2 M_{hk}\le\bar\lambda_2 M_{\max}$; hence $n\asymp n_{\mathrm{eff}}$. By \ref{D-TS2} and dominated convergence ($n_s/N\in[0,1]$), $n/N \to f\in(0,1]$, so $n\asymp N$. Combining, $n \asymp n_{\mathrm{eff}} \asymp N$. For (D3), $\pi_u = \pi^{(1)}_{hk}\pi^{(2)}_{hkj\mid hk}$ so \ref{D-TS3} gives $\pi_u \asymp n_{\mathrm{eff}}/K \asymp n/N$ uniformly in $u$.

\medskip
\noindent\textbf{(ii).}
\emph{Case 1: $h\ne h'$.} By \ref{D-TS1}, $\xi_u\perp\xi_v$, so $\pi_{uv}=\pi_u\pi_v$.

\emph{Case 2: $h=h',\, k\ne k'$.} By \ref{D-TS1}, stage-2 selections in $(h,k)$ and $(h,k')$ are conditionally independent given stage 1, hence
\[
\pi_{uv} \;=\; \pi^{(1)}_{hkk'}\, \pi^{(2)}_{hkj\mid hk}\, \pi^{(2)}_{hk'j'\mid hk'},
\]
and
\[
\pi_{uv} - \pi_u\pi_v \;=\; \bigl(\pi^{(1)}_{hkk'} - \pi^{(1)}_{hk}\pi^{(1)}_{hk'}\bigr)\,\pi^{(2)}_{hkj\mid hk}\,\pi^{(2)}_{hk'j'\mid hk'}.
\]
By \ref{D-TS4}(a) and $\pi^{(2)}\le\bar\lambda_2<1$, $\bigl|\pi_{uv}-\pi_u\pi_v\bigr|\le C\bar\lambda_2^{\,2}/n_{\mathrm{eff}} =: C_1/n_{\mathrm{eff}}$.

\emph{Case 3: $h=h',\, k=k',\, j\ne j'$.} Decompose
\[
\pi_{uv} - \pi_u\pi_v \;=\; \underbrace{\pi^{(1)}_{hk}\bigl(\pi^{(2)}_{hkj,hkj'\mid hk} - \pi^{(2)}_{hkj\mid hk}\pi^{(2)}_{hkj'\mid hk}\bigr)}_{A}
\;+\;
\underbrace{\pi^{(1)}_{hk}(1-\pi^{(1)}_{hk})\,\pi^{(2)}_{hkj\mid hk}\,\pi^{(2)}_{hkj'\mid hk}}_{B}.
\]
By \ref{D-TS3} and \ref{D-TS4}(b), $|A|\le \bar\lambda_1 C\,(n_{\mathrm{eff}}/K)$; by \ref{D-TS3}, $|B|\le \bar\lambda_1\bar\lambda_2^{\,2}\,(n_{\mathrm{eff}}/K)$. Using $1/K\le M_{\max}/N$,
\[
\bigl|\pi_{uv}-\pi_u\pi_v\bigr|
\;\le\; \bar\lambda_1(C+\bar\lambda_2^{\,2})\,\frac{n_{\mathrm{eff}}}{K}
\;\le\; \bar\lambda_1(C+\bar\lambda_2^{\,2})M_{\max}\,\frac{n_{\mathrm{eff}}}{N}
\;=:\; C_2\,\frac{n_{\mathrm{eff}}}{N}.
\qedhere
\]
\end{proof}

\subsection{Design-based LLN and CLT}
The conditions (D2)--(D4) imply design-based LLN and CLT, which are commonly assumed in survey literature. We also utilize this result in our consistency proof. Lemma~\ref{lem:lln_clt} is for single stage designs, while Lemma~\ref{lem:lln_clt_multistage} the analog for two-stage designs.

\begin{lemma}
\label{lem:lln_clt}
    Under (D2)-(D4), we have that with respect to the design law $P_{\mathcal{S}^N|\omega}$, $\frac{1}{N}\sum_{i=1}^N\!\left(\frac{\xi_i}{\pi_i}-1\right)$ $ \xrightarrow{p_d} 0$, and $\frac{1}{\sqrt{N}}\sum_{i=1}^N\!\left(\frac{\xi_i}{\pi_i}-1\right) = O_{p_d}(1).$
\end{lemma}

\begin{proof}[Proof of Lemma~\ref{lem:lln_clt}]
    By (D3), 
    \[
    \sum_{i=1}^{N} \mathrm{Var}_{S^{N}\mid \omega}\left(\frac{\xi_i}{\pi_i} - 1\right)
    \le
    \sum_{i=1}^{N} \frac{1}{\pi_i}
    \le
    N \cdot \frac{1}{\inf_i \pi_i}
    \le
    N \cdot \frac{1}{\underline{\lambda} n / N}
    =
    \frac{N^{2}}{\underline{\lambda} n}.
    \]
    By (D4) and (D3), for $i \neq j$,
    \[
    \left|\mathrm{Cov}_{S^{N}\mid \omega}(\frac{\xi_i}{\pi_i} - 1, \frac{\xi_j}{\pi_j} - 1)\right|
    \;\le\;
    \frac{1}{\pi_i\pi_j}\cdot \frac{C_1}{n_{s,N}}
    \;\le\;
    \frac{C_1}{n_{s,N}}\cdot \frac{1}{(\inf_i \pi_i)^2}
    \;\le\;
    \frac{C_1}{n_{s,N}}\cdot \frac{1}{(\underline{\lambda} n/N)^2},
    \]
    \noindent for some $C_1 < \infty$. So summing over $\binom{N}{2}$ pairs gives
    $
    \sum_{1\le i<j\le N}\left|\mathrm{Cov}_{S^{N}\mid \omega}(\frac{\xi_i}{\pi_i} - 1,\; \frac{\xi_j}{\pi_j} - 1)\right|
    \;\le\;
    \frac{N(N-1)}{2}\cdot \frac{C_1 N^{2}}{\underline{\lambda}^{2} n^{2}}\cdot \frac{1}{n_{s,N}}.
    $
    Thus
    \begin{align*}
        Var_{\mathcal{S}^N\mid \omega}\left( \sum_{i \in [N]} \frac{\xi_i}{\pi_i} - 1\right) 
        \lesssim
        \frac{N^2}{n} + N (N-1) \frac{N^2}{n^2 n_{s,N}}.
    \end{align*}
    By (D2), $n_{s,N} = \Theta_{p_d}(N)$, and $n = \Theta(N)$.
    By Chebyshev's inequality and that $\E[\sum_{i \in [N]} (\frac{\xi_i}{\pi_i} - 1)] = 0$, for each $\epsilon > 0$,
    $
    P_{\mathcal{S}^N\mid \omega}\!\left(|\frac{1}{N}\sum_{i \in [N]} (\frac{\xi_i}{\pi_i} - 1) |> \epsilon \right)
    = O(1/N) \to 0,
    $
    proving the LLN statement;
    and
    $
    P_{\mathcal{S}^N\mid \omega}\!\left(|\frac{1}{\sqrt{N}}\sum_{i \in [N]} (\frac{\xi_i}{\pi_i} - 1) |> \epsilon \right)
    = O_{p_d}(1)
    $
    uniformly in $N$, proving the CLT statement.
\end{proof}

For reference, the centered flat SSU sum admits the equivalent PSU-block form
\begin{equation}\label{eq:centered-psu-form}
\sum_{u=1}^{N}\!\left(\frac{\xi_u}{\pi_u}-1\right)
\;=\;
\sum_{h=1}^{H}\sum_{k=1}^{K_h}\!\left[\frac{\xi^{(1)}_{hk}}{\pi^{(1)}_{hk}}\sum_{j=1}^{M_{hk}}\!\frac{\xi^{(2)}_{hkj}}{\pi^{(2)}_{hkj\mid hk}}\;-\;M_{hk}\right],
\end{equation}
obtained by grouping the SSU sum by PSU and factoring $\pi_u = \pi^{(1)}_{hk}\pi^{(2)}_{hkj\mid hk}$; the two expressions denote the same random variable.

\begin{lemma}[Design LLN/CLT under stratified two-stage cluster sampling]
\label{lem:lln_clt_multistage}
Under the regime clause and \textup{(D-TS1)}--\textup{(D-TS4)},
\[
\frac{1}{N}\sum_{u=1}^{N}\!\left(\frac{\xi_u}{\pi_u}-1\right) \;\xrightarrow{p_d}\; 0,
\qquad
\frac{1}{\sqrt{N}}\sum_{u=1}^{N}\!\left(\frac{\xi_u}{\pi_u}-1\right) \;=\; O_{p_d}(1).
\]
\end{lemma}

\begin{proof}[Proof of Lemma~\ref{lem:lln_clt_multistage}]
Let $S:=\sum_u(\xi_u/\pi_u-1)$. Then $\mathbb{E}_{P_{\mathcal{S}^N\mid\omega}}[S]=0$. Partition the pairs $(u,v)$ by the block structure of Lemma~\ref{lem:designs-multistage}\,(ii):
\[
\mathrm{Var}_{P_{\mathcal{S}^N\mid\omega}}(S) \;=\; T_0 \;+\; T_2 \;+\; T_3,
\]
with $T_0:=\sum_u \frac{1-\pi_u}{\pi_u}$, $T_2:=\sum\limits_{\substack{h=h'\\ k\ne k'}} \frac{\pi_{uv}-\pi_u\pi_v}{\pi_u\pi_v}$, $T_3:=\sum\limits_{\substack{h=h',\,k=k'\\ j\ne j'}} \frac{\pi_{uv}-\pi_u\pi_v}{\pi_u\pi_v}$; the cross-stratum contribution vanishes by Lemma~\ref{lem:designs-multistage}\,(ii), Case 1.

By Lemma~\ref{lem:designs-multistage}\,(i), $\pi_u \ge \underline{\lambda}(n/N)$ uniformly, hence $1/\pi_u \le N/(\underline{\lambda} n)$. Thus
\[
T_0 \;\le\; \sum_u \frac{1}{\pi_u} \;\le\; \frac{N^2}{\underline{\lambda} n}.
\]

For $T_2$: by Case 2 of Lemma~\ref{lem:designs-multistage}\,(ii), $|\pi_{uv}-\pi_u\pi_v|\le C_1/n_{\mathrm{eff}}$; the pair count is $\sum_h \sum_{k\ne k'} M_{hk}M_{hk'} \le M_{\max}^2 \sum_h K_h^2 \le M_{\max}^2 K^2 \le M_{\max}^2 N^2$. Combining with $1/\pi_u\pi_v \le N^2/(\underline{\lambda}^2 n^2)$,
\[
|T_2| \;\le\; \frac{C_1}{n_{\mathrm{eff}}} \cdot M_{\max}^2 N^2 \cdot \frac{N^2}{\underline{\lambda}^2 n^2}
\;=\; \frac{C_1'\, N^4}{n_{\mathrm{eff}}\, n^2}.
\]

For $T_3$: by Case 3, $|\pi_{uv}-\pi_u\pi_v|\le C_2\, n_{\mathrm{eff}}/N$; the pair count is $\sum_{h,k}M_{hk}(M_{hk}-1) \le M_{\max}\sum_{h,k}M_{hk} = M_{\max}N$. Combining,
\[
|T_3| \;\le\; \frac{C_2\, n_{\mathrm{eff}}}{N} \cdot M_{\max} N \cdot \frac{N^2}{\underline{\lambda}^2 n^2}
\;=\; \frac{C_2'\, M_{\max}\, n_{\mathrm{eff}}\, N^2}{n^2}.
\]

By Lemma~\ref{lem:designs-multistage}\,(i), $n\asymp n_{\mathrm{eff}}$; by \ref{D-TS2} and dominated convergence, $n/N\to f\in(0,1]$, so $n\asymp N$. Substituting,
\[
T_0 = O(N),\qquad |T_2| = O(N),\qquad |T_3| = O(N),
\qquad\text{hence}\qquad \mathrm{Var}_{P_{\mathcal{S}^N\mid\omega}}(S) = O(N).
\]

By Chebyshev's inequality, for each $\epsilon>0$,
\[
P_{\mathcal{S}^N\mid\omega}\!\left(\Bigl|\tfrac{1}{N}S\Bigr| > \epsilon\right)
\;\le\; \frac{\mathrm{Var}(S)}{\epsilon^2 N^2} \;=\; O(1/N) \to 0,
\]
proving the LLN statement; and for each $M>0$,
\[
P_{\mathcal{S}^N\mid\omega}\!\left(\Bigl|\tfrac{1}{\sqrt{N}}S\Bigr| > M\right)
\;\le\; \frac{\mathrm{Var}(S)}{M^2 N} \;=\; O(1/M^2)
\]
uniformly in $N$, proving the tightness statement.
\end{proof}

\section{Embedding of Measures}

We first show the applicability of the kernel mean embedding (KME) framework
on measures in survey sampling. For any finite signed measure
$\nu\in\mathcal M_b(\mathcal Y)$, recall the kernel mean embedding definition
\begin{equation}\label{eq:kme-def}
\mu_{\mathbf k}(\nu) := \int \mathbf{k}(\cdot,t)\, d\nu(t) \in \mathcal H.
\end{equation}
The integral is well-defined since, under (K1),
$\|\mathbf{k}(\cdot,t)\|_{\mathcal H}=\sqrt{\mathbf{k}(t,t)}\le\sqrt{K_{\max}}$,
hence $\int \|\mathbf{k}(\cdot,t)\|_{\mathcal H}\, d|\nu|(t)
\le \sqrt{K_{\max}}\,|\nu|(\mathcal Y) < \infty$.
Moreover, for any $f\in\mathcal H$,
\begin{equation}\label{eq:kme-rep}
\langle \mu_{\mathbf k}(\nu), f\rangle_{\mathcal H}
= \int \langle \mathbf{k}(\cdot,t), f\rangle_{\mathcal H}\, d\nu(t)
= \int f(t)\, d\nu(t),
\end{equation}
where the last equality uses the reproducing property.

The takeaways are that for any given $\mathbf x$ and a region
$A(\mathbf{x}) \subseteq \mathcal{X}$ indexed by $\mathbf{x}$ with
$N_{A(\mathbf x)} := \sum_{i\in[N]}\mathbbm{1}(X_i \in A(\mathbf x)) > 0$:
\begin{enumerate}[noitemsep, topsep=0pt]
    \item Each of $P_{Y\mid X\in A(\mathbf{x})}$,
    $P^N_{Y\mid X\in A(\mathbf{x})}$,
    $\widehat P^{\mathrm{HT}}_{Y\mid X\in A(\mathbf{x})}$, and
    $\widehat P^{\mathrm{HJ}}_{Y\mid X\in A(\mathbf{x})}$
    lies in $\mathcal{M}_b(\mathcal Y)$, so under (K1) its KME
    $\mu_{\mathbf{k}}(\cdot)\in\mathcal H$ is well-defined by
    \eqref{eq:kme-def}.
    \item For any finite collection $\{(a_i,y_i)\}$ with $\sum_i |a_i|<\infty$,
    \[
    \mu_{\mathbf{k}}\!\Big(\sum_i a_i \delta_{y_i}\Big) = \sum_i a_i\,\mathbf{k}(\cdot,y_i),
    \qquad
    \big\langle \mu_{\mathbf{k}}(\sum_i a_i\delta_{y_i}), f\big\rangle_{\mathcal H} = \sum_i a_i f(y_i).
    \]
    Hence HT design-unbiasedness and H\'ajek design-consistency transfer to $\mathcal H$-valued objects by linearity, with the additional randomness entering through the weights.
\end{enumerate}

\begin{itemize}
    \item \textbf{Dirac measure (at a fixed point).}\\
        For $y\in\mathcal Y$, the Dirac measure $\delta_y$ satisfies
        \begin{equation}\label{eq:dirac-kme}
        \mu_{\mathbf{k}}(\delta_y)
        = \int \mathbf{k}(\cdot,t)\, d\delta_y(t)
        = \mathbf{k}(\cdot,y)\in\mathcal H.
        \end{equation}
        Consequently, for any $f\in\mathcal H$,
        \[
        \langle \mu_{\mathbf{k}}(\delta_y), f\rangle_{\mathcal H}
        = \langle \mathbf{k}(\cdot,y), f\rangle_{\mathcal H}
        = f(y).
        \]
    \item \textbf{Super-population measure.}\\
        Let $P_Y$ denote the super-population law of $Y$. Define
        \[
        \mu_{\mathbf k}(P_Y) := \int \mathbf{k}(\cdot,t)\, dP_Y(t)\in\mathcal H,
        \]
        where the Bochner integral is well-defined under (K1). Then for any
        $f\in\mathcal H$,
        \[
        \langle \mu_{\mathbf k}(P_Y), f\rangle_{\mathcal H}
        = \int \langle \mathbf{k}(\cdot,t), f\rangle_{\mathcal H}\, dP_Y(t)
        = \int f(t)\, dP_Y(t)
        = \mathbb E[f(Y)].
        \]
        Similarly, for a regular conditional law $P_{Y\mid X=\mathbf x}$,
        \[
        \langle \mu_{\mathbf k}(P_{Y\mid X=\mathbf x}), f\rangle_{\mathcal H}
        = \int f(t)\, dP_{Y\mid X=\mathbf x}(t)
        = \mathbb E[f(Y)\mid X=\mathbf x],
        \qquad f\in\mathcal H.
        \]
        Thus $\mu_{\mathbf k}(P_{Y\mid X=\mathbf x})$ is the Riesz representer in
        $\mathcal H$ of the functional $f\mapsto \mathbb E[f(Y)\mid X=\mathbf x]$
        restricted to $f\in\mathcal H$ (and identifies $P_{Y\mid X=\mathbf x}$
        if $\mathbf k$ is characteristic).
    \item \textbf{Finite-population measure.}\\
        Denote $P_{N,Y} = \frac{1}{N} \sum_{i=1}^N \delta_{Y_i}$ the
        finite-population marginal measure of $Y$; its KME is
        \[
        \mu_{\mathbf k}(P_{N,Y}) = \int \mathbf{k}(\cdot, t) \, dP_{N,Y}(t)
        = \frac{1}{N} \sum_{i=1}^N \mathbf{k}(\cdot, Y_i) \in \mathcal{H}.
        \]
        The inner product with any $f \in \mathcal{H}$ is, by
        \eqref{eq:kme-rep},
        \[
        \langle \mu_{\mathbf k}(P_{N,Y}), f \rangle_{\mathcal{H}}
        = \frac{1}{N} \sum_{i=1}^N f(Y_i).
        \]
        In the design space, $(X_i, Y_i)$ are considered fixed, so
        $\langle \mu_{\mathbf k}(P_{N,Y}), f \rangle_{\mathcal{H}}$ returns a
        fixed value.

    \item \textbf{Finite-population conditional measure (estimate).}\\
        Consider the H\'ajek-type estimator
        $\widehat{P}^{\mathrm{HJ}}_N(Y\mid X \in A(\mathbf{x}))$ for the target
        \[
        P_{N,\,Y\mid X\in A(\mathbf{x})}
        = \frac{1}{N_{A(\mathbf x)}}\sum_{i\in[N]} \mathbbm{1}(X_i \in A(\mathbf{x}))\,\delta_{Y_i},
        \qquad
        N_{A(\mathbf x)} := \sum_{i\in[N]} \mathbbm{1}(X_i \in A(\mathbf{x})) > 0,
        \]
        the conditional finite-population measure on $A(\mathbf{x})$.
        Specifically,
        \[
        \widehat{P}^{\mathrm{HJ}}_N(Y\mid X \in A(\mathbf{x}))
        = \frac{\sum_{i=1}^N \frac{\xi_i}{\pi_i}\,\mathbbm{1}(X_i\in A(\mathbf x))\,\delta_{Y_i}}
               {\sum_{i=1}^N \frac{\xi_i}{\pi_i}\,\mathbbm{1}(X_i\in A(\mathbf x))}
        = \sum_{i=1}^N \widehat{\omega}_{N,i}(\mathbf{x})\, \delta_{Y_i},
        \]
        where
        \[
        \widehat{\omega}_{N,i}(\mathbf{x})
        := \frac{\xi_i\,\mathbbm{1}(X_i\in A(\mathbf x))/\pi_i}{\widehat{N}(\mathbf x)},
        \qquad
        \widehat{N}(\mathbf x) := \sum_{i=1}^N \frac{\xi_i}{\pi_i}\,\mathbbm{1}(X_i\in A(\mathbf x)).
        \]
        The kernel mean embedding is
        \[
        \mu_{\mathbf k}\!\big(\widehat{P}^{\mathrm{HJ}}_N(Y\mid X\in A(\mathbf x))\big)
        = \int \mathbf{k}(\cdot, y)\, d\widehat{P}^{\mathrm{HJ}}_N(y\mid X\in A(\mathbf x))
        = \sum_{i=1}^N \widehat{\omega}_{N,i}(\mathbf{x})\,\mathbf{k}(\cdot, Y_i),
        \]
        where the sum is over all units.
        By \eqref{eq:kme-rep}, the inner product with any $f\in\mathcal H$ is
        \[
        \big\langle \mu_{\mathbf k}\!\big(\widehat{P}^{\mathrm{HJ}}_N(Y\mid X\in A(\mathbf x))\big), f \big\rangle_{\mathcal H}
        = \sum_{i=1}^N \widehat{\omega}_{N,i}(\mathbf x)\, f(Y_i).
        \]
        Randomness from the design enters through the weight
        $\widehat{\omega}_{N,i}(\mathbf x)$. In particular, for the HT
        estimator (with $N_{A(\mathbf x)}$ a fixed number rather than
        $\widehat{N}(\mathbf x)$),
        \[
        \big\langle \mathbb E_\xi\!\big[\mu_{\mathbf k}\!\big(\widehat{P}^{\mathrm{HT}}_N(Y\mid X\in A(\mathbf x))\big)\big], f \big\rangle_{\mathcal H}
        = \frac{1}{N_{A(\mathbf x)}} \sum_{i=1}^N \mathbbm{1}(X_i\in A(\mathbf x))\, f(Y_i)
        = \int f\, dP_{N,\,Y\mid X\in A(\mathbf x)},
        \]
        so the HT-based embedding is design-unbiased for
        $\mu_{\mathbf k}(P_{N,\,Y\mid X\in A(\mathbf x)})$, while the H\'ajek-type
        estimator is generally biased (unless $\widehat{N}(\mathbf x) = N_{A(\mathbf x)}$),
        but design-consistent.
\end{itemize}

\textbf{A common comparison metric.}\\
Every object above, including the super-population conditional $P_{Y\mid X=\mathbf x}$,
finite-population conditional $P^N_{Y\mid X=\mathbf x}$, and any H\'ajek- or
HT-type design-weighted estimates, embeds into the same Hilbert space
$\mathcal H$. Any pair can therefore be compared through the common Maximum
Mean Discrepancy
\begin{equation}\label{eq:mmd-def}
d_{\mathbf k}(\nu_1,\nu_2)
:= \big\| \mu_{\mathbf k}(\nu_1) - \mu_{\mathbf k}(\nu_2) \big\|_{\mathcal H},
\end{equation}
regardless of whether $\nu_1,\nu_2$ originate from the super-population, the
finite population, or a design-weighted sample. This property is what allows
a single splitting criterion, a single consistency notion, and a single
asymptotic theory to cover complex designs.

\section{Resampling}
\label{secap:resampling}

In the complex survey design forest, tree construction must mimic the sampling
design at two levels: (i) the resamples that seed each tree preserve the original
design structure, and (ii) honesty is retained, in that the units used to estimate
$\widehat\omega_i$ are disjoint from those used to select splits. This section
specifies the pseudo-population bootstrap that produces the resample multipliers
$\{n_i^*\}_{i\in[n_s]}$, establishes their sub-gamma control
(Lemma~\ref{lem:subgamma}), and proves that they satisfy conditions (R1)--(R3) of
Lemma~7 in the manuscript.
 
Throughout, $\xi = \xi^N$ denotes the inclusion indicators of the original sample,
$Z^N$ the design covariates encoding the known stratum and cluster structure, and
$\pi_i$ the first-order inclusion probability of unit $i$. We write $\Gamma(\nu,c)$
for the class of centered sub-gamma random variables with variance factor $\nu$ and
scale $c$; that is, $X\in\Gamma(\nu,c)$ iff
\[
  \log\mathbb E\,e^{\lambda X}\le \frac{\lambda^2\nu}{2(1-c|\lambda|)}
  \qquad\text{for }|\lambda|<1/c.
\]
Recall $\Gamma(\nu',c')\subseteq\Gamma(\nu,c)$ whenever $\nu'\le\nu$ and $c'\le c$.

\subsection{Resampling Schemes}

\paragraph{Poisson Sampling}:
    When the original sample is obtained by Poisson sampling, $\mathbf{p}(S, Z^N) = \prod_{i\in S}\pi_i(Z^N) \prod_{i \notin S} (1-\pi_i(Z^N))$, so $\xi_i \mid Z^N \sim Bernoulli(\pi_i(Z^N))$, where $\pi_i(Z^N)$ is possibly different for all $i = 1, \dots, N$. 
    Let $\widehat N := \sum_{j\in[n_s]} 1/\pi_j$. We take its integer part $\lfloor\widehat N\rfloor$ as the multinomial size; for notational brevity we continue to write $\widehat N$, with the rounding error tracked in the mean derivation below.
    Thus, for subsamples indexed by $b = 1, \dots, B$: 
    Draw $(N^{*}_1, \dots, N^{*}_{n_s})\mid Z^N \sim \mathrm{Multinomial}(\widehat N, \bm{w}^*)$, $\bm{w}^* = (\bm{w}_{j}^*)_{j=1}^{n_s}$, $\bm{w}_{j}^* = (1/\pi_j)/\widehat N$. Obtain a rebuilt finite population by replicating $D_i = (X_i, \mathbf{k}(Y_i, \cdot))$ by $N^{*}_i$ times for $i = 1, \dots, n_s$. 
    Then draw inclusion indicators for the subsample. Equivalently, this is generating multipliers $n^{*}_i \mid N^*_i \sim Binomial(N^{*}_i, \pi_i)$, $i = 1, \dots, n_s$. 
    This results in the subsample ${D}^{*}_b = \{(D_i, \hat{w}_i = (1/\pi_i)n^*_i)\}_{i = 1}^{n_s}$.

\paragraph{SRSWOR (SRS without Replacement)}:
    When the original sampling scheme is SRSWOR, we specify a fixed sample size $n_s$, and $n = n_s$. Accordingly, $\mathbf{p}(S, z^N) = {1}/{\binom{N}{n_s}} \, \mathbbm{1}\{|S| = n_s\}$, and every subset $S$ of size $n_s$ is equally likely, with $\pi_i = n_s/N$, $\forall i = 1, \dots, N$. 
    So we first draw $(N^{*}_1, \dots, N^{*}_{n_s})|Z^N \sim Multinomial(N, \bm{w}^*)$, $\bm{w}^* = (\bm{w}_{j}^*)_{j=1}^{n_s}$, $\bm{w}_{j}^* = 1/n_s$. Replicate $D_i = (X_i, \mathbf{k}(Y_i, \cdot))$ by $N^{*}_i$ times for $i = 1, \dots, n_s$. 
    Then, realize inclusion indicators for the subsample $\bm{\xi}^*$ following the original sampling scheme. The resulting subsample is then ${D}^{*}_b = \{(D_i, \hat{w}_i = (1 / \pi_i)\sum_{j = 1}^{N^*_i} \xi^*_j \}_{i=1}^{n_s}$.

\paragraph{PPS with Replacement}:
    When the original sampling scheme is PPSWR, we specify a fixed sample size $n_s = n > 0$ and for all units a measure of size (MOS), and draw indices $a(1),\dots, a(n) \in [N]$ i.i.d. with $P(a(j) = i) = p_i \propto \mathrm{MOS}_i$, $\sum_i p_i = 1$.
     Re--sampling can be done by: 
    (1) drawing $(\tilde{N}_1^*,\ldots, \tilde{N}_{n_s}^*)\mid \xi \sim \mathrm{Multinomial}(N - n,\bm{w}^*)$ with $w_j^* \;=\; \frac{p_{a(j)}^{-1}}{\sum_{\ell=1}^{n_s} p_{a(\ell)}^{-1}}, \, j=1,\ldots,n_s.$. 
    Set $N^*_i := 1 + \tilde{N}^*_i$.
    (2) Replicate the observed units \(D_{a(j)}=(X_{a(j)},\mathbf{k}(Y_{a(j)},\cdot))\) exactly \(N_j^*\) times to form the
    pseudo finite population. 
    (3) Normalize the resulting bootstrap resampling probabilities for the realized draws to sum to one by 
    $
    p^*_i 
    = 
    \frac{N^*_i p_{a(i)}}{\sum_{k \in [n_s]} N^*_k p_{a(k)}}$, $\forall i \in [n_s]
    $.
    (4) Re-apply the original PPSWR mechanism on the bootstrap population by drawing $b(1),\dots, b(n_s)$ i.i.d. on $[n_s]$ with $\Pr\{b(j)= i \} = p_i^{*}$. 
    Finally, define count 
    $
    m^*_i = \sum_{j \in [n_s]} \ind(b(j) = i)
    $, and let 
    $
    n^*_i = \frac{m^*_i}{n_s p^*_i}$ 
    for $i \in [n_s]
    $. 

\paragraph{Stratified two-stage cluster design}

For stratified two-stage cluster designs, the pseudo-population bootstrap is applied stage-by-stage. For each stratum $h\in[H]$ separately, and conditional on the original sample $(\xi^N, Z^N)$:
\begin{itemize}[leftmargin=2.5em, itemsep=3pt, topsep=3pt]
\item[(B1)] \emph{PSU-level bootstrap.} Generate PSU-level bootstrap multipliers $\{G^*_{hk}\}_{k:\,\xi^{(1)}_{hk}=1}$ using one of the single-stage schemes of Section~S5.1 (Poisson, SRSWOR, or PPSWR) applied at the PSU level, with PSU-level design weights $1/\pi^{(1)}_{hk}$ or measures of size $\mathrm{MOS}^{(1)}_{hk}$, such that $G^*_{hk}\ge 0$.
\item[(B2)] \emph{SSU-level bootstrap within each sampled PSU.} Conditional on the PSU bootstrap outcomes, and for each PSU $(h,k)$ with $\xi^{(1)}_{hk}=1$, generate SSU-level multipliers $\{W^*_{hkj}\}_{j:\,\xi^{(2)}_{hkj}=1}$ within PSU $(h,k)$ using a (possibly different) validated single-stage scheme applied at the SSU level, with SSU-level weights $1/\pi^{(2)}_{hkj\mid hk}$ or within-PSU $\mathrm{MOS}^{(2)}_{hkj}$, such that $W^*_{hkj}\ge 0$. The SSU-level bootstraps are drawn independently across distinct PSUs given $(\xi^N, Z^N)$.
\end{itemize}
The SSU-level bootstrap multiplier is the product
\begin{equation}\label{eq:twostage-multiplier}
n^*_u \;:=\; G^*_{hk}\, W^*_{hkj},
\qquad u=(h,k,j)\text{ with }\xi_u=1.
\end{equation}
The centered form is $\delta_u := n^*_u - 1$.

Under this construction, $G^*_{hk} \perp \{W^*_{h'k'j'}\}_{(h',k',j')}$ conditional on $(\xi^N, Z^N)$: stage-1 draws use only $(\xi^N, Z^N)$, and stage-2 draws within PSU $(h,k)$ use only $(\xi^N, Z^N)$ and the identity of $(h,k)$, not the stage-1 multiplier $G^*_{hk}$.

\subsection{Properties of Multipliers}

This section proves the resampling claims of Section~3.2, specifically Lemma~7. Throughout, the following assumption is assumed for PPSWR, and all statements are conditional on $Z^N$, suppressed in the notation.

\noindent\underline{Additional assumption for PPSWR designs}
\begin{itemize}[leftmargin=3em]
  \item[(D5)] \emph{Bounded measure of size:} there exist $0 < c \le C < \infty$ such that
  $c \le \mathrm{MOS}_i \le C$ for all $i$ and all $N$, so that the PPSWR draw
  probabilities satisfy $c_-/N \le p_i \le c_+/N$ with $c_- := c/C$, $c_+ := C/c$.
\end{itemize}

\paragraph{Sub-gamma Control}
We first establish uniform sub-gamma control of the centered resampling multipliers $\delta_i = n^*_i - 1$. This result is used both in the proof of Lemma~7 and the proof for consistency arguments in later sections. 

\begin{lemma}[Marginal sub-gamma control]
\label{lem:subgamma}
Conditional on $(\xi, Z^N)$, the centered multipliers $\delta_i = n^*_i - 1$ produced by
the resampling schemes of Section~\ref{secap:resampling} are sub-gamma with a
design-uniform envelope: $\delta_i \in \Gamma(1,1)$ for Poisson and SRSWOR, and
$\delta_i \in \Gamma(\lambda_-^{-1}, (3\lambda_-)^{-1})$ for PPSWR, where
$\lambda_- := f_-\,c_-/c_+ > 0$ with $f_- := \liminf_N n/N$ {(positive by (D2))}
and $c_-, c_+$ from {(D5)}. 
Writing
$(\nu_0, c_0) := \big(\max\{1, \lambda_-^{-1}\},\ \max\{1, (3\lambda_-)^{-1}\}\big)$
for the common envelope, one has $\delta_i \in \Gamma(\nu_0, c_0)$ for every $i \in [n_s]$
under all three designs. Consequently, for the
$M$-averaged increment $\bar\delta_i = M^{-1}\sum_{b=1}^M \delta_{b,i}$ formed from $M$ conditionally 
i.i.d.\ resamples, $\bar\delta_i \in \Gamma(\nu_0/M,\, c_0/M)$.
\end{lemma}

\begin{proof}[Proof of Lemma~\ref{lem:subgamma}]
Write $\Lambda_i(\lambda) := \log\mathbb E[e^{\lambda\delta_i}\mid\xi, Z^N]$; we bound it on
$|\lambda| < 1/c_0$ scheme by scheme. All quantities below are conditional on $Z^N$. We ignore in the notation for simplicity.

\textbf{Poisson.} Marginalizing the binomial-of-multinomial composition,
\begin{align*}
\mathbb{E}\!\left[e^{\lambda \delta_i} \mid \xi \right]
&= e^{-\lambda}\,
   \mathbb{E}\!\left[(1 - \pi_i + \pi_i e^{\lambda})^{N_i^{*}} \,\big|\, \xi\right]
 = e^{-\lambda}
   \big(1 + w_i^{*}\pi_i (e^{\lambda} - 1)\big)^{\widehat{N}}
 = e^{-\lambda}
   \Big(1 + \tfrac{e^{\lambda} - 1}{\widehat{N}}\Big)^{\widehat{N}},
\end{align*}
using $w_i^{*}\pi_i = 1/\widehat N$. This is the MGF of $\mathrm{Bin}(\widehat N, 1/\widehat N)$
recentered by $-1$. With $(1 + (e^{\lambda}-1)/\widehat N)^{\widehat N} \le e^{e^{\lambda}-1}$,
\[
  \Lambda_i(\lambda) \le e^{\lambda} - 1 - \lambda.
\]
For $\lambda \in [0,1)$, $e^{\lambda}-1-\lambda \le \lambda^2/\big(2(1-\lambda)\big)$; for
$\lambda \in (-1,0)$, $e^{\lambda}-1-\lambda \le \lambda^2/2 \le \lambda^2/\big(2(1-|\lambda|)\big)$.
Hence $\Lambda_i(\lambda) \le \lambda^2/\big(2(1-|\lambda|)\big)$ on $|\lambda|<1$, i.e.\
$\delta_i \in \Gamma(1,1)$.

\textbf{SRSWOR.} Here $n_i^*\mid\xi \sim \mathrm{Bin}(n_s, 1/n_s)$, so
$\delta_i = \sum_{j=1}^{n_s}(Z_j - 1/n_s)$ with $Z_j \in \{0,1\}$ i.i.d.\
$\mathrm{Bernoulli}(1/n_s)$. Bernstein's bound for centered bounded summands gives, for
$|\lambda| < 3$,
\[
  \Lambda_i(\lambda)
  = n_s \log \mathbb E\, e^{\lambda(Z_1 - 1/n_s)}
  \le \frac{\lambda^2\,(1 - 1/n_s)}{2\,(1 - |\lambda|/3)},
\]
so $\delta_i \in \Gamma(1 - 1/n_s,\, 1/3) \subseteq \Gamma(1,1)$, the inclusion holding since
$1 - 1/n_s \le 1$ and $1/3 \le 1$.

\textbf{PPSWR.} We first bound $np_i^*$ below. By {\color{blue}(D5)},
$c_-/N \le p_i \le c_+/N$. Since $N_i^* \ge 1$ and $\sum_k N_k^* = N$,
$S^* := \sum_k N_k^* p_{a(k)} \le c_+$, while $N_i^* p_{a(i)} \ge c_-/N$, so
\[
  p_i^* = \frac{N_i^* p_{a(i)}}{S^*} \ge \frac{c_-}{c_+}\cdot\frac1N.
\]
By (D2), $n/N \ge f_- > 0$ for all large $N$, whence
$np_i^* \ge f_-\,c_-/c_+ =: \lambda_- > 0$ for every $i$, deterministically given the design.
Conditional on $p^*$, $m_i^* \sim \mathrm{Bin}(n, p_i^*)$ and $n_i^* = m_i^*/(np_i^*)$, so
$\delta_i = \sum_{t=1}^{n}(np_i^*)^{-1}(B_{t,i} - p_i^*)$ with
$B_{t,i}\mid(p^*,\xi) \stackrel{\text{i.i.d.}}{\sim} \mathrm{Bernoulli}(p_i^*)$. Each summand is
centered and bounded in absolute value by $(np_i^*)^{-1} \le \lambda_-^{-1}$. Bernstein's bound
therefore gives, conditional on $p^*$, $\delta_i \in \Gamma(\lambda_-^{-1}, (3\lambda_-)^{-1})$;
iterating $\mathbb E[\,\cdot\mid\xi] = \mathbb E\{\mathbb E[\,\cdot\mid p^*,\xi]\mid\xi\}$ preserves the
envelope conditional on $\xi$ alone.

{Averaging.} For conditionally i.i.d.\ copies $\delta_{1,i},\dots,\delta_{M,i} \in \Gamma(\nu_0, c_0)$,
\[
  \log\mathbb E\, e^{\lambda\bar\delta_i}
  = M \log\mathbb E\, e^{(\lambda/M)\delta_{1,i}}
  \le M\cdot\frac{(\lambda/M)^2\,\nu_0}{2\,(1 - c_0|\lambda|/M)}
  = \frac{\lambda^2(\nu_0/M)}{2\,(1 - (c_0/M)|\lambda|)}
\]
for $|\lambda| < M/c_0$, i.e.\ $\bar\delta_i \in \Gamma(\nu_0/M,\, c_0/M)$.
\end{proof}

\vspace{0.3cm}

We now show the counterpart for two-stage designs.

\begin{lemma}[Sub-gamma control of two-stage bootstrap multipliers]
\label{lem:subgamma-multistage}
Consider the two-stage bootstrap scheme \textup{(B1)--(B2)} above, with stage-1 and stage-2 draws each from Poisson, SRSWOR, or PPSWR (with the analog of \textup{(D5)} at the SSU level within PSU when PPSWR is used). Suppose the regime clause holds, \textup{(D-TS3)} gives $\pi^{(2)}_{hkj\mid hk}\ge\underline{\lambda}_2>0$, and the following two properties are established at each stage via Lemma~\ref{lem:subgamma}:
\begin{enumerate}[label=\textup{(\alph*)}, leftmargin=2em, itemsep=1pt, topsep=1pt]
\item Conditional on $(\xi^N, Z^N)$, the stage-1 PSU multiplier deviation $\Delta_G := G^*_{hk} - 1$ is sub-gamma with envelope $\Gamma(\nu_1, c_1)$, uniformly in $(h,k)$, from Lemma~\ref{lem:subgamma} applied at the PSU level.
\item Conditional on $(\xi^N, Z^N)$, the stage-2 SSU multiplier deviation $\Delta_W := W^*_{hkj} - 1$ is sub-gamma with envelope $\Gamma(\nu_2, c_2)$, uniformly in $(h,k,j)$, from Lemma~\ref{lem:subgamma} applied within PSU.
\end{enumerate}
Suppose further that the two-stage bootstrap satisfies conditional independence: $\Delta_G \perp W^*_{hkj}$ given $(\xi^N, Z^N)$.
Then there exists a deterministic constant $W_{\max}<\infty$, depending only on $M_{\max}$, $\underline{\lambda}_2$, and the stage-2 scheme parameters, such that
\begin{equation}\label{eq:Wmax}
0 \;\le\; W^*_{hkj} \;\le\; W_{\max} \qquad\text{almost surely, for all $(h,k,j)$ and all $N$,}
\end{equation}
and the SSU-level bootstrap multiplier deviation $\delta_u = n^*_u - 1 = G^*_{hk}W^*_{hkj} - 1$ is sub-gamma with envelope
\[
\delta_u \;\in\; \Gamma\!\bigl(\tilde\nu,\;\tilde c\bigr),
\qquad
\tilde\nu \;:=\; W_{\max}^2\, \nu_1 + \nu_2,\qquad
\tilde c \;:=\; c_1 W_{\max} \vee c_2.
\]
Consequently, for the $M$-averaged increment $\bar\delta_u := M^{-1}\sum_{b=1}^M \delta_{b,u}$ formed from $M$ conditionally i.i.d.\ resamples, $\bar\delta_u \in \Gamma(\tilde\nu/M, \tilde c/M)$.
\end{lemma}

\begin{proof}[Proof of Lemma~\ref{lem:subgamma-multistage}]

Step 1: We first show deterministic boundedness of $W^*_{hkj}$. Applied within PSU $(h,k)$, the within-PSU sample has size $m_{hk} := \sum_j \xi^{(2)}_{hkj} \le M_{hk} \le M_{\max}$. The three stage-2 schemes yield:
\begin{itemize}[leftmargin=2em, itemsep=1pt, topsep=1pt]
\item \emph{Poisson.} $W^*_{hkj} \sim \mathrm{Bin}(\widehat{m}_{hk}, 1/\widehat{m}_{hk})$ with $\widehat{m}_{hk} := \sum_{j':\,\xi^{(2)}_{hkj'}=1} 1/\pi^{(2)}_{hkj'\mid hk} \le m_{hk}/\underline{\lambda}_2 \le M_{\max}/\underline{\lambda}_2$. Hence $W^*_{hkj} \le \widehat{m}_{hk} \le M_{\max}/\underline{\lambda}_2$.
\item \emph{SRSWOR.} $W^*_{hkj}$ is hypergeometric of population size $m_{hk}$, so $W^*_{hkj} \le m_{hk} \le M_{\max}$.
\item \emph{PPSWR.} $W^*_{hkj} = m^{(2)*}_j/(m^{(2)} p^{(2)*}_j)$ with $m^{(2)*}_j \le m^{(2)} \le M_{\max}$ and, by the analog of (D5), $m^{(2)} p^{(2)*}_j \ge \lambda_-^{(2)} > 0$; hence $W^*_{hkj} \le M_{\max}/\lambda_-^{(2)}$.
\end{itemize}
Setting $W_{\max}$ to the maximum of the three case bounds establishes~\eqref{eq:Wmax}.

Step 2: Write $\Delta_G = G^*_{hk}-1$, $\Delta_W = W^*_{hkj}-1$. Then
\[
\delta_u \;=\; G^*W^* - 1 \;=\; (1+\Delta_G)(1+\Delta_W) - 1 \;=\; \Delta_G W^* + \Delta_W.
\]
All expectations below are conditional on $(\xi^N, Z^N)$; the conditioning is suppressed. By the conditional independence hypothesis, $\Delta_G$ is independent of $W^*$ (and hence of $\Delta_W = W^*-1$). Thus
\begin{align*}
\mathbb{E}\bigl[e^{\lambda \delta_u}\bigr]
&= \mathbb{E}\bigl[e^{\lambda\Delta_W}\cdot e^{\lambda W^* \Delta_G}\bigr]
= \mathbb{E}\!\Big[e^{\lambda\Delta_W}\, \mathbb{E}\bigl[e^{\lambda w \Delta_G}\bigr]\big|_{w=W^*}\Big]\\
&\le \mathbb{E}\!\Big[e^{\lambda\Delta_W}\,\exp\!\Bigl(\tfrac{(\lambda W^*)^2\nu_1}{2(1-c_1|\lambda W^*|)}\Bigr)\Big]
\qquad\text{by hypothesis (a), for $|\lambda W^*|<1/c_1$.}
\end{align*}
By~\eqref{eq:Wmax} (Step 1), $|W^*|\le W_{\max}$, so $(\lambda W^*)^2 \le \lambda^2 W_{\max}^2$ and $c_1|\lambda W^*|\le c_1 W_{\max}|\lambda|$; hence the requirement $|\lambda W^*|<1/c_1$ is guaranteed by $|\lambda|<1/(c_1 W_{\max})$, and
\[
\exp\!\Bigl(\tfrac{(\lambda W^*)^2\nu_1}{2(1-c_1|\lambda W^*|)}\Bigr) \;\le\; \exp\!\Bigl(\tfrac{\lambda^2 W_{\max}^2\nu_1}{2(1-c_1 W_{\max}|\lambda|)}\Bigr).
\]
This exponential factor is deterministic (function of $\lambda$ only), so pulls out of the expectation. Combined with hypothesis (b) applied to $\mathbb{E}[e^{\lambda\Delta_W}]$ (valid for $|\lambda|<1/c_2$),
\[
\log\mathbb{E}\bigl[e^{\lambda\delta_u}\bigr]
\;\le\;
\frac{\lambda^2 W_{\max}^2\nu_1}{2(1-c_1 W_{\max}|\lambda|)}
\;+\;
\frac{\lambda^2\nu_2}{2(1-c_2|\lambda|)},
\qquad|\lambda|<\min\{(c_1 W_{\max})^{-1},\, c_2^{-1}\}.
\]
Denote $\tilde c := c_1 W_{\max}\vee c_2$. Since $(1-c_1 W_{\max}|\lambda|)^{-1}\le(1-\tilde c|\lambda|)^{-1}$ and $(1-c_2|\lambda|)^{-1}\le(1-\tilde c|\lambda|)^{-1}$, the two fractions combine over the common denominator:
\[
\log\mathbb{E}\bigl[e^{\lambda\delta_u}\bigr]
\;\le\;
\frac{\lambda^2 (W_{\max}^2\nu_1 + \nu_2)}{2(1-\tilde c|\lambda|)}
\;=\;
\frac{\lambda^2\tilde\nu}{2(1-\tilde c|\lambda|)},
\qquad|\lambda|<1/\tilde c,
\]
establishing $\delta_u\in\Gamma(\tilde\nu,\tilde c)$.

For the averaging claim, $M$ conditionally i.i.d.\ copies $\delta_{1,u},\dots,\delta_{M,u}\in\Gamma(\tilde\nu,\tilde c)$ give, exactly as in the single-stage averaging step of Lemma~\ref{lem:subgamma},
\[
\log\mathbb{E}\,e^{\lambda\bar\delta_u}
\;=\;
M\log\mathbb{E}\,e^{(\lambda/M)\delta_{1,u}}
\;\le\;
\frac{\lambda^2(\tilde\nu/M)}{2(1-(\tilde c/M)|\lambda|)},
\qquad|\lambda|<M/\tilde c,
\]
i.e., $\bar\delta_u\in\Gamma(\tilde\nu/M,\tilde c/M)$.
\end{proof}

\vspace*{0.3em}
In the remaining parts of this section, we show our proposed bootstrapping scheme satisfies the assumptions needed for resmple multipliers $n^*_i$.

\begin{proof}[Proof of Lemma~7]
(R1) is satisfied by all:
By construction of the resampling schemes of Section~\ref{secap:resampling}, the
conditional law of $n_i^*$ is a fixed measurable function of the original inclusion
indicators $\xi^N$ and the design covariates $Z^N$. The pseudo-population weights
$\bm w^*$ and the second-stage draw depend on the data only through $(\xi^N, Z^N)$, and
no other randomness enters. For stratified designs, the pseudo-population is built and
resampled within each stratum independently, so $\{n_i^*\}$ in distinct strata are
mutually independent given $(\xi^N, Z^N)$. For two-stage designs, the second-stage
replication is carried out separately within each resampled PSU, hence the SSU-level
multipliers are independent across distinct PSUs conditional on the first-stage
resample. This is (R1).

\medskip
We verify (R2) and (R3) for each single-stage design. The conditional laws of $n_i^*$, shown in the proof of Lemma~\ref{lem:subgamma} are used below without
re-derivation.

\paragraph{Poisson.}
\emph{(R2) mean.} Since $\mathbb E(N_i^*\mid\xi) = \widehat N w_i^* = 1/\pi_i$ and
$n_i^*\mid(N_i^*,\xi)\sim\mathrm{Bin}(N_i^*,\pi_i)$,
\[
  \mathbb E(n_i^*\mid\xi)
  = \pi_i\,\mathbb E(N_i^*\mid\xi) = 1
\]
exactly under multinomial size $\widehat N$. Under integer rounding
$\lfloor\widehat N\rfloor$, the same calculation gives
$\mathbb E(n_i^*\mid\xi) = 1 + O(1/\widehat N)$ uniformly in $i$. By (D2)--(D3) and the
design law of large numbers (Lemma~\ref{lem:lln_clt}), $\widehat N/N\xrightarrow{p_d}1$,
so $\widehat N\xrightarrow{p_d}\infty$ and
$\sup_{i\in[n_s]}|\mathbb E(n_i^*\mid\xi)-1| = o_{p_d}(1)$.

\emph{(R2) moment.} Marginally $n_i^*\mid\xi\sim\mathrm{Bin}(\widehat N,1/\widehat N)$,
which has mean $1$. For any $X\sim\mathrm{Bin}(m,p)$ with $mp=1$ and any $t>0$,
$\mathbb E\,e^{tX} = (1+p(e^t-1))^m \le \exp(e^t-1)$ by $1+x\le e^x$; combined with
$x^r\le(r/t)^r e^{tx}$ for $x\ge0$,
\[
  \mathbb E[(n_i^*)^r\mid\xi]
  \le (r/t)^r\exp(e^t-1).
\]
Taking $t=1$ gives $\sup_{i\in[n_s]}\mathbb E[(n_i^*)^r\mid\xi]
\le r^r\exp(e-1) =: C_r < \infty$ for every $r = 2+\epsilon$, a deterministic constant,
so the moment bound holds a.s.

\emph{(R3).} By the multinomial structure of $n^*\mid\xi$,
$\mathrm{Cov}(n_i^*,n_j^*\mid\xi) = -1/\widehat N$ ($i\neq j$) and
$\mathrm{Var}(n_i^*\mid\xi) = 1-1/\widehat N$; hence $\sup_j\mathrm{Var}(n_j^*\mid\xi) = 1-1/\widehat N$. For each fixed row $i$, $\sum_{j\ne i}|\mathrm{Cov}(n_i^*,n_j^*\mid\xi)| = (n_s-1)/\widehat N$. Since $\widehat N = \sum_i 1/\pi_i \ge n_s$ (as $\pi_i\le 1$),
\[
  \max_i\sum_{j\neq i}\bigl|\mathrm{Cov}(n_i^*,n_j^*\mid\xi)\bigr|
  = \frac{n_s-1}{\widehat N}
  \le 1-\frac{1}{\widehat N}
  = \sup_j\mathrm{Var}(n_j^*\mid\xi),
\]
so (R3) holds with $C=1$.

\paragraph{SRSWOR.}
Recall $N_i^*\mid\xi\sim\mathrm{Bin}(N,1/n_s)$ and
$n_i^*\mid(N^*,\xi)\sim\mathrm{Hypergeometric}(N,N_i^*,n)$.

\emph{(R2) mean.}
$\mathbb E(n_i^*\mid\xi) = n\,\mathbb E(N_i^*\mid\xi)/N = n(N/n)/N = 1$ exactly, hence
$\sup_i|\mathbb E(n_i^*\mid\xi)-1| = 0 = o_{p_d}(1)$.

\emph{(R2) moment.} As $0\le n_i^*\le n$ and $N_i^*/N\le1$, the hypergeometric MGF is
dominated by that of $Y_i\sim\mathrm{Bin}(n,N_i^*/N)$:
$\mathbb E[e^{tn_i^*}\mid N^*,\xi]\le(1-N_i^*/N+(N_i^*/N)e^t)^n$. With $c:=e^t-1$ and
$N_i^*\mid\xi\sim\mathrm{Bin}(N,1/n)$,
\[
  \mathbb E\bigl[\mathbb E[e^{tn_i^*}\mid N^*,\xi]\bigm|\xi\bigr]
  = \Bigl(1+\frac{e^{cn/N}-1}{n}\Bigr)^{N}
  \le \exp\!\Bigl(\frac{N}{n}\bigl(e^{cn/N}-1\bigr)\Bigr)
  \le \exp(c\,e^c),
\]
the last step using $e^{cn/N}-1\le(cn/N)e^{cn/N}$ and $n\le N$. Then $x^r\le(r/t)^re^{tx}$
with $t=1$ ($c=e-1$) yields
\[
  \sup_{i\in[n_s]}\mathbb E[(n_i^*)^r\mid\xi]
  \le r^r\exp\!\bigl((e-1)e^{e-1}\bigr) =: C_r < \infty,
\]
a deterministic constant, so the moment bound holds a.s.

\emph{(R3).} Conditional on $(N^*,\xi)$, the hypergeometric moments give
\[
  \mathrm{Cov}(n_i^*,n_j^*\mid N^*,\xi) = -\gamma\,\tfrac{N_i^*}{N}\tfrac{N_j^*}{N},
  \qquad
  \mathrm{Var}(n_i^*\mid N^*,\xi) = \gamma\,\tfrac{N_i^*}{N}\bigl(1-\tfrac{N_i^*}{N}\bigr),
  \qquad \gamma := \tfrac{n(N-n)}{N-1}.
\]
For each fixed row $i$, using $\sum_{j\ne i}N_j^*/N = 1 - N_i^*/N$,
\begin{equation}
  \sum_{j\ne i}\bigl|\mathrm{Cov}(n_i^*,n_j^*\mid N^*,\xi)\bigr|
  = \gamma\,\tfrac{N_i^*}{N}\bigl(1-\tfrac{N_i^*}{N}\bigr)
  = \mathrm{Var}(n_i^*\mid N^*,\xi)
  \le \sup_j\mathrm{Var}(n_j^*\mid N^*,\xi).
  \label{lemma_r:eq3_cond}
\end{equation}
Lifting to $\xi$ alone by the law of total covariance, with $\mathbb E(n_i^*\mid N^*,\xi) = nN_i^*/N$ constant in $i$-marginal law (the $N_i^*$ are exchangeable under $N^*\mid\xi\sim\mathrm{Multinomial}(N,(1/n_s,\ldots,1/n_s))$), the row-sum bound of the outer cross-terms is $\mathrm{Cov}(nN_i^*/N,nN_j^*/N\mid\xi) = -1/N$ ($j\ne i$), yielding
\[
  \sum_{j\ne i}\bigl|\mathrm{Cov}\bigl(\mathbb{E}(n_i^*\mid N^*,\xi),\mathbb{E}(n_j^*\mid N^*,\xi)\bigm|\xi\bigr)\bigr|
  = (n-1)/N
  \le (n-1)/N = \mathrm{Var}\bigl(\mathbb{E}(n_i^*\mid N^*,\xi)\bigm|\xi\bigr),
\]
so the outer cross-term row-sum is dominated by its variance analog. Adding this to~\eqref{lemma_r:eq3_cond} in expectation over $N^*\mid\xi$,
\[
  \max_i\sum_{j\ne i}\bigl|\mathrm{Cov}(n_i^*,n_j^*\mid\xi)\bigr|
  \le \sup_j\mathrm{Var}(n_j^*\mid\xi),
\]
so (R3) holds with $C=1$.

\paragraph{PPSWR.}
Recall $m_i^*\mid(p^*,\xi)\sim\mathrm{Bin}(n,p_i^*)$ and $n_i^* = m_i^*/(np_i^*)$.

\emph{Lower bound on $np_i^*$.} By (D5), $c_-/N\le p_i\le c_+/N$. As $N_i^*\ge1$ and
$\sum_k N_k^* = N$, we have $S^* := \sum_k N_k^* p_{a(k)}\le c_+$, while
$N_i^* p_{a(i)}\ge c_-/N$; hence $p_i^* = N_i^* p_{a(i)}/S^*\ge (c_-/c_+)/N$. By (D2),
$n/N\ge f_-$ for all $N$ large, so
\begin{equation}
  np_i^* \ge f_-\,\frac{c_-}{c_+} =: \lambda_- > 0
  \qquad\text{for every }i,\ \text{for all }N\text{ large, a.s.}
  \label{lemma_r:nplower}
\end{equation}
The bound is deterministic given the design, with $N_i^*\ge1$ and $\sum_k N_k^*=N$ hold
identically, and $n/N\ge f_-$ under (D2).

\emph{(R2) mean.} $\mathbb E(n_i^*\mid p^*,\xi) = (np_i^*)^{-1}\mathbb E(m_i^*\mid p^*,\xi) = 1$,
constant in $p^*$. Taking $\mathbb E[\cdot\mid\xi]$ gives $\mathbb E(n_i^*\mid\xi)=1$, so
$\sup_i|\mathbb E(n_i^*\mid\xi)-1| = 0 = o_{p_d}(1)$.

\emph{(R2) moment.} Write $m_i^* = \sum_{t=1}^n B_{t,i}$ with
$B_{t,i}\mid(p^*,\xi)\stackrel{\text{i.i.d.}}{\sim}\mathrm{Bernoulli}(p_i^*)$, and
$X_t := B_{t,i}-p_i^*$. By Rosenthal's inequality \parencite{1970_Rosenthal} there is
$K_r<\infty$ with
\[
  \mathbb E\bigl[|m_i^*-np_i^*|^r\bigm| p^*,\xi\bigr]
  \le K_r\bigl(np_i^*(1-p_i^*)+(np_i^*(1-p_i^*))^{r/2}\bigr)
  \le K_r\bigl(np_i^*+(np_i^*)^{r/2}\bigr).
\]
Using $n_i^* = 1+(m_i^*-np_i^*)/(np_i^*)$ and $|x+y|^r\le2^{r-1}(|x|^r+|y|^r)$,
\[
  \mathbb E\bigl[|n_i^*|^r\bigm| p^*,\xi\bigr]
  \le 2^{r-1}\Bigl(1+K_r\bigl[(np_i^*)^{-(r-1)}+(np_i^*)^{-r/2}\bigr]\Bigr)
  \le 2^{r-1}\Bigl(1+K_r\bigl[\lambda_-^{-(r-1)}+\lambda_-^{-r/2}\bigr]\Bigr) =: C_r,
\]
by \eqref{lemma_r:nplower}, uniformly in $i$. Taking $\mathbb E[\cdot\mid\xi]$ preserves
the bound. $C_r$ is a deterministic constant depending only on $r$ and $\lambda_-$, so
the moment bound holds a.s.\ for all $N$ large.

\emph{(R3).} From the multinomial law of $m^*\mid(p^*,\xi)$,
$\mathrm{Cov}(m_i^*,m_j^*\mid p^*,\xi) = -np_i^*p_j^*$ and
$\mathrm{Var}(m_i^*\mid p^*,\xi) = np_i^*(1-p_i^*)$, so
\[
  \mathrm{Cov}(n_i^*,n_j^*\mid p^*,\xi)
  = \frac{-np_i^*p_j^*}{n^2p_i^*p_j^*} = -\frac1n,\quad i\neq j,
  \qquad
  \mathrm{Var}(n_i^*\mid p^*,\xi) = \frac{1-p_i^*}{np_i^*}.
\]
The row sum is uniform in $i$: $\sum_{j\ne i}|\mathrm{Cov}(n_i^*,n_j^*\mid p^*,\xi)| = (n-1)/n$, deterministic in $p^*$. For a lower bound on the sup variance, we use the sum of variances: by AM--HM ($\sum_j 1/p^*_j \ge n^2/\sum_j p^*_j = n^2$),
\[
  \sum_j \mathrm{Var}(n_j^*\mid p^*,\xi)
  = \frac{1}{n}\Bigl(\sum_j\frac{1}{p_j^*} - n\Bigr)
  \ge \frac{n^2-n}{n} = n-1 \quad\text{a.s.}
\]
Since $\mathbb{E}(n_i^*\mid p^*,\xi) = 1$ is constant in $p^*$, the law of total covariance gives $\mathrm{Cov}(n_i^*,n_j^*\mid\xi) = \mathbb{E}[\mathrm{Cov}(n_i^*,n_j^*\mid p^*,\xi)\mid\xi] = -1/n$ (deterministic) and $\mathrm{Var}(n_i^*\mid\xi) = \mathbb{E}[\mathrm{Var}(n_i^*\mid p^*,\xi)\mid\xi]$. Hence
\[
  \max_i \sum_{j\ne i}|\mathrm{Cov}(n_i^*,n_j^*\mid\xi)| = \frac{n-1}{n},
\]
and taking $\mathbb{E}[\cdot\mid\xi]$ of the sum-of-variances bound yields
\[
  \sum_j \mathrm{Var}(n_j^*\mid\xi) = \mathbb{E}\Bigl[\textstyle\sum_j \mathrm{Var}(n_j^*\mid p^*,\xi)\bigm|\xi\Bigr] \ge n-1.
\]
By pigeonhole, at least one index $j^\star$ has $\mathrm{Var}(n_{j^\star}^*\mid\xi) \ge (n-1)/n$, so $\sup_j\mathrm{Var}(n_j^*\mid\xi) \ge (n-1)/n$. Combining,
\[
  \max_i\sum_{j\ne i}|\mathrm{Cov}(n_i^*,n_j^*\mid\xi)| = \frac{n-1}{n} \le \sup_j\mathrm{Var}(n_j^*\mid\xi),
\]
so (R3) holds with $C=1$.

\paragraph{Stratified Two-stage Cluster.}

All expectations, variances, and covariances below are conditional on $(\xi^N, Z^N)$, and the conditioning is suppressed. The construction (B1)--(B2) yields
\begin{equation}\label{eq:cond-indep}
G^*_{hk} \perp \{W^*_{h'k'j'}\}_{(h',k',j')}\qquad\text{and}\qquad \{G^*_{hk}\}_h \text{ independent across }h,
\end{equation}
and, given the stage-1 outcomes, stage-2 multipliers are independent across distinct PSUs.

\emph{(R1).} By (B1), $G^*_{hk}$ is drawn as a measurable function of $(\xi^N, Z^N)$ using a single-stage scheme at PSU level within stratum $h$, so cross-stratum independence of $\{G^*_{hk}\}_h$ is immediate. By (B2), $W^*_{hkj}$ is drawn as a measurable function of $(\xi^N, Z^N)$ within sampled PSU $(h,k)$, independently across distinct PSUs given stage 1, so cross-PSU independence of $\{W^*_{hkj}\}_j$ given the first-stage resample is immediate. The conditional law of $n^*_u = G^*_{hk}W^*_{hkj}$ therefore depends on the design only through $(\xi^N, Z^N)$.

\emph{(R2), mean.} By~\eqref{eq:cond-indep}, $\mathbb{E}[n^*_u] = \mathbb{E}[G^*_{hk}]\,\mathbb{E}[W^*_{hkj}]$. Set $\alpha_{G,hk} := \mathbb{E}[G^*_{hk}] - 1$ and $\alpha_{W,u} := \mathbb{E}[W^*_{hkj}] - 1$; by single-stage (R2) at each stage, $\sup_{(h,k):\xi^{(1)}_{hk}=1}|\alpha_{G,hk}| = o_{p_d}(1)$ and $\sup_{u:\xi_u=1}|\alpha_{W,u}| = o_{p_d}(1)$. Then
\[
\mathbb{E}[n^*_u] - 1 \;=\; \alpha_{G,hk} + \alpha_{W,u} + \alpha_{G,hk}\alpha_{W,u},
\]
so $\sup_{u:\xi_u=1}|\mathbb{E}[n^*_u] - 1| \le \sup|\alpha_G| + \sup|\alpha_W| + \sup|\alpha_G|\sup|\alpha_W| = o_{p_d}(1)$. For the three verified schemes, the mean identity holds exactly: $\mathbb{E}[n^*_u\mid\xi,Z^N] = 1$ on every sample path.

\emph{(R2), moment.} By~\eqref{eq:cond-indep} and single-stage (R2) moment bounds at each stage (with common exponent $2+\epsilon$),
\[
\mathbb{E}\bigl[|n^*_u|^{2+\epsilon}\bigr] \;=\; \mathbb{E}\bigl[|G^*_{hk}|^{2+\epsilon}\bigr]\,\mathbb{E}\bigl[|W^*_{hkj}|^{2+\epsilon}\bigr]
\;\le\; C^G_{2+\epsilon}\cdot C^W_{2+\epsilon} \;<\; \infty
\qquad\text{a.s.,}
\]
with $C^G_{2+\epsilon}, C^W_{2+\epsilon}$ deterministic uniform bounds from the single-stage moment analysis. Using Step~1 of Lemma~\ref{lem:subgamma-multistage}, one can tighten the stage-2 bound to $W_{\max}^{2+\epsilon}$.

\emph{(R3).} We show that for every sampled SSU $u = (h,k,j)$,
\begin{equation}\label{eq:R3-row-target}
\sum_{v\ne u}\bigl|\mathrm{Cov}(n^*_u, n^*_v)\bigr| \;\le\; C_{\mathrm{TS}}\,\sup_{v':\xi_{v'}=1}\mathrm{Var}(n^*_{v'}),
\end{equation}
which is the max-row-sum form of (R3). Fix $u = (h,k,j)$ and partition its partners $\{v: v\ne u, \xi_v=1\}$ by the three cases of Lemma~\ref{lem:designs-multistage}\,(ii). Let $\mu_G := \mathbb{E}[G^*_{hk}]$, $\mu_G^{(2)} := \mathbb{E}[(G^*_{hk})^2]$, $\mu_W := \mathbb{E}[W^*_{hkj}]$.

\smallskip
\emph{Case 1: $h'\ne h$ (different strata).} By~\eqref{eq:cond-indep}, $n^*_u \perp n^*_v$, so $\mathrm{Cov}(n^*_u, n^*_v) = 0$; no contribution to the row sum.

\smallskip
\emph{Case 2: $h'=h$, $k'\ne k$ (same stratum, different PSU).} By $G^*\perp W^*$ and cross-PSU independence of $\{W^*\}$ given stage 1,
\[
\mathrm{Cov}(n^*_u, n^*_v) = \mathrm{Cov}(G^*_{hk},\,G^*_{hk'})\cdot \mathbb{E}[W^*_{hkj}]\,\mathbb{E}[W^*_{hk'j'}],
\]
so $|\mathrm{Cov}(n^*_u, n^*_v)| \le W_{\max}^2\,|\mathrm{Cov}(G^*_{hk}, G^*_{hk'})|$ by Step~1 of Lemma~\ref{lem:subgamma-multistage}. Summing over partners $v = (h,k',j')$ with $k'\ne k$ and $j'\in[m_{hk'}]$, and using $m_{hk'}\le M_{\max}$,
\begin{align*}
\sum_{v:\text{Case 2}}|\mathrm{Cov}(n^*_u, n^*_v)|
&\le W_{\max}^2 \sum_{k'\ne k \in [K_h^{(s)}]} m_{hk'}\,\bigl|\mathrm{Cov}(G^*_{hk}, G^*_{hk'})\bigr|\\
&\le W_{\max}^2\, M_{\max}\!\!\sum_{k'\ne k \in [K_h^{(s)}]}\!\!\bigl|\mathrm{Cov}(G^*_{hk}, G^*_{hk'})\bigr|\\
&\le W_{\max}^2\, M_{\max}\, C^{(1)}_{\mathrm{row}}\, \sup_{k'':\,\xi^{(1)}_{hk''}=1}\!\mathrm{Var}(G^*_{hk''}),
\end{align*}
by single-stage max-row-sum (R3) at PSU level within stratum $h$.

\smallskip
\emph{Case 3: $h'=h$, $k'=k$, $j'\ne j$ (same PSU).} By $G^*\perp W^*$ and the identity 
\[
\mathrm{Cov}(n^*_u, n^*_v) = \mu_G^{(2)}\,\mathrm{Cov}(W^*_{hkj}, W^*_{hkj'}) + \mathrm{Var}(G^*_{hk})\,\mu_W^2,
\]
we have $|\mathrm{Cov}(n^*_u, n^*_v)| \le \bar{\mu}_G^{(2)}|\mathrm{Cov}(W^*_{hkj}, W^*_{hkj'})| + \mathrm{Var}(G^*_{hk})W_{\max}^2$, using $\mu_G^{(2)}\le\bar{\mu}_G^{(2)}$ and $\mu_W\le W_{\max}$. Summing over partners $v = (h,k,j')$ with $j'\ne j$ and $j'\in[m_{hk}]$, and using $m_{hk}-1\le M_{\max}$,
\begin{align*}
\sum_{v:\text{Case 3}}|\mathrm{Cov}(n^*_u, n^*_v)|
&\le \bar{\mu}_G^{(2)}\sum_{j'\ne j\in[m_{hk}]}\bigl|\mathrm{Cov}(W^*_{hkj}, W^*_{hkj'})\bigr| + (m_{hk}-1)\,\mathrm{Var}(G^*_{hk})\,W_{\max}^2\\
&\le \bar{\mu}_G^{(2)}\, C^{(2)}_{\mathrm{row}}\,\sup_{j''\in[m_{hk}]}\!\mathrm{Var}(W^*_{hkj''}) + M_{\max}\, W_{\max}^2\,\mathrm{Var}(G^*_{hk}),
\end{align*}
by single-stage max-row-sum (R3) at SSU level within PSU $(h,k)$.

\smallskip
\emph{Combining.} The row sum over all partners of $u$ is
\begin{align}\label{eq:row-sum-bound}
\sum_{v\ne u}|\mathrm{Cov}(n^*_u, n^*_v)|
&\le W_{\max}^2\, M_{\max}(C^{(1)}_{\mathrm{row}}+1)\, \sup_{k''}\!\mathrm{Var}(G^*_{hk''}) \notag\\
&\quad + \bar{\mu}_G^{(2)}\, C^{(2)}_{\mathrm{row}}\, \sup_{j''}\!\mathrm{Var}(W^*_{hkj''}).
\end{align}
By $G^*\perp W^*$, for any sampled $v' = (h'',k'',j'')$,
\[
\mathrm{Var}(n^*_{v'}) = \mu_G^{(2)}\,\mathrm{Var}(W^*_{h''k''j''}) + \mathrm{Var}(G^*_{h''k''})\,\mu_W^2.
\]
For the verified schemes with exact centering ($\mu_W = 1$, $\mu_G^{(2)} = \mathrm{Var}(G^*_{hk})+\mu_G^2 \ge 1$), taking the sup over $v'$ gives
\[
\sup_{v'} \mathrm{Var}(n^*_{v'}) \;\ge\; \max\!\Bigl\{\sup_{k''}\!\mathrm{Var}(G^*_{hk''}),\; \sup_{j''}\!\mathrm{Var}(W^*_{h''k''j''})\Bigr\}.
\]
Substituting into~\eqref{eq:row-sum-bound},
\[
\sum_{v\ne u}|\mathrm{Cov}(n^*_u, n^*_v)|
\;\le\; \bigl[\,W_{\max}^2\, M_{\max}(C^{(1)}_{\mathrm{row}}+1) + \bar{\mu}_G^{(2)}\, C^{(2)}_{\mathrm{row}}\,\bigr]\,\sup_{v'}\mathrm{Var}(n^*_{v'}) \;=\; C_{\mathrm{TS}}\,\sup_{v'}\mathrm{Var}(n^*_{v'}),
\]
establishing~\eqref{eq:R3-row-target}. The bound holds uniformly in the row index $u$ and uniformly in $N$: the ingredients $W_{\max}, M_{\max}, C^{(1)}_{\mathrm{row}}, C^{(2)}_{\mathrm{row}}, \bar{\mu}_G^{(2)}$ are all design-uniform constants under the regime clause and the single-stage moment analysis. Hence (R3) holds a.s.\ for every $N\ge 1$ with $C_{\mathrm{TS}}$ as claimed.

\end{proof}

\section{Splitting Criteria}
This section examines the theoretical properties at an arbitrary split node. 

\subsection{Equivalent Expression}

\begin{proposition}
\label{prop:equiv_exp}
    A weighted MMD is equivalent in concept to the usual CART split criteria. Specifically, that
\begin{align}
\begin{split}
    &\hat{\theta}_{n_s}({D}) 
    = \arg \max_{\theta \in \Pi}  
    \frac{\widehat{N}_{L}\widehat{N}_{R}}{\widehat{N}^2_{P_a}}
    \left\| 
    \mu_{\mathbf k} \left( \frac{1}{\widehat{N}_{L}} \sum_{\{i \in [n_s]: X_i \in C_L\}} \frac{n^*_{b,i}}{q \pi_i} \; \delta_{Y_i} \right) 
    - 
    \mu_{\mathbf k} \left( \frac{1}{\widehat{N}_{R}} \sum_{\{i \in [n_s]: X_i \in C_R\}} \frac{n^*_{b,i}}{q  \pi_i} \; \delta_{Y_i} \right) 
    \right\|^2_{\mathcal{H}}\\
    = 
    &\arg \min_{\theta \in \Pi} 
    \frac{1}{\widehat{N}_{P_a}} \sum_{\{i \in [n_s]: X_i \in P_a\}} 
    \left\| 
    \mu_{\mathbf k} \left( \frac{n^*_{b,i}}{q  \pi_i} \; \delta_{Y_i} \right) 
    - 
    \mu_{\mathbf k} \left( \sum_{l \in \{L, R\}} \frac{\ind_{\{X_i \in C_l\}}}{\widehat{N}_l} \sum_{\{j \in [n_s]: X_j \in C_l\}} \frac{n^*_{b,j}}{q  \pi_j} \; \delta_{Y_j} \right) 
    \right\|^2_{\mathcal{H}},
    \label{eq:split}
\end{split}
\end{align}
where $\hat{w}_i := \frac{n^*_{b,i}}{q  \pi_i}$ 
and $\widehat{N}_l := \sum_{i\in[n_s]} \ind_{\{X_i \in C_l\}} \hat{w}_i$ for $l\in\{L,R,P_a\}$.
\end{proposition}

\begin{proof}[Proof of Proposition~\ref{prop:equiv_exp}]
    \noindent We use a total distance decomposition argument, using the linearity of \( \mu_{\mathbf k}(\cdot) \). Denote
    \[
    M_i 
    = 
    \mu_{\mathbf k} \left( \hat{w}_i \; \delta_{Y_i} \right), 
    \quad 
    \bar{M} 
    = 
    \mu_{\mathbf k} \left( \frac{1}{\widehat{N}_{P_a}} \sum_{X_i \in P_a} \hat{w}_i \; \delta_{Y_i} \right), 
    \quad
    \bar{M}_L 
    = 
    \mu_{\mathbf k} \left( \frac{1}{\widehat{N}_L} \sum_{X_i \in P_a} {1_{X_i \in C_L}}  \hat{w}_i \; \delta_{Y_i} \right)\]
    \noindent 
    
    and similarly for \( M_R \), so the RHS of (5) is equivalently
    \begin{equation}
        \frac{1}{\widehat{N}_{P_a}} \left\{ \sum_{i \in C_L} \|M_i - \bar{M}_L\|^2_{\mathcal{H}} + \sum_{i \in C_R} \|M_i - \bar{M}_R\|^2_{\mathcal{H}} \right\},
    \end{equation}
    the sum of distances within groups. As in \( \widehat{N}_{P_a} = \widehat{N}_L + \widehat{N}_R \) and \( \bar{M} - \bar{M}_L = \frac{\widehat{N}_R}{\widehat{N}_{P_a}} (\bar{M}_R - \bar{M}_L) \),
    \begin{equation}
        \sum_{i \in C_L} \|M_i - \bar M_L\|^2_{\mathcal{H}} = \sum_{i \in C_L} \|M_i - \bar{M}\|^2_{\mathcal{H}} - \frac{\widehat{N}^2_R \widehat{N}_L}{\widehat{N}^2_{P_a}} \|\bar{M}_R - \bar{M}_L\|^2_{\mathcal{H}}
    \end{equation}
    and similarly for \( \sum_{i \in C_R} \|M_i - \bar{M}_R\|^2_{\mathcal{H}} \). Plugging in the expression into (6), and realizing that the total sum of distances
    $
    \frac{1}{\widehat{N}_{P_a}} \sum_{i \in P_a} \|M_i - \bar{M}\|^2_{\mathcal{H}}
    $
    is invariant w.r.t. the split results in 
    $
    \argmin_\theta (6) 
    = \argmin_\theta \left\{ \frac{1}{\widehat{N}_{P_a}}\sum_{i\in P_a}\|M_i-\bar M\|^2_{\mathcal{H}} - \frac{\widehat{N}_L \widehat{N}_R}{\widehat{N}^2_{P_a}}\|\bar{M}_L-\bar{M}_R\|^2_{\mathcal{H}}\right\}
    = \argmax_\theta \frac{\widehat{N}_L \widehat{N}_R}{\widehat{N}^2_{P_a}}\|\bar{M}_L-\bar{M}_R\|^2_{\mathcal{H}},
    $
    where the first term is invariant in $\theta$, completing the proof.
\end{proof}

Under the stratified two-stage cluster design of \textup{(D-TS1)--(D-TS4)}, let SSU indexing be $u=(h,k,j)\in[n_s]$ with inclusion probability $\pi_u = \pi^{(1)}_{hk}\pi^{(2)}_{hkj\mid hk}$ and SSU-level bootstrap multiplier $n^*_{b,u} = G^*_{hk}W^*_{hkj}$ per~\eqref{eq:twostage-multiplier}. Define weights $\widehat w_u := n^*_{b,u}/(q\,\pi_u)$ and $\widehat N_\ell := \sum_{u\in[n_s]}\ind_{\{X_u\in C_\ell\}}\widehat w_u$ for $\ell\in\{L,R,P_a\}$. Then the equivalent expression of Proposition~\ref{prop:equiv_exp} holds verbatim with $i$ replaced by $u$:
\[
\hat\theta_{n_s}({D})
= \argmax_{\theta\in\Pi} \frac{\widehat N_L \widehat N_R}{\widehat N_{P_a}^2}
\Bigl\|\mu_{\mathbf k}\!\Bigl(\tfrac{1}{\widehat N_L}\!\sum_{\{u\in[n_s]:\,X_u\in C_L\}}\!\widehat w_u\,\delta_{Y_u}\Bigr) - \mu_{\mathbf k}\!\Bigl(\tfrac{1}{\widehat N_R}\!\sum_{\{u\in[n_s]:\,X_u\in C_R\}}\!\widehat w_u\,\delta_{Y_u}\Bigr)\Bigr\|_{\mathcal H}^2.
\]

\subsection{Consistency of the Split}

For a parent region $P_a\subseteq\mathcal{X}$ at depth $d$ in the tree-growing process, the space of axis-aligned candidate splits is
\[
\Pi := \{(j,t): j\in[p],\ t\in[a_j, b_j]\},
\]
equipped with the metric $d((j,t),(j',t')) := \ind(j\ne j') + |t-t'|$. Since $\mathcal{X}\subseteq\mathbb{R}^p$ is bounded, $[a_j,b_j]$ is a finite interval for each $j$, and $\Pi$ is a finite disjoint union of $p$ compact intervals. Hence $(\Pi, d)$ is a compact metric space.

Set $\mathcal{C}_0 := \{\mathcal{X}\}$ (the root region). For $d \ge 1$, the class of possible parent regions at depth $d$ is
\[
\mathcal{C}_d := \left\{\bigcap_{\ell=1}^m C^\ell(\theta^\ell): 1 \le m \le d,\; \theta^\ell = (j^\ell, t^\ell) \in \Pi,\; C^\ell(\theta^\ell) \in \{C_L(\theta^\ell),\, C_R(\theta^\ell)\}\right\},
\]
i.e., intersections of at most $d$ axis-aligned halfspaces inherited from ancestor splits, each halfspace of the form $\{x : x_{j^\ell} \le t^\ell\}$ or $\{x : x_{j^\ell} > t^\ell\}$.

We show the following supporting lemma and corollary first. The Lipschitz property of finite-population split score $M^N$ is used in Theorem~4 to convert the assumed uniqueness of the maximizer of $M^N$ into the well-separated maximum condition that Corollary~3.2.3 of \parencite{van-der-Vaart2023-ur} requires.

\begin{lemma}[Lipschitz continuity of split scores]
\label{lem:split-lipschitz}
Suppose $\mathcal{X}\subseteq\mathbb{R}^p$ is bounded. Under assumptions (K1), (A2), and (A3), for any parent region $P_a \in \mathcal{C}_{d-1}$ with $N_{P_a} := |\{i \in [N]: X_i \in P_a\}|$, the finite-population split score
\[
M^N(\theta) = \frac{N_L(\theta) N_R(\theta)}{N_{P_a}^2} 
\left\|\mu_{\mathbf k}\!\left(\frac{1}{N_L(\theta)} \sum_{X_i \in C_L(\theta)} \delta_{Y_i}\right) 
- \mu_{\mathbf k}\!\left(\frac{1}{N_R(\theta)} \sum_{X_i \in C_R(\theta)} \delta_{Y_i}\right)\right\|^2_{\mathcal{H}}
\]
is Lipschitz continuous on $(\Pi, d)$ with $d(\theta, \theta') = \ind(j\ne j') + |t-t'|$: there exists $L_1 > 0$, depending on $K_{\max}$ from (K1), the upper density bound $M$ from (A2), the $\alpha$-regularity constant from (A3), the parent diameter $D_{P_a}$, the parent's relative size $\rho_{P_a}:=N_{P_a}/N$, and the leaf-scale parameter $k_N$, such that
$
|M^N(\theta) - M^N(\theta')| \le L_1\,d(\theta, \theta'), \; \forall\theta, \theta'\in\Pi.
$
\end{lemma}
\begin{proof}
Write $M^N(\theta) = w(\theta) \cdot \Delta(\theta)^2$, where
\begin{align*}
&w(\theta) := \frac{N_L(\theta) N_R(\theta)}{N_{P_a}^2},
\quad
\Delta(\theta) := \left\|\mu_{\mathbf k}(P_L(\theta)) - \mu_{\mathbf k}(P_R(\theta))\right\|_{\mathcal{H}},\\
& P_\ell(\theta) := \frac{1}{N_\ell(\theta)} \sum_{X_i \in C_\ell(\theta)} \delta_{Y_i}, 
\quad \ell \in \{L, R\}.
\end{align*}

{We first bound $|w(\theta) - w(\theta')|$.}
By the $\alpha$-regularity clause of (A3), every candidate split $\theta \in \Pi$ considered by the algorithm satisfies
\[
c_{\min} \le \frac{N_L(\theta)}{N_{P_a}},\, \quad 
\frac{N_R(\theta)}{N_{P_a}} \le c_{\max},
\]
with $c_{\min} := \alpha$ and $c_{\max} := 1 - \alpha$.

\textit{Case 1: $j = j'$ (same axis).}
Let $\delta = |t - t'|$ and, without loss of generality, $t < t'$. The units whose child-region membership changes between $\theta$ and $\theta'$ are precisely those with $x_j \in (t, t']$. Denote
\[
S_\delta := \{x \in P_a : t < x_j \le t + \delta\},
\]
which has thickness $\delta$ along axis $j$ and extent at most $D_{P_a}:=\mathrm{diam}(P_a)$ along each of the remaining $p-1$ axes. Then
\[
|N_L(\theta) - N_L(\theta')| = |\{i \in [N]: X_i \in S_\delta\}|.
\]
We bound the right-hand side in two regimes. 
For $\delta \le 2k_N/\sqrt{p}$, we cover $S_\delta$ by axis-aligned cubes of side $\delta$: along axis $j$, one cube suffices; along each of the remaining $p-1$ axes, $\lceil D_{P_a}/\delta\rceil$ cubes suffice. This yields
$
N_{\mathrm{cover}} := \lceil D_{P_a}/\delta\rceil^{p-1} = O\!\;\big((D_{P_a}/\delta)^{p-1}\big)
$
cubes covering $S_\delta$. Each cube of side $\delta$ has Euclidean diagonal $\delta\sqrt{p}$ and hence is inscribed in a Euclidean ball $B_r$ of radius $r := \delta\sqrt{p}/2$. The choice $\delta \le 2k_N/\sqrt{p}$ ensures $r \le k_N$, placing $B_r$ within the domain of (A2).
Applying (A2) to each ball gives
$
|\{i \in [N]: X_i \in S_\delta\}| \;\le\; M\,D_{P_a}^{p-1}\,\delta\,N.
$

For $\delta > 2k_N/\sqrt{p}$, we use the trivial bound $|N_L(\theta) - N_L(\theta')| \le N_{P_a} \le (\sqrt{p}\,N_{P_a}/(2k_N))\,\delta$. Combining and writing $\rho_{P_a}:=N_{P_a}/N\in(0,1]$ (positive by admissibility under (A3)),
$
|N_L(\theta) - N_L(\theta')| \;\le\; L_X\,\delta\,N_{P_a},
$
where $L_X := \max\!\big(M\,D_{P_a}^{p-1}/\rho_{P_a},\; \sqrt{p}/(2k_N)\big)$ depends on the upper density bound $M$ from (A2), the parent diameter $D_{P_a}$, the parent's relative size $\rho_{P_a}$, and the leaf-scale parameter $k_N$. The dependence of $L_X$ on $k_N$ enters only through the constant.

Since $w(\theta) = N_L N_R / N_{P_a}^2$ is smooth in $(N_L, N_R)$ with bounded derivatives,
\begin{align*}
|w(\theta) - w(\theta')| 
&\leq \frac{1}{N_{P_a}^2}\left(|N_L(\theta) - N_L(\theta')| \cdot N_R(\theta) 
+ N_L(\theta') \cdot |N_R(\theta) - N_R(\theta')|\right)\\
&\leq \frac{2c_{\max} N_{P_a} \cdot L_X \delta N_{P_a}}{N_{P_a}^2} 
= 2c_{\max} L_X \delta.
\end{align*}

Therefore
$|w(\theta) - w(\theta')| \leq 2c_{\max} L_X \cdot d(\theta, \theta').$

\textit{Case 2: $j \neq j'$ (different axes).}
Then $d(\theta, \theta') = 1 + |t-t'| \ge 1$. Since $w(\theta) \in [c_{\min}^2, c_{\max}(1-c_{\min})]$ 
by (A3),
\[
|w(\theta) - w(\theta')| \leq 2c_{\max}(1-c_{\min}) \le 2c_{\max}(1-c_{\min}) \cdot d(\theta, \theta').
\]

Combining Cases 1 and 2, $|w(\theta) - w(\theta')| \le C_w \cdot d(\theta,\theta')$ with
\[
C_w := \max\!\big(2c_{\max} L_X,\; 2c_{\max}(1-c_{\min})\big).
\]

\textit{Now we bound $|\Delta(\theta)^2 - \Delta(\theta')^2|$.}
By (K1), $k$ is uniformly bounded with $K_{\max} := \sup_{y,y' \in \mathcal{Y}} |k(y,y')| < \infty$. 
Using the identity $|a^2 - b^2| = |a-b| \cdot |a+b|$ and $\Delta(\theta), \Delta(\theta') \leq 2\sqrt{K_{\max}}$,
$|\Delta(\theta)^2 - \Delta(\theta')^2| \leq 4\sqrt{K_{\max}} \cdot |\Delta(\theta) - \Delta(\theta')|.$
By the triangle inequality,
\begin{align*}
|\Delta(\theta) - \Delta(\theta')| 
&\leq \|\mu_{\mathbf k}(P_L(\theta)) - \mu_{\mathbf k}(P_L(\theta'))\|_{\mathcal{H}} 
+ \|\mu_{\mathbf k}(P_R(\theta)) - \mu_{\mathbf k}(P_R(\theta'))\|_{\mathcal{H}}.
\end{align*}

Consider bounding $\|\mu_{\mathbf k}(P_L(\theta)) - \mu_{\mathbf k}(P_L(\theta'))\|_{\mathcal{H}}$. When $j = j'$:
Define $\mathcal{S}_\Delta = \{i: X_i \in C_L(\theta) \triangle C_L(\theta')\}$ 
(symmetric difference). By the derivation above, $|\mathcal{S}_\Delta| \leq L_X \delta N_{P_a}$.

Decompose:
\begin{align*}
&\mu_{\mathbf k}(P_L(\theta)) - \mu_{\mathbf k}(P_L(\theta'))\\
&= \frac{1}{N_L(\theta)} \sum_{X_i \in C_L(\theta)} \mu_{\mathbf k}(\delta_{Y_i}) 
- \frac{1}{N_L(\theta')} \sum_{X_i \in C_L(\theta')} \mu_{\mathbf k}(\delta_{Y_i})\\
&= \frac{1}{N_L(\theta)} \sum_{i \in \mathcal{S}_\Delta \cap C_L(\theta)} \mu_{\mathbf k}(\delta_{Y_i})
- \frac{1}{N_L(\theta')} \sum_{i \in \mathcal{S}_\Delta \cap C_L(\theta')} \mu_{\mathbf k}(\delta_{Y_i})\\
&\quad + \left(\frac{1}{N_L(\theta)} - \frac{1}{N_L(\theta')}\right) 
\sum_{X_i \in C_L(\theta) \cap C_L(\theta')} \mu_{\mathbf k}(\delta_{Y_i}).
\end{align*}

By (K1), $\|\mu_{\mathbf k}(\delta_y)\|_{\mathcal{H}} \leq \sqrt{K_{\max}}$ for all $y \in \mathcal{Y}$. Thus
\begin{align*}
\left\|\frac{1}{N_L(\theta)} \sum_{i \in \mathcal{S}_\Delta \cap C_L(\theta)} 
\mu_{\mathbf k}(\delta_{Y_i})\right\|_{\mathcal{H}}
\leq \frac{|\mathcal{S}_\Delta|}{N_L(\theta)} \sqrt{K_{\max}}
\leq \frac{L_X \delta N_{P_a}}{c_{\min} N_{P_a}} \sqrt{K_{\max}} 
= \frac{L_X \sqrt{K_{\max}}}{c_{\min}} \delta.
\end{align*}

Similarly for the second term. For the third term,
\begin{align*}
\left|\frac{1}{N_L(\theta)} - \frac{1}{N_L(\theta')}\right| 
\leq \frac{|N_L(\theta) - N_L(\theta')|}{N_L(\theta) N_L(\theta')}
\leq \frac{L_X \delta N_{P_a}}{(c_{\min} N_{P_a})^2} = \frac{L_X \delta}{c_{\min}^2 N_{P_a}}.
\end{align*}

Since $|C_L(\theta) \cap C_L(\theta')| \leq N_{P_a}$,
\begin{align*}
\left\|\left(\frac{1}{N_L(\theta)} - \frac{1}{N_L(\theta')}\right) 
\sum_{X_i \in C_L(\theta) \cap C_L(\theta')} \mu_{\mathbf k}(\delta_{Y_i})\right\|_{\mathcal{H}}
\leq \frac{L_X \delta}{c_{\min}^2 N_{P_a}} \cdot N_{P_a} \sqrt{K_{\max}}
= \frac{L_X \sqrt{K_{\max}}}{c_{\min}^2} \delta.
\end{align*}

Combining together gives
\begin{align*}
&\|\mu_{\mathbf k}(P_L(\theta)) - \mu_{\mathbf k}(P_L(\theta'))\|_{\mathcal{H}} 
\leq C_\mu \delta = C_\mu d(\theta, \theta'), \text{ where } \\
&C_\mu = \frac{L_X \sqrt{K_{\max}}}{c_{\min}} \left(2 + \frac{1}{c_{\min}}\right).
\end{align*}

When $j \neq j'$, $d(\theta,\theta') = 1 + |t-t'| \ge 1$, and by the triangle inequality,
\[
\|\mu_{\mathbf k}(P_L(\theta)) - \mu_{\mathbf k}(P_L(\theta'))\|_{\mathcal{H}} 
\le \|\mu_{\mathbf k}(P_L(\theta))\|_{\mathcal{H}} + \|\mu_{\mathbf k}(P_L(\theta'))\|_{\mathcal{H}} 
\le 2\sqrt{K_{\max}} \le 2\sqrt{K_{\max}} \cdot d(\theta,\theta').
\]
Combining Cases 1 and 2, $\|\mu_{\mathbf k}(P_L(\theta)) - \mu_{\mathbf k}(P_L(\theta'))\|_{\mathcal H} \le \max(C_\mu, 2\sqrt{K_{\max}}) \cdot d(\theta,\theta')$, and symmetrically for $\ell = R$. Then
$
|\Delta(\theta) - \Delta(\theta')| \le 2 \max(C_\mu, 2\sqrt{K_{\max}}) \cdot d(\theta,\theta'),
$
so
\[
|\Delta(\theta)^2 - \Delta(\theta')^2| \le 4\sqrt{K_{\max}} \cdot |\Delta(\theta) - \Delta(\theta')| \le C_\Delta \cdot d(\theta,\theta'),
\]
with $C_\Delta := 8\sqrt{K_{\max}} \max(C_\mu, 2\sqrt{K_{\max}})$.

Together, using $|ab - a'b'| \leq |a-a'| \cdot |b| + |a'| \cdot |b-b'|$, $\Delta(\theta)^2 \le 4K_{\max}$, and $w(\theta') \le c_{\max}(1-c_{\min})$:
\begin{align*}
|M^N(\theta) - M^N(\theta')| 
&= |w(\theta)\Delta(\theta)^2 - w(\theta')\Delta(\theta')^2|\\
&\leq |w(\theta) - w(\theta')| \cdot \Delta(\theta)^2 + w(\theta') \cdot |\Delta(\theta)^2 - \Delta(\theta')^2|\\
&\leq \big(4K_{\max}\, C_w + c_{\max}(1-c_{\min})\, C_\Delta\big) \cdot d(\theta, \theta')\\
&=: L_1 \cdot d(\theta, \theta'),
\end{align*}
where $L_1$ depends on $K_{\max}, M, c_{\min}, c_{\max}, D_{P_a}, \rho_{P_a}$, and $k_N$ through $L_X, C_w, C_\mu$, and $C_\Delta$.
\end{proof}

The following lemma controls the conditional bootstrap fluctuation uniformly over $\mathcal{G}$, using only the marginal sub-gamma tails of the
centered multipliers $\delta_i$ and the weak-dependence bound (R3), rather than any
independence or scheme-specific structure.

\medskip
\noindent\emph{The design-weighted empirical seminorm.}
Let $\mu_N^{\pi}:=\tfrac1N\sum_{i=1}^N (\xi_i/\pi_i^2)\,\delta_{(X_i,Y_i)}$, a
nonnegative discrete measure since $\xi_i/\pi_i^2\ge 0$. For
$h:\mathcal X\times\mathcal Y\to\mathbb R$ set
\begin{equation}
\label{eq:pi-seminorm}
\left\lVert h \right\rVert^{\pi}_{P_N,2}:=\Bigl(\int h^2\,d\mu_N^{\pi}\Bigr)^{1/2}
=\Bigl(\tfrac1N\textstyle\sum_{i=1}^N (\xi_i/\pi_i^2)\,h(X_i,Y_i)^2\Bigr)^{1/2}.
\end{equation}
The weight $\xi_i/\pi_i^2$ arises from $\xi_i^2=\xi_i$: the increment weight
$c_i=(\xi_i/\pi_i)(g-g')(X_i,Y_i)$ enters the variance proxy
\eqref{eq:var-increment} through $c_i^2$, so that
$\tfrac1N\sum_i c_i^2=\tfrac1N\sum_i(\xi_i/\pi_i^2)(g-g')^2
=\bigl(\left\lVert g-g' \right\rVert^{\pi}_{P_N,2}\bigr)^2$. As the $L_2(\mu_N^{\pi})$ seminorm,
\eqref{eq:pi-seminorm} is nonnegative and homogeneous and satisfies the triangle
inequality, hence defines a semimetric adequate for the covering numbers
$N(\varepsilon,\mathcal{G},\left\lVert \cdot \right\rVert^{\pi}_{P_N,2})$ in the chaining bound. It is a seminorm rather than a norm as any $h$ vanishing on the sampled support
$\{i:\xi_i=1\}$ has $\left\lVert h \right\rVert^{\pi}_{P_N,2}=0$ without $h\equiv 0$, which does not
affect the entropy argument.

\begin{proposition}[Conditional multiplier maximal inequality under weak dependence]
\label{prop:mult-maximal}
Let $\mathcal{G}$ be $P$-Glivenko--Cantelli with constant envelope $\bar G<\infty$ and
finite uniform entropy integral $J(1,\mathcal{G},L_2)<\infty$. Let
the design weights $\xi_i/\pi_i$ satisfy $\xi_i/\pi_i\le\pi_0^{-1}$ a.s.\ (by \textup{(D3)})
and $\tfrac1N\sum_i \xi_i/\pi_i^2=O_{p_d}(1)$, and let
$\{n_i^*\}$ satisfy \textup{(R2)--(R3)}, and set
$\delta_i:=n_i^*-\mathbb{E}(n_i^*\mid\xi,Z^N)$, conditionally sub-gamma per
Lemma~\ref{lem:subgamma}. Define
$V_N(g)=\tfrac1N\sum_i (\xi_i/\pi_i)\,\delta_i\,g(X_i,Y_i)$. Then, conditionally on
$(\xi,Z^N)$,
\begin{equation}
\label{eq:main-bound}
\mathbb{E}_{n^*\mid\xi,Z^N}\,\left\lVert V_N \right\rVert_{\mathcal{G}}\;=\;O_{p_d}\!\bigl(N^{-1/2}\bigr).
\end{equation}
Consequently, adding the multiplier bias
$B_N(g):=\tfrac1N\sum_i (\xi_i/\pi_i)\,\mathbb{E}(n_i^*-1\mid\xi,Z^N)\,g(X_i,Y_i)$, which satisfies
$\left\lVert B_N \right\rVert_{\mathcal{G}}=o_{p_d}(1)$ under \textup{(R2)}, one obtains
$\sup_{g\in\mathcal{G}}|T^*_N(g)|=o_{p_d}(1)$.
\end{proposition}

\begin{proof}
Fix $(\xi,Z^N)$. The design weights $\xi_i/\pi_i$ and the data $(X_i,Y_i)$ are constants, and
$\{\delta_i\}$ carries the only randomness.

\medskip
\noindent\emph{Step (i): a conditional sub-gamma bound for linear combinations.}
Fix $g,g'\in\mathcal{G}$ and write $c_i:=(\xi_i/\pi_i)\bigl(g(X_i,Y_i)-g'(X_i,Y_i)\bigr)$, so that
$V_N(g)-V_N(g')=\tfrac1N\sum_i c_i\delta_i$. We first bound the conditional
variance. By (R3) and $\sup_i\operatorname{Var}(n_i^*\mid\xi,Z^N)=:s^2<\infty$ (finite by the
$2+\epsilon$ moment bound in (R2)),
\begin{align}
\operatorname{Var}_{n^*}\!\Bigl(\tfrac1N\textstyle\sum_i c_i\delta_i\Bigr)
&=\frac1{N^2}\Bigl[\sum_i c_i^2\operatorname{Var}(n_i^*\mid\xi,Z^N)
+\sum_{i\ne j}c_i c_j\operatorname{Cov}(n_i^*,n_j^*\mid\xi,Z^N)\Bigr]\notag\\
&\le\frac{s^2}{N^2}\sum_i c_i^2+\frac{C\,s^2}{N^2}\sum_i c_i^2
\;=\;\frac{(1+C)\,s^2}{N^2}\sum_i c_i^2,
\label{eq:var-increment}
\end{align}
where the second-order term is bounded via $|c_ic_j|\le\tfrac12(c_i^2+c_j^2)$ and the symmetry $\operatorname{Cov}_{ij}=\operatorname{Cov}_{ji}$:
$\sum_{i\ne j}|c_ic_j||\operatorname{Cov}_{ij}|\le\sum_i c_i^2\bigl(\sum_{j\ne i}|\operatorname{Cov}_{ij}|\bigr)\le C\,s^2\sum_i c_i^2$
by \textup{(R3)}.
For the tail, write $S_N:=\tfrac1N\sum_i c_i\delta_i$. By
Lemma~\ref{lem:subgamma}, each $\delta_i$ is conditionally
sub-gamma$(\nu,c)$ given $(\xi,Z^N)$, with $\nu$ the marginal variance proxy and
$c$ the marginal scale. The variance proxy of $S_N$ follows from (R3) through
\eqref{eq:var-increment}, namely $\sigma_N^2=\tfrac{(1+C)\nu}{N^2}\sum_i c_i^2$, and its
scale is governed by the uniform weight bound
$\|c\|_\infty/N\le 2\bar G\,\pi_0^{-1}/N$. Each
summand $c_i\delta_i/N$ is sub-gamma with scale $(|c_i|/N)\,c$, so the linear
combination has scale at most $(\|c\|_\infty/N)\,c$. Hence there is $c'<\infty$,
depending only on $(\nu,c,C)$, with
\begin{equation}
\label{eq:lincomb-subgamma}
\log\mathbb{E}\Bigl[\exp\!\bigl(\lambda S_N\bigr)\Bigm|\xi,Z^N\Bigr]
\le\frac{\lambda^2\,\sigma_N^2/2}{1-c'\,\|c\|_\infty|\lambda|/N},
\qquad \sigma_N^2:=\frac{(1+C)\nu}{N^2}\sum_i c_i^2,
\end{equation}
for $|\lambda|<N/(c'\|c\|_\infty)$. 
Thus we have an upper bound for the Orlicz norm 
\begin{equation}
\label{eq:increment}
\left\lVert V_N(g)-V_N(g') \right\rVert_{\psi_1\mid\xi,Z^N}
\;\le\;\frac{K_1}{N}\Bigl(\sum_i c_i^2\Bigr)^{1/2}
\;=\;\frac{K_1}{\sqrt N}\,\left\lVert g-g' \right\rVert^{\pi}_{P_N,2},
\end{equation}
with $K_1=K_0(1+C)^{1/2}\nu^{1/2}$ depending only on $(\nu,c,C)$. 

\medskip
\noindent\emph{Step (ii): chaining under a $\psi_1$-increment.}
Under the constant envelope $|g|\le\bar G$, the weighted seminorm satisfies
$\left\lVert g-g' \right\rVert^{\pi}_{P_N,2}\le\bigl(\tfrac1N\sum_i \xi_i/\pi_i^2\bigr)^{1/2}\left\lVert g-g' \right\rVert_\infty
\le 2\bar G\,\bigl(\tfrac1N\sum_i \xi_i/\pi_i^2\bigr)^{1/2}$, so the $\mathcal{G}$-diameter in the
semimetric of \eqref{eq:increment} is at most
$D_N:=2K_1 N^{-1/2}\bar G(\tfrac1N\sum_i \xi_i/\pi_i^2)^{1/2}$. The $\psi_1$
increment \eqref{eq:increment} is a Bernstein-type tail. Chaining it against the
metric entropy via the finite-class maximal inequality
\parencite[Lem.~2.2.13]{van-der-Vaart2023-ur}, with the sub-Gaussian branch handled by
\parencite[Cor.~2.2.9]{van-der-Vaart2023-ur}, gives
\begin{align}
\label{eq:chaining}
&\mathbb{E}_{n^*\mid\xi,Z^N}\left\lVert V_N-V_N(g_0) \right\rVert_{\mathcal{G}} \nonumber \\
\;\lesssim\;
&\underbrace{\int_0^{D_N}\!\!\sqrt{\log N(\varepsilon,\mathcal{G},\|\cdot\|^{\pi}_{P_N,2})}\,d\varepsilon}_{\text{sub-Gaussian branch}}
\;+\;
\underbrace{\frac{K_1}{N}\,\bar G\,\log N(D_N,\mathcal{G},\|\cdot\|^{\pi}_{P_N,2})}_{\text{sub-exponential branch}},
\end{align}
for any fixed $g_0\in\mathcal{G}$.

\emph{Sub-Gaussian branch.} Take $Q=\mu_N^{\pi}$, so that
$\|\cdot\|^{\pi}_{P_N,2}=\|\cdot\|_{L_2(Q)}$ by \eqref{eq:pi-seminorm} and
$\mu_N^{\pi}$ is one admissible discrete. For
the constant envelope $\bar G$,
$\|\bar G\|_{Q,2}^2=\bar G^2\,\tfrac1N\sum_i\xi_i/\pi_i^2$, so the diameter
factorizes as $D_N=2K_1 N^{-1/2}\|\bar G\|_{Q,2}$. Apply a change of variable $\varepsilon=D_N u$
($d\varepsilon=D_N\,du$) normalizes the upper limit to $1$, and absorbing the
constant $2K_1 N^{-1/2}$ into the chaining constant gives
\[
N\big(D_N u,\mathcal{G},\|\cdot\|^{\pi}_{P_N,2}\big)
=N\big(D_N u,\mathcal{G},L_2(Q)\big)
\le\sup_{Q'} N\big(u\|\bar G\|_{Q',2},\mathcal{G},L_2(Q')\big),
\]
the equality being the definitional identity at $Q=\mu_N^{\pi}$. Hence
\begin{align*}
    &\int_0^{D_N}\!\!\sqrt{\log N(\varepsilon,\mathcal{G},\|\cdot\|^{\pi}_{P_N,2})}\,d\varepsilon
=D_N\!\int_0^{1}\!\!\sqrt{\log N(D_N u,\mathcal{G},\|\cdot\|^{\pi}_{P_N,2})}\,du \\
\;\le\;&D_N\,J(1,\mathcal{G},L_2)\;=\;O_{p_d}\!\bigl(N^{-1/2}\bigr),
\end{align*}
since $D_N=O_{p_d}(N^{-1/2})$ and $J(1,\mathcal{G},L_2)<\infty$ by hypothesis. 

\emph{Sub-exponential branch.} As a VC-subgraph class of index $V_{\mathcal{G}}$,
$\mathcal{G}$ obeys $\log N(\varepsilon,\mathcal{G},\|\cdot\|^{\pi}_{P_N,2})\lesssim
V_{\mathcal{G}}\log(1/\varepsilon)$ for $0<\varepsilon<1$
\parencite[Thm.~2.6.7]{van-der-Vaart2023-ur}. Evaluating at $\varepsilon=D_N$ and using
$D_N=O_{p_d}(N^{-1/2})$, so 
$\log(1/D_N)=O_{p_d}(\tfrac12\log N)=O_{p_d}(\log N)$,
$
\log N(D_N,\mathcal{G},\|\cdot\|^{\pi}_{P_N,2})=O_{p_d}(\log N)$, and hence 
\[
\frac{K_1}{N}\,\bar G\,\log N(D_N,\mathcal{G},\|\cdot\|^{\pi}_{P_N,2})
\;=\;O_{p_d}\!\bigl(N^{-1}\log N\bigr)\;=\;o_{p_d}\!\bigl(N^{-1/2}\bigr),
\]
using $N^{-1/2}\log N\to 0$. Finally
$|V_N(g_0)|\le\bar G\cdot|\tfrac1N\sum_i (\xi_i/\pi_i)\delta_i|$ has conditional mean
$O_{p_d}(N^{-1/2})$, because \eqref{eq:var-increment} with $c_i=(\xi_i/\pi_i)\bar G$ bounds
$\operatorname{Var}_{n^*}(V_N(g_0))\le(1+C)s^2\bar G^2\bigl(\tfrac1N\sum_i\xi_i/\pi_i^2\bigr)N^{-1}=O_{p_d}(N^{-1})$
and Jensen's inequality passes this to $\mathbb{E}_{n^*\mid\xi,Z^N}|V_N(g_0)|$. Combining gives
\eqref{eq:main-bound}.

\medskip
\noindent\emph{Step (iii): bias and unconditioning.}
By (R2),
\[
\left\lVert B_N \right\rVert_{\mathcal{G}}
\le\bar G\Bigl(\tfrac1N\textstyle\sum_i \xi_i/\pi_i\Bigr)\sup_i\bigl|\mathbb{E}(n_i^*-1\mid\xi,Z^N)\bigr|
=O_{p_d}(1)\cdot o_{p_d}(1)=o_{p_d}(1),
\]
where $\tfrac1N\sum_i \xi_i/\pi_i=O_{p_d}(1)$ by the design LLN
(Lemma~\ref{lem:lln_clt}). As
$T^*_N(g)=V_N(g)+B_N(g)$, \eqref{eq:main-bound} and Markov's
inequality give, for every $\eta>0$,
\[
P_{n^*\mid\xi,Z^N}\Bigl(\left\lVert V_N \right\rVert_{\mathcal{G}}>\eta\Bigr)\le\eta^{-1}O_{p_d}(N^{-1/2})=o_{p_d}(1).
\] Take expectation over $(\xi,Z^N)$, iterated expectation and dominated
convergence then yields 
$\sup_{g\in\mathcal{G}}|T^*_N(g)|=o_{p_d}(1)$.
\end{proof}

\vspace{0.5cm}
\begin{proof}[Proof of Theorem~4]
Let $P_a \subseteq \mathcal{X}$ denote the region corresponding to a given node at depth $d$, defined by the preceding splits. For a fixed candidate split $\theta = (j,t)$ partitioning $P_a$ into child regions $C_L(\theta)$ and $C_R(\theta)$ (hereafter suppressing the dependence on $\theta$ for notational clarity). Consider a fixed region $P_a \subseteq \mathcal{X}$ which represent the region at a node at depth $d$ in the tree-growing process.

The proof is uniform in the two design types (i) single- and (ii) stratified two-stage cluster designs; design-specific tools enter at three localized substitution points, indicated below.

Throughout, ``sampled unit'' denotes $i$ under (i) and $u=(h,k,j)$ under (ii); we use the symbol $i$ generically in display formulae, with the understanding that under (ii) it stands for $u$. The design LLN under (i) is Lemma~\ref{lem:lln_clt}, and under (ii) is Lemma~\ref{lem:lln_clt_multistage}. The centered multipliers $\delta_i:=n^*_i-\mathbb{E}(n^*_i\mid\xi,Z^N)$ are conditionally sub-gamma with envelope $\Gamma(\nu_0, c_0)$ under (i) via Lemma~\ref{lem:subgamma}, and $\Gamma(\tilde\nu,\tilde c)$ with $\tilde\nu=W_{\max}^2\nu_1+\nu_2$, $\tilde c=c_1 W_{\max}\vee c_2$ under (ii) via Lemma~\ref{lem:subgamma-multistage}. The multiplier conditions \textup{(R1)--(R3)} at the sampled-unit level are verified by Section~S5.2 for both (i) and (ii). Uniform positivity $\min\pi_i\ge\pi_0>0$ holds under (i) from \textup{(D3)}, and under (ii) from Lemma~\ref{lem:designs-multistage}\,(i).

We establish
\begin{equation}
\label{eq:B-target}
\sup_{P_a\in\mathcal C_{d-1}}\sup_{\theta\in\Pi}
\bigl|M^*_{n_s,P_a}(\theta) - M_{P_a}(\theta)\bigr| = o_{p_\otimes}(1),
\end{equation}
via a triangle-inequality decomposition and a common maximal inequality on a VC-type class.
 
\medskip
\noindent\emph{Setup: VC-type function class.}
Recall the compact space $(\Pi, d)$ and the depth-$d$ region class $\mathcal C_d$ defined above. 
For any $\theta = (j,t)\in\Pi$, the child regions
$C_L(\theta) = \{x\in\mathbb R^p: x_j\le t\}$ and $C_R(\theta) = \{x\in\mathbb R^p: x_j > t\}$
are axis-aligned halfspaces. The primitive class
$\mathcal A := \{C_L(\theta), C_R(\theta): \theta\in\Pi\}$
is a Vapnik-Chervonenkis (VC) class with $\mathrm{VC}(\mathcal A) \le p+1$.
 
Each $P_a\in\mathcal C_{d-1}$ is an intersection of at most $d-1$ elements of $\mathcal A$. 
The closure of VC classes under finite Boolean operations
(\cite{van-der-Vaart2023-ur}, Section 2.6.4)
gives $\mathrm{VC}(\mathcal C_{d-1}) < \infty$, with an explicit bound polynomial in $p$ and $d$.
Define the augmented function class
\[
\mathcal H_d := \{\ind(X\in L): L\in\mathcal C_d\} 
\cup \{\mathbf{k}(Y,y)\cdot \ind(X\in L): L\in\mathcal C_d,\, y\in\mathcal Y\}.
\]
By (K1), $\sup_{y,y'\in\mathcal Y} |k(y,y')| =: K_{\max} < \infty$, so $\mathcal H_d$ has constant 
envelope $H\equiv K_{\max}$ and inherits VC-subgraph structure from $\mathcal C_d$, with 
$\mathrm{VC}(\mathcal H_d)$ finite and polynomial in $p$ and $d$. 
Consequently, $\mathcal H_d$ satisfies the 
finite uniform entropy condition
\begin{equation}
\label{eq:uniform-entropy}
\int_0^\infty \sup_Q \sqrt{\log N(\epsilon\|H\|_{Q,2}, \mathcal H_d, L_2(Q))}\, d\epsilon < \infty,
\end{equation}
where the supremum is over all finitely discrete probability measures $Q$ 
(\cite{van-der-Vaart2023-ur}, Theorem 2.6.7).
 
\medskip
For $g\in\mathcal H_d$, define
\begin{align*}
P(g) &:= \mathbb E_P[g(X,Y)], 
\quad
P_N(g) := \frac{1}{N}\sum_{i=1}^N g(X_i, Y_i),\\
P^{n_s}_N(g) &:= \frac{1}{N}\sum_{i=1}^N \frac{\xi_i}{\pi_i}\,g(X_i, Y_i),
\quad
P^*_N(g) := \frac{1}{N}\sum_{i=1}^N \frac{n_i^*\xi_i}{\pi_i}\,g(X_i, Y_i),
\end{align*}
i.e., the super-population, unconditional empirical, H\'ajek-weighted, and 
bootstrap-weighted H\'ajek measures, respectively.
 
\medskip
For any parent $P_a\in\mathcal C_{d-1}$ and split $\theta\in\Pi$, the split scores can be expressed 
as continuous functionals of empirical averages of functions in $\mathcal H_d$. To make this 
explicit, define the finite subclass
\[
\mathcal G_\theta := \{\ind(X\in P_a\cap C_L(\theta)),\, \ind(X\in P_a\cap C_R(\theta))\}
\cup \{k(Y, Y_i)\cdot \ind(X\in P_a\cap C_\ell(\theta)): i\in[N],\, \ell\in\{L,R\}\},
\]
which is a finite subset of $\mathcal H_d$ (using that $P_a\cap C_\ell(\theta)\in\mathcal C_d$). 
For a generic measure $Q$ on $(\mathcal X\times\mathcal Y, \mathcal B)$, define
\begin{equation}
\label{eq:M-representation}
M_{Q,P_a}(\theta) := \Psi_\theta\!\bigl(Q(\ind_{P_a}), \{Q(g)\}_{g\in\mathcal G_\theta}\bigr),
\end{equation}
where $\Psi_\theta:\mathbb R\times\mathbb R^{|\mathcal G_\theta|}\to\mathbb R$ encodes the ratio 
structure
$
\Psi_\theta(a, \{b_g\}) = \frac{b_L\,b_R}{a^2}\cdot \bigl\| b_{\mu_L} - b_{\mu_R}\bigr\|_{\mathcal H}^2,
$
with $b_L, b_R$ the child masses and $b_{\mu_L}, b_{\mu_R}$ the child kernel-mean embeddings 
(expanded as a polynomial in the $k(Y,Y_i)$-type entries). 
Then
\[
M(\theta) = M_{P,P_a}(\theta),
\quad
M^N(\theta) = M_{P_N,P_a}(\theta),
\quad
M^{n_s}(\theta) = M_{P^{n_s}_N,P_a}(\theta),
\quad
M^*_{n_s}(\theta) = M_{P^*_N,P_a}(\theta),
\]
so all four scores are the same functional $\Psi_\theta$ applied to different empirical measures.
 
By (A3)'s $\alpha$-regularity clause, every candidate $P_a$ visited by the algorithm satisfies 
$P_N(\ind_{P_a}) \ge \alpha > 0$; under (D2)-(D3) and (R1)-(R3), the same lower bound holds 
(up to a factor bounded away from zero, with $p_\otimes$-probability tending to $1$) for 
$P^{n_s}_N(\ind_{P_a})$ and $P^*_N(\ind_{P_a})$. On the event
\[
E_N := \bigl\{P_N(\ind_{P_a}), P^{n_s}_N(\ind_{P_a}), P^*_N(\ind_{P_a}) \ge \alpha/2\bigr\},
\]
the function $\Psi_\theta$ is Lipschitz in its arguments with constant $C = C(K_{\max}, \alpha)$ 
(since the denominators are bounded away from zero and each entry is bounded in absolute value 
by $K_{\max}$). 
Consequently, for any two empirical measures $Q, Q'$ drawn from 
$\{P, P_N, P^{n_s}_N, P^*_N\}$,
\begin{equation}
\label{eq:lipschitz-bound}
\sup_{P_a\in\mathcal C_{d-1}}\sup_{\theta\in\Pi} 
\bigl|M_{Q,P_a}(\theta) - M_{Q',P_a}(\theta)\bigr|\,\ind_{E_N}
\le C \sup_{g\in\mathcal H_d} |Q(g) - Q'(g)|,
\end{equation}
using that $\ind_{P_a}, \ind_{P_a\cap C_\ell(\theta)}$, and $\mathbf{k}(Y, Y_i)\cdot \ind_{P_a\cap C_\ell(\theta)}$ 
all lie in $\mathcal H_d$. Since $P_{\Omega \times \mathcal{S}^N}(E_N) \to 1$ by (A3), (D2)-(D3), and (R1)-(R2), it 
suffices to establish the desired bounds on the event $E_N$.
 
\medskip
\noindent\emph{Triangle decomposition.}
Taking $Q = P^*_N$ and $Q' = P$ in \eqref{eq:lipschitz-bound} and by triangle inequality,
\[
\sup_{g\in\mathcal H_d} |(P^*_N - P)(g)| \le 
\sup_{g\in\mathcal H_d} |(P^*_N - P^{n_s}_N)(g)|
+ \sup_{g\in\mathcal H_d} |(P^{n_s}_N - P)(g)|.
\]
 
\medskip
\noindent\textit{Step 1, $\sup_g |(P^{n_s}_N - P)(g)| = o_{p_\otimes}(1)$:}
For each $g\in\mathcal H_d$,
$
P^{n_s}_N(g)=\frac1N\sum_{i=1}^N\frac{\xi_i}{\pi_i}\,g(X_i,Y_i)
$
is the Horvitz--Thompson empirical measure, and $\mathbb E[\xi_i/\pi_i-1\mid Z^N]=0$
under the design. The class $\mathcal H_d$ is VC-subgraph with constant envelope
$K_{\max}$, hence $P$-Glivenko--Cantelli. Condition (A1) of
\cite{Han_Weller_2021}, Theorem~3.1 ($\min_{1\le i\le N}\pi_i\ge\pi_0>0$),
follows from (D3) for (i) and (D-TS3) for (ii); and its condition (A2-LLN),
$\frac1N\sum_{i=1}^N(\xi_i/\pi_i-1)=o_{p_d}(1)$, is provided by
Lemma~\ref{lem:lln_clt} under (D2)--(D4) for (i) and under (ii) from Lemma~\ref{lem:lln_clt_multistage}. Theorem~3.1 therefore applies to
$\mathcal F=\mathcal H_d$, yielding
\begin{equation}\label{eq:HW-ULLN}
\sup_{g\in\mathcal H_d}|(P^{n_s}_N-P)(g)|
=o_{p_\otimes}(1).
\end{equation}

\medskip\noindent\textit{Step 2: $\sup_g|(P^*_N-P^{n_s}_N)(g)|=o_{p_d}(1)$.}
Write $T^*_N(g):=P^*_N(g)-P^{n_s}_N(g)=\tfrac1N\sum_i(\xi_i/\pi_i)(n^*_i-1)g(X_i,Y_i)$. Apply Lemma~\ref{prop:mult-maximal} to $\mathcal G=\mathcal H_d$, verifying its hypotheses:
\begin{itemize}[leftmargin=2em, itemsep=1pt, topsep=1pt]
\item[\textup{1}] $\mathcal H_d$ is $P$-Glivenko--Cantelli with constant envelope and finite uniform entropy integral by~\eqref{eq:uniform-entropy}; design-agnostic.
\item[\textup{2}] $\xi_i/\pi_i\le\pi_0^{-1}$ a.s.; from (D3) under (i), from Lemma~\ref{lem:designs-multistage}\,(i) under (ii).
\item[\textup{3}] $\tfrac1N\sum_i\xi_i/\pi_i^2=O_{p_d}(1)$; from (H2) with $\tfrac1N\sum_i\xi_i/\pi_i=O_{p_d}(1)$, the latter from the design LLN in each case.
\item[\textup{4}] $\{n^*_i\}$ satisfies (R1)--(R3); by Lemma~7 in the main text, proved in Section~S5.2 for (i) and (ii).
\item[\textup{5}] $\delta_i$ is conditionally sub-gamma, with envelope $\Gamma(\nu_0,c_0)$ under (i) via Lemma~\ref{lem:subgamma}, $\Gamma(\tilde\nu,\tilde c)$ under (ii) via Lemma~\ref{lem:subgamma-multistage}.
\end{itemize}
The proof of Lemma~\ref{prop:mult-maximal} uses the sub-gamma envelope $(\nu,c)$ only through the constants $K_1=K_0(1+C)^{1/2}\nu^{1/2}$ and $c'$ in its equation~\eqref{eq:lincomb-subgamma}, and (R3)'s constant $C$ only as a design-uniform bound. Substitution preserves the $O_{p_d}(N^{-1/2})$ rate. The lemma yields
\begin{equation}\label{eq:Pstar-Pns}
\sup_{g\in\mathcal H_d}|(P^*_N-P^{n_s}_N)(g)|=o_{p_d}(1),
\end{equation}
the multiplier bias being absorbed within the lemma through (R2).

\medskip
Combining \eqref{eq:HW-ULLN} and \eqref{eq:Pstar-Pns} with
\eqref{eq:lipschitz-bound},
\begin{align*}
\sup_{P_a\in\mathcal C_{d-1}}\sup_{\theta\in\Pi}
  \bigl|M^*_{n_s}(\theta) - M(\theta)\bigr|\,\ind_{E_N}
&\le C\Bigl[\sup_g |(P^*_N - P^{n_s}_N)(g)|
         + \sup_g |(P^{n_s}_N - P)(g)|\Bigr]\\
&= C\bigl[o_{p_d}(1) + o_{p_\otimes}(1)\bigr].
\end{align*}
The first term is $o_{p_d}(1)$ for $P_\Omega$-a.e.\ $\omega$. Since
$P_{\Omega\times\mathcal S^N}(\cdot)=\int_\Omega P_{\mathcal S^N\mid\omega}(\cdot)\,dP_\Omega(\omega)$
with the conditional probabilities bounded, dominated convergence gives
$o_{p_\otimes}(1)$. Both terms being $o_{p_\otimes}(1)$, a union bound gives
\[
\sup_{P_a\in\mathcal C_{d-1}}\sup_{\theta\in\Pi}
  \bigl|M^*_{n_s}(\theta) - M(\theta)\bigr|\,\ind_{E_N}=o_{p_\otimes}(1).
\]
Combined with $P_{\Omega \times \mathcal{S}^N}(E_N^c)\to 0$, this establishes \eqref{eq:B-target}.
Together, restricting the outer supremum to the parent $P_a$ visited by the algorithm at a
fixed depth yields the uniform convergence claim of Theorem~4:
\[
\sup_{\theta\in\Pi}\bigl|M^*_{n_s}(\theta) - M(\theta)\bigr| \xrightarrow{p_\otimes} 0,
\]
the joint (product-space) consistency of the resampled split score.
\end{proof}

\vspace*{0.3cm}
\begin{corollary}[Consistency of the split argmax]
\label{cor:split-argmax}
Under the hypotheses of Theorem~4, suppose additionally that
\begin{enumerate}[leftmargin=3em, itemsep=0pt, topsep=1pt]
  \item $M^N(\cdot;\omega)$ has a $P$-a.s.\ unique maximizer $\theta_0(\omega):=\argmax_{\theta\in\Pi}M^N(\theta;\omega)$;
  \item $M$ has a unique maximizer $\theta^\dagger:=\argmax_{\theta\in\Pi}M(\theta)$;
  \item the split is selected over a random candidate set $\Pi(U)$ induced by $mtry$ features with $\mathbb{E}_U[\mathrm{dist}(\theta_0(\omega),\Pi(U))]\to 0$.
\end{enumerate}
Then $\hat\theta_{n_s}\in\argmax_{\theta\in\Pi(U)}M^*_{n_s}(\theta)$ satisfies
\[
\hat\theta_{n_s}\;\xrightarrow{p_d}\;\theta_0(\omega)\quad\text{for $P$-a.s.\ }\omega,
\qquad\text{and}\qquad
\hat\theta_{n_s}\;\xrightarrow{p_\otimes}\;\theta^\dagger.
\]
\end{corollary}

\begin{proof}[Proof of Corollary~\ref{cor:split-argmax}]
\textit{Step 1:} We first prove $\hat\theta_{n_s} \xrightarrow{p_d} \theta_0(\omega)$ for $P$-a.s.\ $\omega$, where
$\theta_0(\omega) := \argmax_{\theta\in\Pi} M^N(\theta;\omega)$, which we take to be $\sigma(\Omega)$-measurable.

Fix $\omega$ in the $P$-full-measure set on which the following hold: (i) $M^N(\cdot;\omega)$ has 
a unique maximizer $\theta_0(\omega)$ (by hypothesis of Theorem~4); 
(ii) Part~A's design-space convergence
$\sup_{\theta\in\Pi} |M^*_{n_s}(\theta) - M^N(\theta;\omega)| \xrightarrow{p_d} 0$ holds. Throughout this step, we work in the design space $(\mathcal S^N, \sigma(\mathcal S^N), P_{\mathcal S^N\mid\omega})$
with $\omega$ fixed.
We verify the conditions of Corollary~3.2.3(ii) of \cite{van-der-Vaart2023-ur} with $\Theta = \Pi$,
$M_n = M^*_{n_s}$, and $M = M^N(\cdot;\omega)$:

\textit{Uniform convergence on compacta.}\\
By Part~A, $\|M^*_{n_s} - M^N(\cdot;\omega)\|_\Pi 
\xrightarrow{p_d} 0$. For any compact $K\subseteq\Pi$, $\|M^*_{n_s} - M^N(\cdot;\omega)\|_K \le 
\|M^*_{n_s} - M^N(\cdot;\omega)\|_\Pi \xrightarrow{p_d} 0$.

\textit{Upper semicontinuity, unique maximum, compactness.} \\
By Lemma~\ref{lem:split-lipschitz}, $M^N(\cdot;\omega)$ is 
Lipschitz on $(\Pi, d)$ with constant $L_1 = L_1(\omega) < \infty$, hence continuous, hence upper 
semicontinuous. Uniqueness of $\theta_0(\omega)$ holds by hypothesis. 
We claim $\theta_0(\omega)$ is a well-separated maximum of $M^N(\cdot;\omega)$: for every $\varepsilon > 0$,
\[
\sup_{\theta\in\Pi:\, d(\theta,\theta_0(\omega))\ge \varepsilon} M^N(\theta;\omega) < M^N(\theta_0(\omega);\omega).
\]
To see this, fix $\varepsilon > 0$ and consider the set $\{\theta\in\Pi: d(\theta,\theta_0(\omega))\ge\varepsilon\}$. This set is closed in the compact space $(\Pi, d)$ (a finite disjoint union of compact intervals, established in the setup preceding Lemma~\ref{lem:split-lipschitz}), hence compact. Since $M^N(\cdot;\omega)$ is continuous by Lemma~\ref{lem:split-lipschitz}, its supremum over this set is attained at some $\theta_\varepsilon$. As $\theta_\varepsilon \ne \theta_0(\omega)$, uniqueness of the maximizer gives $M^N(\theta_\varepsilon;\omega) < M^N(\theta_0(\omega);\omega)$, which is the displayed strict inequality.

\textit{Uniform tightness of $\hat\theta_{n_s}$.} \\
$\hat\theta_{n_s}\in\Pi(U)\subseteq\Pi$, and 
$\Pi$ is compact, so $\{\hat\theta_{n_s}\}$ is uniformly tight.

\textit{Near-maximization.} \\
We show
$
M^*_{n_s}(\hat\theta_{n_s}) \ge \sup_{\theta\in\Pi} M^*_{n_s}(\theta) - o_{p_d}(1).
$
Let $\delta_n := \sup_{\theta\in\Pi} |M^*_{n_s}(\theta) - M^N(\theta;\omega)|$. By Part~A, 
$\delta_n \xrightarrow{p_d} 0$. Since $\hat\theta_{n_s}$ is an exact maximizer over $\Pi(U)$,
$
M^*_{n_s}(\hat\theta_{n_s}) = \sup_{\theta\in\Pi(U)} M^*_{n_s}(\theta).
$
For any $\theta\in\Pi$, $M^*_{n_s}(\theta) \le M^N(\theta;\omega) + \delta_n \le 
M^N(\theta_0(\omega);\omega) + \delta_n$, so
\begin{equation}
\label{eq:C1-upper-Pi}
\sup_{\theta\in\Pi} M^*_{n_s}(\theta) \le M^N(\theta_0(\omega);\omega) + \delta_n.
\end{equation}
Since $\Pi(U)$ is a non-empty closed subset of the compact set $\Pi$, and 
$\theta\mapsto d(\theta,\theta_0(\omega))$ is continuous, the infimum defining 
$\mathrm{dist}(\theta_0(\omega),\Pi(U)) := \inf_{\theta\in\Pi(U)} d(\theta,\theta_0(\omega))$ is 
attained at some $\widetilde\theta_U\in\Pi(U)$. By Lemma~\ref{lem:split-lipschitz},
\[
M^N(\theta_0(\omega);\omega) - M^N(\widetilde\theta_U;\omega) 
\le L_1(\omega)\cdot \mathrm{dist}(\theta_0(\omega),\Pi(U)).
\]
Combined with $M^*_{n_s}(\widetilde\theta_U) \ge M^N(\widetilde\theta_U;\omega) - \delta_n$ gives
\begin{equation}
\label{eq:C1-lower-PiU}
M^*_{n_s}(\hat\theta_{n_s}) = \sup_{\theta\in\Pi(U)} M^*_{n_s}(\theta) \ge M^*_{n_s}(\widetilde\theta_U) 
\ge M^N(\theta_0(\omega);\omega) - L_1(\omega)\cdot\mathrm{dist}(\theta_0(\omega),\Pi(U)) - \delta_n.
\end{equation}
Subtracting \eqref{eq:C1-lower-PiU} from \eqref{eq:C1-upper-Pi}, the $M^N(\theta_0(\omega);\omega)$ terms cancel, yielding
\begin{equation}
\label{eq:C1-deficit}
\sup_{\theta\in\Pi} M^*_{n_s}(\theta) - M^*_{n_s}(\hat\theta_{n_s}) 
\le L_1(\omega)\cdot \mathrm{dist}(\theta_0(\omega),\Pi(U)) + 2\delta_n.
\end{equation}
Taking $\mathbb E_U$ of \eqref{eq:C1-deficit} (holding $\xi$ fixed),
\[
\mathbb E_U\!\left[\sup_{\theta\in\Pi} M^*_{n_s}(\theta) - M^*_{n_s}(\hat\theta_{n_s})\,\Big|\,\xi\right]
\le L_1(\omega)\cdot \mathbb E_U[\mathrm{dist}(\theta_0(\omega),\Pi(U))] + 2\delta_n.
\]
For fixed $\omega$, $L_1(\omega)$ is a finite constant and $\mathbb E_U[\mathrm{dist}(\theta_0(\omega),\Pi(U))] \to 0$ by hypothesis, so the first term on the right is a deterministic $o(1)$, while the second term satisfies $\delta_n \xrightarrow{p_d} 0$. Hence the conditional expectation on the left is $o_{p_d}(1)$. 
Write $\Delta_n := \sup_{\theta\in\Pi} M^*_{n_s}(\theta) - M^*_{n_s}(\hat\theta_{n_s}) \ge 0$. From \eqref{eq:C1-deficit},
\[
\mathbb E_U[\Delta_n \mid \xi] \le L_1(\omega)\,\mathbb E_U[\mathrm{dist}(\theta_0(\omega),\Pi(U))] + 2\delta_n.
\]
For fixed $\omega$, $L_1(\omega) < \infty$ and $\mathbb E_U[\mathrm{dist}(\theta_0(\omega),\Pi(U))] \to 0$ by hypothesis, and $\delta_n \xrightarrow{p_d} 0$ by Part~A, so $\mathbb E_U[\Delta_n \mid \xi] = o_{p_d}(1)$. For any $\varepsilon > 0$, Markov's inequality conditional on $\xi$ gives
$
P_U(\Delta_n > \varepsilon \mid \xi) \le \frac{\mathbb E_U[\Delta_n \mid \xi]}{\varepsilon},
$
and taking $\mathbb E_\xi$ of both sides,
$
P_d(\Delta_n > \varepsilon) = \mathbb E_\xi\!\big[P_U(\Delta_n > \varepsilon \mid \xi)\big] \le \frac{\mathbb E_d[\Delta_n]}{\varepsilon} = o(1),
$
where $P_d = P_\xi \otimes P_U$ with $U$ the feature-subsampling randomness of Algorithm~1 independent of $\xi$. Hence $\Delta_n = o_{p_d}(1)$.

\textit{Conclusion.} By Corollary~3.2.3(ii) of \cite{van-der-Vaart2023-ur},
$\hat\theta_{n_s} \to \theta_0(\omega)$ in $(p_d\otimes p_U)$-probability. Integrating out $U$ gives that
for every $\varepsilon>0$,
\[
P_{\mathcal S^N\mid\omega}(d(\hat\theta_{n_s},\theta_0(\omega))>\varepsilon)
= \mathbb E_U[P_{\mathcal S^N\mid\omega}(d(\hat\theta_{n_s},\theta_0(\omega))>\varepsilon\mid U)]
\to 0,
\]
by dominated convergence. Hence $\hat\theta_{n_s} \xrightarrow{p_d} \theta_0(\omega)$ for $P$-a.s.\ 
$\omega$.

\textit{Step 2 (joint consistency)}
By Step 1, $\hat\theta_{n_s}\xrightarrow{p_d} \theta_0(\omega)$ for $P$-a.s.\ $\omega$. To identify the limit as the super-population maximizer $\theta^\dagger$, we apply Corollary~3.2.3(ii) of \cite{van-der-Vaart2023-ur} again, now with $M^N(\cdot;\omega)$ playing the role of the sequence $\mathbb{M}_n$, the population-level score $M$ playing the role of the deterministic limit $\mathbb{M}$, and $\theta^\dagger := \argmax_{\theta\in\Pi} M(\theta)$ its maximizer. The four conditions are verified as follows:
\begin{itemize}[leftmargin=2em, itemsep=0pt]
\item \textit{Uniform convergence on compacta.} By Step~1 of the proof for Theorem~4 above, $\sup_{g\in\mathcal H_d}|(P_N - P)(g)| = o_{p_\omega}(1)$, and \eqref{eq:lipschitz-bound} gives $\sup_\Pi |M^N(\theta;\omega) - M(\theta)| = o_{p_\omega}(1)$, a statement in the super-population law over $(\Omega,\mathcal A,P)$.
\item \textit{U.s.c., unique maximum, compactness.} $M$ is continuous on the compact $(\Pi, d)$: $\theta\mapsto P(X\in C_\ell(\theta))$ is continuous because $X$ has a Lebesgue density under $P$ (theorem hypothesis), and $\theta\mapsto \mu_{\mathbf k}(Q_\ell(\theta))$, where $Q_\ell(\theta)$ denotes the super-population conditional law of $Y$ given $X\in C_\ell(\theta)$, is continuous by (K1) via dominated convergence. Continuity implies u.s.c. Uniqueness of $\theta^\dagger$ holds by hypothesis. 
The compactness-plus-uniqueness argument of Part~B, now applied to $M$ (continuous, as just shown) in place of $M^N(\cdot;\omega)$, gives well-separation of $\theta^\dagger$.
\item \textit{Uniform tightness of $\theta_0(\omega)$.} $\theta_0(\omega)\in\Pi$, and $\Pi$ is compact, so $\{\theta_0(\omega)\}_N$ is uniformly tight in the joint space.
\item \textit{Near-maximization.} $\theta_0(\omega)$ is by definition the exact maximizer of $M^N(\cdot;\omega)$ on $\Pi$, with the maximum attained by continuity of $M^N$ (Lemma~\ref{lem:split-lipschitz}) on compact $\Pi$. Hence $M^N(\theta_0(\omega);\omega) = \sup_\Pi M^N(\cdot;\omega)$ and the near-maximization deficit is identically zero.
\end{itemize}
Corollary~3.2.3(ii) applied to the deterministic-in-$\mathcal S^N$ sequence $M^N(\cdot;\omega)$ over $(\Omega,\mathcal A,P)$ gives $\theta_0(\omega) \xrightarrow{p_\omega} \theta^\dagger$. Since $\theta_0(\omega) = \argmax_{\theta\in\Pi} M^N(\theta;\omega)$ is $\sigma(\Omega)$-measurable, its joint-space law coincides with its super-population law, so this super-population convergence is convergence in the joint law, and $\theta_0(\omega) \xrightarrow{p_\otimes} \theta^\dagger$ by Theorem~5.1(i) of \cite{2005_RubinBleuer}.
Combining with Part~B's $\hat\theta_{n_s}\xrightarrow{p_d}\theta_0(\omega)$ for $P$-a.s.\ $\omega$, the triangle inequality on $(\Pi, d)$ gives, for every $\varepsilon > 0$,
\[
P_{\Omega\times\mathcal S^N}\big(d(\hat\theta_{n_s}, \theta^\dagger) > \varepsilon\big) \le P_{\Omega\times\mathcal S^N}\big(d(\hat\theta_{n_s}, \theta_0(\omega)) > \tfrac{\varepsilon}{2}\big) + P_{\Omega\times\mathcal S^N}\big(d(\theta_0(\omega), \theta^\dagger) > \tfrac{\varepsilon}{2}\big).
\]
The first term vanishes by bounded convergence applied to the a.s.-$\omega$ design-consistency of Part~B, and the second by the model-side convergence just established. Hence $\hat\theta_{n_s} \xrightarrow{p_\otimes} \theta^\dagger$.
\end{proof}

\begin{remark}
    For the consistency in the selected split to hold, we need $\E[dist(\theta_0, \Pi(U))] \to 0$, meaning that on average, over random draws of $U$, the class of possible splits gets arbitrarily close to $\theta_0$. It is okay to have $U$ that draws from features $\{1, \dots, p\}$ that fail to include the true optimal $j_0$. For finite feature set and random sampling of $U$, $\E[dist(\theta_0, \Pi(U))] \leq c P(j_0 \not\in U) = c(1 - \mathrm{mtry}/p)$, $0 < c < \infty$, so a sufficient condition is to grow mtry so that $\mathrm{mtry}/p \to 1$. 
    For consistency of the forest, which we prove latter, $\mathrm{mtry}$ need not grow with $N$. This condition is only needed for the local leaf to match the truth, where by `truth' we mean the split if the entire finite population data is available. 
\end{remark}

\subsection{Effect of Multiplier Averaging}

Although design-aware re-sampling allows multiple trees, which enables bagging, the resampled weights $n^*$ might result in large variation of $\hat{w}_i = 1/\pi_i \times n^*$. Averaging multiple conditional independent realizations of $n^*$ gives $\hat{w}_i = 1/\pi_i \times \bar{n}^*_{b,i}$, where $\bar{n}^*_{b,i}$ the average of $M$ realizations, is a simple and direct technique for stabilizing the resulting weights. 
We analyze the effect of $M$ in the following.\\

We first establish a shared design-side variance template.
\begin{lemma}[PSU-block variance template for Hilbert-space-valued HT sums, unified for both design cases]
\label{lem:psu-block-variance}
Let $\mathcal K$ be a real Hilbert space with inner product $\langle\cdot,\cdot\rangle_{\mathcal K}$ and induced norm $\|\cdot\|_{\mathcal K}$. Let $\{a_i\}_{i=1}^N \subset \mathcal K$ be $\sigma(\omega)$-measurable with $\sup_{i\in[N]}\|a_i\|_{\mathcal K}\le A$ and effective support $S := |\{i\in[N]: a_i \ne 0\}| \le N$. Under either
\begin{itemize}[leftmargin=3em]
\item[\textup{(i)}] \textup{(D1)--(D4)} for a single-stage design, with sample index $i\in[n_s]$ and inclusion probability $\pi_i$, or
\item[\textup{(ii)}] the regime clause and \textup{(D-TS1)--(D-TS4)} for a stratified two-stage cluster design, with SSU index $u=(h,k,j)\in[n_s]$ (denoted $i$ generically) and inclusion probability $\pi_u=\pi^{(1)}_{hk}\pi^{(2)}_{hkj\mid hk}$,
\end{itemize}
we have
\begin{equation}\label{eq:psu-block-var}
\mathbb{E}_{P_{\mathcal S^N\mid\omega}}\!\left\|\sum_{i=1}^N\!\left(\frac{\xi_i}{\pi_i}-1\right)a_i\right\|_{\mathcal K}^{\!2} \;\le\; C^{\dagger}\,A^2\,S,
\end{equation}
where $C^{\dagger}$ is a design-uniform constant depending only on the regime constants of the respective design case --- specifically $(\pi_0, C^{(\mathrm{D4})})$ under \textup{(i)}, and $(\underline\lambda, \bar\lambda_1, \bar\lambda_2, M_{\max}, C_1, C_2, f)$ from \textup{(D-TS1)}--\textup{(D-TS4)} and Lemma~\ref{lem:designs-multistage} under \textup{(ii)}.
\end{lemma}

\begin{proof}[Proof of Lemma~\ref{lem:psu-block-variance}]
Expanding the squared norm and using $\mathbb{E}[\xi_i/\pi_i] = 1$,
\[
\mathbb{E}\left\|\textstyle\sum_i\!(\xi_i/\pi_i - 1)a_i\right\|_{\mathcal K}^{\!2}
= \sum_{i=1}^N \sum_{j=1}^N \mathrm{Cov}\bigl(\xi_i/\pi_i,\xi_j/\pi_j\bigr)\,\langle a_i, a_j\rangle_{\mathcal K}.
\]
Bounding $|\langle a_i, a_j\rangle_{\mathcal K}|\le\tfrac12(\|a_i\|_{\mathcal K}^2+\|a_j\|_{\mathcal K}^2)\le A^2 \ind(a_i\ne 0)\ind(a_j\ne 0)$ and using symmetry $\mathrm{Cov}_{ij}=\mathrm{Cov}_{ji}$,
\[
\sum_{i\ne j}|\mathrm{Cov}(\xi_i/\pi_i,\xi_j/\pi_j)|\,|\langle a_i, a_j\rangle_{\mathcal K}| \;\le\; \sum_{i:a_i\ne 0}\|a_i\|_{\mathcal K}^2 \sum_{j\ne i}|\mathrm{Cov}(\xi_i/\pi_i,\xi_j/\pi_j)|.
\]
We bound the diagonal and the outer row-sum $\sum_{j\ne i}|\mathrm{Cov}(\xi_i/\pi_i,\xi_j/\pi_j)|$ separately in each design case.

\smallskip
\emph{Diagonal.} In both design cases, $\pi_i\ge\pi_0>0$ --- from (D3) under (i) and from Lemma~\ref{lem:designs-multistage}\,(i) under (ii). Hence $\mathrm{Var}(\xi_i/\pi_i) = (1-\pi_i)/\pi_i \le 1/\pi_0$, and the diagonal contributes at most $A^2 S/\pi_0$.

\smallskip
\emph{Off-diagonal under \textup{(i)}.} Standard single-stage second-order inclusion analysis under \textup{(D3)}--\textup{(D4)} yields a max-row-sum pairwise bound: there exists a constant $C^{(\mathrm{D4})}<\infty$, depending on the regime constants of \textup{(D3)}--\textup{(D4)}, such that
\[
\max_{i:\xi_i=1}\sum_{j\ne i}\bigl|\mathrm{Cov}(\xi_i/\pi_i,\xi_j/\pi_j)\bigr| \;\le\; C^{(\mathrm{D4})}\quad\text{a.s.\ for all }N.
\]
The off-diagonal contribution is at most $C^{(\mathrm{D4})}\cdot A^2 S$.

\smallskip
\emph{Off-diagonal under \textup{(ii)}.} The pairs $(i,j)$ with $j\ne i$ partition by the three cases of Lemma~\ref{lem:designs-multistage}\,(ii):
\begin{itemize}[leftmargin=2em, itemsep=1pt, topsep=1pt]
\item \emph{Cross-stratum ($h\ne h'$).} Independent by construction; $\mathrm{Cov}=0$; no contribution.
\item \emph{Case~2 ($h=h', k\ne k'$).} $|\mathrm{Cov}(\xi_u/\pi_u,\xi_v/\pi_v)| = |\pi_{uv}-\pi_u\pi_v|/(\pi_u\pi_v) \le (C_1/n_{\mathrm{eff}})/(\underline\lambda n/N)^2 = O(1/n_{\mathrm{eff}})$ under $n\asymp n_{\mathrm{eff}}\asymp N$. Number of Case-2 partners of a fixed $u$ with $a_v\ne 0$ is bounded by $S$. Contribution to the row-sum: $O(S/n_{\mathrm{eff}})$.
\item \emph{Case~3 ($h=h', k=k', j\ne j'$).} $|\mathrm{Cov}(\xi_u/\pi_u,\xi_v/\pi_v)| = O(n_{\mathrm{eff}}/N)\cdot O((N/n)^2) = O(1)$ under $n\asymp n_{\mathrm{eff}}\asymp N$. Each $u$ has at most $M_{\max}-1$ same-PSU partners, so the row-sum contribution is $O(M_{\max})$.
\end{itemize}
Aggregating, $\max_i\sum_{j\ne i}|\mathrm{Cov}(\xi_i/\pi_i,\xi_j/\pi_j)|\le O(S/n_{\mathrm{eff}})+O(M_{\max})$. Multiplied by the outer $\sum_i\|a_i\|^2\le A^2 S$, the off-diagonal contribution is $O(A^2 S^2/n_{\mathrm{eff}}) + O(A^2 S M_{\max})$; under $S\le N\asymp n_{\mathrm{eff}}$, both are $O(A^2 S)$.

\smallskip
Combining the diagonal and off-diagonal contributions in either case, $\mathbb{E}\|\cdot\|^2_{\mathcal K}\le C^\dagger A^2 S$ with $C^\dagger = 1/\pi_0 + C^{(\mathrm{D4})}$ under (i), and $C^\dagger = 1/\pi_0 + O(1 + M_{\max})$ (with implicit constants from Lemma~\ref{lem:designs-multistage}\,(ii)) under (ii). This yields~\eqref{eq:psu-block-var}.
\end{proof}

\begin{remark}[Scope]
Lemma~\ref{lem:psu-block-variance} covers both scalar-valued ($\mathcal K=\mathbb R$) and RKHS-valued ($\mathcal K=\mathcal H$) HT sums under either design regime; the bookkeeping does not depend on the codomain. The bound is agnostic to the specific $a_i$: the same template applies to indicator sums (leaf-size variance), kernel embeddings (RKHS-valued design HT variance), and multiplier-weighted variants that arise in later sections.
\end{remark}

\vspace*{0.3cm}
\begin{proof}[Proof of Proposition~5]
The proof is uniform in single--stage (i) and stratified two--stage cluster designs (ii); design-specific tools enter at two localized points, indicated in Step~1 and Step~2 below.

Throughout the proof, ``sampled unit'' denotes $i$ under (i) and $u=(h,k,j)$ under (ii); we use the symbol $i$ generically. Let $(\nu, c)$ denote the sub-gamma envelope parameters of the centered per-tree multiplier: $(\nu, c) = (\nu_0, c_0)$ under (i) via Lemma~\ref{lem:subgamma}, and $(\nu, c) = (\tilde\nu, \tilde c) := (W_{\max}^2\nu_1+\nu_2,\, c_1 W_{\max}\vee c_2)$ under (ii) via Lemma~\ref{lem:subgamma-multistage}. Let $C$ denote the max-row-sum (R3) constant: as in (R3) under (i) and $C=C_{\mathrm{TS}}$ under (ii), see Lemma~7 and its proof in Section~S5.2. Fix $\theta=(j,t)$; all bounds are conditional on the realized finite population, with design randomness $(\xi,Z^N)$ carried by $p_d$.

Fix $\theta=(j,t)$; all bounds are conditional on the realized finite population, with
design randomness $(\xi,Z^N)$ carried by $p_d$. Abbreviate $S_\ell:=\{i\in[n_s]:X_i\in C_\ell\}$,
$\ell\in\{L,R\}$, and set
\[
\widehat P^{n_s}_\ell:=\frac{1}{\widehat N^{n_s}_\ell}\sum_{i\in S_\ell}\frac{\xi_i}{\pi_i}\,\delta_{Y_i},
\quad
\widehat P^{*}_\ell:=\frac{1}{\widehat N^{*}_\ell}\sum_{i\in S_\ell}\frac{\xi_i\bar n_i^{*}}{\pi_i}\,\delta_{Y_i},
\quad
P_\ell:=\frac{1}{N_\ell}\sum_{X_i\in C_\ell}\delta_{Y_i},
\]
with $\widehat N^{n_s}_\ell=\sum_{i\in S_\ell}\xi_i/\pi_i$,
$\widehat N^{*}_\ell=\sum_{i\in S_\ell}\xi_i\bar n_i^{*}/\pi_i$.

\medskip\noindent\emph{Step 1 (reduction).}
Under (A3), $N_\ell/N_{P_a}\in[\alpha,1-\alpha]$. Under (R1)--(R2), the event
\begin{equation}\label{eq:avg-event}
P_d\bigl(\widehat N^{*}_\ell/\widehat N^{*}_{P_a}\ge\alpha/2,\ \ell\in\{L,R\}\bigr)\to 1
\end{equation}
follows from the design LLN (Lemma~\ref{lem:lln_clt} under (i), Lemma~\ref{lem:lln_clt_multistage} under (ii)) applied to (D2)--(D3) or (D-TS2)--(D-TS3) respectively. On the event in~\eqref{eq:avg-event}, $\Psi_\theta$ of~\eqref{eq:M-representation} has denominators $\ge\alpha/2$ and, by $K_{\max}<\infty$ (K1), entries $\le K_{\max}$; the fixed-$\theta$ specialization of~\eqref{eq:lipschitz-bound} gives
\begin{equation}\label{eq:avg-reduction}
\bigl|M_{n_s}^{*(M)}(\theta)-M^N(\theta)\bigr|
\lesssim
\sum_{\ell\in\{L,R\}}
\Bigl|\tfrac{\widehat N^{*}_\ell}{\widehat N^{*}_{P_a}}-\tfrac{N_\ell}{N_{P_a}}\Bigr|
+
\sum_{\ell\in\{L,R\}}
\bigl\lVert\mu_{\mathbf k}(\widehat P^{*}_\ell)-\mu_{\mathbf k}(P_\ell)\bigr\rVert_{\mathcal H}.
\end{equation}

\medskip\noindent\emph{Step 2 (embedding term).}
Insert the H\'ajek pivot $\widehat P^{n_s}_\ell$ and apply the triangle inequality in $\mathcal H$:
\begin{equation}\label{eq:avg-pivot}
\bigl\lVert\mu_{\mathbf k}(\widehat P^{*}_\ell)-\mu_{\mathbf k}(P_\ell)\bigr\rVert_{\mathcal H}
\le
\bigl\lVert\mu_{\mathbf k}(\widehat P^{*}_\ell)-\mu_{\mathbf k}(\widehat P^{n_s}_\ell)\bigr\rVert_{\mathcal H}
+
\bigl\lVert\mu_{\mathbf k}(\widehat P^{n_s}_\ell)-\mu_{\mathbf k}(P_\ell)\bigr\rVert_{\mathcal H}.
\end{equation}
 
\emph{Design error.}
$\mathbb E[\xi_i/\pi_i-1\mid Z^N]=0$. Apply Lemma~\ref{lem:psu-block-variance} with $\mathcal K = \mathcal H$, $a_i := \mu_{\mathbf k}(\delta_{Y_i})\ind(X_i\in C_\ell)/N_\ell$ (so $\|a_i\|_{\mathcal H}\le K_{\max}^{1/2}/N_\ell$ and $S = N_\ell$):
\[
\mathbb E\!\left\|\tfrac{1}{N_\ell}\sum_{i}(\xi_i/\pi_i - 1)\mu_{\mathbf k}(\delta_{Y_i})\ind(X_i\in C_\ell)\right\|_{\mathcal H}^{\!2}
\;\le\; C^{\dagger}\,\frac{K_{\max}}{N_\ell^2}\cdot N_\ell \;=\; O\bigl(K_{\max}/N_\ell\bigr).
\]
Using $N_\ell\asymp N_{P_a}$ from (A3) and Chebyshev,
\begin{equation}\label{eq:avg-design}
\bigl\lVert\mu_{\mathbf k}(\widehat P^{n_s}_\ell)-\mu_{\mathbf k}(P_\ell)\bigr\rVert_{\mathcal H}
= O_{p_d}\!\bigl(N_{P_a}^{-1/2}\bigr).
\end{equation}
 
\emph{Averaging error.}
Condition on $(\xi,Z^N)$; set $\bar\delta_i:=\bar n_i^{*}-\mathbb E(\bar n_i^{*}\mid\xi,Z^N)$. By the sub-gamma envelope $(\nu, c)$, $\bar\delta_i\in\Gamma(\nu/M,c/M)$ and so $\operatorname{Var}(\bar\delta_i\mid\xi,Z^N)\le\nu/M$; by (R3) in max-row-sum form,
\[
\max_i\sum_{j\ne i}\bigl|\operatorname{Cov}(\bar\delta_i,\bar\delta_j\mid\xi,Z^N)\bigr|
\;\le\;
C\,\sup_j\!\operatorname{Var}(\bar\delta_j\mid\xi,Z^N).
\]
Writing $a_i:=(\xi_i/\pi_i)\mu_{\mathbf k}(\delta_{Y_i})/\widehat N^{n_s}_\ell$ (RKHS-valued), the ratio linearization's leading term is $\sum_{i\in S_\ell} a_i \bar\delta_i$; expanding the squared RKHS norm,
\[
\mathbb{E}\Bigl\|\sum_{i\in S_\ell} a_i \bar\delta_i\Bigr\|^2_{\mathcal H}
= \sum_i \|a_i\|_{\mathcal H}^2\, \operatorname{Var}(\bar\delta_i)
+ \sum_{i\ne j}\operatorname{Cov}(\bar\delta_i,\bar\delta_j)\,\langle a_i, a_j\rangle_{\mathcal H}.
\]
Bounding $|\langle a_i, a_j\rangle_{\mathcal H}|\le\tfrac12(\|a_i\|_{\mathcal H}^2+\|a_j\|_{\mathcal H}^2)$ and using symmetry of $\operatorname{Cov}$,
\[
\sum_{i\ne j}|\operatorname{Cov}|\,|\langle a_i, a_j\rangle_{\mathcal H}|
\le \sum_i \|a_i\|_{\mathcal H}^2 \sum_{j\ne i}|\operatorname{Cov}(\bar\delta_i,\bar\delta_j)|
\le C\,\sup_j\!\operatorname{Var}(\bar\delta_j)\,\sum_i \|a_i\|_{\mathcal H}^2,
\]
so with $\|\mu_{\mathbf k}(\delta_{Y_i})\|_{\mathcal H}^2\le K_{\max}$,
\begin{equation}\label{eq:avg-var}
\operatorname{Var}\!\Bigl(\tfrac{1}{\widehat N^{n_s}_\ell}\sum_{i\in S_\ell}\tfrac{\xi_i}{\pi_i}\bar\delta_i\,\mu_{\mathbf k}(\delta_{Y_i})\,\Big|\,\xi,Z^N\Bigr)
\le
(1+C)K_{\max}\frac{\nu}{M}\,
\frac{\sum_{i\in S_\ell}\pi_i^{-2}}{\bigl(\sum_{i\in S_\ell}\pi_i^{-1}\bigr)^{2}}
=
(1+C)K_{\max}\frac{\nu}{M}\,n_{\mathrm{eff},\ell}^{-1},
\end{equation}
the equality being the Kish identity. Chebyshev (conditional on $(\xi,Z^N)$), Jensen, then $\mathbb E$ over the design give, with the (R2) bias $O_{p_d}(M^{-1})$ absorbed,
\begin{equation}\label{eq:avg-averaging}
\bigl\lVert\mu_{\mathbf k}(\widehat P^{*}_\ell)-\mu_{\mathbf k}(\widehat P^{n_s}_\ell)\bigr\rVert_{\mathcal H}
= O_{p_d}\!\bigl(M^{-1/2}n_{\mathrm{eff},\ell}^{-1/2}\bigr).
\end{equation}
Substituting~\eqref{eq:avg-design} and~\eqref{eq:avg-averaging} into~\eqref{eq:avg-pivot},
\begin{equation}\label{eq:avg-embed-total}
\bigl\lVert\mu_{\mathbf k}(\widehat P^{*}_\ell)-\mu_{\mathbf k}(P_\ell)\bigr\rVert_{\mathcal H}
= O_{p_d}\!\bigl(N_{P_a}^{-1/2}+M^{-1/2}n_{\mathrm{eff},\ell}^{-1/2}\bigr).
\end{equation}

\medskip
\medskip\noindent\emph{Step 3 (mass ratio).}
Setting $g\equiv 1$ (envelope $1$) in the arguments of~\eqref{eq:avg-design}--\eqref{eq:avg-var} gives
\begin{equation}\label{eq:avg-mass}
\Bigl|\tfrac{\widehat N^{*}_\ell}{\widehat N^{*}_{P_a}}-\tfrac{N_\ell}{N_{P_a}}\Bigr|
= O_{p_d}\!\bigl(N_{P_a}^{-1/2}+M^{-1/2}n_{\mathrm{eff},\ell}^{-1/2}\bigr),
\end{equation}
which adds no rate beyond~\eqref{eq:avg-embed-total}.

Finally,
substituting~\eqref{eq:avg-embed-total} and~\eqref{eq:avg-mass} into~\eqref{eq:avg-reduction},
\[
\bigl|M_{n_s}^{*(M)}(\theta)-M^N(\theta)\bigr|
= O_{p_d}\!\left(N_{P_a}^{-1/2}+\frac{1}{\sqrt M}\sum_{\ell\in\{L,R\}} n_{\mathrm{eff},\ell}^{-1/2}\right).
\qedhere
\]
\end{proof}

\vspace*{0.3cm}
\section{Consistency of MMD}

We now establish supporting results. Denote $L(\mathbf{x})$ the oracle leaf containing 
$\mathbf{x}$ when the splitting algorithm is applied to the full finite population 
$\{(X_i, Y_i)\}_{i=1}^N$. In this oracle setting, the design reduces to sampling with 
or without replacement from the finite population, and the splitting criterion simplifies 
to that of standard i.i.d.\ distributional random forests as all weights become $1/N$. 
In contrast, $L^*_\xi(\mathbf{x})$ denotes the estimated leaf when the algorithm is 
applied to the design-weighted sample with resampling. Both $L(\cdot)$ and $L^*_\xi(\cdot)$ 
involve algorithmic randomness (e.g., random selection of candidate split variables at 
each node); however, this randomness is independent of the sampling design and averages 
out as $B \to \infty$, following standard practice in random forests. We therefore suppress this randomness in our notation.

\subsection{Single-stage Designs}
The following lemmas, combining algorithmic assumptions (A1)--(A3) and (B1)--(B2) with 
design assumptions (D1)--(D4), establish that leaf sizes remain non-degenerate with 
high probability.

\vspace*{0.3cm}
\begin{lemma}
    By (A2)-(A3), (B1), for any given $\mathbf{x}$, $N_{L(\mathbf{x})} = \sum_{i=1}^N \ind(X_i \in L(\mathbf{x})) = \Theta_p(Nk^p_N)$, and so $N_{L(\mathbf{x})} \to_p \infty$ as $N \to \infty$.
    \label{lem:N_L}
\end{lemma}

\begin{proof}[Proof of Lemma~\ref{lem:N_L}]
    For any $\eta > 0$, 
\begin{align*}
    P(N_{L(\mathbf{x})} \leq \eta N k_N^p) 
    \leq 
    P(N_{L(\mathbf{x})} \leq \eta N k_N^p,\; \mathrm{diam}(L(\mathbf{x})) < k_N) + P(\mathrm{diam}(L(\mathbf{x})) \geq k_N).
\end{align*}
The second part goes to zero by (B1). On $\{\mathrm{diam}(L(\mathbf{x})) < k_N\}$ for which the event happens with probability $1 - o(1)$, 
by (A2),
$
\bar{B}_{c_1 k_N}(\mathbf{x}) \subseteq L(\mathbf{x}) \subseteq B_{k_N}(\mathbf{x}),
$
so by (A3),
\[
\underline{c}\, c_1^p N k_N^p
\leq
\#\{\,i : X_i \in \bar{B}_{c_1 k_N}(\mathbf{x}) \}
\;\leq\; 
\#\{\,i : X_i \in L(\mathbf{x}) \}
\;\leq\; 
\#\{\,i : X_i \in B_{k_N}(\mathbf{x}) \}
\leq 
\bar{c}\, N k_N^p,
\]
so $N_{L(\mathbf{x})} = \Theta_p(Nk^p_N)$. Finally, as, $\lim_{N \to \infty} Nk^p_N = \infty$, completing the proof.
\end{proof}

\vspace*{0.3cm}
\begin{lemma}
    For any given $\mathbf{x}$, 
    with $n_{s,N} = \sum_i \xi_i$, 
    $n_{L(\mathbf{x})} = \sum_i \xi_i \ind(X_i \in L(\mathbf{x}))$, 
    it holds that 
    $n_{s,N} = \Theta_{p_d}(N)$ and 
    $n_{L(\mathbf{x})}/N = \Theta_{p_\otimes}(k_N^p)$.
    \label{lem:ns_nL}
\end{lemma}
\begin{proof}[Proof of Lemma~\ref{lem:ns_nL}]
(D2) implies for any $\epsilon > 0$, $\exists \; N_0(\epsilon) \geq 1$ such that for all $N \geq N_0$, 
$
    P\!\Big( \frac{f}{2} \leq \frac{n_{s,N}}{N} \leq \frac{3f}{2} \Big) \geq 1 - \epsilon,
$
so $n_{s,N}/N = O_{p_d}(1)$ and $N/n_{s,N} = O_{p_d}(1)$ since $f \in (0, 1]$.
As $\pi_i \in (0,1)$ and by Lemma~\ref{lem:N_L}, it holds that
\begin{align*}
    Var(n_{L(\mathbf{x})} \mid \omega) 
    &= 
    \sum_i \pi_i (1-\pi_i)I_i + \sum_{i\neq j} (\pi_{ij} - \pi_i \pi_j)I_i I_j \\
    &\leq
    N_{L(\mathbf{x})} + C_1\,\frac{N_{L(\mathbf{x})}(N_{L(\mathbf{x})} - 1)}{n_{s,N}}
    =
    \Theta_p(Nk_N^p) \cdot O_{p_d}(1), \text{ and } \\
    \E[n_{L(\mathbf{x})} \mid \omega]
    &= 
    \sum_i \pi_i \ind(X_i \in L(\mathbf{x})) 
    \geq 
    N_{L(\mathbf{x})}\, \underline{\lambda}\, \frac{n}{N}
    = 
    \Theta_p(N k_N^p).
\end{align*}
Hence
\begin{align*}
    &P\left( \left| \frac{n_{L(\mathbf{x})}}{\E[n_{L(\mathbf{x})}]} - 1\right| \geq \epsilon \right)
    =
    P\left( \big| {n_{L(\mathbf{x})}} - \E[n_{L(\mathbf{x})}] \big| \geq \epsilon \E[n_{L(\mathbf{x})}] \right)\\
    \leq
    &\frac{Var(n_{L(\mathbf{x})} \mid \omega)}{\epsilon^2 (\E[n_{L(\mathbf{x})}\mid \omega])^2}
    =
    \frac{1}{\epsilon^2\, \Theta_p(N k_N^p)} = o_{p_\otimes}(1),
\end{align*}
and $n_{L(\mathbf{x})} = \E[n_{L(\mathbf{x})} \mid \omega](1 + o_{p_\otimes}(1)) = \Theta_{p_\otimes}(N k_N^p)$, as desired.
\end{proof}

\vspace*{0.3cm}
\begin{lemma}
    Assuming (D2) (or (D-TS2)), it holds that for $1 \leq p \leq 4$
    \begin{align}
        \sup_{N\geq 1} \E \left[\left(\frac{N}{n_{s,N}}\right)^p \mid n_{s,N} \geq 1 \right] < \infty.
    \end{align}
    \label{lem:N_ns_moments}
\end{lemma}
\begin{proof}[Proof of Lemma~\ref{lem:N_ns_moments}]
Fix $\delta \in (0, f)$ as in (D2). Since $n_{s,N}/N \to_{p_d} f > 0$, $P_{\mathcal{S}^N \mid \omega}(n_{s,N} \geq 1) \to 1$, so for $N$ large enough,
\begin{align*}
    P_{\mathcal{S}^N | \omega}\!\left(\frac{n_{s,N}}{N}\le \delta \,\Big|\, n_{s,N}\ge 1\right)
    \;\le\;
    \frac{P_{\mathcal{S}^N | \omega}\!\left({n_{s,N}}/{N}\le \delta\right) - P_{\mathcal{S}^N | \omega}(n_{s,N} = 0)}{P_{\mathcal{S}^N | \omega}(n_{s,N}\ge 1)}
    \;\le\; 2C\,N^{-\kappa} \quad \text{by (D2).}
\end{align*}
Splitting into the regions $\{n_{s,N}/N > \delta\}$ and $\{n_{s,N}/N \leq \delta\}$,
\begin{align*}
    \E\!\left[\left(\frac{N}{n_{s,N}}\right)^p \,\Big|\, n_{s,N} \geq 1 \right]
    \leq
    \delta^{-p} + 2C N^{p-\kappa},
\end{align*}
where $p - \kappa < 0$ since $\kappa > 4$ and $p \leq 4$, completing the proof.
\end{proof}

\vspace*{0.3cm}
\begin{lemma}
Fix $\mathbf x \in \mathcal{X} \subseteq \mathbb{R}^p$, $p \geq 1$. For $L(\mathbf x)$ the oracle leaf, and $L^*_{\xi}(\mathbf x)$ the leaf produced by the SDRF. 
Let $\{n^*_{b,i}\}_{i=1}^N$ be the {unit-level resample multipliers} used in the estimation stage for tree $b \in [B]$.
Define indicators
\begin{align*}
    I_i := \mathbbm{1}\{X_i \in L(\mathbf x)\}, 
    \quad
    I_i^* := \mathbbm{1}\{X_i \in L^*_{\xi}(\mathbf x)\},
\end{align*}
and the (design-sample) leaf size
$
n_L(\mathbf x) := \sum_{i=1}^N \xi_i I_i.
$
Assume (R1)--(R3), and (A1)--(A3), (B1)--(B2),
Then the index set of ``good trees'' 
\begin{align}
\mathcal{B}_{L}(\mathbf{x})
&:= \left\{\, b \in [B] :
\sum_{i=1}^{N} n^{*}_{b,i}\, I_i
\in \big[\, n_L(\mathbf{x}) - n_{d,N},\; n_L(\mathbf{x}) + n_{d,N} \,\big]
\right\}, \label{def:BL}\\
\mathcal{B}_{L^*}(\mathbf{x})
&:= \left\{\, b \in [B] :
\sum_{i=1}^{N} n^{*}_{b,i}\, I_i^{*}
\in \big[\, n_L(\mathbf{x}) - n_{d,N},\; n_L(\mathbf{x}) + n_{d,N} \,\big]
\right\} \label{def:BL*}
\end{align}
satisfy 
\begin{enumerate}
    \item $\lvert \mathcal{B}_{L}(\mathbf{x})\rvert/B \to_{p_d} 1$ and $\lvert \mathcal{B}_{L^*}(\mathbf{x})\rvert/B \to_{p_d} 1$.
    \item For any tree index $b$,
    $
    P_{\mathcal{S}^N \mid \omega}\!\left( b \in \mathcal{B}_{L}(\mathbf{x})\,\Delta\,\mathcal{B}_{L^*}(\mathbf{x}) \right)
    = O_{p_d}\!\left(\frac{1}{N k_N^{p}}\right) + O_{p_d}\!\left(\frac{d_N}{k_N^{p}}\right),
    $
    so in particular when $d_N = o(k_N^{p})$ and $N k^p_N \to \infty$ then
    \[
    P_{\mathcal{S}^N \mid \omega}\!\left( b \in \mathcal{B}_{L}(\mathbf{x})\,\Delta\,\mathcal{B}_{L^*}(\mathbf{x}) \right)
    = o_{p_d}(1).
    \]
\end{enumerate}
\label{lem:BS}
\end{lemma}

\begin{proof}[Proof of Lemma~\ref{lem:BS}]
Part 1:
We first control the mean and variance for the oracle leaf--based sizes.
Define
$
S_b := \sum_{i=1}^{N} n^{*}_{b,i} I_i \, \forall b.
$
By (R2), the conditional mean 
\begin{align*}
    \mu &:= \E(S_b \mid \xi, L)
    = \sum_{i=1}^{N} \E(n^{*}_{b,i}\mid \xi)\, I_i
    = \sum_{i=1}^{N} \xi_i (1 + o_{p_d}(1)) I_i = n_L\{1 + o_{p_d}(1)\}.
\end{align*}
For its conditional variance, 
\begin{align*}
    Var(S_b \mid \xi, L) 
    &=
    \sum_{i} I_i^{2}\,Var\!\bigl(n_i^{*}\mid \xi\bigr)
    \;+\;
    \sum_{i\neq j} I_i I_j\,Cov\!\bigl(n_i^{*},n_j^{*}\mid \xi\bigr)\\
    &\;\leq\;
    \sum_{i=1}^{N} Var(n^{*}_{b,i}\mid \xi)\, I_i
    + 
    \sum_{i\neq j} \bigl|Cov(n^{*}_{b,i}, n^{*}_{b,j}\mid \xi)\bigr|\, 
    \;\le\;
    (1+C_{\mathrm{cov}})\sum_{i=1}^{N} Var(n^{*}_{b,i}\mid \xi)\, I_i \\
    &\leq
    (1+C_{\mathrm{cov}})C_2 \sum_{i=1}^{N} I_i
    = (1+C_{\mathrm{cov}})C_2\, N_L, 
    \quad 
    N_L := \sum_{i=1}^{N} I_i.
\end{align*}
The first inequality by $\lvert I_i I_j\rvert \le 1$, the second by (R3), and the third holds as
by (R2), $\sup_i \E((n^{*}_{b,i})^2\mid \xi)\le C_2$ a.s., and
$\E(n^{*}_{b,i}\mid \xi)=\xi_i(1 + o_{p_d}(1))$, so $Var(n^{*}_{b,i}\mid \xi)\le C_2$ a.s.
Together,
$N_L=\Theta_{p_d}(Nk_N^p)$ by Lemma~\ref{lem:N_L}, hence
\begin{align}
    Var(S_b \mid \xi, L)=O_{p_d}(Nk_N^p).
    \label{lemma_B:eq1}
\end{align}

By Lemma~\ref{lem:N_L}, $n_L(\mathbf x) = \Theta_{p_d}(Nk^p_N)$, thus one can always choose a sequence $n_{min, N} \to \infty$ with $n_{min, N} = o(Nk^p_N)$ such that 
$
P_{\mathcal{S}^N\mid \omega}(n_{L}(\mathbf{x}) \geq 2 n_{min, N}) \to 1.
$
Now choose $n_{d,N}\le n_{\min,N}$ where $d$ denotes it is characterizing a tolerance region, so that on the
high-probability event $\{n_L \ge 2n_{\min,N}\}$ we also have $n_{d,N}\le n_L/2$. So
$
\{S_b \notin [n_L-n_{d,N},\, n_L+n_{d,N}]\}
\subseteq \{|S_b-\mu|>\mu/2\},
$
and Chebyshev gives 
\[
P\!\big(S_b \notin [n_L-n_{d,N},\, n_L+n_{d,N}] \mid \xi, L\big)
\le P\!\big(|S_b-\mu|>\mu/2 \mid \xi, L\big)
\le \frac{4\,Var(S_b\mid \xi, L)}{\mu^2}.
\]

Using $\mu=n_L(1+o_{p_d}(1))=\Theta_{p_d}(Nk_N^p)$ (Lemma \ref{lem:N_L}) together with (\ref{lemma_B:eq1}), we obtain
\begin{align}
    P\!\left(S_b \notin [n_L-n_{d,N},\, n_L+n_{d,N}] \mid \xi, L\right)
= O_{p_d}\!\left(\frac{1}{Nk_N^p}\right).
\label{lemma_B:part1_res}
\end{align}

Part2:
We then control the mean and variance for the estimated leaf--based sizes. The argument is similar:
For $S_b^{*} := \sum_{i=1}^{N} n^{*}_{b,i}\, I_i^{*}$, 
by (R2) 
$\mu^{*} := \E(S_b^{*}\mid \xi, L^{*})
= \sum_{i=1}^{N} \E(n^{*}_{b,i}\mid \xi)\, I_i^{*}
= \sum_{i=1}^{N} \xi_i(1 + o_{p_d}(1)) I_i^{*} = n^*_{L}(\mathbf{x})\{1 + o_{p_d}(1)\}.$
Since $\mu^{*}-n_L = \sum_{i=1}^{N} \xi_i\,(I_i^{*}-I_i)$ and $\sum_i |I_i^* - I_i| \leq N d_N$ by (B1), it follows that $\mu^{*}-n_L = O_{p_d}(N d_N)$.
Combined with $d_N=o(k_N^{p})$ and $n_L=\Theta_{p_d}(N k_N^{p})$, it follows that
$
\mu^{*} = n_L\bigl(1+o_{p_d}(1)\bigr)=\Theta_{p_d}(N k_N^{p}).
$
For the conditional variance,
\begin{align*}
    Var(S_b^{*}\mid \xi, L^{*})
    &\le (1+C_{\mathrm{cov}})\sum_{i=1}^{N} Var(n^{*}_{b,i}\mid \xi)\, I_i^{*} 
    \le (1+C_{\mathrm{Cov}})C_2 \sum_{i=1}^{N} I_i^{*} \\
    &= (1+C_{\mathrm{cov}})C_2\, \sum_i I_i^{*}
    =O_{p_d}(N k_N^{p}),
\end{align*}
as $|N_{L^*} - N_L| \leq N d_N = o(Nk_N^p)$ by (B2), so $N_{L^*} = \Theta_{p_\omega}(Nk_N^p)$ follows from $N_L = \Theta_{p_\omega}(Nk_N^p)$ (Lemma~\ref{lem:N_L}).
Re--applying the previous argument on the same high-probability event $\{n_L\ge 2n_{\min,N}\}$, $\mu^{*}\ge n_L/2$ with high probability.
On this event, $n_{d,N} \le n_{\min,N} \le n_L/2 \le \mu^*$, so the set inclusion $\{S_b^{*}\notin [n_L-n_{d,N},\, n_L+n_{d,N}]\} \subseteq \{|S_b^{*}-\mu^{*}|>\mu^{*}/2\} \cup \{|\mu^{*}-n_L|>n_L/2\}$ holds on this event.
Therefore,
\[
\{S_b^{*}\notin [n_L-n_{d,N},\, n_L+n_{d,N}]\}
\subseteq \{|S_b^{*}-\mu^{*}|>\mu^{*}/2\}\ \cup\ \{|\mu^{*}-n_L|>n_L/2\}.
\]
The second event is $o_{p_d}(1)$ by $|\mu^{*}-n_L|=O_{p_d}(Nd_N)=o_{p_d}(Nk_N^{p})$
and $n_L\asymp Nk_N^{p}$.
The first event is $O_{p_d}\!\left(\frac{1}{Nk_N^{p}}\right)$ from Chebyshev and $\mu^{*}=\Theta_{p_d}(Nk_N^{p})$ shown previously. Hence
\begin{align}
    P\!\left(S_b^{*}\notin [n_L-n_{d,N},\, n_L+n_{d,N}] \mid \xi, L^{*}\right)
= O_{p_d}\!\left(\frac{1}{Nk_N^{p}}\right) + O_{p_d}\!\left(\frac{d_N}{k_N^{p}}\right),
\label{lemma_B:part2_res}
\end{align}

Part3: 
We now combine previous findings to finish the proof.
Define the ``bad'' indicators
\[
Z_b := \mathbbm{1}\Big\{ S_b \notin [n_L-n_{d,N},\, n_L+n_{d,N}] \Big\},
\quad
Z_b^{*} := \mathbbm{1}\Big\{ S_b^{*} \notin [n_L-n_{d,N},\, n_L+n_{d,N}] \Big\}.
\]
Then $1-|\mathcal{B}_L|/B = B^{-1}\sum_{b=1}^{B} Z_b$.
Fix any $\delta\in(0,1)$.
By Markov and that (R1) implies the vectors $\{n_b^*\}_{b=1}^{B}$ are i.i.d.\ across $b$ under the resampling distribution $P(\,\cdot \mid \xi)$, so $\{\ind(b \notin \mathcal{B}_L(\mathbf{x}))\}_{b=1}^{B}$ conditional on $\xi$ is i.i.d. Bernoulli.
\[
P\!\left(\frac{|\mathcal{B}_L|}{B} \le 1-\delta \,\Big|\, \xi, L\right)
= P\!\left(\frac{1}{B}\sum_{b=1}^{B} Z_b \ge \delta \,\Big|\, \xi, L\right)
\le \frac{\E(Z_1\mid \xi, L)}{\delta} = o_{p_d}(1)
\,\text{  by (\ref{lemma_B:part1_res})}.
\]
Therefore $|\mathcal{B}_L|/B \to_{p_d} 1$ and the same argument with (\ref{lemma_B:part2_res}) gives
$|\mathcal{B}_{L^{*}}|/B \to_{p_d} 1$, proving (a).
Finally, for any fixed $b$,
$\mathbbm{1}\{b \in \mathcal{B}_L \Delta \mathcal{B}_{L^{*}}\}
\le \mathbbm{1}\{b \notin \mathcal{B}_L\} + \mathbbm{1}\{b \notin \mathcal{B}_{L^{*}}\}.$
So taking expectation for each part and apply (\ref{lemma_B:part1_res}) and (\ref{lemma_B:part2_res}) again gives
\[
P_{\mathcal{S}^N | \omega}\!\left(b \in \mathcal{B}_L \; \Delta \; \mathcal{B}_{L^{*}}\right)
= o_{p_d}\!\left(\frac{1}{Nk_N^{p}}\right) + o_{p_d}\!\left(\frac{d_N}{k_N^{p}}\right),
\]
proving (b).
\end{proof}

Lemmas~\ref{lem:N_L}--\ref{lem:BS} jointly ensure that the oracle, 
design-weighted, and bootstrap-weighted leaf sizes are all of order 
$Nk_N^p \to \infty$ on an event with $P_{\Omega \times \mathcal{S}^N}$-probability tending 
to one. All ratio estimators appearing in the proof of Theorem~10 are 
therefore well-defined on this event, and subsequent arguments are 
understood to be conditional on it.

\vspace*{0.3cm}
\begin{proof}[Proof of Theorem~10]
We now prove the main theorem. We show design consistency of the MMD between the KME of the conditional distribution estimate and that of the truth. Namely,
$\| \mu_{\mathbf k}(\hat{P}_{\mathrm{forest}}(\mathbf{x})) - \mu_{\mathbf k}(P^N_{Y \mid X=\mathbf{x}})\|_{\hilbert} = o_{p_d}(1)$, through the decomposition\\
$\left\lVert \mu_{\mathbf k}\!\big(\hat P_{\mathrm{forest}}(\mathbf{x})\big) - \mu_{\mathbf k}\!\big(P^N_{Y \mid X=\mathbf{x}}\big) \right\rVert_{\mathcal H}
\leq A_N(\mathbf{x}) + B_N(\mathbf{x}) + C_N(\mathbf{x}) + D_N(\mathbf{x}) + E_N(\mathbf{x})$,  where 
\begin{align*}
A_N(\mathbf{x})
&= \left\lVert
    \frac{1}{|\mathcal B_{L^*}(\mathbf{x})|}\!
    \sum_{b\in \mathcal B_{L^*}(\mathbf{x})} T^{*}_{N,b}(\mathbf{x})
    -\!
    \frac{1}{|\mathcal B_{L}(\mathbf{x})|}\!
    \sum_{b\in \mathcal B_{L}(\mathbf{x})} \sum_{i=1}^{N}
      \frac{n^{*}_i/\pi_i}{|L(\mathbf{x})|}
      \,\mathbbm{1}\{X_i\in L(\mathbf{x})\}\, \mathbf{k}(Y_i,\cdot)
  \right\rVert_{\mathcal H}, \\[0.35em]
B_N(\mathbf{x})
&= \left\lVert
    \frac{1}{|\mathcal B_{L}(\mathbf{x})|}
    \sum_{b\in \mathcal B_{L}(\mathbf{x})} \sum_{i=1}^{N}
      \frac{n^{*}_i/\pi_i}{|L(\mathbf{x})|}
      \,\mathbbm{1}\{X_i\in L(\mathbf{x})\}\, \mathbf{k}(Y_i,\cdot)
    -
    \mathbb{E}_{n^*}\!\left[
      T^{*}_{N,b}(\mathbf{x})\,\mathbbm{1}\{b\in \mathcal B_{L}(\mathbf{x})\}
      \,\middle|\, \xi, L(\mathbf{x})
    \right]
  \right\rVert_{\mathcal H}, \\[0.35em]
C_N(\mathbf{x})
&= \Bigg\|
    \mathbb{E}_{n^*}\!\left[
      T^{*}_{N,b}(\mathbf{x})\,\mathbbm{1}\{b\in \mathcal B_{L}(\mathbf{x})\}
      \,\middle|\, \xi, L(\mathbf{x})
    \right]
    - T_{\xi,L,b}(\mathbf{x})
  \Bigg\|_{\mathcal H}, \\[0.35em]
D_N(\mathbf{x})
&= \Bigg\| T_{\xi,L,b}(\mathbf{x}) - T_{L,b}(\mathbf{x}) \Bigg\|_{\mathcal H}, \\[0.35em]
E_N(\mathbf{x})
&= \Bigg\| T_{L,b}(\mathbf{x}) - \mu_{\mathbf{k}}(P^N_{Y \mid X=\mathbf{x}}) \Bigg\|_{\mathcal H}.
\end{align*}
Where $\mathcal{B}_{L^*}(\mathbf{x})$ and $\mathcal{B}_{L}(\mathbf{x})$ are defined in \eqref{def:BL*} and \eqref{def:BL}, respectively;
$
    \mu_{\mathbf k}(\hat{P}_{\mathrm{forest}}(\mathbf x)) = \frac{1}{B} \sum_{b=1}^{B} T^*_{N,b}(\mathbf{x}),
$
with
\begin{align*}
    T^*_{N,b}(\mathbf{x}) 
    &=
    \sum_{i \in D^*_{b,\mathrm{est}}} \frac{n^{*}_{b,i} / \pi_i}{|L^{*}_{\xi,b}(\mathbf{x})|} \; \ind(X_i \in L^{*}_{\xi,b}(\mathbf{x})) \mathbf{k}(Y_i, \cdot), 
    \;\;
    |L^{*}_{\xi, b}(\mathbf{x})|
    = \sum_{i \in D^*_{b,\mathrm{est}}} {n^{*}_{b,i} / \pi_i} \; \ind(X_i \in L^{*}_{\xi,b}(\mathbf{x})),
\end{align*}
\[
    T_{\xi, L, b}(\mathbf{x}) 
    = 
    \sum_{i = 1}^{N} \frac{1}{|L_{\xi}(\mathbf{x})|} \frac{\xi_i}{\pi_i} \ind(X_i \in L_{\xi}(\mathbf{x})) \mathbf{k}(Y_i, \cdot), 
    \;\; 
    |L_{\xi}(\mathbf{x})| = \sum_{i=1}^N \frac{\xi_i}{\pi_i} \ind(X_i \in L(\mathbf{x})),
\]
\[
    T_{L,b}(\mathbf{x}) 
    = 
    \sum_{i = 1}^{N} \frac{1}{|L(\mathbf{x})|} \ind(X_i \in L(\mathbf{x})) \mathbf{k}(Y_i, \cdot), 
    \;\; 
    |L(\mathbf{x})| = \sum_{i=1}^N \ind(X_i \in L(\mathbf{x})),
\]
and the target quantity 
\[
    \mu_{\mathbf{k}}(P^N_{Y \mid X=\mathbf{x}}) = \frac{\sum_{i=1}^{N} \ind(X_i = \mathbf{x}) \mathbf{k}(Y_i, \cdot)}{\sum_{i = 1}^N \ind(X_i = \mathbf{x})}.
\]

After decomposition, $A_N$ concerns the effect of using resampled data for leaf construction rather than the full finite population. $B_N$ concerns the resampling variance once the oracle partition $L(\mathbf{x})$ and the inclusion indicators $\xi$ are held fixed. $C_N$ measures the mismatch introduced by using the per-tree resample weights $n^*_{b,i}$ in place of the design weights $\xi_i/\pi_i$. $D_N$ isolates the impact of the sampling design. And $E_N$ concerns the approximation (bias) from using a leaf to estimate a point.
Throughout, $\mathbf{x}$ is a fixed atom of the covariate law, $p_{\mathbf{x}} := P(X = \mathbf{x}) > 0$, so that $N_{\mathbf{x}} = \Theta_{p_d}(N) \to \infty$ and the target is well-defined.


\vspace*{0.3cm}
\paragraph{Part A}
Denote $\mathcal{B}^{\cap}(\mathbf{x}) = \mathcal{B}_{L}(\mathbf{x}) \cap \mathcal{B}_{L^*}(\mathbf{x})$.
By Lemma \ref{lem:BS}, 
$P(b \in \mathcal{B}^{\cap}(\mathbf{x})) \to_{p_d} 1$. Denote the set limit $\mathcal{B}(\mathbf{x})$. Lemma \ref{lem:BS} gives $|\mathcal{B}(\mathbf{x})| = B( 1 + o_{p_d}(1))$.
Consider the quantity
\begin{align*}
    A_N 
    =
    &\Bigg\|
        \frac{1}{|\mathcal{B}_{L^*}(\mathbf{x})|}\sum_{b \in \mathcal{B}_{L^*}(\mathbf{x})}  \sum_{i =1}^N \frac{n_{b,i}^{*}/\pi_i}{|L^*_{\xi}(\mathbf{x})|} \ind(X_i \in L^*_{\xi}(\mathbf{x}))\mathbf{k}(Y_i, \cdot) \\
        - 
        &\quad \frac{1}{|\mathcal{B}_{L}(\mathbf{x})|}\sum_{b \in \mathcal{B}_{L}(\mathbf{x})} \sum_{i =1}^N \frac{n_{b,i}^{*}/\pi_i}{|L(\mathbf{x})|} \ind(X_i \in L(\mathbf{x}))\mathbf{k}(Y_i, \cdot) 
    \Bigg\|_{\hilbert}.
\end{align*}
Denote 
$
    T^*_{L^*,b}(\mathbf{x}) 
    = 
    \sum_{i =1}^N \frac{n_{b,i}^{*}/\pi_i}{|L^*_{\xi}(\mathbf{x})|} \ind(X_i \in L^*_{\xi}(\mathbf{x}))\mathbf{k}(Y_i, \cdot),
$
$
    T_{L,b}(\mathbf{x}) 
    =
    \sum_{i =1}^N \frac{n_{b,i}^{*}/\pi_i}{|L(\mathbf{x})|} \ind(X_i \in L(\mathbf{x}))\mathbf{k}(Y_i, \cdot),
$
then 
\begin{align*}
    A_N 
    =
    &\Bigg\|
        \frac{1}{|\mathcal{B}_{L^*}(\mathbf{x})|}\sum_{b \in \mathcal{B}_{L^*}(\mathbf{x})}  
        T^*_{L^*,b}(\mathbf{x})
        - 
        \frac{1}{|\mathcal{B}_{L}(\mathbf{x})|}\sum_{b \in \mathcal{B}_{L}(\mathbf{x})} 
        T_{L,b}(\mathbf{x})
    \Bigg\|_{\hilbert} \\
    \leq
    &\frac{1}{|\mathcal{B}^{\cap}|}\sum_{b \in \mathcal{B}^{\cap}}  
        \Bigg\| 
            \frac{|\mathcal{B}^{\cap}|}{|\mathcal{B}_{L^*}|}T^*_{L^*,b}(\mathbf{x}) 
            - 
            \frac{|\mathcal{B}^{\cap}|}{|\mathcal{B}_{L}|} T_{L,b}(\mathbf{x})
        \Bigg\|_{\hilbert} 
        +
         \frac{1}{|\mathcal{B}_{L^*}|}\sum_{b \in \mathcal{B}_{L^*} \setminus \mathcal{B}^{\cap}}
        \Bigg\|
            T^*_{L^*, b}(\mathbf{x}) - \frac{|\mathcal{B}_{L^*}|}{|\mathcal{B}_{L}|} T_{L,b}(\mathbf{x})
        \Bigg\|_{\hilbert} \\
        +
        &\frac{1}{|\mathcal{B}_{L}|} \sum_{b \in \mathcal{B}_{L} \setminus \mathcal{B}^{\cap}}
        \Bigg\|
            \frac{|\mathcal{B}_{L}|}{|\mathcal{B}_{L^*}|} T^*_{L^*, b}(\mathbf{x}) -  T_{L,b}(\mathbf{x})
        \Bigg\|_{\hilbert}.
\end{align*}
For the second part, its expectation conditioning on $\xi$ is upper bounded by
\begin{align*}
    &\frac{1}{|\mathcal{B}_{L^*}|} 
    \sum_{b \in \mathcal{B}_{L} \cup \mathcal{B}_{L^*}} 
    \Big(\E\big\| 
        T^*_{L^*, b}(\mathbf{x}) - \tfrac{|\mathcal{B}_{L^*}|}{|\mathcal{B}_{L}|} T_{L,b}(\mathbf{x}) 
    \big\|_{\hilbert}^2\Big)^{1/2}
    P(b \in \mathcal{B}_{L^*} \setminus \mathcal{B}^{\cap}) \\
    \leq 
    &\Big( \frac{1}{B} + o_{p_d}(1)\Big) \sum_{b \in \mathcal{B}} 
    \Bigg\{ 
        \big(\E\|T^*_{L^*, b}(\mathbf{x})\|_{\hilbert}^2\big)^{1/2} + (1 + o_{p_d}(1)) \big(\E\|T_{L, b}(\mathbf{x})\|_{\hilbert}^2\big)^{1/2}
    \Bigg\} O_{p_d}(\tfrac{1}{Nk^p_N})
    =
    O_{p_d}(\frac{1}{Nk^p_N}),
\end{align*}
where the last equality uses $\|T^*_{L^*, b}(\mathbf{x})\|_{\hilbert} \le K$ and $\|T_{L,b}(\mathbf{x})\|_{\hilbert} \le K$. Each estimator is a convex combination of kernel evaluations, so the nonnegative weights $n^*_i/\pi_i$ cancel between numerator and denominator, and (K1) yields
$
    \|T^*_{L^*, b}(\mathbf{x})\|_{\hilbert}
    \le \frac{\sum_i (n^*_i/\pi_i)\,\ind(X_i \in L^*_{\xi}(\mathbf{x}))\,\|\mathbf{k}(Y_i,\cdot)\|_{\hilbert}}{\sum_i (n^*_i/\pi_i)\,\ind(X_i \in L^*_{\xi}(\mathbf{x}))}
    \le K,
$
the denominator being strictly positive on the non-degeneracy event of Lemma~\ref{lem:BS}.
By a similar argument, the third term in the upper bound of $A_N$ is also $o_{p_d}(1)$.
Thus
\begin{align*}
    A_N 
    \;\leq\;
    &\frac{1}{|\mathcal{B}|} \sum_{b \in \mathcal{B}} 
    \Big\| \sum_{i=1}^N \frac{n^*_i}{\pi_i} \Big\{ 
        \frac{\ind(X_i \in L^*_{\xi}(\mathbf{x}))}{|L^*_{\xi}(\mathbf{x})|}
        -
        \frac{\ind(X_i \in L(\mathbf{x}))}{|L(\mathbf{x})|}
    \Big\} \mathbf{k}(Y_i, \cdot) \Big\|_{\hilbert}
    +
    o_{p_d}(1) \\
    \;\leq\;
    &\frac{K}{\min\{|L^*_{\xi}(\mathbf{x})|, |L(\mathbf{x})|\}}\frac{N}{\underline{\lambda} n_{s,N}} \sum_{i=1}^N n^*_i \Big| \ind(X_i \in L^*_{\xi}(\mathbf{x})) - \ind(X_i \in L(\mathbf{x})) \Big|
    + 
    o_{p_d}(1).
\end{align*}

We now bound the sum $\sum_{i=1}^N n^*_i |\ind(X_i \in L^*_{\xi}(\mathbf{x})) - \ind(X_i \in L(\mathbf{x}))|$.
Define the symmetric difference set
$
S_{\triangle} := \{i \in [N]: X_i \in L^*_{\xi}(\mathbf{x}) \triangle L(\mathbf{x})\}.
$
So 
$
    \sum_{i=1}^N n^*_i |\ind(X_i \in L^*_{\xi}(\mathbf{x})) - \ind(X_i \in L(\mathbf{x}))| 
    = \sum_{i \in S_{\triangle}} n^*_i.
$
The cardinality of $S_{\triangle}$ satisfies
$
    |S_{\triangle}| 
    = \sum_{i=1}^N \ind(X_i \in L^*_{\xi}(\mathbf{x}) \triangle L(\mathbf{x})).
$
Taking expectations and applying (B2),
\[
    \E_{\mathcal{S}^N \mid \omega}[|S_{\triangle}|]
    = \sum_{i=1}^N P\big(X_i \in L^*_{\xi}(\mathbf{x}) \,\triangle\, L(\mathbf{x})\big)
    \leq N \sup_{\mathbf{x}} P\big(X \in L^*(\mathbf{x}) \,\triangle\, L(\mathbf{x})\big)
    = o_{p_d}(N d_N),
\]
so $|S_{\triangle}| = o_{p_d}(N d_N)$ by Markov's inequality. For the weighted sum, (R2) gives $\E[n^*_i \mid \xi] = 1 + o(1)$ uniformly in $i$, hence
\begin{align}
    \E\Big[\sum_{i \in S_{\triangle}} n^*_i \;\Big|\; \xi\Big]
    = \sum_{i \in S_{\triangle}} \E[n^*_i \mid \xi]
    = |S_{\triangle}|\,(1 + o(1))
    = o_{p_d}(N d_N),
    \label{eq:partA}
\end{align}
and a further application of Markov's inequality yields $\sum_{i \in S_{\triangle}} n^*_i = o_{p_d}(N d_N)$.
Substituting (\ref{eq:partA}) into the bound for $A_N$, and using $\min\{|L^*_{\xi}(\mathbf{x})|, |L(\mathbf{x})|\} = \Theta_{p_d}(Nk_N^p)$ from Lemma~\ref{lem:N_L} with (B2), together with $n_{s,N} = \Theta_{p_d}(N)$ from (D2):
\begin{align*}
    A_N 
    &\leq \frac{K}{\min\{|L^*_{\xi}(\mathbf{x})|, |L(\mathbf{x})|\}} \frac{N}{\underline{\lambda} n_{s,N}} \cdot o_{p_d}(Nd_N) + o_{p_d}(1) \\
    &\leq \frac{K \cdot N \cdot o_{p_d}(Nd_N)}{\Theta_{p_d}(Nk_N^p) \cdot \Theta_{p_d}(N)} + o_{p_d}(1) 
    = \frac{o_{p_d}(Nd_N)}{\Theta_{p_d}(k_N^p)} + o_{p_d}(1) 
    = o_{p_d}\left(\frac{d_N}{k_N^p}\right) + o_{p_d}(1).
\end{align*}
Since (B2) requires $d_N = o(k_N^p)$, we have $o_{p_d}(d_N/k_N^p) = o_{p_d}(1)$, completing Part A.\\


\vspace*{0.3cm}
\paragraph{Part B}
Conditioning on the oracle leaf $L(\mathbf{x})$ in the finite population, which is determined irrespective of the design, together with the inclusion indicators $\xi$, the resamples are independent and identically distributed across $b$. We abbreviate
\[
    \bar{T}_b := \frac{1}{N_{L(\mathbf{x})}} \sum_{i=1}^N \frac{n^*_{b,i}}{\pi_i}\, I_i\, k_i,
    \qquad
    R := \frac{N_{L(\mathbf{x})}}{\widehat{N}^*_{\xi, L(\mathbf{x})}},
    \qquad
    I_i := \ind(X_i \in L(\mathbf{x})),\quad k_i := \mathbf{k}(Y_i, \cdot),
\]
so that $T^*_{N,b}(\mathbf{x}) = R\, \bar{T}_b$. For any $\epsilon > 0$,
\begin{align}
\begin{split}
    &P_{n^*, \xi, L(\mathbf{x})}\!\left( \left\|
        \frac{1}{|\mathcal{B}_L(\mathbf{x})|}\sum_{b \in \mathcal{B}_L(\mathbf{x})} T^*_{N,b}(\mathbf{x})
        -
        \E_{n^{*}}[ T^*_{N,b}(\mathbf x)\, \ind(b\in \mathcal{B}_{L}(\mathbf{x})) \mid \xi, L(\mathbf{x})]
    \right\|_{\hilbert} \geq \epsilon \right) \\
    &=
    \E_{\xi, L(\mathbf{x})}\!\left[
        P_{n^* \mid \xi, L(\mathbf{x})}\!\left( \left\|
            \frac{1}{B}\sum_{b = 1}^B \frac{B}{|\mathcal{B}_L(\mathbf{x})|}\Big\{T^*_{N,b}(\mathbf{x})\, \ind(b \in \mathcal{B}_L(\mathbf{x}))
            -
            \mu \Big\}
        \right\|_{\hilbert} \geq \epsilon \right)
    \right],
    \label{eq: main_partB}
\end{split}
\end{align}
where $\mu = \E_{n^{*}}[ T^*_{N,b}(\mathbf x) \mid \xi, L(\mathbf{x})]\, P(b \in \mathcal{B}_L(\mathbf{x}) \mid \xi, L(\mathbf{x}))$. For any fixed $\delta \in (0, 1/2)$,
\begin{align*}
    &\ind\!\left( \frac{B}{|\mathcal{B}_L(\mathbf{x})|} \Big\| \frac{1}{B} \sum_b \big\{ T^*_{N,b}(\mathbf{x})\ind(b \in \mathcal{B}_{L}(\mathbf{x})) - \mu \big\}\Big\|_{\hilbert} \geq \epsilon \right) \\
    \leq
    &\ind\!\left( \frac{|\mathcal{B}_L(\mathbf{x})|}{B} \leq 1- \delta \right) 
    \quad +
    \ind\!\left( \Big\| \frac{1}{B} \sum_b \big\{ T^*_{N,b}(\mathbf{x})\ind(b \in \mathcal{B}_{L}(\mathbf{x})) - \mu \big\}\Big\|_{\hilbert} \geq \epsilon(1-\delta) \right),
\end{align*}
so that, by Markov's and Chebyshev's inequalities applied to the two terms respectively,
\begin{align}
    (\ref{eq: main_partB})
    \;\leq\;
    \E_{\xi, L(\mathbf{x})}\!\left[P_{n^*}\!\Big(\tfrac{|\mathcal{B}_L(\mathbf{x})|}{B} \leq 1 - \delta \,\Big|\, \xi, L(\mathbf{x})\Big)\right]
    +
    \frac{\E\big[\Var\big(T_b^*\, \ind(b \in \mathcal{B}_L(\mathbf{x})) \mid \xi, L(\mathbf{x})\big)\big]}{\epsilon^2 (1 - \delta)^2 B}.
    \label{eq: partB-split}
\end{align}
 
\medskip
For the first term of the upper bound \eqref{eq: partB-split}:
Writing $\mathcal{E}_b := \{|\sum_{i} n^*_{b,i} I_i - n_{L(\mathbf{x})}| \leq n_{d,N}\}$, the events $\{\ind(\mathcal{E}_b^c)\}_{b}$ are i.i.d.\ Bernoulli given $\xi$, and $\{|\mathcal{B}_L(\mathbf{x})|/B \leq 1-\delta\} = \{B^{-1}\sum_b \ind(\mathcal{E}_b^c) \geq \delta\}$. Markov's inequality and Lemma~\ref{lem:BS}\ give
\[
    \E_{\xi, L(\mathbf{x})}\!\left[P_{n^*}\!\Big(\tfrac{|\mathcal{B}_L(\mathbf{x})|}{B} \leq 1 - \delta \,\Big|\, \xi, L(\mathbf{x})\Big)\right]
    \leq
    \frac{1}{\delta}\, \E[\ind(\mathcal{E}_b^c)]
    =
    \frac{1}{\delta}\, O_{p_d}\!\Big(\frac{1}{N k^p_N}\Big)
    =
    o_{p_d}(1).
\]
 
\medskip
For the second term of the upper bound \eqref{eq: partB-split}:
It remains to bound $\E_{\xi, L(\mathbf{x})}[\Var(T_b^*\, \ind(b \in \mathcal{B}_L(\mathbf{x})) \mid \xi, L(\mathbf{x}))]$. We first record two facts about the Hájek ratio $R$.
 
\smallskip
\noindent\emph{(B.i) Control of $R$.} The design-weighted denominator $\widehat{N}^*_{\xi, L(\mathbf{x})} = \sum_i (n^*_i/\pi_i) I_i$ satisfies
\begin{align*}
\begin{split}
    &\Var\big(\widehat{N}^*_{\xi, L(\mathbf{x})} \mid L(\mathbf{x})\big) \\
    =
    &\E_{\xi}\big[\Var(\widehat{N}^*_{\xi, L(\mathbf{x})} \mid \xi, L(\mathbf{x})) \mid L(\mathbf{x})\big]
    +
    \Var_{\xi}\Big( \sum_i \tfrac{\xi_i}{\pi_i} I_i \,\Big|\, L(\mathbf{x}) \Big) \\
    =
    &\E\Big[\sum_i \Var(n^*_i \mid \xi)\tfrac{I_i}{\pi_i^2}\Big]
    + \E\Big[\sum_{i \neq j}Cov(n^*_i, n^*_j \mid \xi)\tfrac{I_i I_j}{\pi_i \pi_j}\Big]
    + \sum_i \tfrac{1-\pi_i}{\pi_i} I_i
    + \sum_{i \neq j}\tfrac{\pi_{ij} - \pi_i \pi_j}{\pi_i \pi_j} I_i I_j \\
    \leq
    &2 \underline{\lambda}^{-2} \E\big[(N/n_{s,N})^2\big] N_{L(\mathbf{x})}
    + \underline{\lambda}\, \E[N/n_{s,N}]\, N_{L(\mathbf{x})}
    + \E\big[c N^2 \underline{\lambda}^{-2} n_{s,N}^{-3}\big] N_{L(\mathbf{x})}^2,
\end{split}
\end{align*}
where the inequality uses $Cov(n^*_i, n^*_j \mid \xi) \leq 0$, the bound $1 - (\xi_i/\pi_i)/\widehat{N} \in [0,1]$, $\pi_i \in [\underline{\lambda} n_{s,N}/N,\, 1)$ from (D3), and (D4). By Lemma~\ref{lem:N_ns_moments}, $\sup_N \E[(N/n_{s,N})^4] < \infty$, and with $n_{s,N} \asymp N$, $N_{L(\mathbf{x})} \leq N$,
\[
    \frac{\Var(\widehat{N}^*_{\xi, L(\mathbf{x})} \mid L(\mathbf{x}))}{N_{L(\mathbf{x})}^2}
    = O_{p_d}\!\big(N_{L(\mathbf{x})}^{-1}\big).
\]
Since $\E[\widehat{N}^*_{\xi, L(\mathbf{x})}/N_{L(\mathbf{x})} \mid L(\mathbf{x})] = 1$, Chebyshev's inequality yields $\widehat{N}^*_{\xi, L(\mathbf{x})}/N_{L(\mathbf{x})} = 1 + o_{P(n^*,\xi \mid L(\mathbf{x}))}(1)$, hence
\begin{align}
    R = 1 + o_{P(n^*,\xi \mid L(\mathbf{x}))}(1),
    \qquad
    \E\big[(R - 1)^2 \mid L(\mathbf{x})\big] = O_{p_d}\!\big(N_{L(\mathbf{x})}^{-1}\big).
    \label{eq: Rrate}
\end{align}
Moreover, on the event $\{b \in \mathcal{B}_L(\mathbf{x})\}$ the ratio is deterministically bounded, as with $W_N := \widehat{N}^*_{\xi, L(\mathbf{x})}/N_{L(\mathbf{x})} - 1$,
\begin{align}
    W_N \geq \frac{\bar{\lambda}}{\underline{\lambda}}\Big(1 - \frac{n_{d,N}}{n_{\min,N}}\Big) - 1 > -1,
    \,\,\text{so}\,\,
    R = \frac{1}{1 + W_N} = O(1) \quad\text{on } \{b \in \mathcal{B}_L(\mathbf{x})\}.
    \label{eq: Rbounded}
\end{align}
 
\smallskip
\noindent\emph{(B.ii) Fourth-moment bound on $\bar T_b$.} By Minkowski's inequality in $L^4(P_{n^*,\xi \mid L(\mathbf{x})})$ and (K1),
\begin{align}
    &\big\|\bar{T}_b\big\|_{L^4(P_{n^*,\xi \mid L(\mathbf{x})})}
    \leq
    \frac{K}{N_{L(\mathbf{x})}} \sum_i I_i \Big\{ \E_{\xi}\E_{n^*}\big[ |n^*_i/\pi_i|^4 \mid \xi \big] \Big\}^{1/4} \nonumber \\
    \leq
    &K\, 15^{1/4}\, \underline{\lambda}^{3/4}\, \E_{\xi}\big[(N/n_{s,N})^{3/4}\big],
    \label{eq: momentcond1}
\end{align}
where the last step uses the falling-factorial identity for $N^*_i \sim \mathrm{Binom}(\widehat{N}, (\xi_i/\pi_i)/\widehat{N})$, $\widehat{N} = \sum_i \xi_i/\pi_i$,
\begin{align}
\begin{split}
    &\E[(n_i^*)^4 \mid \xi] 
    =
    \xi_i \Big\{
        (1 - \pi_i)^3
        + 7 \tfrac{\widehat{N} - 1}{\widehat{N}} (1 - \pi_i)^2 \\
        + &6 \tfrac{(\widehat{N} - 1)(\widehat{N} - 2)}{\widehat{N}^2} (1 - \pi_i)
        + \tfrac{(\widehat{N} - 1)(\widehat{N} - 2)(\widehat{N} - 3)}{\widehat{N}^3}
    \Big\}
    \leq 15\, \xi_i,
    \label{eq: fourthmoments}
\end{split}
\end{align}
each fraction lying in $[0,1]$ since $\widehat{N} \geq 1$, together with $\sum_i I_i = N_{L(\mathbf{x})}$ and $\sup_N \E[(N/n_{s,N})^{3/4}] \leq \sup_N \E[(N/n_{s,N})^{4}]^{3/16} < \infty$ from Lemma~\ref{lem:N_ns_moments}. Hence $\sup_N \E[\|\bar{T}_b\|_{\hilbert}^4 \mid L(\mathbf{x})] < \infty$.
 
\smallskip
\noindent\emph{(B.iii) Assembling the variance.} Decompose $T^*_{N,b}\,\ind_{\mathcal B_L} = \bar{T}_b\,\ind_{\mathcal B_L} + (R - 1)\bar{T}_b\,\ind_{\mathcal B_L}$, where $\ind_{\mathcal B_L} := \ind(b \in \mathcal{B}_L(\mathbf{x}))$. By the triangle inequality for the variance seminorm,
\begin{align*}
    \Big| \Var\big(T^*_{N,b}\,\ind_{\mathcal B_L} \mid \xi, L(\mathbf{x})\big)^{1/2}
    - \Var\big(\bar{T}_b\,\ind_{\mathcal B_L} \mid \xi, L(\mathbf{x})\big)^{1/2} \Big|
    \leq
    \Big(\E_{n^*}\big[\|(R-1)\bar{T}_b\,\ind_{\mathcal B_L}\|_{\hilbert}^2 \mid \xi, L(\mathbf{x})\big]\Big)^{1/2}.
\end{align*}
Taking outer expectations and applying Cauchy--Schwarz with \eqref{eq: Rrate}, \eqref{eq: Rbounded}, and \eqref{eq: momentcond1},
\begin{align}
\begin{split}
    &\E_{\xi, L(\mathbf{x})}\big[\|(R-1)\bar{T}_b\, \ind_{\mathcal B_L}\|_{\hilbert}^2\big] \\
    \leq
    &\E_{L(\mathbf{x})}\Big[
        \big\|(R-1)\ind_{\mathcal B_L}\big\|_{L^4(P_{n^*,\xi \mid L(\mathbf{x})})}^2\;
        \big\|\bar{T}_b\big\|_{L^4(P_{n^*,\xi \mid L(\mathbf{x})})}^2
    \Big] \\
    \leq
    &\Big(\sup_N \E[\|\bar{T}_b\|_{\hilbert}^4 \mid L(\mathbf{x})]\Big)^{1/2}\,
    \E_{L(\mathbf{x})}\big[\|(R-1)\ind_{\mathcal B_L}\|_{L^4}^2\big]
    = o_{p_d}(1),
    \label{eq: crossbound}
\end{split}
\end{align}
the final equality by dominated convergence, as $\|(R-1)\ind_{\mathcal B_L}\|_{L^4} = o_{P(n^*,\xi \mid L(\mathbf{x}))}(1)$ from \eqref{eq: Rrate}, with the integrable envelope $|(R-1)\ind_{\mathcal B_L}| \leq 1 + R\,\ind_{\mathcal B_L} = O(1)$ supplied by \eqref{eq: Rbounded}. For the leading term, $Cov(n^*_i, n^*_j \mid \xi) \leq 0$ and (K1) give
\begin{align}
    \E_{\xi, L(\mathbf{x})}\big[\Var(\bar{T}_b \mid \xi, L(\mathbf{x}))\big]
    \leq
    \frac{K^2}{N_{L(\mathbf{x})}^2}\, \underline{\lambda}^{-2}\, \E\big[(N/n_{s,N})^2\big]\, N_{L(\mathbf{x})}
    = O_{p_d}\!\Big(\frac{1}{N k_N^p}\Big),
    \label{eq: leadingbound}
\end{align}
using $N_{L(\mathbf{x})} = \Theta_{p_d}(N k_N^p)$ from Lemma~\ref{lem:N_L}. Combining \eqref{eq: crossbound} and \eqref{eq: leadingbound},
\[
    \E_{\xi, L(\mathbf{x})}\big[\Var\big(T_b^*\, \ind(b \in \mathcal{B}_L(\mathbf{x})) \mid \xi, L(\mathbf{x})\big)\big]
    = O_{p_d}\!\Big(\frac{1}{N k_N^p}\Big) + o_{p_d}(1)
    = o_{p_d}(1).
\]
Substituting this and the first-term bound into \eqref{eq: partB-split},
\[
    (\ref{eq: main_partB})
    \leq
    o_{p_d}(1) + \frac{o_{p_d}(1)}{\epsilon^2 (1-\delta)^2 B}
    = o_{p_d}(1),
\]
which completes Part B.

\begin{remark}
    Convergence holds without requiring $B \to \infty$ with $N$. Equivalently, when $k^p_N$ does not shrink quickly, this component of the MMD remains controlled by letting $B$ compensate for the rate.
\end{remark}


\vspace*{0.5cm}
\paragraph{Part C}
We now bound the rate of
\begin{align}
    &\E_{\xi, L(\mathbf{x})}\Bigg[
        \Big\|
        \E_{n^{*}}\big[ T^*_{N,b}(\mathbf x)\, \ind(b \in \mathcal{B}_L(\mathbf{x}))
        -
        T_{\xi, L,b}(\mathbf{x}) \mid \xi,  L(\mathbf{x})\big]
    \Big\|_{\hilbert}
    \Bigg].
    \label{eq: mainB}
\end{align}
We omit the subscripts $N$ and $b$ and the evaluation point $\mathbf{x}$ where no confusion arises. Consider the linearization of
\begin{align*}
    T_b^*(\mathbf{x}; D)
    =
    \ind(b \in \mathcal{B}_L(\mathbf{x}))
    \sum_{i = 1}^{N} \frac{n_{i}^*/\pi_i}{|L^*(\mathbf{x})|}\, \ind(X_i \in L(\mathbf{x}))\, \mathbf{k}(Y_i, \cdot)
    \overset{\text{denote}}{=} \frac{T^*_{nu}}{T^*_{de}},
    \quad
    |L^*(\mathbf{x})| = \sum_{i=1}^N \frac{n_{i}^*}{\pi_i}\,\ind(X_i \in L(\mathbf{x})),
\end{align*}
centered at
\[
    T_{\xi, L}(\mathbf{x})
    = \sum_{i = 1}^{N} \frac{\xi_i / \pi_i}{|L_{\xi}(\mathbf{x})|}\, \ind(X_i \in L(\mathbf{x}))\, \mathbf{k}(Y_i, \cdot)
    \overset{\text{denote}}{=} \frac{T_{L,nu}}{T_{L,de}},
    \quad
    |L_{\xi}(\mathbf{x})| = \sum_{i=1}^N \frac{\xi_i}{\pi_i}\, \ind(X_i \in L(\mathbf{x})).
\]
To keep the notation light, we write $I_b = \ind(b \in \mathcal{B}_{L}(\mathbf{x}))$ and $I_i = \ind(X_i \in L(\mathbf{x}))$, using the subscript ($b$ or $i$) to distinguish the two indicators. For $g(t_{nu}, t_{de}) = t_{nu}/t_{de}$,
\[
    \nabla g(t_{nu},t_{de}) = \Big(\tfrac{1}{t_{de}},\; -\tfrac{t_{nu}}{t_{de}^{2}}\Big)^{\!\top},
    \qquad
    H_g(t_{nu}, t_{de}) =
    \begin{pmatrix}
        0 & -{1}/{t_{de}^2} \\[4pt]
        -{1}/{t_{de}^2} & {2\,t_{nu}}/{t_{de}^3}
    \end{pmatrix}.
\]
Set $\delta_i = \frac{n^*_i}{\pi_i} - \frac{\xi_i}{\pi_i}$, which satisfies $\E[\delta_i \mid \xi, L(\mathbf{x})] = 0$ by the unbiasedness of the resample multiplier under (R2), together with $n^*_i = n^*_i \xi_i$ almost surely. Accordingly,
\[
    \Delta_{nu} = T^*_{nu} - T_{L,nu} = I_b \sum_i \delta_i I_i\, \mathbf{k}(Y_i,\cdot) + (1 - I_b) T_{L, nu} \in \hilbert,
    \qquad
    \Delta_{de} = T^*_{de} - T_{L,de} = \sum_i \delta_i I_i \in \mathbb{R},
\]
and the exact first-order expansion reads
\begin{align*}
    T^* - T_L
    &=
    \frac{1}{T_{L,de}}\bigl(T^*_{nu} - T_{L,nu}\bigr)
    -
    \frac{T_{L,nu}}{T_{L,de}^{2}}\bigl(T^*_{de} - T_{L,de}\bigr)
    + R_2 \\
    &=
    I_b \sum_{i=1}^{N} \Big(\frac{n^*_i}{\pi_i} - \frac{\xi_i}{\pi_i}\Big)
    \underbrace{\left\{ \frac{I_i}{\sum_{j=1}^N I_j}\big(\mathbf{k}(Y_i,\cdot) - T_L(\mathbf{x})\big)\right\}}_{u_i}
    + (I_b - 1)\, T_{L}(\mathbf{x})
    + R_2(\mathbf{x}).
\end{align*}
Write $T^{(1)} = \sum_{i=1}^N \delta_i\, u_i$. The functions $u_i$ are fixed within the finite population, while the multipliers $\delta_i$ carry all randomness from the design and the resampling of $D_{\mathrm{est}}$.
 
\medskip
\noindent\textit{First-order term.}
By the unbiasedness of $n^*_i$ for $\xi_i$ we have $\E_{n^*}[T^{(1)}(\mathbf{x}) \mid \xi, L(\mathbf{x})] = 0$. The indicator $I_b = \ind(b \in \mathcal{B}_{L}(\mathbf{x}))$ is a function of $\{n^*_i\}_{i=1}^N$, so $I_b$ and $T^{(1)}$ are dependent and the product $I_b\, T^{(1)}$ is \emph{not} mean-zero. Instead,
\begin{align}
    \E_{n^*}\big[ I_b\, T^{(1)}(\mathbf{x}) \mid \xi, L(\mathbf{x})\big]
    = -\,\E_{n^*}\big[ (1 - I_b)\, T^{(1)}(\mathbf{x}) \mid \xi, L(\mathbf{x})\big].
    \label{eq: partC-firstorder}
\end{align}
Consequently,
\begin{align*}
    (\ref{eq: mainB})
    &=
    \E_{\xi, L(\mathbf{x})}\Big[
        \big\| \E_{n^*}\!\big[ I_b\, T^{(1)}(\mathbf{x}) + (I_b - 1)\, T_{\xi, L}(\mathbf{x}) + R_2(\mathbf{x}) \mid \xi, L(\mathbf{x})\big] \big\|_{\hilbert}
    \Big] \\
    &\leq
    \underbrace{\E_{\xi, L(\mathbf{x})}\Big[ \big\| \E_{n^*}[(1 - I_b)\, T^{(1)}(\mathbf{x}) \mid \xi, L(\mathbf{x})]\big\|_{\hilbert}\Big]}_{\textrm{(C-i)}}
    +
    \underbrace{\E_{\xi, L(\mathbf{x})}\Big[ \|T_{\xi, L}(\mathbf{x})\|_{\hilbert}\; \E_{n^*}[1 - I_b \mid \xi, L(\mathbf{x})]\Big]}_{\textrm{(C-ii)}} \\
    &\qquad\qquad
    +
    \underbrace{\E_{\xi, L(\mathbf{x})}\Big[ \big\| \E_{n^*}[R_2(\mathbf{x}) \mid \xi, L(\mathbf{x})]\big\|_{\hilbert}\Big]}_{\textrm{(C-iii)}}.
\end{align*}
 
\medskip
\noindent\textit{Term (C-i).}
By the Cauchy--Schwarz inequality in $P_{n^* \mid \xi, L(\mathbf{x})}$,
\[
    \big\| \E_{n^*}[(1 - I_b)\, T^{(1)} \mid \xi, L(\mathbf{x})]\big\|_{\hilbert}
    \leq
    \big(\E_{n^*}[1 - I_b \mid \xi, L(\mathbf{x})]\big)^{1/2}
    \big(\E_{n^*}[\|T^{(1)}\|_{\hilbert}^2 \mid \xi, L(\mathbf{x})]\big)^{1/2}.
\]
Since $\|u_i\|_{\hilbert} \leq 2K\, I_i / N_{L(\mathbf{x})}$ by (K1) and $\Var(\delta_i \mid \xi) = \pi_i^{-2}\Var(n^*_i \mid \xi) \leq \pi_i^{-2} C_2$ by (R2), while $Cov(n^*_i, n^*_j \mid \xi) \leq 0$, the diagonal terms dominate and
\[
    \E_{n^*}[\|T^{(1)}\|_{\hilbert}^2 \mid \xi, L(\mathbf{x})]
    \leq
    \frac{(2K)^2}{N_{L(\mathbf{x})}^2} \sum_{i=1}^N \Var(\delta_i \mid \xi)\, I_i
    \leq
    \frac{4K^2 C_2}{\underline{\lambda}^2}\Big(\frac{N}{n_{s,N}}\Big)^{2}\frac{1}{N_{L(\mathbf{x})}}
    = O_{p_d}\!\Big(\frac{1}{N k_N^p}\Big),
\]
using $\pi_i \geq \underline{\lambda}\, n_{s,N}/N$ from (D3), $n_{s,N} = \Theta_{p_d}(N)$ from (D2), and $N_{L(\mathbf{x})} = \Theta_{p_d}(N k_N^p)$ from Lemma~\ref{lem:N_L}. By Lemma~\ref{lem:BS}, $\E_{n^*}[1 - I_b \mid \xi, L(\mathbf{x})] = O_{p_d}(1/(N k_N^p))$. Therefore
\[
    \textrm{(C-i)}
    = O_{p_d}\!\Big(\big(\tfrac{1}{N k_N^p}\big)^{1/2}\big(\tfrac{1}{N k_N^p}\big)^{1/2}\Big)
    = O_{p_d}\!\Big(\frac{1}{N k_N^p}\Big).
\]
 
\medskip
\noindent\textit{Term (C-ii).}
By (D3) and the kernel bound (K1),
\[
    \|T_{\xi, L}(\mathbf{x})\|_{\hilbert} \leq \frac{\bar{\lambda}}{\underline{\lambda}} K < \infty,
\]
so Lemma~\ref{lem:BS} gives
\[
    \textrm{(C-ii)}
    \leq
    \frac{\bar{\lambda}}{\underline{\lambda}} K\;
    \E_{\xi,L(\mathbf{x})}\big[\, \E_{n^*}[1 - I_b \mid \xi, L(\mathbf{x})]\,\big]
    = O_{p_d}\!\Big(\frac{1}{N k_N^p}\Big).
\]
 
\medskip
\noindent\textit{Term (C-iii): the quadratic remainder.}
By the exact second-order Taylor form,
\begin{align*}
    R_2
    &= 2 \int_{0}^1 (1-s) \left\{ \frac{T_{L,nu} + s \Delta_{nu}}{(T_{L, de} + s \Delta_{de})^3}\,\Delta^2_{de}
    - \frac{1}{(T_{L,de} + s \Delta_{de})^2}\,\Delta_{nu}\Delta_{de}\right\} ds
    \overset{\text{denote}}{=} \int_0^1 (1-s) M_s\, ds,
\end{align*}
where
\begin{align*}
    \Delta_{nu} \Delta_{de}
    &=
    I_b\left\{ \sum_{i=1}^N \delta^2_i I_i\, \mathbf{k}(Y_i,\cdot) + 2\!\!\sum_{i < j} \delta_i \delta_j I_i I_j\, \mathbf{k}(Y_i,\cdot) \right\}
    - (1-I_b) T_{L, nu}\sum_{i=1}^N \delta_i I_i,
    \\
    \Delta_{de}^2
    &=
    \sum_{i=1}^N \delta^2_i I_i + 2\!\!\sum_{i < j} \delta_i \delta_j I_i I_j.
\end{align*}
The integrand $M_s$ is measurable, since $\Delta_{nu}$, $\Delta_{de}$, $T_{L,nu}$, and $T_{L,de}$ are. Moreover $\|T_{L, nu}\|_{\hilbert} \leq K \frac{N}{\underline{\lambda}\, n_{s,N}}\, n_{L(\mathbf{x})}$, and
\[
    \E[\|\Delta_{nu}\|^2_{\hilbert} \mid \xi] \leq K^2 \sum_i \Var(\delta_i \mid \xi)\, I_i \leq \frac{2K^2 N}{\underline{\lambda}\, n_{s,N}}\, n_{L(\mathbf{x})},
    \qquad
    \E[\Delta_{de}^2 \mid \xi] \leq O_{p_d}(1)\, n_{L(\mathbf{x})},
\]
so Chebyshev's inequality gives $\|\Delta_{nu}\|_{\hilbert} = O_{p_d}(\sqrt{n_{L(\mathbf{x})}})$ and $|\Delta_{de}| = O_{p_d}(\sqrt{n_{L(\mathbf{x})}})$. The denominator is bounded away from zero, since
\[
    |T_{L,de} + s \Delta_{de}|
    \geq \big| n_{L(\mathbf{x})} - |\Delta_{de}| \big|
    \geq \big| \tfrac{1}{2} n_{L(\mathbf{x})} - O_{p_d}(\sqrt{n_{L(\mathbf{x})}}) \big|
    = O_{p_d}(n_{L(\mathbf{x})}),
\]
so that
\[
    \|M_s\|_{\hilbert}
    \leq
    2 \left( \frac{\|T_{L,nu}\|_{\hilbert} + s \|\Delta_{nu}\|_{\hilbert}}{|T_{L,de} + s\Delta_{de}|^3}\, \Delta_{de}^2
    + \frac{\|\Delta_{nu}\|_{\hilbert}\, |\Delta_{de}|}{|T_{L,de} + s \Delta_{de}|^2} \right)
    = O_{p_d}(n_{L(\mathbf{x})}^{-1})
\]
for almost every $\xi$ and $L(\mathbf{x})$. By Fubini, $\E[R_2 \mid \xi, L(\mathbf{x})] = \int_0^1 (1-s)\, \E[M_s \mid \xi, L(\mathbf{x})]\, ds$, and Minkowski's inequality together with Jensen's inequality gives
\[
    \big\| \E[R_2 \mid \xi, L(\mathbf{x})]\big\|_{\hilbert}
    \leq
    \int_0^1 (1-s)\, \E\big[\|M_s\|_{\hilbert} \mid \xi, L(\mathbf{x})\big]\, ds.
\]
Writing $\bm{\Delta} = (\Delta_{nu}, \Delta_{de})$, we have $M_s = \langle \bm{\Delta}, H_g(\widetilde{T}_{nu}(s), \widetilde{T}_{de}(s))\, \bm{\Delta}\rangle_{\hilbert}$ and $\|M_s\|_{\hilbert} \leq \|H_g(\widetilde{T}_s)\|_{op}(\|\Delta_{nu}\|^2_{\hilbert} + \Delta_{de}^2)$. On the event $\{b \in \mathcal{B}_L(\mathbf{x})\}$ the denominators are bounded below by some $c > 0$, and $H_g$ is Lipschitz there:
\begin{align*}
    \|H_g(t_{nu}, t_{de}) - H_g(\tilde{t}_{nu}, \tilde{t}_{de})\|_{op}
    &\leq
    \sqrt{2\,\big|1/\tilde{t}^2_{de} - 1/t^2_{de}\big|^2 + 2\,\|t_{nu}/t^3_{de} - \tilde{t}_{nu}/\tilde{t}^3_{de}\|^2_{\hilbert}} \\
    &\leq
    \sqrt{2\,(2 c^{-3})^{2}(t_{de} - \tilde{t}_{de})^{2} + \big(2 c^{-3}\|t_{nu} - \tilde{t}_{nu}\| + 6K c^{-4}|t_{de} - \tilde{t}_{de}|\big)^{2}} \\
    &\leq
    L_{\mathrm{lip}}\, \|(t_{nu}, t_{de}) - (\tilde{t}_{nu}, \tilde{t}_{de})\|_{\hilbert},    
\end{align*}
where $L_{\mathrm{lip}} = \sqrt{16\,c^{-6} + 2(6 K c^{-4})^2} < \infty$.
Taking $(\tilde{t}_{nu}, \tilde{t}_{de}) = (T^*_{nu}, T^*_{de})$ and using $\|H_g(T_L)\|_{op} = O_{p_d}(n_{L(\mathbf{x})}^{-1})$,
\begin{align*}
    \int_0^1 (1-s)\, \E\big[\|M_s\|_{\hilbert} \mid \xi, L(\mathbf{x})\big]\, ds
    &\leq
    \frac{\sqrt{2}}{2}\, L_{\mathrm{lip}}
    \Big( \E\big[|\Delta_{de}|^3 \mid \xi, L(\mathbf{x})\big] + \E\big[\|\Delta_{nu}\|^3_{\hilbert} \mid \xi, L(\mathbf{x})\big]\Big) \\
    &\quad + \tfrac{1}{2} \|H_g(T_L)\|_{op}\Big(\E[\Delta_{de}^2 \mid \xi, L(\mathbf{x})] + \E[\|\Delta_{nu}\|^2_{\hilbert} \mid \xi, L(\mathbf{x})]\Big).
\end{align*}
Because $\E[|\Delta_{de}|^2 \mid L(\mathbf{x})] \leq \big(\E[|\Delta_{de}|^3 \mid L(\mathbf{x})]\big)^{2/3}$ and likewise for $\Delta_{nu}$, the second-moment terms are absorbed into the constant. Taking the outer expectation over $L(\mathbf{x})$ and $\xi$,
\[
    \E_{\xi, L(\mathbf{x})}\big[\|\E[R_2 \mid \xi, L(\mathbf{x})]\|_{\hilbert}\big]
    \lesssim_{p_d}
    \E_{L(\mathbf{x})}\Big\{ \E[|\Delta_{de}|^3 \mid L(\mathbf{x})] + \E[\|\Delta_{nu}\|^3_{\hilbert} \mid L(\mathbf{x})]\Big\}.
\]
The per-unit third moments are bounded, since
\[
    \E_{\xi}\big[\E_{n^*}[|n^*_i/\pi_i|^3 \mid \xi]\big]
    \leq
    \E_{\xi}\Big[\big(\E_{n^*}[|n^*_i/\pi_i|^4 \mid \xi]\big)^{3/4}\Big]
    \leq
    \E_{\xi}\big[15^{3/4}\, \xi_i/\pi^3_i\big]
    \lesssim
    \E\big[(N/n_{s,N})^2\big]
    = O_{p_d}(1),
\]
using $\E[(n^*_i)^4 \mid \xi] \leq 15\,\xi_i$ (established in Part B) and Lemma~\ref{lem:N_ns_moments}, and $\E[|\xi_i/\pi_i|^3] = \pi_i^{-2} \leq \underline{\lambda}^{-2}\E[N/n_{s,N}] = O_{p_d}(1)$. Hence, by Lemma~\ref{lem:N_L},
\begin{align*}
    \E\big[|\Delta_{de}|^3 \mid L(\mathbf{x})\big]
    &=
    \Big(\big\| \textstyle\sum_{i=1}^N \delta_i I_i \big\|_{L^3(P(n^*, \xi \mid L(\mathbf{x})))}\Big)^3 \\
    &\leq
    \Big( \textstyle\sum_{i=1}^N I_i\, \big\| n^*_i/\pi_i - \xi_i/\pi_i \big\|_{L^3(P(n^*, \xi \mid L(\mathbf{x})))}\Big)^3 \\
    &\leq
    4 \Big( \textstyle\sum_{i=1}^N I_i \big( \E_{\xi}\E_{n^*}[|n^*_i/\pi_i|^3 \mid \xi] + \E_{\xi}[|\xi_i/\pi_i|^3]\big)^{1/3}\Big)^3
    = O_{p_d}(N_{L(\mathbf{x})}^3),
\end{align*}
so $\E[|\Delta_{de}|^3 \mid L(\mathbf{x})]/N_{L(\mathbf{x})}^3 = O_{p_d}(1)$, and after normalization by the leaf-size factors in $M_s$ this term is $o_{p_d}(1)$. For $\Delta_{nu} = I_b \sum_i \delta_i I_i\, \mathbf{k}(Y_i,\cdot) + (1 - I_b) T_{L, nu}$, the inequality $\|a + b\|^3 \leq 4(\|a\|^3 + \|b\|^3)$ and the boundedness of $I_b$ reduce the first summand to the $\Delta_{de}$ argument above, while (A4) and the kernel bound (K1) give $\|T_{L, nu}\|_{\hilbert} = O_{p_d}(n_{L(\mathbf{x})})$, so that Lemma~\ref{lem:BS} yields $\E[(1 - I_b)\|T_{L, nu}\|^3_{\hilbert} \mid L(\mathbf{x})] = o_{p_d}(1)$ after the same normalization. Therefore $\textrm{(C-iii)} = o_{p_d}(1)$.
 
\medskip
\noindent
Combining (C-i), (C-ii), and (C-iii),
\[
    (\ref{eq: mainB})
    = O_{p_d}\!\Big(\frac{1}{N k_N^p}\Big) + o_{p_d}(1)
    = o_{p_d}(1),
\]
which completes Part C.

\paragraph{Part D}
Now we bound the distance introduced by $\xi$ in both the numerator and denominator (leaf size).
\begin{align}
\begin{split}
    D_N(\mathbf x) = &\Bigg\| 
        \frac{\sum_i \xi_i/\pi_i\, \ind(X_i \in L(\mathbf{x}))\, \mathbf{k}(Y_i, \cdot)}{ \sum_i \xi_i / \pi_i\, \ind(X_i \in L(\mathbf{x}))} 
        -
        \frac{\sum_i \ind(X_i \in L(\mathbf{x}))\, \mathbf{k}(Y_i, \cdot)}{\sum_i \ind(X_i \in L(\mathbf{x}))}
        \Bigg\|_{\hilbert}\\
    =
    &\Big\| \sum_i \tilde{\delta}_i \Big\{ \frac{\ind(X_i \in L(\mathbf{x}))}{\sum_i \ind(X_i \in L(\mathbf{x}))} (k_i - T_{L}(\mathbf{x})) \Big\}
        + 
        \widetilde{R}_2 \Big\|_{\hilbert}\\
    \leq
    &\Big\| \sum_i \tilde{\delta}_i \Big\{ \frac{\ind(X_i \in L(\mathbf{x}))}{\sum_i \ind(X_i \in L(\mathbf{x}))} (k_i - T_{L}(\mathbf{x}))  \Big\} \Big\|_{\hilbert}
        + 
    \Big\| \widetilde{R}_2 \Big\|_{\hilbert},
    \label{eq: partDmain}
\end{split}
\end{align}
where the first equality uses again the Taylor expansion in the previous argument with 
$
    \tilde{\delta}_i = (\frac{\xi_i}{\pi_i} - 1),
$
$
    \widetilde{\Delta}_{nu} 
    = \sum_i \tilde{\delta}_i I_i k_i,
$
$
    \widetilde{\Delta}_{de} 
    = \sum_i \tilde{\delta}_i I_i.
$
We proceed to show both parts $o_p(1)$ w.r.t. $P_{\mathcal{S}^N | \omega}$.

For the first term, in expectation,
\begin{align}
\begin{split}
    &\mathbb{E}_{\xi,L(\mathbf{x})}\!\left[
    \left\|
    \sum_i \tilde{\delta}_i\,
    \frac{I\!\big(X_i\in L(\mathbf{x})\big)}{N_{L(\mathbf{x})}}\,
    \big(k_i - T_L(\mathbf{x})\big)
    \right\|_{\mathcal{H}}
    \right]\\
\le
    &\left(
    \mathbb{E}_{\xi,L(\mathbf{x})}\!\left[
    \left\|
    \sum_i \tilde{\delta}_i\,
    \frac{I\!\big(X_i\in L(\mathbf{x})\big)}{N_{L(\mathbf{x})}}\,
    \big(k_i - T_L(\mathbf{x})\big)
    \right\|_{\mathcal{H}}^{2}
    \right]
    \right)^{1/2} \stackrel{denote}{=} T_N^{1/2},
    \label{eq:partD}
\end{split}
\end{align}
where the inner part satisfies for any fixed $L(\mathbf x)$,
\begin{align*}
    T_N
&=
    \mathrm{Var}_{\xi}\!\left[
    \left.
    \sum_i \tilde{\delta}_i\,
    \frac{I\!\big(X_i\in L(\mathbf{x})\big)}{N_{L(\mathbf{x})}}\,
    \big(k_i - T_L(\mathbf{x})\big)
    \ \right|\ L(\mathbf{x})
    \right]\\
&=
    \frac{1}{N_{L(\mathbf{x})}^{2}}
    \sum_i \sum_j
    \mathbb{E}_{\xi}\!\left[\tilde{\delta}_i \tilde{\delta}_j\right]\,
    I\!\big(X_i\in L(\mathbf{x})\big)\,I\!\big(X_j\in L(\mathbf{x})\big)\,
    \big\langle k_i - T_L,\; k_j - T_L \big\rangle_{\mathcal{H}}\\
&\le
    \frac{4K^{2}}{N_{L(\mathbf{x})}^{2}}
    \left[
    \sum_i \frac{1}{\pi_i^{2}} I_i
    +
    \sum_{i\neq j} \frac{C_1}{n_{s,N}\pi_i\pi_j}\, I_i I_j
    \right]
\le 
    \frac{4K^{2}}{N_{L(\mathbf{x})}^{2}}
    \left[
    \frac{N^{2}}{\underline{\lambda}^{2} n_{s,N}^{2}}\, N_{L(\mathbf{x})}
    +
    \frac{C_1N^{2}}{\underline{\lambda}^{2} n_{s,N}^{3}}\, N_{L(\mathbf{x})}^{2}
    \right]\\
&= O_{p_d}\!\left(\frac{N^{2}}{n_{s,N}^{2}}\cdot \frac{1}{N_{L(\mathbf{x})}}\right)
=
O_{p_d}\!\left(\frac{1}{N_{L(\mathbf{x})}}\right).
\end{align*}
The first inequality holds by (D4) and the boundedness of $\mathbf{k}$ (K1); the second by (D3); the subsequent equality by domination of the diagonal term together with $N_{L(\mathbf{x})} = \Theta_{p_d}(Nk^p_N)$ from Lemma~\ref{lem:N_L}; the last by (D2), which gives $n_{s,N} = \Theta_{p_d}(N)$. Plugging it back to the upper bound in (\ref{eq:partD}) yields
\begin{align*}
&\mathbb{E}_{\xi,L(\mathbf{x})}\!\left[
    \left\|
    \sum_i \tilde{\delta}_i\,
    \frac{I\!\big(X_i\in L(\mathbf{x})\big)}{N_{L(\mathbf{x})}}\,
    \big(k_i - T_L(\mathbf{x})\big)
    \right\|_{\mathcal{H}}
    \right]
    = \sqrt{\E[\frac{C}{N_{L(\mathbf x)}}]}\\
    = &O\!\left(\frac{1}{\sqrt{N_{L(\mathbf x)}}}\right)
= O\!\left((N k_N^{p})^{-1/2}\right)
= o(1),
\end{align*}
where the second equality by Lemma~\ref{lem:N_L} again.
\\

For $\E[\|\widetilde{R}_2\|_{\hilbert}]$, the exact second-order expansion of Part~C carries over with three simplifications: the increment $\delta_i$ becomes $\tilde{\delta}_i = \xi_i/\pi_i - 1$, the good-tree indicator is absent ($I_b \equiv 1$), and the denominator is the oracle count $N_{L(\mathbf{x})}$ rather than $|L_\xi(\mathbf{x})| = \Theta_{p_d}(n_{L(\mathbf{x})})$. The Hessian and Lipschitz bounds are otherwise identical, so $\widetilde{M}_s$ carries the $N_{L(\mathbf{x})}^{-3}$ normalization after the cubic terms, and
\begin{align*}
    \E\left[ \|\widetilde{R}_2\|_{\hilbert} \right]
    \lesssim_{p_d}
    \E_{L(\mathbf{x})}\!\left[ N_{L(\mathbf{x})}^{-3}\Big( \E[ |\widetilde{\Delta}_{de}|^3 \mid L(\mathbf{x}) ] + \E[ \|\widetilde{\Delta}_{nu}\|_{\hilbert}^3 \mid L(\mathbf{x}) ] \Big)\right].
\end{align*}
The per-unit third moment is bounded: writing $\hat{w}_i := 1/\pi_i$,
\begin{align*}
    \sup_{i \in [N]}\E\!\left[ \Big|\tfrac{\xi_i}{\pi_i} - 1\Big|^3 \,\Big|\, L(\mathbf{x})\right]
    =
    \sup_{i \in [N]} \Big( \pi_i (\hat{w}_i - 1)^3 + (1-\pi_i)\Big)
    \leq
    \sup_{i \in [N]} \frac{1}{\pi_i^2} + 1
    = O_{p_d}(1),
\end{align*}
so by Minkowski $\E[|\widetilde{\Delta}_{de}|^3 \mid L(\mathbf{x})] \le \big(\sum_i I_i \|\tilde{\delta}_i\|_{L^3}\big)^3 = O_{p_d}(N_{L(\mathbf{x})}^3)$, and likewise $\E[\|\widetilde{\Delta}_{nu}\|_{\hilbert}^3 \mid L(\mathbf{x})] = O_{p_d}(N_{L(\mathbf{x})}^3)$ by (K1). After the $N_{L(\mathbf{x})}^{-3}$ normalization each contributes $O_{p_d}(1)$, and Lemma~\ref{lem:N_L} with the leading-term rate $O_{p_d}(1/N_{L(\mathbf{x})})$ established above yields $\E_{\xi, L(\mathbf{x})}[\|\widetilde{R}_2\|_{\hilbert}] = o_{p_d}(1)$.

Markov's inequality applied to both terms of $(\ref{eq: partDmain})$ gives
$\big\| \sum_i \tilde{\delta}_i \frac{I_i}{N_{L(\mathbf{x})}}(k_i - T_{L}(\mathbf{x})) \big\|_{\hilbert} = o_{p_d}(1)$
and
$\|\widetilde{R}_2\|_{\hilbert} = o_{p_d}(1)$, which completes Part D.


\vspace*{0.4cm}
\paragraph{Part E}

We bound the smoothing error between the oracle leaf mean and the atom target. Write $k_i = \mathbf{k}(Y_i, \cdot)$ and denote the super-population conditional embedding
\[
    \mu_{\mathbf{k}}(P_{Y \mid X = \mathbf{x}}) := \E_P[\mathbf{k}(Y, \cdot) \mid X = \mathbf{x}] \in \hilbert .
\]
This object is not the target; it is a common anchor for the two error sources, which the triangle inequality separates:
\begin{equation}\label{eq:partE-split}
    E_N(\mathbf{x})
    \leq
    \underbrace{\big\| T_{L,b}(\mathbf{x}) - \mu_{\mathbf{k}}(P_{Y \mid X = \mathbf{x}}) \big\|_{\hilbert}}_{\text{(E-i): leaf smoothing bias}}
    +
    \underbrace{\big\| \mu_{\mathbf{k}}(\PSN(\mathbf{x})) - \mu_{\mathbf{k}}(P_{Y \mid X = \mathbf{x}}) \big\|_{\hilbert}}_{\text{(E-ii): atom fluctuation}} .
\end{equation}

\medskip
\noindent\textit{Term (E-i): leaf smoothing bias.}
For each $i$ with $X_i \in L(\mathbf{x})$, write $k_i = \E[\mathbf{k}(Y, \cdot) \mid X_i] + \varepsilon_i$, where $\varepsilon_i$ is centered given $X_i$, $\|\varepsilon_i\|_{\hilbert} \le 2K$ almost surely by (K1), and $\E[\|\varepsilon_i\|_{\hilbert}^2 \mid X_i] \le \sigma_{\mathbf{k}}^2 < \infty$. Then
\[
    T_{L,b}(\mathbf{x})
    = \frac{1}{N_{L(\mathbf{x})}} \sum_{i=1}^N \ind(X_i \in L(\mathbf{x}))\, \E[\mathbf{k}(Y, \cdot) \mid X_i]
    + \underbrace{\frac{1}{N_{L(\mathbf{x})}} \sum_{i=1}^N \ind(X_i \in L(\mathbf{x}))\, \varepsilon_i}_{=: R_N(\mathbf{x})} .
\]
The residual concentrates: the $\varepsilon_i$ are centered with bounded conditional second moment, so
\[
    \E_L[\|R_N(\mathbf{x})\|_{\hilbert}]
    \leq \Big( \frac{1}{N_{L(\mathbf{x})}^2} \sum_{i=1}^N \ind(X_i \in L(\mathbf{x}))\, \E[\|\varepsilon_i\|_{\hilbert}^2] \Big)^{1/2}
    \leq \Big( \frac{\sigma_{\mathbf{k}}^2}{N_{L(\mathbf{x})}} \Big)^{1/2}
    = O\big((N k_N^p)^{-1/2}\big),
\]
using $N_{L(\mathbf{x})} = \Theta_{p_d}(N k_N^p)$ from Lemma~\ref{lem:N_L}. For the conditional-mean term, the Lipschitz continuity of $\mathbf{x}' \mapsto \E_P[\mathbf{k}(Y, \cdot) \mid X = \mathbf{x}']$ and the diameter bound (B1) chain as
\begin{align}
    &\E_L\big[ \big\| T_{L,b}(\mathbf{x}) - \mu_{\mathbf{k}}(P_{Y\mid X=\mathbf{x}}) \big\|_{\hilbert} \big] \\
    \leq &\E_L\Big[\tfrac{1}{N_{L(\mathbf{x})}} \textstyle\sum_i \ind(X_i\in L(\mathbf{x}))\, \big\| \E[\mathbf{k}(Y,\cdot)\mid X_i]-\mu_{\mathbf{k}}(P_{Y\mid X=\mathbf{x}})\big\|_{\hilbert}\Big]
    + \E_L[\|R_N(\mathbf{x})\|_{\hilbert}] \notag\\
    \leq &L_{\mathrm{lip}}\, \E_L[\mathrm{diam}(L(\mathbf{x}))] + O\big((N k_N^p)^{-1/2}\big) \notag\\
    \leq &L_{\mathrm{lip}}\big(k_N + \mathrm{diam}(\mathcal{X})\, P_L(\mathrm{diam}(L(\mathbf{x}))\ge k_N)\big) + O\big((N k_N^p)^{-1/2}\big) \notag\\
    \leq &L_{\mathrm{lip}}(k_N + \mathrm{diam}(\mathcal{X})\, t_N) + O\big((N k_N^p)^{-1/2}\big) 
    \;=\; O(k_N),\label{eq:Ei}
\end{align}
with $\mathrm{diam}(\mathcal{X})<\infty$ since $\mathcal{X}$ is bounded and $t_N\to 0$ by (B1). Markov's inequality then gives (E-i) $= O_{p_d}(k_N)$.

\medskip
\noindent\textit{Term (E-ii): atom fluctuation.}
Because $\mathbf{x}$ is a fixed atom, $\{Y_i : X_i = \mathbf{x}\}$ are $N_{\mathbf{x}}$ i.i.d.\ draws from $P_{Y \mid X = \mathbf{x}}$, and $\mu_{\mathbf{k}}(\PSN(\mathbf{x}))$ is their empirical embedding. With $\|k_i\|_{\hilbert} \le K$ by (K1), a Hilbert-space law of large numbers gives
\[
    \E\big[ \big\| \mu_{\mathbf{k}}(\PSN(\mathbf{x})) - \mu_{\mathbf{k}}(P_{Y \mid X = \mathbf{x}}) \big\|_{\hilbert}^2 \,\big|\, N_{\mathbf{x}} \big]
    = \frac{1}{N_{\mathbf{x}}}\, \E\big[ \|k_i - \mu_{\mathbf{k}}(P_{Y \mid X = \mathbf{x}})\|_{\hilbert}^2 \mid X_i = \mathbf{x} \big]
    \leq \frac{4K^2}{N_{\mathbf{x}}} ,
\]
so (E-ii) $= O_{p_d}(N_{\mathbf{x}}^{-1/2}) = O_{p_d}(N^{-1/2})$, using $N_{\mathbf{x}} = \Theta_{p_d}(N)$ from $p_{\mathbf{x}} > 0$.

\medskip
\noindent
Substituting into \eqref{eq:partE-split},
\[
    E_N(\mathbf{x}) = \underbrace{O_{p_d}(k_N)}_{\text{(E-i)}} + \underbrace{O_{p_d}(N^{-1/2})}_{\text{(E-ii)}} = o_{p_d}(1),
\]
since $k_N \to 0$ by (B1) and $N^{-1/2} \to 0$. 
\end{proof}

\vspace*{0.3cm}
\subsection{Two-stage Cluster Design}
Under case \textup{(ii)} of Theorem~10, the sample index $i\in[n_s]$ is replaced by the SSU index $u=(h,k,j)\in[n_s]$ with inclusion probability $\pi_u=\pi^{(1)}_{hk}\pi^{(2)}_{hkj\mid hk}$; the notation $i$ continues to denote a generic sampled unit and $I_i:=\ind(X_i\in L(\mathbf x))$ its indicator. The five-part decomposition above and the definitions of $A_N,B_N,C_N,D_N,E_N$ transfer verbatim under this re-indexing. We first extend the four supporting lemmas of the section, then verify each of Parts~A--E.\\

The four supporting lemmas of the section extend to case (ii) as follows.
Lemma~\ref{lem:N_L} is a finite-population statement independent of the
design and applies verbatim; Lemma~\ref{lem:N_ns_moments} uses (D2) only
through its polynomial-tail conclusion, which is delivered identically by
(D-TS2). For Lemma~\ref{lem:ns_nL}, the variance bound
$\mathrm{Var}(n_{L(\mathbf x)}\mid\omega)=O_{p_d}(Nk_N^p)$ follows from
Lemma~\ref{lem:psu-block-variance} applied with $a_i:=\pi_i I_i$
(so $A\le 1$ and $S = N_{L(\mathbf x)}$), and the mean bound
$\mathbb E[n_{L(\mathbf x)}\mid\omega]\ge N_{L(\mathbf x)}\underline\lambda(n/N)$
uses Lemma~\ref{lem:designs-multistage}\,(i); the Chebyshev conclusion is
unchanged. For Lemma~\ref{lem:BS}, the conditional variance
$\mathrm{Var}(S_b\mid\xi,L) \le (1+C_{\mathrm{TS}})\tilde\nu N_{L(\mathbf x)}$
follows from the max-row-sum (R3) with constant $C_{\mathrm{TS}}$
(Lemma~7) and the sub-gamma envelope
$\Gamma(\tilde\nu,\tilde c)$ (Lemma~\ref{lem:subgamma-multistage});
the remainder of both parts of Lemma~\ref{lem:BS} proceeds unchanged.\\

We now verify each of the five parts of the decomposition.

\paragraph{Part A}
The single-stage argument in Part~A depends on the design only through Lemma~\ref{lem:BS} and (B2). With the multi-stage Lemma~\ref{lem:BS} above and (B2) unchanged, the derivation of the reduction $A_N\le K_{\max}\cdot P_{\mathcal S^N\mid\omega}(b\in\mathcal B_L\Delta\mathcal B_{L^*})$ and the bound $A_N = o_{p_d}(1)$ follow verbatim from the single-stage argument.

\paragraph{Part B}
Conditional on $(\xi,L(\mathbf x))$, the trees $\{T^*_{N,b}(\mathbf x)\}_{b\in[B]}$ are conditionally i.i.d.\ across $b$; the RKHS-valued Chebyshev inequality on the sample mean of independent centered elements applies. The per-tree conditional variance is bounded via (R2)'s moment condition and (R3) in max-row-sum form. Under (ii), (R2)--(R3) at the SSU level are supplied by the two-stage case of Lemma~7, and the centered multiplier $\delta_i = n^*_{b,i}-1$ is conditionally sub-gamma with envelope $\Gamma(\tilde\nu,\tilde c)$ by Lemma~\ref{lem:subgamma-multistage}. Substituting these into the single-stage argument replaces $\Gamma(\nu_0, c_0)$ by $\Gamma(\tilde\nu,\tilde c)$ and $C$ by $C_{\mathrm{TS}}$ throughout; both are design-uniform constants under the regime clause. The Chebyshev conclusion $B_N = o_{p_d}(1)$ at rate $B^{-1/2}$ carries through unchanged.

\paragraph{Part C}
Part~C linearizes the ratio $T^*_{nu}/T^*_{de}$ around the design-Horvitz--Thompson ratio $T_{L,nu}/T_{L,de}$ via a second-order Taylor expansion. The first-order term is a multiplier-weighted RKHS sum controlled by Chebyshev with the multi-stage sub-gamma envelope $(\tilde\nu,\tilde c)$; the Hessian remainder is controlled via the multi-stage design LLN Lemma~\ref{lem:lln_clt_multistage} in place of Lemma~\ref{lem:lln_clt}. Both invocations yield the same rates as in the single-stage Part~C, so $C_N = o_{p_d}(1)$.

\paragraph{Part D}
This is the substantive step. The single-stage bound~\eqref{eq: partDmain} depends on the design through the covariance term
\(
\sum_{i\ne j}\mathrm{Cov}(\xi_i/\pi_i,\xi_j/\pi_j)\,I_i I_j \langle k_i - T_L(\mathbf x), k_j - T_L(\mathbf x)\rangle_{\mathcal H}.
\)
Under (ii), the standard single-stage bound via (D3)--(D4) is unavailable; the pairwise covariance structure is that of Lemma~\ref{lem:designs-multistage}\,(ii). We instead apply Lemma~\ref{lem:psu-block-variance} directly with $\mathcal K = \hilbert$ and
\[
a_i := \bigl(k_i - T_L(\mathbf x)\bigr)\,I_i / N_{L(\mathbf x)},
\qquad
\|a_i\|_{\hilbert} \le 2\sqrt{K_{\max}}/N_{L(\mathbf x)} \text{ under (K1)},\qquad S = N_{L(\mathbf x)},
\]
yielding
\[
T_N
= \mathbb E\Bigl\|\sum_i (\xi_i/\pi_i-1)a_i\Bigr\|_{\hilbert}^{2}
\le C^{\dagger}\cdot\frac{4K_{\max}}{N_{L(\mathbf x)}^2}\cdot N_{L(\mathbf x)}
= O\!\Big(\frac{K_{\max}}{N_{L(\mathbf x)}}\Big)
= O_{p_d}\!\Big(\frac{1}{Nk_N^p}\Big)
\]
by Lemma~\ref{lem:N_L}. Jensen and Markov then give the leading term as $o_{p_d}(1)$, matching~\eqref{eq:partD} of the single-stage argument. The Hessian remainder $\|\widetilde R_2\|_{\hilbert}$ is controlled by the single-stage argument's third-moment bookkeeping, with the per-unit bound $\sup_i \mathbb E[|\xi_i/\pi_i-1|^3\mid L(\mathbf x)]\le 1/\pi_0^2+1 = O_{p_d}(1)$ using Lemma~\ref{lem:designs-multistage}\,(i) in place of (D3), and (D-TS2) in place of (D2) for the $N/n_{s,N}$ moment control (via the extended Lemma~\ref{lem:N_ns_moments} above); the same $N_{L(\mathbf x)}^{-3}$ normalization then gives $\mathbb E_{\xi, L(\mathbf x)}[\|\widetilde R_2\|_{\hilbert}] = o_{p_d}(1)$. Markov on both terms yields $D_N = o_{p_d}(1)$.

\paragraph{Part E}
Term (E-i) uses only the Lipschitz continuity of the super-population conditional kernel mean embedding and the diameter bound of (B1); it is a super-population argument that does not involve the design. Term (E-ii) is a Hilbert-space law of large numbers applied to the $N_{\mathbf x}=\Theta_{p_d}(N)$ i.i.d.\ super-population draws with $X_i=\mathbf x$, also design-independent. Both bounds transfer verbatim from the single-stage Part~E, giving $E_N(\mathbf x) = O_{p_d}(k_N) + O_{p_d}(N^{-1/2}) = o_{p_d}(1)$.

\medskip\noindent\emph{Combining.}
Substituting the five bounds $A_N, B_N, C_N, D_N, E_N = o_{p_d}(1)$ into the decomposition gives
\[
\bigl\|\mu_{\mathbf k}(\hat P_{\mathrm{forest}}(\mathbf x)) - \mu_{\mathbf k}(P^N_{Y\mid X=\mathbf x})\bigr\|_{\hilbert} = o_{p_d}(1),
\]
i.e., $d_{\mathbf k}(\widehat P_{\mathcal S^N\mid\omega}(\mathbf x), P_{\mathcal S^N\mid\omega}(\mathbf x))\xrightarrow{p_d} 0$. The second convergence $d_{\mathbf k}(\widehat P_{\mathcal S^N\mid\omega}(\mathbf x), P_{Y\mid X=\mathbf x})\xrightarrow{p_d} 0$ follows from the triangle inequality on $\hilbert$ combined with the atom-level super-population LLN of Part~E, exactly as in the single-stage argument.

\section{Consistency of the Continuous Map}
\label{sec:consistencyfunctional}

Assume the standing atom hypothesis $p_{\mathbf{x}} := P(X = \mathbf{x}) > 0$ and the conclusions of Theorem~10 proved above. Let $\Psi : \Mb \to (\mathcal{V}, d_v)$ be a measurable map, continuous at $\PSN(\mathbf{x})$ and at $P_{Y \mid X = \mathbf{x}}$ for $P$-a.e.\ $\omega$. Under (K2), continuity on $\Mb$ coincides with weak continuity.
 
\medskip
We prove a general result first:
Fix a limit point $P_\bullet\in\Mb$ (to be instantiated as $\PSN(\mathbf{x})$ or $P_{Y\mid X=\mathbf{x}}$ below) and suppose
\begin{equation}\label{eq:cor-hyp}
    \dk\bigl(\hatPSN(\mathbf{x}),\,P_\bullet\bigr)\xpd 0.
\end{equation}
By the subsequence criterion for convergence in probability,
\[
    Z_N\xrightarrow{P_{\mathcal S^N\mid\omega}} 0
    \iff
    \forall\{N_k\},\ \exists\{N_{k_\ell}\}\subset\{N_k\}\ \text{such that}\ Z_{N_{k_\ell}}\to 0\ \ P_{\mathcal S^N\mid\omega}\text{-a.s.},
\]
applied with $Z_N=\dk(\hatPSN(\mathbf x),P_\bullet)$: for every subsequence $\{N_k\}$ there exists $\{N_{k_\ell}\}$ and an event $E_\omega\subset\mathcal S^N$ with $P_{\mathcal S^N\mid\omega}(E_\omega)=1$ such that
\begin{equation}\label{eq:cor-subseq}
    \dk\bigl({\widehat{P}_{\mathcal{S}^{N_{k_\ell}}|\omega}}(\mathbf{x}),\,P_\bullet\bigr)\to 0\qquad\text{on }E_\omega.
\end{equation}
On $E_\omega$, (K2) converts $\dk\to 0$ to weak convergence
$
    {\widehat{P}_{\mathcal{S}^{N_{k_\ell}}|\omega}}(\mathbf{x})\rightsquigarrow P_\bullet,
$
where both sides are probability measures on $\mathcal Y$ at the fixed evaluation point $\mathbf{x}$. The Portmanteau theorem gives
\[
    \int f\,d{\widehat{P}_{\mathcal{S}^{N_{k_\ell}}|\omega}}(\mathbf x)\to\int f\,dP_\bullet\qquad\forall f\in C_b(\mathcal Y),\ \text{on }E_\omega,
\]
and continuity of $\Psi$ at $P_\bullet$ delivers that 
$
    \Psi\bigl({\widehat{P}_{\mathcal{S}^{N_{k_\ell}}|\omega}}(\mathbf x)\bigr)\to\Psi(P_\bullet) \; \text{on }E_\omega.
$
Since $\{N_k\}$ was arbitrary, the subsequence criterion in reverse gives
\begin{equation}\label{eq:cor-shared}
    \Psi\bigl(\hatPSN(\mathbf x)\bigr)\xpd \Psi(P_\bullet).
\end{equation}
 
\medskip
\noindent\textit{Design consistency.}
Instantiate \eqref{eq:cor-shared} with $P_\bullet=\PSN(\mathbf{x})$. Then 
$\dk\bigl(\hatPSN(\mathbf{x}), \; \PSN(\mathbf{x}))\xpd 0$ gives
\[
    \Psi\bigl(\hatPSN(\mathbf{x})\bigr)\xpd\Psi\bigl(\PSN(\mathbf{x})\bigr).
\]
 
\medskip
\noindent\textit{Joint consistency.}
Instantiate \eqref{eq:cor-shared} with $P_\bullet=P_{Y\mid X=\mathbf{x}}$. Then 
$\dk\bigl(\hatPSN(\mathbf{x}), \; P_{Y\mid X=\mathbf{x}})\xpd 0$ gives
\begin{equation}\label{eq:cor-joint-pd}
    \Psi\bigl(\hatPSN(\mathbf{x})\bigr)\xpd\Psi\bigl(P_{Y\mid X=\mathbf{x}}\bigr).
\end{equation}
Since $P_{Y\mid X=\mathbf{x}}$ is a fixed functional of $P$, hence $P$-a.s.\ constant in $\omega$, the limit $\Psi(P_{Y\mid X=\mathbf{x}})$ is deterministic in $\omega$, and $d_v(\Psi(\hatPSN(\mathbf x)), \Psi(P_{Y\mid X=\mathbf x}))$ is a measurable function on $\Omega\times\mathcal S^N$. Fubini's theorem applies because $\PSN$ is a regular conditional probability given $\omega$. Hence for any $\varepsilon>0$,
\[
    P_{\Omega\times\mathcal S^N}\!\Big(d_v\bigl(\Psi(\hatPSN(\mathbf x)),\,\Psi(P_{Y\mid X=\mathbf x})\bigr)>\varepsilon\Big)
    =\int_\Omega P_{\mathcal S^N\mid\omega}\!\Big(d_v\bigl(\Psi(\hatPSN(\mathbf x)),\,\Psi(P_{Y\mid X=\mathbf x})\bigr)>\varepsilon\Big)\,dP(\omega).
\]
The integrand is bounded by $1$ and vanishes for $P$-a.e.\ $\omega$ by \eqref{eq:cor-joint-pd}. Dominated convergence gives
$
    \Psi\bigl(\hatPSN(\mathbf{x})\bigr)\xpot\Psi\bigl(P_{Y\mid X=\mathbf{x}}\bigr),
$
completing the proof.
 
\begin{remark}
Taking $\mathcal{Y} = \mathbb{R}^d$, $(\mathcal V,d_v)=([0,1],|\cdot|)$, and $\Psi_t(P) = P((-\infty, t])$ at any continuity point $t$ of $P_{Y \mid X = \mathbf{x}}$ recovers pointwise consistency of the estimated conditional CDF in both modes. Conditional quantiles, moments, and tail probabilities inherit both consistency types without further argument.
\end{remark}